\documentclass[%
 aip,
 jap,
 citeautoscript,
 amsmath,
 amssymb,
 reprint
]{revtex4-1}

\usepackage{graphicx}
\usepackage{dcolumn}
\usepackage{bm}
\usepackage{color}

\newcommand{\kk}[1]{}

\usepackage[utf8]{inputenc}
\usepackage[T1]{fontenc}
\usepackage{mathptmx}

\begin{document}

\preprint{AIP/123-QED}

\title{Perspective on Metallic Antiferromagnets}

\author{Saima A. Siddiqui}
\affiliation{ 
Department of Materials Science and Engineering, University of Illinois at Urbana-Champaign
}%
\affiliation{Materials Research Laboratory, University of Illinois at Urbana-Champaign, Urbana, IL 61801, USA}
\author{Joseph Sklenar}
\affiliation{%
Department of Physics and Astronomy, Wayne State University
}%
\author{Kisung Kang}%
\affiliation{ 
Department of Materials Science and Engineering, University of Illinois at Urbana-Champaign
}%
\author{Matthew J. Gilbert}%
\affiliation{ 
Department of Electrical and Computer Engineering, University of Illinois at Urbana-Champaign
}%
\author{André Schleife}%
\affiliation{ 
Department of Materials Science and Engineering, University of Illinois at Urbana-Champaign
}%
\affiliation{Materials Research Laboratory, University of Illinois at Urbana-Champaign, Urbana, IL 61801, USA}
\affiliation{National Center for Supercomputing Applications, University of Illinois at Urbana-Champaign, Urbana, IL 61801, USA}

\author{Nadya Mason}%
\affiliation{Materials Research Laboratory, University of Illinois at Urbana-Champaign, Urbana, IL 61801, USA}
\affiliation{%
Department of Physics, University of Illinois at Urbana-Champaign
}%
\author{Axel Hoffmann}%
\email{axelh@illinois.edu.}
\affiliation{ 
Department of Materials Science and Engineering, University of Illinois at Urbana-Champaign
}%
\affiliation{ 
Department of Electrical and Computer Engineering, University of Illinois at Urbana-Champaign
}%
\affiliation{Materials Research Laboratory, University of Illinois at Urbana-Champaign, Urbana, IL 61801, USA}
\affiliation{%
Department of Physics, University of Illinois at Urbana-Champaign
}%

\date{\today}

\begin{abstract}
Antiferromagnet materials have recently gained renewed interest due to their possible use in spintronics technologies, where spin transport is the foundation of their functionalities.
In that respect metallic antiferromagnets are of particular interest, since they enable complex interplays between electronic charge transport, spin, optical, and magnetization dynamics.
Here we review phenomena where the metallic conductivity provides unique perspectives for the practical use and fundamental properties of antiferromagnetic materials. 
\end{abstract}

\maketitle

\section{\label{sec:Introduction}Introduction}

As conventional ferromagnetic (FM) digital storage devices reach the end of scaling,\cite{Meena2014} interest has burgeoned in exploring antiferromagnetic (AFM) materials for information storage and manipulation.
This interest is largely motivated by the robustness of antiferromagnetic order to moderate external magnetic fields, zero net magnetization that does not produce stray fields, and switching time-scales that correspond to switching rates in the THz regime.
In particular, the precession frequency of antiferromagnetic order is set by the geometric mean of the anisotropy and exchange energies,\cite{Bossini2016,GomonayNatPhys2018} leading to antiferromagnetic switching that is up to two orders of magnitude faster than FM switching. 
While research on antiferromagnets has been ongoing for decades, antiferromagnets had proven difficult to manipulate and read.
However, the field has now been newly motivated by recent experiments and theory that seem to show that antiferromagnetic order can be manipulated, possibly by spin-orbit-torques generated by charge currents,\cite{WadleyScience2016} staggered local relativistic fields induced by electrical currents,\cite{olejnikNatComm2017} domain wall motion,\cite{baldrati2019mechanism} and optical excitation by circularly polarized light.\cite{Kirilyuk2010}
Yet many aspects regarding manipulation of antiferromagnetic order are still unknown, including the influence of thermal effects,\cite{chiang2019absence} the time-scale of antiferromagnetic switching and manipulation, and even the mechanism and robustness of the switching itself.\cite{chiang2019absence,matallawagner2019resistive,grzybowski2017imaging}
These unknowns motivate further experimental and theoretical study of antiferromagnetic materials.

A large number of antiferromagnetic materials are available in nature.
Insulating antiferromagnets, which are mostly oxides such as NiO and halides such as MnF$_2$, have been well-studied recently, because of their potential to carry chargeless spin waves (magnons).
Conducting antiferromagnets, while regularly used as sources of exchange bias in magnetic spin-valve and tunnel junction-based devices,\cite{Nogues1999JMMM} have been less considered as spintronic devices.
Conducting antiferromagnets also have great potential for fundamental studies and applications due to their high electrical and thermal conductivities, and the strong interactions of electrons, spin, phonons, and photons.
In this perspective, we focus on conducting antiferromagnets. which include materials that are of high current research focus such, as CuMnAs, Mn$_3$Sn, Mn$_2$Au, and FeRh.

A detailed review on antiferromagnetic spintronics has been published previously \cite{BaltzRMP2018}.
In this perspective, we discuss recent progress and understanding of key properties of antiferromagnetic metals: charge transport, dynamics, and optical effects.
We first discuss the appearance of anisotropic magnetoresistance (AMR) effects in AFM metals, where the electrical resistance depends on the relative orientation of the current and N\'{e}el vector; we also mention the related spin-Hall magnetoresistance (SMR) effect.
We then discuss how antiferromagnetic metals can be used to generate spin currents, via spin Hall or anomalous Hall effects, as well as absorb spin currents, possibly leading to tunable sub-THz frequency oscillators.
The mechanisms behind all of these "charge-related" effects are active areas of investigation.
Thus, we discuss how understanding these phenomena may depend on an interplay between antiferromagnetism and topology, and further show how tunable magnetism may be derived from first-principles calculations.
Beyond charge transport, we discuss progress in layered AFM materials and coupling of light to AFM materials, including studies of the AFM structure, phases, and dynamics using linear, quadratic, and non-linear magneto-optical techniques.

\section{\label{sec:ChargeTransport}Charge Transport in Metallic Antiferromagnets}

\subsection{\label{sec:AMR}Magnetoresistance}

Anisotropic magnetoresistance (AMR) is a long-studied electrical property of ferromagnetic metals where resistivity depends upon the relative orientation between current and magnetization.\cite{mcguire1975anisotropic}
In spintronics research involving ferromagnetic materials, AMR is used in a variety of contexts.
The mixing of AMR with microwave currents in spintronic devices leads to a rectification effect\cite{JuretschkeJAP1960,JuretschkeJAP1963} that can be used to detect high-frequency magnetization dynamics excited by spin torques;\cite{liu2011spin, sklenar2017unidirectional} this effect is often used to quantify spin torque symmetries and torque magnitudes.\cite{macneill2017control,nan2019controlling}
AMR also enables the emission of microwave radiation in spin-torque oscillators that are driven by electric currents.\cite{langenfeld2016exchange,awad2017long,safranski2019spin,zahedinejad2019two}
AMR is not exclusive to ferromagnets; in antiferromagnet metals, AMR refers to the dependence of the electrical resistance upon the relative orientation between the current and the N\'{e}el vector. 
As we will discuss, AMR is a useful way to read out a magnetic memory state stored within an antiferromagnetic metal.\cite{marti2014room,WadleyScience2016,BodnarNatComm2018}
As interest in these materials evolves, it remains to be seen if the above noted applications of AMR in ferromagnets will have analogs in antiferromagnets.  

In ferromagnetic materials, a straightforward way to characterize the AMR of a sample is to apply a large enough magnetic field that overcomes internal demagnetization fields and anisotropies of the system.\cite{oh2019angular}
By saturating the magnetization and rotating the applied magnetic field, the resistivity can be measured as a function of the angle between the current and the field/magnetization of a given sample.
Similar measurements can be made in antiferromagnets, provided the magnetocrystalline anisotropies of the material do not restrict rotation of the N\'{e}el vector.
A good example of characterizing AMR with rotating fields can be found in alloys of CoGd.\cite{moriyama2018spin} 
In bilayers of CoGd films with different compositions, the temperature can be adjusted to a "compensation" point where the bilayer behaves as two antiferromagnetically coupled macrospins.
At the compensation temperature, a phase shift of 90$^\circ$ in the AMR signal occurs relative to the AMR traces at higher (or lower) temperatures when there is a net moment.
The phase shift arises from competition between the Zeeman and antiferromagnetic exchange interaction, leading to a $90^\circ$ angular offset between the N\'{e}el vector and the external field.
Examples of rotating magnetic fields leading to an AMR signal have been reported in antiferromagnetic materials, including Sr$_2$IrO$_4$,\cite{wang2014anisotropic,fina2014anisotropic} MnTe,\cite{kriegner2016multiple} and EuTiO$_3$.\cite{ahadi2019anisotropic}
In Sr$_2$IrO$_4$ and EuTiO$_3$, a field-dependent AMR signal was reported where both the angular period and the AMR amplitude were shown to depend on the magnitude of the rotating field.
The origin of the field-dependent AMR in these materials is still under active investigation.    

\begin{figure}[b]
\centering
\includegraphics[width=3.5 in]{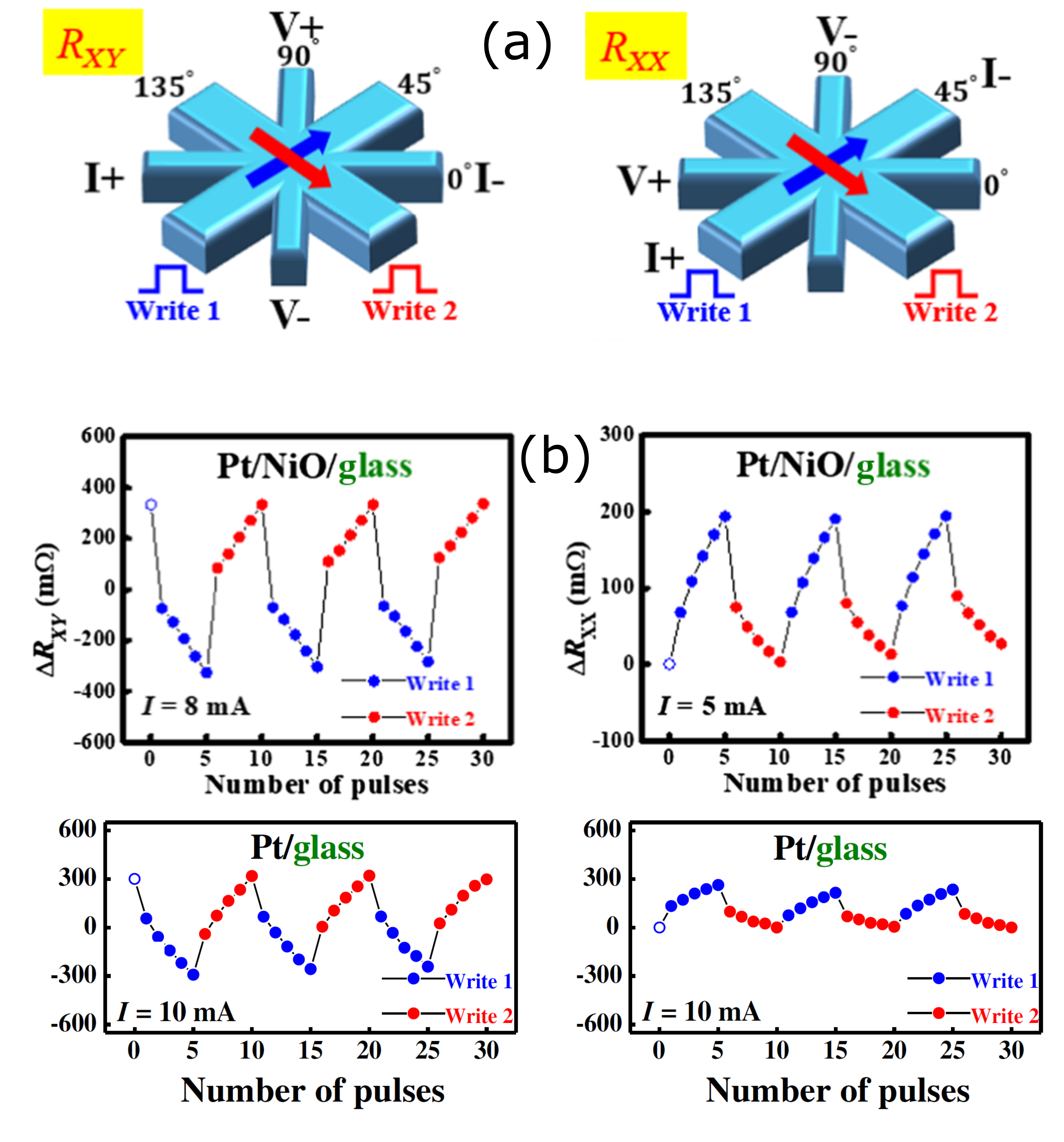}
\caption{\label{figAMR}
(a) Illustration of a multi-terminal device with electrical leads for both the ``writing'' and ``reading'' of information into a heterostruture containing Pt with or without NiO.  (b) Switching behavior in both the longitudinal and transverse ``read'' configuration is observed both with and without NiO.  Reprinted figure with permission from Ref.~\onlinecite{chiang2019absence}.
}
\end{figure}

In antiferromagnetic spintronics, AMR and related effects can be used to ``read'' out memory states, e.g.\ of metallic antiferromagnets or heterostructures incorporating antiferromagnetic layers.
An early example was the usage of AMR in FeRh.\cite{marti2014room,moriyama2015sequential} 
FeRh possesses a first order phase transition from ferro- to antiferromagnetism which occurs near room temperature.\cite{mcgrath2019self}
By applying a field in the ferromagnetic phase, and cooling to the antiferromagnetic phase, the N\'{e}el vector can be initialized perpendicular to the field-cooling orientation.
The field-cooling process ``writes'' information into the FeRh, and  AMR is used to ``read'' out the memory state.
Using X-ray linear magnetic dichroism, combined with the AMR measurement, a higher (lower) resistance state was found when N\'{e}el order was parallel (perpendicular) to the current.\cite{marti2014room}
In Section~\ref{sec:SOT} we discuss how AFM materials such as CuMnAs and Mn$_2$Au can have memory states ``written'' by electrical means.\cite{WadleyScience2016,BodnarNatComm2018} 
Although the writing process differs, the read-out mechanism in these in CuMnAs and Mn$_2$Au was attributed to their intrinsic AMR.

Phenomenologically similar to AMR is the spin Hall magnetoresistance (SMR) effect.
SMR was first discovered in bilayers  of a ferromagnetic insulator adjacent to a spin Hall metal.\cite{nakayama2013spin}
Here, an anisotropic resistance is endowed into the spin Hall metal, that depends upon the relative orientation between the spin polarization in the spin Hall metal and the magnetic order.
SMR also exists in all-metallic bilayers.\cite{avci2015unidirectional,kim2016spin} 
More recently, SMR has been discovered in bilayers consisting of an antiferromagnetic insulator adjacent to a spin Hall metal.\cite{hoogeboom2017negative, fischer2018spin, baldrati2018full,ji2017spin} 
The resistance depends on the orientation of the N\'{e}el order relative to the spin Hall effect induced spin polarization.
Using spin torque effects from the spin Hall metal, switching experiments that used SMR to read out the memory state were reported in NiO/Pt.\cite{chen2018antidamping} 
The switching experiments in NiO/Pt were qualitatively quite similar to the experiments in CuMnAs and Mn$_2$Au.

However, recently there have been a series of experiments in NiO/Pt\cite{chiang2019absence,churikova2020non} and Fe$_2$O$_3$/Pt\cite{zhang2019quantitative, cheng2020electrical} suggesting that the AMR/SMR signal, {\em i.e.}, the read-out in electrical switching experiments, can be a thermal or electromigration artifact arising from the high current densities needed to switch the magnetic state (see Fig.~\ref{figAMR}).
In the Fe$_2$O$_3$/Pt bilayer system, intentional thermal annealing of the sample was used to distinguish two distinct types of switching in the magnetoresistance.
A saw-tooth magnetoresistance shape was attributed to the thermal artifact, while a smaller amplitude step-like change in the resistance was identified as antiferromagnetic switching.
The implications of these new SMR switching experiments have not been fully reconciled yet with the earlier AMR switching experiments.
Clearly, a future goal in the field of antiferromagnetic spintronics will be to identify and separate magnetoresistance effects arising from thermal or electromigration artifacts in both SMR and AMR based systems and devices. 

\subsection{\label{sec:SpinCurrent}Spin current generation}

In the previous section, we discussed general features of the charge transport, and how the charge transport interacts with the spin structure within an antiferromagnet.
Another important question is how metallic antiferromagnets can be used to generate spin currents that can be injected into other adjacent materials.
These spin currents may originate from charge currents, temperature gradients, or magnetization dynamics.

\begin{figure}[b]
\centering
\includegraphics[width=3.5 in]{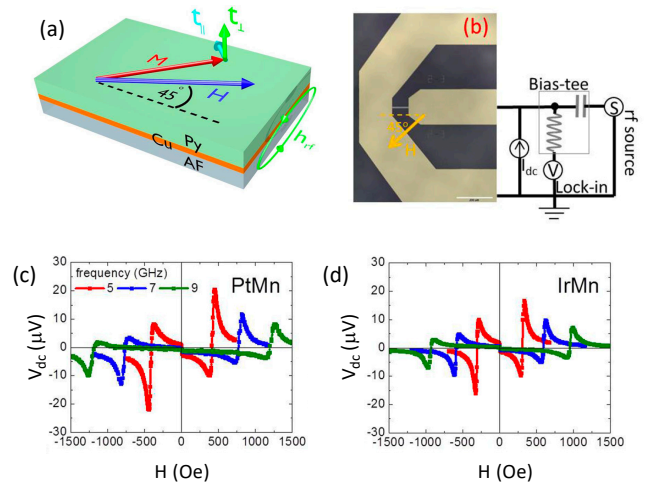}
\caption{\label{figSHE}
(a) Schematic of torques generated due to charge current flow in a Ni$_{80}$Fe$_{20}$ (Py)/Cu/antiferromagnet (AF) multilayer sample.  The torques generated from Oersted fields ($\tau_\perp$) are perpendicular to the damping-like spin-orbit torques ($\tau_\parallel$). (b) Optical image of a sample integrated into a terminated coplanar waveguide together with a schematic of a spin-torque ferromagnetic resonance measurement (ST-FMR).  (c) and (d) ST-FMR measurements at different resonance frequencies for PtMn and IrMn.  Reprinted figure with permission from Ref.~\onlinecite{ZhangPRB2015}.
}
\end{figure}

The generation of spin currents from charge currents in the bulk of conducting materials is known as spin Hall effects.\cite{Hoffmann2013IEEETM}
These exist in any conducting materials and are a simple consequence of spin-orbit coupling. 
Therefore, one can naively expect that spin Hall effects are more pronounced for materials with heavier elements, and, in fact, a systematic study of different CuAu-I-type metallic antiferromagnets based on Mn demonstrated such a dependence, both experimentally and theoretically.\cite{ZhangPRL2014} 
Using spin pumping and inverse spin Hall effect measurements it was shown that the spin Hall angles, which are material-specific parameters describing the efficiency of the charge- to spin-current conversion, follow the relationship PtMn > IrMn > PdMn > FeMn.
In fact, the experimentally observed spin Hall conductivities are reasonably well explained by first-principles calculations of the intrinsic spin Hall effects in these alloys.
Similar measurements were subsequently performed by spin-torque ferromagnetic resonance,\cite{Li2011PRL} which has become one of the standard approaches for quantifying spin Hall effects.
When an \emph{rf} current is passed through a bilayer of the spin Hall material and a ferromagnetic metallic layer [such as Ni$_{80}$Fe$_{20}$, permalloy (Py)] then this will result in different torques acting on the magnetization of the ferromagnet, as is shown schematically in Fig.~\ref{figSHE}(a).
An optical image of a sample integrated in a terminated coplanar waveguide is shown in Fig.~\ref{figSHE}(b).
Here, the current passing through the antiferromagnetic layer may generate on Oersted field $h_{rf}$, which results in a torque $\tau_\perp \propto M \times h_{rf}$, and via the spin Hall effect also generates a damping-like torque $\tau_\parallel$, which is perpendicular to the torque from the Oersted field.
Note that for investigating spin Hall effects in metallic antiferromagnets, one commonly inserts a non-magnetic layer (typically copper) in between the ferromagnet and antiferromagnet to avoid additional spurious magnetic interactions.
Since the two torques are perpendicular to each other, they drive the magnetization dynamics in the ferromagnet with different phases.  Consequently, the magnetization dynamics in the ferromagnet results through its anisotropic magnetoresistance in a time-varying resistance change, which via mixing with the original {\em rf} charge current results in a phase sensitive detection of the ferromagnetic magnetization dynamics.\cite{JuretschkeJAP1960,JuretschkeJAP1963}  Therefore, analyzing the resonance lineshape allows the determination of the magnitude of the spin Hall angle.\cite{liu2011spin}  This is shown in Figs.~\ref{figSHE}(c) and (d), which show voltage spectra as a function of applied magnetic field measured for fixed \emph{rf} frequency for PtMn and IrMn, respectively.\cite{ZhangPRB2015} 
In both cases the lineshape is a combination of an antisymmetric and a symmetric Lorentzian, which correspond to the Oersted and spin Hall torques, respectively.  Thus the larger symmetric component in Fig.~\ref{figSHE}(c) compared to Fig.~\ref{figSHE}(d) indicates a larger spin Hall angle for PtMn compared to IrMn.

The initial experimental results\cite{ZhangPRL2014} were obtained for polycrystalline films of the metallic antiferromagnets.
At the same time, first-principles calculations\cite{ZhangPRL2014} suggested pronounced anisotropies of the spin Hall conductivities, which reflect the different broken symmetries due to the antiferromagnetic order.
Subsequent measurements for epitaxial films grown along different directions confirmed indeed pronounced anisotropies.
\emph{E.g.}, PtMn films grown with an $a$-axis orientation has a spin Hall conductivity about twice as large as PtMn films grown with a $c$-axis orientation.\cite{ZhangPRB2015}
Similarly, large anisotropies for different crystalline orientations have also been observed for other antiferromagnetic systems, such as IrMn$_3$.\cite{ZhangSciAdv2016}
Generally, these strong anisotropies directly reflect the possibility of additional spin Hall contributions, once the antiferromagnetic order reduced the symmetry.\cite{CulcerPRL2007}
Therefore, the question arises in general what the role of the antiferromagnetic spin structure is with respect to the spin Hall effects in metallic antiferromagnets.\cite{ZhangPRL2014,SklenarAIPAdv2016}
A first attempt to investigate this question was pursued by assuming that different exchange bias configurations also reflect different microscopic spin configurations in the antiferromagnets. 
However, so far experiments show that the spin Hall effects are mostly independent of exchange bias.\cite{SaglamPRB2018,KhodadadiPRB2019} 
Nevertheless, these experiments were performed for polycrystalline films and more conclusive investigations may require measurements for epitaxial systems.
Another open question is how spin Hall effects in antiferromagnetic metallic alloys depend on composition and doping.

Interestingly, the spin Hall effects in antiferromagnetic metals that incorporate heavier elements, such as Ir or Pt, are comparable in efficiency to the spin Hall effects that are used in other commonly used non-magnetic metals.
Thus, antiferromagnetic metals can be used for switching magnetization via spin-orbit torques.
At the same time, they may provide an effective field via exchange bias on the magnetization in an adjacent ferromagnetic layer.
This turns out to be useful for switching magnetization in ferromagnetic layers with perpendicular anisotropies.
For many magnetic memory devices, it is often beneficial to have information stored in perpendicular magnetized layers.
However, in order to have deterministic electric switching of perpendicular magnetizations via spin-orbit torques, an additional symmetry breaking in-plane magnetic field is required.
Thus, if the symmetry breaking magnetic field is provided via exchange bias, field-free switching of perpendicular magnetizations can be achieved even without any additional externally applied magnetic field.\cite{FukamiNatMat2016,OhNatNano2016} 
Furthermore, inhomogeneities of exchange bias in polycrystalline metallic antiferromagnets may result in magnetization switching via complex intermediate magnetization states, which can be exploited for memristive behavior\cite{FukamiNatMat2016} that has already been used for implementing associative memory devices.\cite{BordersAPE2017}

In addition to the ordinary spin Hall effects discussed so far, the magnetic structure in metallic antiferromagnets may also give rise to spin current generation with unusual symmetries.
In particular, antiferromagnets with chiral non-collinear spin structures are expected to show anomalous Hall effects.
This was first discussed theoretically for strained $\gamma$-FeMn\cite{ShindouPRL2001}, which has a $3Q$ spin-structure where spins are arranged on the corner of a tetrahedron and either point towards or away from the center of the tetrahedron.
More recently, similar effects have been theoretically predicted for metallic antiferromagnets with spins arranged on a Kagome-lattice\cite{ChenPRL2014,KueblerEPL2014} and indeed corresponding anomalous Hall effects have been observed experimentally.\cite{Nakatsuji2015,Nayak2016}
Note that in ferromagnets it has already been demonstrated that the anomalous Hall effect is accompanied by a transverse spin current, which can give rise to spin accumulations and spin-orbit torques.\cite{QinPRB2017,DasPRB2017,GibbonsPRAppl2018}
Thus, an open question is whether the anomalous Hall effects in antiferromagnets can similarly give rise to concomitant transverse spin currents.
Towards this end, it has already been demonstrated that the response of the non-collinear spin structure to an applied field can give rise to a magnetic spin Hall effect, which is even in magnetic fields.
This has been observed for Mn$_3$Sn\cite{KimataNature2019} and Mn$_3$Ir\cite{HolandaArxiv2019}.
Furthermore, it was shown that for [001] oriented Mn$_3$Ir films the generated spin current can have a significant polarization in the out-of-plane direction,\cite{LiuPRAppl2019} which provides interesting new perspectives for manipulating magnetization of ferromagnets with perpendicular anisotropies. 
In addition, the close relationship between the broken symmetries due to antiferromagnetic spin structure and the resultant spin currents\cite{WinklerArxiv2019} opens up entirely new perspectives for reconfigurable spin-orbit torques.

Aside from using charge currents in antiferromagnets for generating spin currents, it is also known that heat current due to thermal gradients can inject spin current from antiferromagnets into ferromagnets,\cite{SekiPRL2015,WuPRL2019} a phenomenon known as spin Seebeck effect.
Note that due to the compensated nature of the spin structure there can be degenerate magnon modes with opposite spins in antiferromagnets.\cite{RezendeJAP2019} 
Therefore, a magnetic field is required to lift this degeneracy in order to generate a net spin Seebeck signal. 
Although it is known that magnons can contribute to spin transport in metallic antiferromagnets,\cite{SaglamPRB2016,GladiiPRB2018} there has been so far no reports of spin Seebeck effect in metallic antiferromagnets.
Note that unlike for metallic ferromagnets, where spin Seebeck effects are hard to distinguish from anomalous Nernst effects,\cite{HuangPRL2011} this is generally not an issue with colinear antiferromagnets, where anomalous Hall effects are absent.
At the same time non-colinear, chiral antiferromagnets may give rise to anomalous Nernst effects,\cite{Ikhlas2017} and therefore, just as with the above discussed anomalous Hall effects, the question arises, whether these anomalous Nernst effects also give rise to concomitant spin currents.

Another possibility for generating spin currents from antiferromagnets is via spin pumping from antiferromagnetic magnetization dynamics,\cite{ChengPRL2014} in analogy to the well-established spin pumping from ferromagnetic resonance.\cite{TserkovnyakPRL2002,HeinrichPRL2003}
As already mentioned, magnons in antiferromagnets may carry two opposite directions of angular momentum, and thus in principle it is possible to have two different spin polarizations pumped from an antiferromagnet. 
Very recently, it has been shown that such spin pumping is indeed possible from insulating antiferromagnets,\cite{Li2020Nature} but similar measurements with metallic antiferromagnets are still missing.

\subsection{\label{sec:SOT}Spin torques}

\begin{figure}[b]
\centering
\includegraphics[width=3.5 in]{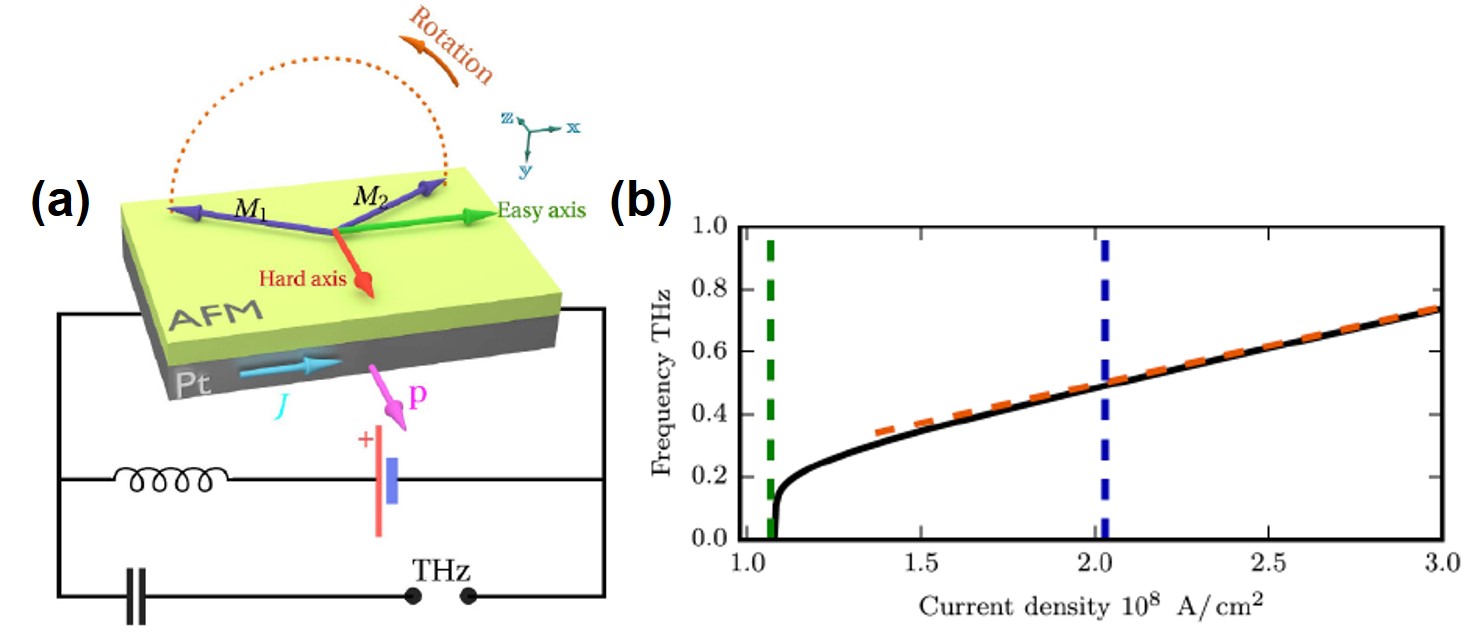}
\caption{\label{figTHz}
(a) Schematic of antiferromagnetic magnetization dynamics induced via spin-orbit torques in a bi-axial antiferromagnet. (b) Numerically calculated rotation frequency of the N\'eel vector as a function of driving current. The orange dashed line corresponds to an analytical approximation, while the blue and green dashed lines indicate the minimum current densities required to initiate and maintain the dynamics, respectively. Reprinted figure with permission from Ref.~\onlinecite{KhymynSciRep2017}.}
\end{figure}

Just as metallic antiferromagnets can be utilized for generating spin currents, they may also absorb spin currents.
The absorption of spin currents becomes of particular interest when the angular momentum associated with the spin currents gets absorbed into the antiferromagnetic spin structure and results in changes or dynamic excitations of the spin structures through spin transfer torques. 
For ferromagnetic systems with a net magnetization $\mathbf{M}$, one generally distinguishes between field-like torques $\mathbf{\tau_{fl}} \propto \mathbf{M} \times \mathbf{\sigma}$ and damping-like torques $\mathbf{\tau_{dl}} \propto \mathbf{M} \times (\mathbf{\sigma} \times \mathbf{M})$, where $\mathbf{\sigma}$ is the polarization direction of the injected spin current. 
If one adopts the same torques due to spin current injections to antiferromagnetic systems, then it is easy to see\cite{GomonayNatPhys2018} that the field-like torques cancel each other out due to the opposite direction of the two antiferromagnetic spin sublattices. 
However, since the damping-like torque is even in the magnetization direction, it creates identical torques for both sublattices.  
Thus if a spin current is injected into the antiferromagnet with a polarization perpendicular to Néel vector, then the spin-torque induced canting of the two sublattices should lead to a rotation of the Néel vector via the torques from the antiferromagnetic exchange interactions [see Fig.~\ref{figTHz}(a)].\cite{GomonayJMSJ2008,ChengPRB2015,KhymynSciRep2017} 
As shown by theoretical calculations, see Fig.~\ref{figTHz}(b), this may then give rise to tunable oscillators in the sub-THz frequency range.
So far an experimental demonstration of such \emph{dc} current driven oscillations is still missing.
One issue might be that the spin diffusion lengths in metallic antiferromagnets are very short,\cite{AcharuyyaJAP2011,MerodioAPL2014,ZhangPRB2015} and so far are reported to be mostly below 2~nm.\cite{BaltzRMP2018} 
Nevertheless, measurements in ultrathin 1-nm thick IrMn films may suggest some spin-torque related magnetization changes in the metallic antiferromagnet.\cite{ReichlovaPRB2015}

Another possible way to electrically manipulate the antiferromagnetic spin structure is via N\'eel spin-orbit torques, as will be discussed further in section~\ref{sec:Dirac}. 
The basic idea is that if the crystal structure of the antiferromagnet is such that each crystal site for the two antiferromagnetic spin sublattices has locally opposite inversion symmetry, then this may result in local staggered (N\'eel) spin accumulation with opposite signs for each antiferromagnetic spin, and therefore result in identical field-like torques for both antiferromagnetic sublattices.  
This idea was first theoretically proposed for Mn$_2$Au,\cite{ShickPRB2010,ZeleznyPRL2014,BarthemNatComm2013} but experimental observation of sublattice switching via N\'eel spin-orbit torques was first demonstrated for CuMnAs.\cite{WadleyScience2016} 
Subsequently, similar experimental results were obtained for Mn$_2$Au films.\cite{BodnarNatComm2018,MeinertPRAppl2018} 
Since there are specific symmetry requirements for the crystal structure of the metallic antiferromagnet in order to observe N\'eel spin orbit torques, this effect has only been reported for CuMnAs and Mn$_2$Au. 
Another experimental complication is that all-electrical measurements of the current induced switching of antiferromagnetic spin structures generally require measurements with currents applied in several different directions with respect to the crystalline orientation, and with current densities that are close to the damaging threshold of the devices.
Thus extrinsic effects due to electromigration may very often be mistaken for changes of the antiferromagnetic spin-structure.\cite{chiang2019absence,zhang2019quantitative,churikova2020non}  
Therefore, further exploration of N\'eel torque related effects will benefit from detailed direct experimental detection of the antiferromagnetic order in these devices.

\section{\label{sec:NewMaterials}New Materials}

\subsection{\label{sec:LayeredMAterials}Layered Materials}
\begin{figure}
\centering
\includegraphics[width=3.5 in]{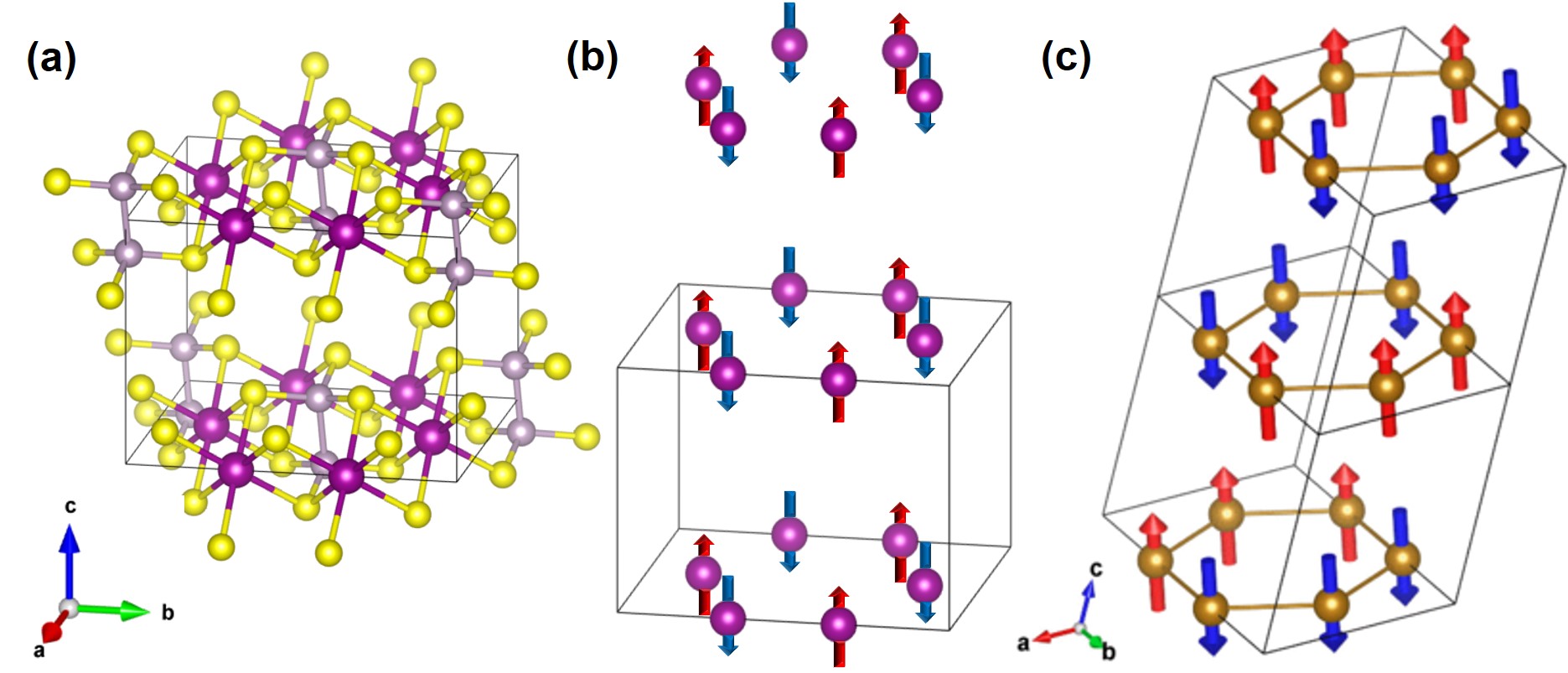}
\caption{\label{figLayerd}
(a) Crystal structure of $X$PS$_3$ ($X$ = Mn and Fe). Spin orientation in (b) MnPS$_3$ and (c) FePS$_3$ from Ref.~\onlinecite{lancon_PRB_2016}.}
\end{figure}

Layered ferromagnetic and antiferromagnetic materials have been studied since the early sixties.
Only recently, due to the advances in exfoliation processes, it has been possible to separate van der Waals materials into layers with thicknesses corresponding to a single unit cell.
One example is Cr$X_3$ ($X$ = Cl, Br and I), which has magnetic ordering down to monolayers at low temperature.
Among the trihalides, CrI$_3$ is the most studied antiferromagnetic insulator and has an out-of-plane anisotropy.
The intralayer Cr$^{3+}$ ions in CrI$_3$ are coupled ferromagnetically , while the interlayers are coupled antiferromagnetically.
CrCl$_3$ has similar magnetic coupling but with an in-plane anisotropy and CrBr$_3$ is identified as a Heisenberg ferromagnetic insulator.\cite{kimPNAS2019}
Other layered materials with antiferromagnetic ordering include Cr$_2$Ge$_2$Te$_6$,\cite{gong2017discovery} $X$PS$_3$ ($X$ = Mn and Fe),\cite{wang_2Dmaterials_2016,kim_2Dmaterials_2019} V$Y_2$ ($Y$ = S and Se), \cite{ma_ACSNano_2012} and RuCl$_3$\cite{weber_NanoLett_2016,du_2Dmaterials_2018,zhou_JPCS_2019}.
In MnPS$_3$, all nearest-neighbor interactions within a layer are antiferromagnetic \cite{kurosawa_JPSJ_1983} [see Fig.~\ref{figLayerd}(b)], whereas in FePS$_3$, Fe$^{2+}$ is coupled ferromagnetically to two of the nearest neighbors and antiferromagnetically to the third so that within the layer the Fe$^{2+}$ moments appear as ferromagnetic chains coupled antiferromagnetically to each other [see Fig.~\ref{figLayerd}(c)].\cite{lancon_PRB_2016}
For both compounds the magnetic moments are perpendicular to the layer.
Tunable magnetism has also been identified in many other novel materials from first-principles calculations.\cite{LvPRB2015,HemantACSNano2017,LiPRB2018}
Only recently, the antiferromagnetic van der Waals metal GdTe$_3$ has been identified.\cite{LeiSciAdv2020}
Lei {\em et al.}\ have experimentally shown for low temperatures that the antiferromagnetic order of GdTe$_3$ persists down to three monolayers.
Other layered antiferromagnetic metals have been predicted from first-principles calculations to have high spin-orbit torque and N\'eel temperatures well above room temperature.\cite{jiao_Nanoscale_2019}
However, the experimental demonstration of such materials yet has to be explored.
One limitation is the air sensitivity of layered antiferromagnetic materials, which adds to the challenges for their technological applications.

\subsection{\label{sec:Weyl}Antiferromagnetic Weyl Metals}

In addition to layered materials, the advent of topological materials has cast a new light on many different aspects of the properties of materials that were, heretofore, considered to be well-understood.
More specifically, there is growing experimental and theoretical evidence that there is a strong connection between magnetism and topology within condensed matter and materials\cite{Smejkal2017}, though the connection is not understood.
The lack of understanding represents a unique opportunity to explore antiferromagnetic semimetals for signatures of the presence of topology.
In this context, we discuss the current understanding of both the theoretical and experimental search for antiferromagnetic Weyl and Dirac semimetals in an effort to not only uncover the origin of the coexistence of these two seemingly disparate orders, but to assess their potential usefulness for future spintronic technologies. 

Before descending to the current state of research in topological antiferromagnetic semimetals, we briefly review some of the important topological concepts that are needed to understand the developments.
To date, the vast majority of observed topological phases in non-magnetic metals are stabilized as a result of the presence of time-reversal ($T$) and inversion ($P$)  symmetry.\cite{QiRMP,HasanRMP}
When either time-reversal symmetry or inversion symmetry is broken, the resultant non-degenerate conduction and valence bands may touch at discrete points or lines within the Brillouin zone.
The low-energy quasiparticle excitations around the non-degenerate band touching points are two-fold degenerate and described by Weyl fermions whose Hamiltonian is 
\begin{equation}
H(\mathbf{k}) = \sum_{i,j=x,y,z}v_{ij}k_{i}\sigma_{j}, 
\end{equation}
where $\sigma_{i=x,y,z}$ are the Pauli matrices and $v_{ij}$ is the Fermi velocity assuming that $\det[v_{ij}] \neq 0$.
Materials that possess such bandstructures are referred to as Weyl semimetals.\cite{ArmitageRMP}
The band touching points, or Weyl nodes, in Weyl semimetals may not be removed by perturbations due to the fact that there are no remaining Pauli matrices that may be added to the Hamiltonian.
Weyl nodes come in pairs and act as monopoles of Berry curvature with one Weyl node acting as a source of Berry curvature and the other as the sink.
The locations of the Weyl nodes within the Brillouin zone is determined by the preserved symmetry present in the material.
Considering the two aforementioned canonical symmetries individually, we note that when $T$ is present, then a Weyl node located at $\mathbf{k}$ must have a time-reversed partner located at $\mathbf{-k}$ that carries the same topological charge.
Therefore, to avoid having a non-zero topological charge within a given material, there must be two additional Weyl nodes present that both carry oppositely compensating topological charge to ensure that the total topological charge remains zero.
On the other hand, when $P$ is present, a Weyl node located at $\mathbf{k}$ must have a partner of opposite topological charge located at $\mathbf{-k}$.

The theory of charge transport in Weyl semimetals is well established and many of the predictions revolve around manifestations of axion electrodynamics.
To be more precise, axion electrodynamics refers to the addition of an axion term to the traditional electromagnetic Lagrangian where the action is $S_{\theta} = \frac{e^{2}}{2\pi h}\int dt dr\,\theta(\mathbf{r},t)\, \mathbf{E}\cdot \mathbf{B}$.
In the preceding equation, $\theta$ is the axion background, or magnetoelectric polarization, and $\mathbf{E}$ and $\mathbf{B}$ are the electric and magnetic fields, respectively.
The presence of this magnetoelectric term in the crystal produces prominent charge transport responses such as the chiral anomaly and the anomalous Hall effect (AHE).\cite{Zyuzin2012A,Zyuzin2012B}

\begin{figure}[b]
\centering
\includegraphics[width=3.4 in]{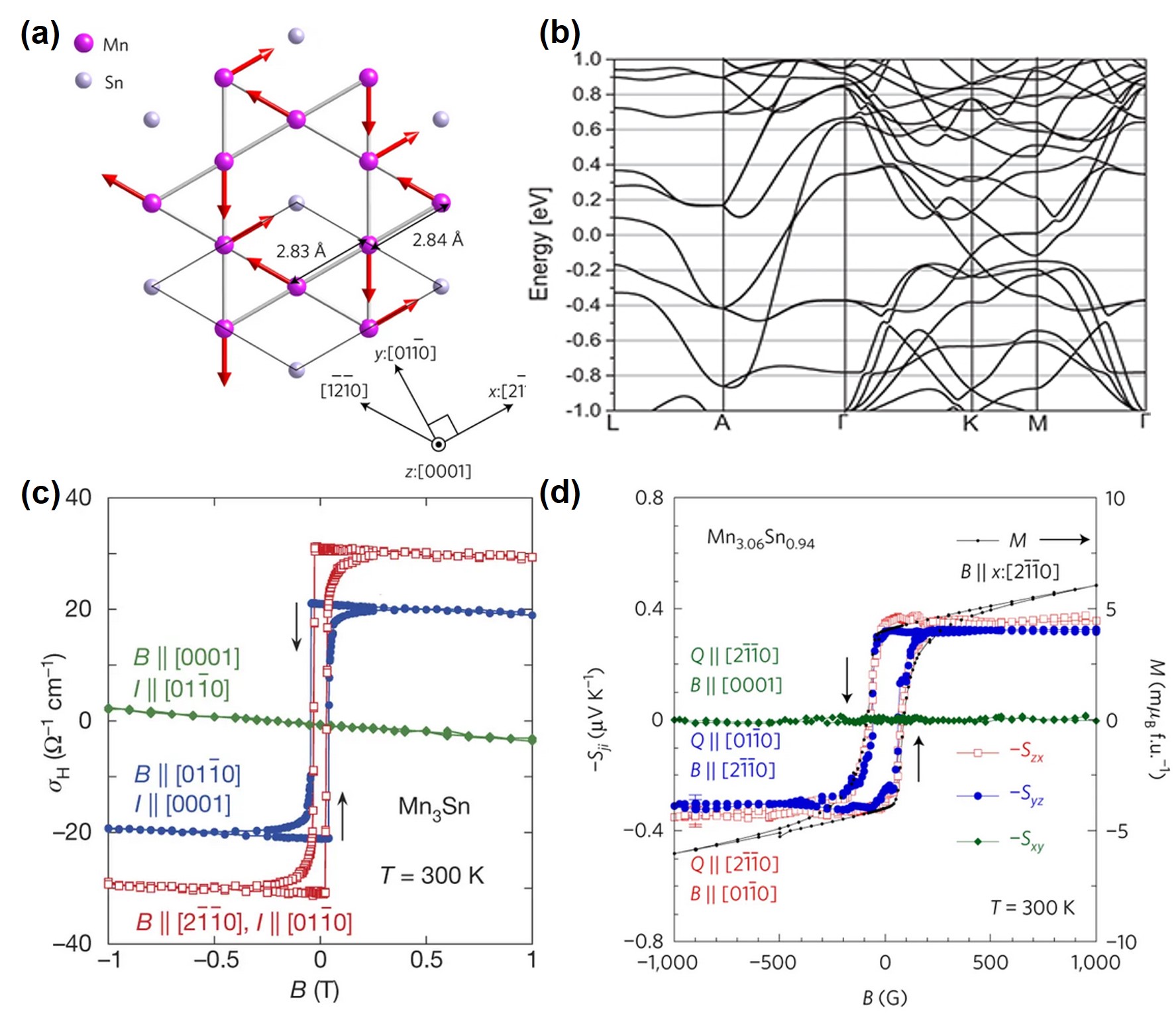}
\caption{\label{fig:weyl}
(a) Schematic representation of the $ab$-plane of the Mn$_3$Sn crystal lattice. The connections between the atoms consisting of alternating large and small triangles illustrate the breathing modes of the Kagome lattice.
Reprinted figure with permission from Ref.~\onlinecite{Ikhlas2017}. (b) Electronic bandstructure of Mn$_3$Sn from {\em ab-initio} calculations using the spin-density functional approximation. Reprinted figure with permission from Ref.~\onlinecite{Felser2017}. (c) Measured AHE in Mn$_3$Sn at room temperature along two different crystal directions. Reprinted figure with permission from Ref.~\onlinecite{Nakatsuji2015}. (d) Measured Nernst signal in Mn$_3$Sn at room temperature showing clear hysteresis as the in-plane magnetic field is varied. Reprinted figure with permission from Ref.~\onlinecite{Ikhlas2017}.}
\end{figure}

The presence of the AHE arises from contributions that are both intrinsic to the crystal, such as broken $T$ and spin-orbit coupling, and extrinsic, such as defect scattering.\cite{NagaosaRMP}
It is well understood that the size of the AHE in a given ferromagnetic crystal will be proportional to the magnetization of the system and thus, the AHE in an antiferromagnetic metal must be zero.
In opposition to this assertion, recent theoretical work has shown that antiferromagnetic metals that have broken time-reversal symmetry and non-collinear spin order will have a non-zero Berry curvature and, consequently, possess an AHE.\cite{MacDonald2014}
To illustrate the AHE in an antiferromagnetically ordered Weyl semimetal, consider the Heusler compound Mn$_3$Sn.
In Fig.~\ref{fig:weyl}(a), we show the crystal structure of Mn$_3$Sn.
While it is known that Mn$_3$Sn\cite{Kren1975} may crystallize in both tetragonal and hexagonal structures, Fig.~\ref{fig:weyl}(a) depicts the more common hexagonal form of the Mn$_3$Sn crystal structure with magnetic ordering temperature that is well above room temperature.\cite{Nakatsuji2015,Ikhlas2017}
The crystal consists of a Kagome lattice formed by the Mn atoms within the $ab$-plane that are subsequently stacked vertically along the $c$-axis to form a tube of face-sharing octahedra.
In Fig.~\ref{fig:weyl}(b), we show the {\em ab-initio} calculated bandstructure of Mn$_3$Sn to illustrate the existence of Weyl nodes near the Fermi energy, $E_{F} = 0$~eV, giving rise to a large Berry curvature at the Fermi surface, where the adiabatic motion of the quasiparticles in the Berry curvature leads to the AHE.\cite{Haldane2004}
In Fig.~\ref{fig:weyl}(c), the AHE is plotted for Mn$_3$Sn at room temperature along two distinct crystal directions producing an anomalous Hall conductivity\cite{Nakatsuji2015} of 20~$\Omega^{-1}$cm$^{-1}$.
In the closely related material, {\em i.e.} Mn$_3$Ge, that value is 50~$\Omega^{-1}$cm$^{-1}$, and the predicted spin Hall conductivity\cite{Nayak2016} is 1100~$(\hbar/e)~ \Omega^{-1}$cm$^{-1}$, which is comparable to that of platinum.\cite{Guo2008}

In close association to the Fermi surface properties of the AHE, measurements on the anomalous Nernst effect, shown in Fig.~\ref{fig:weyl}(d), similarly demonstrate that Mn$_3$Sn possesses a large Nernst effect resulting in a Seebeck coefficient of $\approx 0.35~\mu$VK$^{-1}$ without an externally applied magnetic field at room temperature.\cite{Ikhlas2017}

\subsection{\label{sec:Dirac} Antiferromagnetic Dirac Metals}

In addition to the existence of Weyl fermions in antiferromagnetic semimetals, it is possible to find additional fermionic excitations.
To begin to see how one may find Dirac fermions, consider that in order for the bandstructure of a Weyl semimetal to remain two-fold degenerate, the material may not possess both $P$ and $T$.
The resulting energy spectrum places Weyl nodes of opposite topological charge at the same point in momentum space resulting in a four-fold degenerate band crossing that is not topologically stable.
However, if there is an additional symmetry present in the semimetallic crystal structure, then Weyl nodes of opposite topological charge may be stabilized at the same point in momentum space.
Fortunately, the presence of additional crystalline point group or space group symmetries is capable of constraining the Hamiltonian such that the mixing of Weyl nodes is forbidden, leading to a stable four-fold degenerate band crossing.
The stable merger of different Weyl nodes realizes a $(3+1)$-D Dirac vacuum and materials containing such four-fold degenerate nodes are referred to as Dirac semimetals.\cite{ArmitageRMP}  

\begin{figure}[b]
\centering
\includegraphics[width=3.4 in]{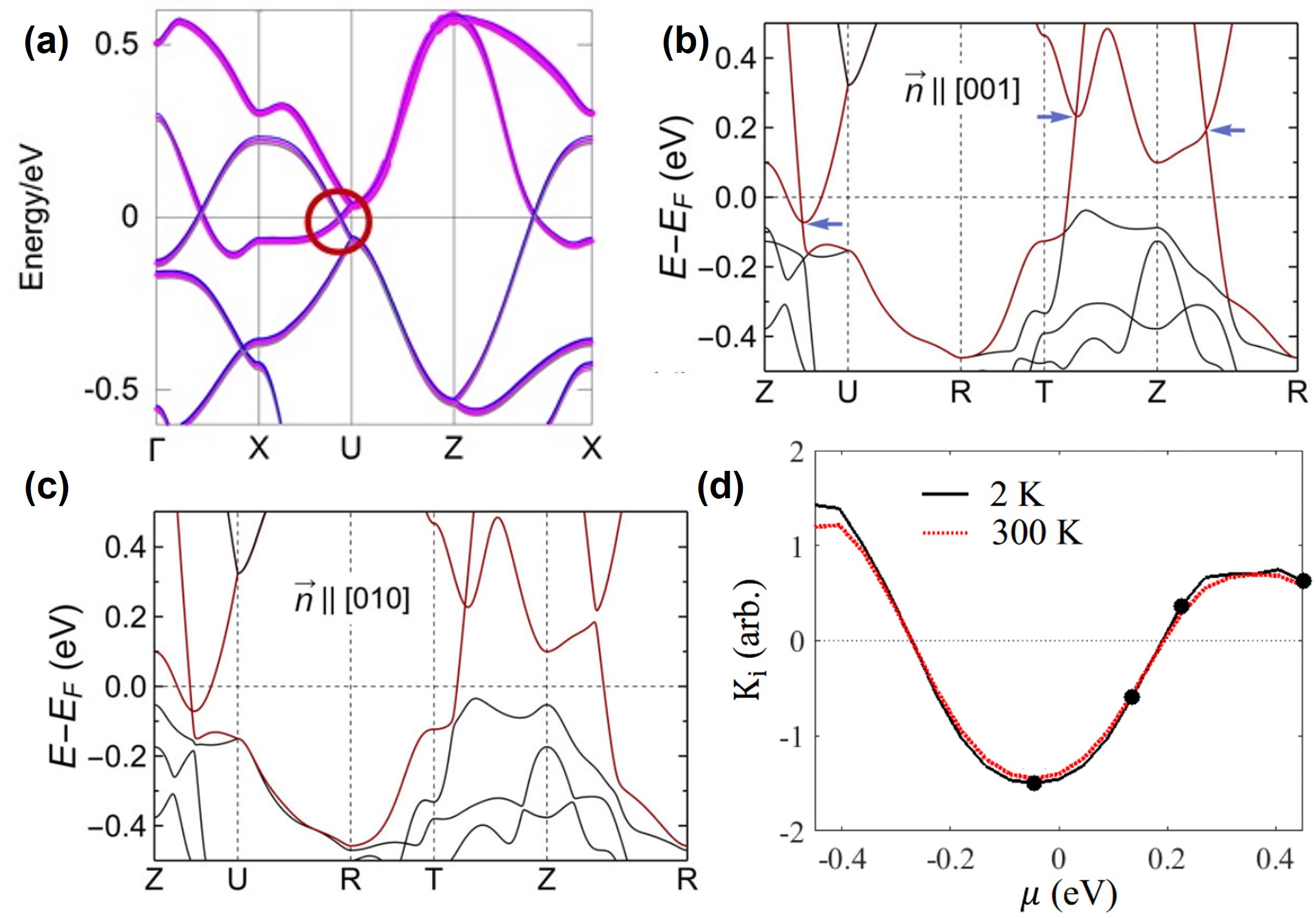}
\caption{\label{fig:dirac}
(a) Calculated bandstructure for orthorhombic CuMnAs showing the clear Dirac bandcrossings within the Brillouin zone. Reprinted figure with permission from Ref.~\onlinecite{Smejkal2017}. Calculated electronic bandstructure for MnPd$_2$ with the N\'{e}el vector aligned along the (b) [001]-direction and (c) [010]-direction. Reprinted figure with permission from Ref.~\onlinecite{Shao2019}. (d) Calculated anisotropy energy for CuMnAs as a function of the chemical potential where a positive value indicates that the system prefers a gapless phase and a negative value a gapped phase. Reprinted figure with permission from Ref.~\onlinecite{Kim2018}.}
\end{figure}

In antiferromagnetic metals, the inherent magnetism breaks either one or both $P$ and $T$ symmetries. Therefore, in order for a Dirac semimetal to preserve the 4-fold band degeneracy in the presence of antiferromagnetism, it must possess an emergent antiunitary symmetry that serves to stabilize relativistic band crossings within the metal.\cite{FangAFM2013}
Such conditions are satisfied in CuMnAs in which $P$ and $T$ symmetries are individually broken but the combination $PT$ is preserved in the presence of an additional nonsymmorphic space group $D_{2h}$.
The resulting bandstructure of this material, shown in Fig.~\ref{fig:dirac}(a), shows several degenerate band crossings within the Brillouin zone that are protected by the expansion of the little group resulting from the presence of the nonsymmorphic crystalline symmetries.
In groundbreaking experimental work, charge transport measurements have demonstrated that in CuMnAs one may indeed manipulate the antiferromagnetic order using the N\'{e}el spin-orbit torque.\cite{WadleyScience2016}

The work on utilizing the N\'{e}el spin-orbit torque clearly points to the potential for both reading and writing states in antiferromagnets via manipulation of the position of the N\'{e}el vector between gapped and gapless phases of the topological antiferromagnet.
Clearly, there are additional materials that possess a similar nonsymmorphic crystal structure, noncollinear antiferromagnetism, and augmented antiunitary symmetries that may be capable of operating at higher temperatures than for CuMnAs.
In Figs.~\ref{fig:dirac}(b) and (c), we plot the bandstructure calculated via {\em ab-initio} methods for MnPd$_2$ for two different orientations of the N\'{e}el vector.
In Fig.~\ref{fig:dirac}(b), the N\'{e}el vector is aligned along the [001]-direction where several topologically protected degenerate band crossings are observed, denoted by the arrows.\cite{Shao2019}
Using the N\'{e}el spin-orbit torque, the N\'{e}el vector may be reoriented to the [010]-direction that results in the magnetic orientation breaking the underlying crystal symmetries that serve to protect the gapless nature of the topological phase, creating gaps in spectrum. Figure ~\ref{fig:dirac}(b) and ~\ref{fig:dirac}(c) show biaxial anisotropy energy corresponding to the gapped and gapless phases, respectively.  

The principle behind the N\'{e}el spin-orbit torque is that charge transport reorients the N\'{e}el vector and underlying antiferromagnetic order.
While it is certain that the N\'{e}el spin-orbit torque provides sufficient torque to reorient the magnetism within the topological Dirac semimetal when the phase is initially gapless, it is unclear if sufficient torque is produced when the semimetal is in the gapped phase.
Another methodology to reorient the antiferromagnetic order within a topological Dirac semimetal is to manipulate the location of the chemical potential within the material using electrostatic gating.\cite{Kim2018}
In Fig.~\ref{fig:dirac}(d), we show the calculated anisotropy energy for CuMnAs as a function of the location of the chemical potential.
The anisotropy energy is defined as the energetic difference between the gapless phase when the N\'{e}el vector is aligned along the [100]-direction and the gapped phase when the N\'{e}el vector is aligned along the [001]-direction with positive values indicating the system prefers the gapless phase.
We see clearly that by simply manipulating the chemical potential we are able to change the preferred state of the system without the need for a charge current.
Therefore, one is able to move between the two bistable phases with less energy.

\section{\label{sec:Dynamics}Dynamics}

Ferromagnetic resonance (FMR) and the associated magnon mode spectrum are the fundamental dynamic excitations of magnetization, and are ubiquitous across many areas of magnetism such as spintronics \cite{hoffmann2015opportunities} and magnonics.\cite{chumak2015magnon}
FMR can be thought of as an infinite wavelength magnon mode.
For long wavelength excitations, precession frequencies are in the GHz range and are set by the external magnetic field, magnetic anisotropies, and the dipolar interaction.\cite{damon1961magnetostatic, sklenar2012generating} 
Short wavelength magnons in ferromagnets can have much higher frequencies that are set by the exchange interaction.\cite{kalinikos1986theory}
In antiferromagnets, there are both acoustic and optical antiferromagnetic resonance (AFMR) dynamic modes.\cite{keffer1952theory} 
The two modes are distinguished by a phase difference in the precession of the antiferromagnetically coupled magnetic sub-lattices. 
The energy scale of optical magnon modes is set by the exchange interaction \textit{across all length scales}, and optical AFMR can have frequencies in the THz range.\cite{gomonay2018antiferromagnetic, kampfrath2011coherent,tzschaschel2017ultrafast}
Because spatially uniform modes are easier to access experimentally, long wavelength THz modes in antiferromagnets represent a unique difference compared with ferromagnets. 
These modes are actively being considered for their technological potential in terms of ultrafast switching of magnetic memories or as potential sources for THz electromagnetic radiation. 

Basic research into the dynamics of antiferromagnets can be more readily enabled if the antiferromagnetic exchange interaction is reduced, since this shifts the resonance frequencies into a range that is more readily experimentally accessible. 
Synthetic antiferromagnets are an artificial material system typically comprised of multiple magnetic layers that are weakly coupled.\cite{duine2018synthetic}
A simple example is the insertion of a non-magnetic spacer layer between two ferromagnets which faciliates an antiferromagnetic interaction via the Ruderman--Kittel--Kasuya--Yosida interaction.
Using the low damping ferromagnetic material, permalloy (Ni$_{80}$Fe$_{20}$, Py), trilayers of Py/Ru/Py have been used to directly examine acoustic and optical AFMR\cite{liu2014interlayer}.
More recently, magnetic garnet materials known to have exceptionally low magnetic damping have been used to create synthetic antiferromagnets.\cite{gomez2018synthetic}
Low damping ferromagnetic insulators, such as yttrium iron garnet (YIG), have  previously enabled unique coherent magnon phenomena like the Bose-Einstein condensation of magnons.\cite{demokritov2006bose}
There has been previous interest exploring Bose-Einstein condensation of magnons within antiferromagnets as well.\cite{radu2005bose,fjaerbu2017electrically} 
It will be intriguing to see if synthetic antiferromagnets, employing low-damping insulators, can bridge these two areas of interest.  

\begin{figure}[b]
\centering
\includegraphics[width=3.5 in]{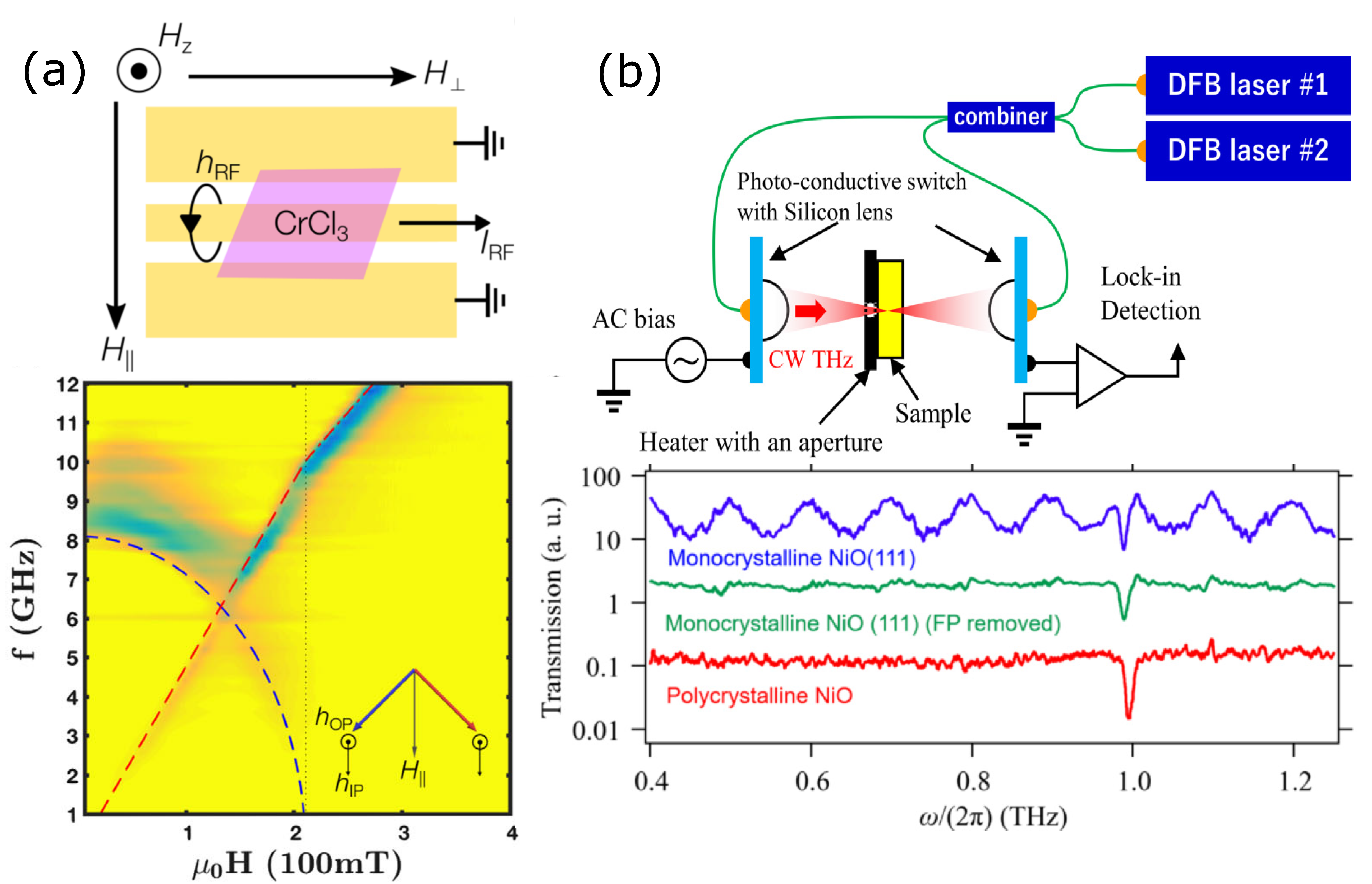}
\caption{\label{figAFMR}
(a) CrCl$_3$ platelets are fixed onto a co-planar wave guide and absorption of microwaves in the wave guide, at a fixed frequency, is measured as a function of external field.  Both an optical and acoustic AFMR mode are observed at less than 10 GHz.  (b) An experimental set-up for measuring AFMR in the frequency domain up to 1.2 THz is illustrated.  The transmission of THz radiation passing through the NiO sample is measured as a function of frequency, and an AFMR is identified near 1.0 THz.  Reprinted figure with permission from Ref.~\onlinecite{macneill2019gigahertz} and Ref.~\onlinecite{moriyama2019intrinsic}.}
\end{figure}

From the perspective of magnetization dynamics, two dimensional magnets based upon van der Waals materials\cite{gong2017discovery,huang2017layer} (see also Sec. \ref{sec:LayeredMAterials}) are quite similar to synthetic magnets.
In insulating CrI$_3$ and CrCl$_3$ individual atomic layers are ferromagnetic, but there is also a weaker antiferromagnetic interlayer coupling\cite{mcguire2017magnetic}.
Thick platelets of CrCl$_3$ have been been used to study both optical and acoustic AFMR at GHz frequencies [see Fig.~\ref{figAFMR}(a)].\cite{macneill2019gigahertz}
After the observation of GHz-AFMR in CrCl$_3$ it was reported that thinner samples near the monolayer limit have an increased interlayer exchange coupling.\cite{klein2019enhancement}
This observation may help explain very recent experiments where magnons are optically detected in CrI$_3$ with reported frequencies varying from tens of GHz\cite{zhang2020gate} to the THz regime.\cite{cenker2020direct} 
In the thick platelet limit, out-of-plane magnetic fields have been used to hybridize optical and acoustic magnon modes in CrCl$_3$.
This is appealing because parallel efforts involving ferromagnetic materials have identified magnon-photon\cite{huebl2013high, tabuchi2014hybridizing, bai2015spin, li2019strong} and magnon-magnon\cite{klingler2018spin, chen2018strong, li2019coherent} hybridized modes as being promising platforms for quantum information processing.
Antiferromagnetic materials may have unique potential in these hybrid quantum systems simply because of the separate optical and acoustic modes which can be independently targeted for hybridization with each other or with microwave photons.
 
The discovery of intrinsic N\'eel spin-orbit torques and the associated current-induced switching of memory states in antiferromagnets possessing these torques\cite{WadleyScience2016,BodnarNatComm2018} raises the question of whether current-induced switching at THz speeds is possible\cite{vzelezny2014relativistic, roy2016robust}.
In materials with N\'eel spin-orbit torques, like CuMnAs and Mn$_2$Au, the state that is switched is comprised of many magnetic domains.\cite{grzybowski2017imaging}
The switching process itself involves a redistribution of magnetic domains through domain wall motion.\cite{baldrati2019mechanism, gray2019spin}
A current induced switching process which exploits THz dynamics in an antiferromagnetic metal has thus far not been demonstrated. 
Other promising work indicates that a pulse train of THz pulses can lead to a switching process in CuMnAs similar to how switching occurs after a series of electrical current pulses.\cite{olejnik2018terahertz} 
 
Looking ahead, research into the magnetization dynamics of antiferromagnets will benefit from experimental techniques, {\em i.e.} measuring antiferromagnetic resonance in the frequency domain.
Recently, terahertz spectroscopy techniques have been used to electrically detect AFMR via spin pumping and the inverse spin Hall effect in Cr$_2$O$_3$.\cite{Li2020Nature}
In addition, broadband techniques working within the frequency domain have helped to study the origin of magnetic damping in both polycrystalline and single crystalline NiO [see Fig.~\ref{figAFMR}(b)].\cite{moriyama2019intrinsic} 
By measuring the antiferromagnetic resosnance in the frequency domain, and quantifying the linewidth of AFMR as a function of field and temperature, damping mechanisms may be partially elucidated.
Future devices, such as antiferromagnetic spin-torque oscillators, will greatly benefit from identifying materials that have low damping;
theoretical progress is being made in this area.\cite{simensen2020magnon}  
From this standpoint, FeRh becomes an intriguing material.
Magnetic damping in the ferromagnetic phase of FeRh has been reported,\cite{mancini2013magnetic} and it is a relatively low damping material similar to permalloy.
It remains to be seen if the low damping observed in the ferromagnetic phase has any implications for damping within the antiferromagnetic phase, and this would appear to be an interesting direction to pursue.

\section{\label{sec:Optical}Optical and magneto-optical properties}

\begin{figure}[htbp]
\includegraphics[width=0.9\columnwidth]{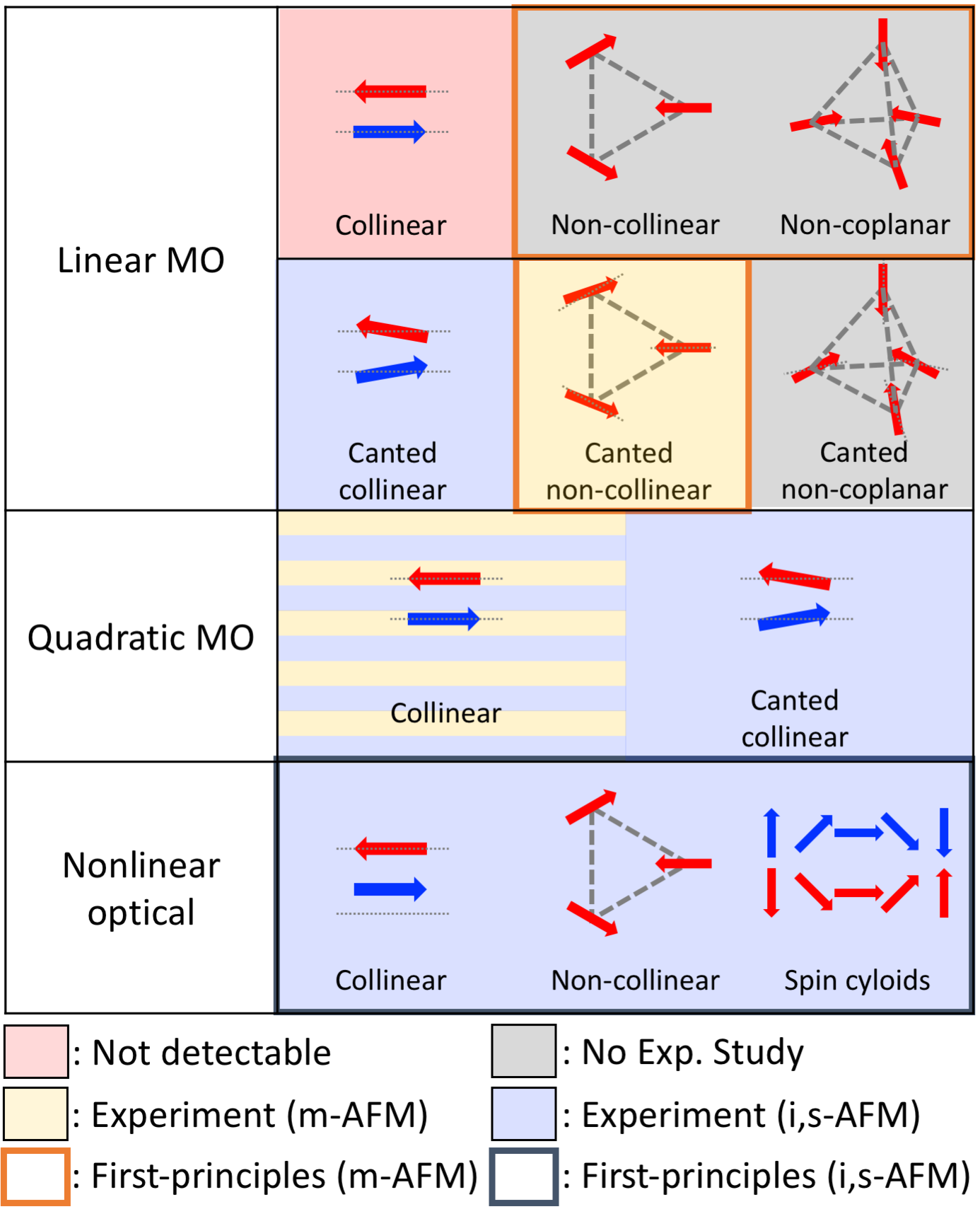}
\caption{\label{figure1}
Schematic categorization of optical and magneto-optical effects in antiferromagnets (AFMs).
Each color label represents a magnetic ordering, studied by different optical or magneto-optical methods.
i,s,m-AFM correspond to insulating, semiconducting, and metallic antiferromagnets, respectively.
}
\end{figure}

Interaction of materials with light provides rich information both statically, and also dynamically with femto-second time-resolution.
For magnetically ordered materials, this includes visualization of details of the magnetic structure and magnetic domains, {\em i.e.}, fundamental material properties that are also essential for applications.
For the specific example of antiferromagnetic materials, the N\'eel vector orientation of domains can be determined optically, which turns out difficult to probe otherwise due to the lack of net magnetization \cite{Nemec2018,Cheong2020}.
For device applications, magneto-optical effects are discussed in the literature, {\em e.g.}, for reading and writing information through manipulation of the magnetic ordering.

This constitutes a challenge especially for antiferromagnetic \emph{metals}:
While neutron diffraction \cite{Wadley2015, BarthemNatComm2013} and synchrotron X-ray techniques \cite{Wadley2017, Sapozhnik2018} analyze the magnetic structure with high resolution, such measurements require large-scale experimental facilities.
Transmission electron diffraction through Lorentz microscopy was successfully implemented for antiferromagnetic NiO \cite{Loudon2012}, but suffers from the same problem, compared to much more easily accessible optical or magneto-optical probes.
We now discuss experimental and theoretical results for accessing fundamental optical and magneto-optical effects in metallic antiferromagnets \cite{McCord2015}, many of which rely on relatively simple, laboratory-scale experimental setups \cite{Oppeneer2017,Saidl2017}.
Our discussion will be divided into linear optical and non-linear optical effects (see Fig.\ \ref{figure1}), where the former refer to optical processes that merely affect light polarization, but do not change the frequency of the light, and the latter allow for such frequency changes \cite{Cheong2020}.

In particular, linear (see Sec.\ \ref{sec:linmo}) and quadratic magneto-optical effects (see Sec.\ \ref{sec:quadmo}), both of which are linear optical, are suitable for reading magnetic configurations of different domains.
Nonlinear optical effects (see Sec.\ \ref{sec:nonlinear}), such as second-harmonic generation, require strong electromagnetic fields but can provide direct information of the antiferromagnetic order \cite{Cheong2020}.
Manipulation of the magnetic order on time scales of about 100 fs has been achieved by excitation with short laser pulses (see Sec.\ \ref{sec:dynamic}).
In addition, throughout we will point out magneto-optical effects that so far were only observed in semiconducting or insulating antiferromagnets, but that also have high potential for yielding important insight into metallic antiferromagnets.
For a more detailed introduction and more comprehensive overview we refer to the excellent reviews in Refs.~\onlinecite{Nemec2018, Cheong2020}.

\subsection{\label{sec:linmo}Linear magneto-optical effects}

Magneto-optical effects describe the change of polarization of light upon interaction with the magnetic configuration of a material.
Depending on the magnetic symmetry of the specific material, linear and quadratic magneto-optical effects can occur (see Fig.~\ref{figure1}).
In particular, for two-sublattice collinear antiferromagnets the N{\'e}el vector is a good magnetic order parameter and magneto-optical effects can be expressed through an expansion of the complex, frequency-dependent dielectric tensor \cite{tzschaschel2017ultrafast,Iida2011,Eremenko2012,Yang2019}
\begin{equation}
\label{eq:exp}
\epsilon_{ij}=\epsilon_{ij}^{(0)}+K_{ijk}M_{k}+G_{ijkl}^{MM}M_{k}M_{l}+G_{ijkl}^{LL}L_{k}L_{l}+G_{ijkl}^{ML}M_{k}L_{l}+\dots
\end{equation}
Here $M$ is the net magnetization ($M=M_1+M_2$), $L$ is the N\'eel vector ($L=M_1-M_2$), $\epsilon_{ij}^{(0)}$ is the magnetization-independent dielectric tensor, $K_{ijk}$ is the linear magneto-optic tensor, and $G_{ijkl}$ is the quadratic magneto-optic tensor.

Linear magneto-optical effects, as the first-order term ($K_{ijk}M_k$) in the expansion in Eq.~\eqref{eq:exp}, are proportional to the net magnetization $M$.
Examples for these include the magneto-optical Kerr effect (MOKE)\cite{John1877} that is measured in reflected light and the Faraday effect\cite{Faraday1855}, measured in transmitted light.
Due to the zero net magnetization of collinear antiferromagnets the off-diagonal components of the dielectric tensor vanish, precluding observation of linear magneto-optical effects in these materials.

Contrary to ferromagnets, antiferromagnetic materials also comprise of systems with non-collinear or non-coplanar magnetic ordering (see Fig.~\ref{figure1}), for which the N{\'e}el vector cannot be defined and Eq.~\eqref{eq:exp} is not applicable.
For these, symmetry analysis and first-principles simulations recently lead to the prediction of anomalous Hall conductivity \cite{MacDonald2014} and magneto-optical Kerr effect (MOKE) \cite{Feng2015}.
These seminal works illustrate that magneto-optical effects are not simply linked to the net magnetization, but instead to the underlying magnetic and crystalline symmetries as represented in the off-diagonal elements of the dielectric tensor $\epsilon_{ij}$ \cite{Oppeneer2001}.
This insight, along with potential applications for visualizing antiferromagnetic order, triggered large interest in magneto-optical effects also for antiferromagnetic metals and in particular, materials with (i) non-collinear/non-coplanar orderings and (ii) canted collinear orderings, or combinations thereof (see Fig.\ \ref{figure1}).

In 2015, Feng \emph{et al.}\ were the first to conclude from their first-principles simulations that three non-spinpolarized, non-magnetic metals Mn$_3$X (with X=Rh, Ir, Pt) show large MOKE \cite{Feng2015}.
They attributed this to strong spin-orbit interaction and degeneracy-breaking band splitting, arising from non-collinear antiferromagnetic ordering.
The first experimental observation of MOKE in an antiferromagnetic metal was reported shortly after for Mn$_3$Sn, which shows large zero-field Kerr rotation, comparable in its magnitude to ferromagnets\cite{Higo2018}.
Mn$_3$Sn also shows non-collinear ordering, with an inverse triangular spin structure and uniform negative vector chirality of the in-plane Mn magnetic moments.
While the authors also note that the magnetic moments are slightly canted, causing a small net ferromagnetic moment, they discuss that the large Hall resistivity and field-dependent MOKE measurements indicate that this ferromagnetic moment is not responsible for the large MOKE signal they observed\cite{Higo2018}.

This claim is supported by symmetry analyses and cluster multipole moments, that are suggested as an order parameter to measure symmetry breaking for commensurate non-collinear magnetic order \cite{Suzuki2017}.
Cluster multipole moments work similar to ferromagnets and can generate large linear MOKE and anomalous Hall effect.
Higo \emph{et al.}\ specifically invoke magnetic octupole domains \cite{Higo2018} in their work to explain their large MOKE signals.
In addition, Ref.\ \onlinecite{Higo2018} reports first-principles calculations that discuss the fully \emph{compensated} antiferromagnetic state of Mn$_3$Sn, also confirming large MOKE signals in the absence of any ferromagnetic contributions.

More recently, other works followed up on these results and investigated MOKE in non-collinear as well as non-coplanar antiferromagnetic metals:
Wimmer \emph{et al.}\ \cite{Wimmer2019} use symmetry arguments to discuss magneto-optical phenomena and non-zero off-diagonal elements of the frequency-dependent conductivity tensor of coplanar, non-collinear Mn$_3$Ir and Mn$_3$Ge.
Zhou \emph{et al.}\ investigate different non-collinear antiferromagnetic orderings of Mn$_3$XN (X=Ga, Zn, Ag, Ni) and report strong MOKE signals as well as their dependence on the specific magnetic ordering.
In addition to these coplanar antiferromagnetic metals, Feng \emph{et al.}\ recently identified compensated non-coplanar orderings as candidates for strong MOKE signals and illustrate this using first-principles results for Kerr rotation angles of $\gamma$-Fe$_{0.5}$Mn$_{0.5}$ \cite{Feng2020}.

Most of the above results focus on polar MOKE, i.e., MOKE for surfaces perpendicular to the direction characterizing magnetic ordering, e.g.\ that of weak magnetization.
In addition, Higo \emph{et al.}\ also reported longitudinal MOKE for Mn$_3$Sn, where weak ferromagnetism lies within the surface plane \cite{Higo2018}.
Also Balk \emph{et al.}\ recently measured longitudinal MOKE for non-collinear antiferromagnetic Mn$_3$Sn with an extremely small in-plane magnetic moment  \cite{Balk2019}.
They studied three different antiferromagnetic orderings by increasing the temperature above the N{\'e}el temperature of about 420 K.
Their results further point to a difference of surface and bulk magnetism that requires a more detailed investigation.

Finally, spontaneously canted \emph{collinear} antiferromagnets with a weak ferromagnetic contribution show linear magneto-optical effects.
Such spontaneous canting of magnetic moments is typically on the order of $1^{\circ}$ and can arise, for instance, due to the Dzyaloshinskii-Moriya interaction \cite{Dzyaloshinsky1958, moriya1960, Nemec2018}.
This was observed early on for the spontaneously canted antiferromagnetic insulator $\alpha$-Fe$_2$O$_3$ using the Faraday effect \cite{Williams1958}.
While orthoferrites are also not metallic, they show canting and are another example that illustrates observation of Faraday rotation without any external stimulation \cite{Tabor1970,Schmool1999}.
Providing direct evidence of magneto-optical effects also in canted collinear antiferromagnetic metals, and understanding its magnitude quantitatively, is a promising but outstanding goal.

\begin{figure}[htbp]
\includegraphics[width=0.99\columnwidth]{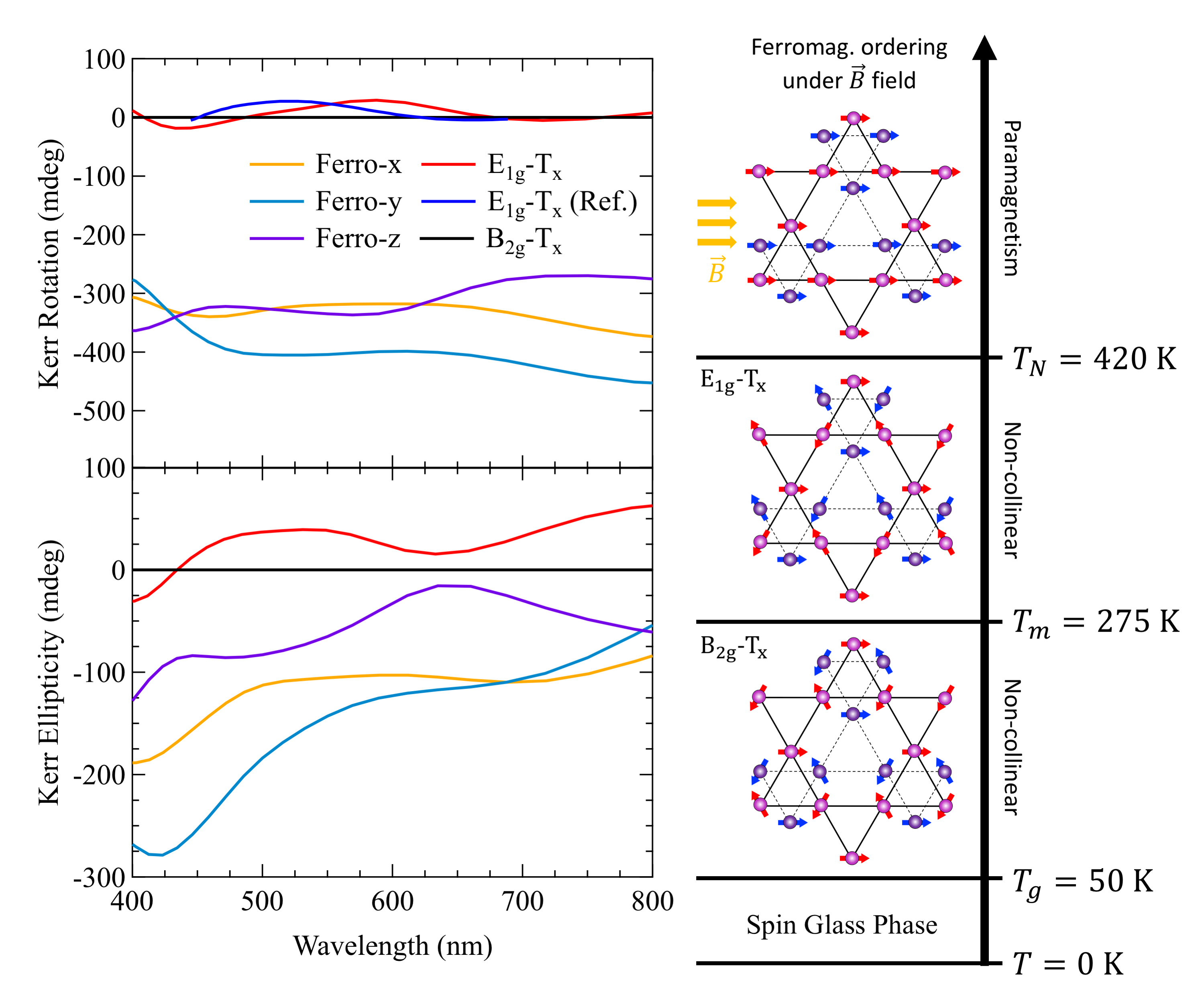}
\caption{\label{fig:mn3sn-2}
Detecting magnetic phase transitions through switching of magneto-optical effects for Mn$_3$Sn.
Left: Polar magneto-optical Kerr effect signals for different magnetic phases.
The blue curve shows a computational result by Higo \emph{et al.}\cite{Higo2018}
Right: Magnetic phase transitions and corresponding temperatures.
Above the N\'eel temperature, Mn$_3$Sn becomes paramagnetic, but can exhibit ferromagnetic ordering under strong external magnetic fields.
}
\end{figure}

In addition to using optical and magneto-optical measurements for visualizing domains, they can also be used to identify magnetic ordering and magnetic phase transitions of a material.
Saidl \emph{et al.}\ illustrated this for the transition from antiferromagnetic to ferromagnetic FeRh using reflectivity and transmittance measurements \cite{Saidl2016v2}.
While the non-collinear antiferromagnetic metal Mn$_3$Sn has three different magnetic phase transitions (see Fig.\ \ref{fig:mn3sn-2}), their optical detection was studied near the magnetic transition temperature $T_m$ \cite{Balk2019}:
Balk \emph{et al.}\ reported that longitudinal MOKE and anomalous Hall effect are almost zero below $T_m$, while finite signals arise when the material is heated right above $T_m$ \cite{Balk2019}.
Motivated by earlier studies, in which we showed that the combination of accurate first-principles simulations and experiments can successfully identify signatures in optical spectra that correlate directly with \emph{crystal} structure \cite{Lim2017, Zhang2018}, we aimed to explore this also for the \emph{magnetic} structure, e.g.\ of Mn$_3$Sn.
To this end, we performed first-principles density functional theory calculations within the Vienna \textit{Ab Initio} Simulation Package (VASP) \cite{Kresse:1996,Kresse:1999}, using the projector-augmented wave method to describe electron-ion interaction \cite{Blochl:1994}.
Kohn-Sham states were expanded into plane waves with a cutoff energy of 600 eV.
Relaxed atomic geometries and electronic and optical properties \cite{Gajdos:2006} were computed using a $13\times13\times13$ Monkhorst-Pack \cite{Monkhorst:1976} $\mathbf{k}$-point grid to sample the Brillouin zone.
Exchange and correlation were described using the generalized-gradient approximation by Perdew, Burke, and Ernzerhof \cite{Perdew:1997}.

From our first-principles calculations of the frequency-dependent complex dielectric tensor across the visible spectral range we find similar results for polar MOKE around $T_m$ as discussed from experiment \cite{Balk2019}, allowing the distinction of the $B_{2g}$ and the $E_{1g}$ phases (see Fig.\ \ref{fig:mn3sn-2}).
We also note that our result for $E_{1g}$ agrees to within 0.2 eV with that of Higo \emph{et al.}\cite{Higo2018}
We computed polar MOKE for a total of seven non-collinear antiferromagnetic configurations ($E_{1g}$-$T_x$, $E_{1g}$-$T_y$, $B_{2g}$-$T_x$, B$_{1g}$-$T_y$, $A_{2g}$-$T_z$, $E_{2g}$-$T_{xyz}$, and $E_{2g}$-$T_z$) and three collinear antiferromagnetic configurations (Anti-x, Anti-y, Anti-z).
Of these, only $E_{1g}$-$T_x$ and $E_{1g}$-$T_y$ show non-vanishing MOKE signals that are very similar in their magnitude, but exhibit different directionality:
$E_{1g}$-$T_x$ and $E_{1g}$-$T_y$ can be detected from different surface orientations, i.e., along the $x$ axis and $y$ axis, respectively.
Further increasing the temperature from $T_m$ eventually turns Mn$_3$Sn paramagnetic above the N\'eel temperature $T_N$ (see Fig.\ \ref{fig:mn3sn-2}).
No polar MOKE signal is expected for paramagnetic Mn$_3$Sn, indicating that both magnetic phase transitions can be distinguished by magneto-optical detection.
However, magnetic moments of the paramagnetic state can align in the presence of a strong enough external field, leading to a ferromagnetic configuration.
For three different orientations of this phase our results show sizable polar MOKE with different spectral behavior, enabling optical distinction of these orientations.

\subsection{\label{sec:quadmo}Quadratic magneto-optical effects}

While linear magneto-optical effects occur only in antiferromagnets with certain magnetic orderings, most antiferromagnetic metals exhibit collinear ordering.
Even though quadratic magneto-optical effects are typically weaker than their linear counterparts, they enable studying such collinear antiferromagnets for which linear magneto-optical properties vanish.
For collinear ordering with a N{\'e}el vector larger than the magnetization the second-order term ($G_{ijkl}^{LL}L_{k}L_{l}$) of the expansion in Eq.~\eqref{eq:exp} dominates and is proportional to the square of the N\'eel vector.

Quadratic magneto-optical effects include magnetic linear dichroism \cite{Pisarev1972, Kharchenko2005} and magnetic linear birefringence of reflected \cite{Ferre1984, Yang2019} (also called quadratic MOKE or Hubert-Sch{\"a}fer effect) or transmitted light \cite{Saidl2017} (also called Voigt effect or Cotton-Mouton effect).
Magnetic linear birefringence arises from a contribution to the dielectric tensor that is separate from crystal-structure driven terms.
It can be measured indirectly through large changes of the birefringence near the N{\'e}el or Curie temperature of magnetic phase transitions.
In this case, changes of the birefringence are typically attributed exclusively to magnetic contributions.
The magnetic contribution can also be measured directly, {\em e.g.}, in cubic systems with vanishing structure-driven birefringence, in which magnetism reduces the symmetry from cubic to uniaxial, leading to magnetic birefringence \cite{Silber2019}.

In the context of collinear antiferrogmagnetic metals, Saidl \emph{et al.}\ used the Voigt effect to \emph{optically} determine the orientation of the N{\'e}el vector in thin films of CuMnAs on a GaP substrate \cite{Saidl2017}.
They used a pump-probe setup to accomplish separating the small polarization rotation due to the Voigt effect from all other changes of polarization in experiment, such as strain.
Interestingly, measurements with a laser pump-probe system also allow studying the connection of the temperature dependence of the magnetic heat capacity and magnetic linear birefringence \cite{Ferre1984}:
For metallic antiferromagnetic Fe$_2$As it was shown that this connection is mediated by the exchange interaction \cite{Yang2019}.
We also note that magnetic linear birefringence was studied in the \emph{canted} collinear antiferromagnets DyFeO$_3$ \cite{Gnatchenko1989} and $\alpha$-FeO$_3$\cite{LeGall1976}, however, these two materials are not metallic.

Finally, Pisarev \emph{et al.}\ reported linear dichroism for antiferromagnetic KNiF$_3$, attributed it to a purely magneto-optical origin, and disentangled this contribution from strain effects \cite{Pisarev1972}.
Kharchenko \emph{et al.}\ report observation of magnetic linear dichroism in MnF$_2$ with the magnitude of the effect being large enough to visually observe antiferromagnetic domains in the material \cite{Kharchenko2005}.
While both of these materials are non-metallic, the potential of linear magnetic dichroism for visualizing antiferromagnetic domains renders this effect of interest also for metallic antiferromagnets.

\subsection{\label{sec:nonlinear}Non-linear optical effects}

While the previous two sections discussed linear-optical effects, {\em i.e.}, processes that do not change the frequency of the light, also non-linear optical processes couple to magnetic properties and magnetic ordering of materials.
Due to their low efficiency, these require high electromagnetic field strengths, making their experimental realization more involved and, thus, less common.
To the best of our knowledge, non-linear optical effects were not studied in metallic antiferromagnets so far;
instead, we now highlight important examples of semiconductors and insulators to illustrate the potential of non-linear optics for antiferromagnetic metals.
Nonlinear \emph{magneto-optical} effects have also been discussed, especially for ferromagnetic materials, in the literature \cite{Dahn1996, Zvezdin1999}.

The most common non-linear optical technique in the present context is second-harmonic generation.
Similar to the above discussion of magnetic linear birefringence, there are also crystal-structure and magnetic-structure driven contributions to second-harmonic generation \cite{Fiebig2005}, that manifest themselves in the nonlinear susceptibility tensor of a given material.
It is reported that second-harmonic generation is particularly well-suited for studying magnetic ordering with broken inversion symmetry \cite{Cheong2020} and it is sensitive to the direction of the N{\'e}el vector or net magnetization.
This can be measured using the difference between left and right circularly polarized light, nonlinear rotation and ellipticity of linearly polarized light, or studying the temperature dependence of the second-harmonic signal near the N{\'e}el temperature \cite{Fiebig1994}.
Furthermore, second-harmonic spectra can distinguish $180^{\circ}$ N{\'e}el vector domains, which cannot be achieved using linear-optical methods \cite{Fiebig2005}.
The effect was used to distinguish the sign change under the time-reversal operation, which allows to investigate different domains, {\em e.g.}, in antiferromagnetic Cr$_2$O$_3$ \cite{Fiebig1994}.
Theoretical predictions exist that second-harmonic generation can be used to probe antiferromagnetism at surfaces and in thin films of NiO \cite{Dahn1996,Trzeciecki1999} and experimental results were presented for the model systems CoO and NiO \cite{Fiebig2005}.
Second-harmonic generation was also studied from first principles for NiO \cite{Satitkovitchai2003}.
Finally, this effect was used to study non-collinear antiferromagnets such as $R$MnO$_3$ ($R$ = Sc, Y, In, Ho, Er, Tm, Yb, Lu) \cite{Frohlich1998, Fiebig2000, Degenhardt2001, Fiebig2002, Fiebig2004, Fiebig2005}.
For $R$MnO$_3$, Fiebig \emph{et al.}\ report that they can distinguish the different magnetic phases corresponding to different non-collinear antiferromagnetic configurations \cite{Fiebig2005}. 
Manz \emph{et al.}\ identified antiferromagnetic spin cycloids in TbMnO$_3$ by the helicity of the structure \cite{Manz2016}.

Higher order non-linear optical effects are even more rare, however, one example is the use of the inverse Faraday effect, as a third-order nonlinear optical effect, to induce a magnetization in antiferromagnetic NiO that was subsequently probed by means of the Faraday effect \cite{Satoh2010}.

The above examples represent studies of insulating or semiconducting systems, {\em i.e.}, materials with a spectral region of optical transparency.
While a comprehensive discussion of the experimental feasibility of non-linear optical and magneto-optical effects in antiferromagnetic \emph{metals} is beyond the scope of this paper, we note that second-harmonic generation has been accomplished in ferromagnetic metals \cite{Bennemann1998} and third-order nonlinear optics was studied for metallic thin films \cite{Liao1998} and, thus, we envision that it can also be a powerful tool to study antiferromagnetic metals.

\subsection{\label{sec:dynamic}Laser-induced dynamics}

\begin{figure}[htbp]
\includegraphics[width=0.9\columnwidth]{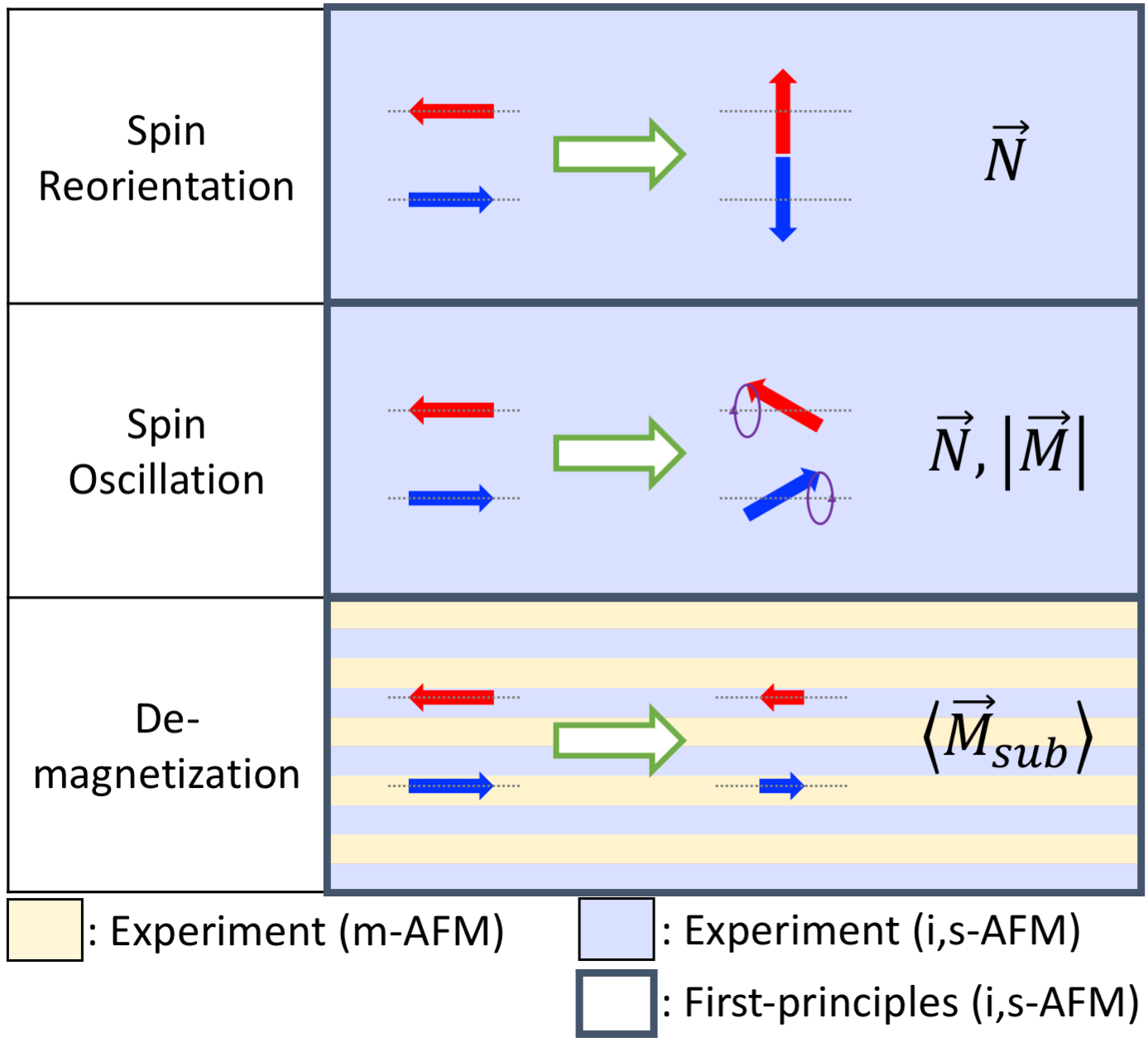}
\caption{\label{figure3}
Schematic categorization of laser-induced dynamics in antiferromagnets (AFMs).
Color labels represent magnetic structures studied with different magneto-optical methods.
i,s,m-AFM correspond to insulating, semiconducting, and metallic antiferromagnets, respectively.
}
\end{figure}

One exciting potential application of antiferromagnetic materials in general and metals in particular, is information storage, since antiferromagnets are expected to show orders of magnitude faster spin dynamics, compared to ferromagnets \cite{Kimel2004}.
Reading and writing of information is a prerequisite for such applications and has, for instance, been achieved in non-collinear, non-metallic antiferromagnets \cite{Satoh2015}.
Electrical switching of metallic antiferromagnets has indeed been reported, {\em e.g.}, in CuMnAs \cite{WadleyScience2016} and Mn$_2$Au \cite{MeinertPRAppl2018}.
In addition, magneto-optical effects are effective in reading magnetic information from metallic antiferromagnets, and they may also be utilized to write magnetic information optically by reorienting spins.
This triggered interest in the question of whether ultrafast switching can be achieved \emph{optically} in metallic antiferromagnets \cite{Cheong2020}.
So far, this question was investigated experimentally only for semiconducting or insulating antiferromagnets, and computionally for metallic antiferromagnets (see Fig.~\ref{figure3}).
The prospect of applications and the fundamental interest in ultrafast phenomena \cite{Nemec2018,Cheong2020} are the reason why laser-induced dynamics remains an interesting, rapidly evolving research direction and below we provide a current overview.
A review of laser-induced phenomena can be found in Ref.~\onlinecite{Kirilyuk2010}.

Laser-induced dynamical phenomena include demagnetization \cite{Kimel2002, Trzeciecki2001, Lefkidis2007, Gomez-Abal2004}, N\'eel vector reorientation \cite{Kimel2004, Kimel2009}, and spin oscillations \cite{Satoh2010v2,tzschaschel2017ultrafast, Kalashnikova2007,Iida2011,Kimel2006}.
Most of these optical techniques to manipulate magnetic order were applied to insulating and semiconducting antiferromagnets (see Fig.~\ref{figure3}) and below we briefly discuss key insights, since these are promising research directions also for metallic antiferromagnets.
To the best of our knowledge, only optical-pump induced demagnetization was realized experimentally for a metallic antiferromagnet \cite{Yang2019}:
Fe$_2$As was studied using a pump-probe technique and the observed change of the magnetic birefringence signal near the N{\'e}el temperature was attributed to laser-induced demagnetization \cite{Yang2019}.

In 2001, Trzeciecki \emph{et al.}\ developed a theoretical description of ultrafast spin dynamics in antiferromagnetic NiO and reported femto-second time scales for dephasing-rephasing dynamics \cite{Trzeciecki2001}.
Shortly thereafter, laser-induced demagnetization was shown experimentally via the optically induced phase transition from antiferromagnetic to paramagnetic FeBO$_3$ \cite{Kimel2002} and a time scale of 700 ps was reported.
The work on NiO was followed up later 
to investigate demagnetization and magneto-optical switching for bulk and the (001) surface of antiferromagnetic NiO \cite{Gomez-Abal2004,Lefkidis2007}.

Spin reorientation was triggered by Kimel \emph{et al.}\ in antiferromagnetic TmFeO$_3$ using a short, 100 fs full width at half maximum laser pulse and detected using time-resolved linear magnetic birefringence \cite{Kimel2004}.
They report a reorientation of spins by several tens of degrees within a few picoseconds and explain the underlying mechanism by an optical excitation that subsequently causes a change of the magnetocrystalline anisotropy axis via electron-phonon and phonon-phonon coupling.
Their measurement also relies on an optical approach, making the entire process of influencing and detecting the spin orientation an all-optical technique.

More recently, Kimel \emph{et al.}\ used a magnetic field pulse, generated from a 100 fs circularly polarized optical pump pulse by means of the inverse Faraday effect, to demonstrate an inertia-mediated spin switching mechanism in antiferromagnetic HoFeO$_3$ \cite{Kimel2009}.
They report switching between the $\Gamma_{12}$ magnetic state, where the N{\'e}el vector is in the $zy$ plane, and the $\Gamma_{24}$ state with the N{\'e}el vector in the $xz$ plane.
This spin dynamics was monitored via the Faraday effect in the probe pulse.
The inertia-driven mechanism allows spin switching with extremely short laser pulses, since it circumvents the use of very strong fields that are typically needed for ultrashort pulses, but that are detrimental because they destroy the magnetic order.
It also decouples addressing a given bit from actually switching it, which has large potential for writing large amounts of data.

Optically induced spin and N{\'e}el vector oscillations in antiferromagnets were reported early on using the antiferromagnetic resonance \cite{Keffer1961, Brunner1970} or magnon generation \cite{Jongerden1990}.
The first report of optically induced coherent spin oscillations leading to a net magnetization is for MnF$_2$ \cite{Holzrichter1971}.
Later, building on a study by Satoh \emph{et al.}\cite{Satoh2010v2}, Tzschaschel \emph{et al.}\ use optical pumping (90 fs, 0.98 eV) and probing (50 fs, 1.55 eV) with linearly and circularly polarized light to study the excitation of two optical magnon modes in antiferromagnetic NiO \cite{tzschaschel2017ultrafast}.
They disentangle in-plane and out-of-plane magnon modes by observing the Faraday effect and magnetic linear birefringence, in agreement with their theoretical predictions, including selection rules for both modes.
Essenberger \emph{et al.}\ further study the magnon dispersion of the antiferromagnetic transition-metal oxides NiO, FeO, MnO, and CoO from first principles \cite{Essenberger:2011}.

Earlier experimental studies of optically induced spin oscillations showed that linearly polarized 150 fs light pulses excite coherent spin precession in antiferromagnetic FeBO$_3$ \cite{Kalashnikova2007} and DyFeO$_3$ \cite{Iida2011}. 
These were performed in the transparent regime of the antiferromagnetic materials, which prevented heating of the sample and allowed non-thermal excitation;
this may constitute a difficulty when applied to antiferromagnetic metals.
In an earlier work on TmFeO$_3$, Kimel \emph{et al.}\ showed that thermal excitations can still excite antiferromagnetic resonances, however, at a different frequency than the resonance that is excited non-thermally \cite{Kimel2006}.

Recently, first-principles techniques are increasingly applied for simulating real-time dynamics of magnetic order.
Different techniques are available to study (de-)magnetization dynamics of ferromagnetic metals \cite{Krieger2015,Elliott2016v2,Simoni2017,Chen2019}, but also Heusler compounds \cite{Elliott2016} and nanoclusters \cite{Simoni2017}.
The time scales found in these studies generally agree with experiment.
In addition, the simulations provide valuable insight into the underlying mechanisms:
This is illustrated, for instance, by the phase diagram of all-optical spin switching in Ref.\ \onlinecite{Zhang2017}, by attributing demagnetization of Ni and Co to spin flips \cite{Shokeen2017}, and by distinguishing mechanisms for demagnetization in bulk from those at surfaces of Ni \cite{Krieger2017}.
Spin selective charge transfer between magnetic sublattices was identified as the underlying mechanism for ultrafast switching of magnetic order in Fe-Mn and Co-Mn multilayers and antiferromagnetically ordered NiO \cite{Dewhurst2018} and was also identified as an important mechanism near the Co/Cu interface of a ferromagnetic heterostructure \cite{Chen2019v2}.
This mechanism was also shown to be important in FeNi alloys \cite{Hofherr2020}.
Real-time propagation was also applied to study magnons in Fe, Co, and Ni \cite{Tancogne-Dejean2020}.

\section{\label{sec:Outlook}Outlook and Conclusions}

Over the past five years the interest in metallic antiferromagnets has significantly increased due to the realization that charge transport and magnetic spin structures can have very complex interactions. 
While these interactions often were inspired by phenomena that have already been well studied with respect to spintronics based on ferromagnets, it turns out that the different symmetries of antiferromagnetic materials enable new types of phenomena. 
In particular, antiferromagnets with non-collinear chiral or non-coplanar spin structures do not have easy corresponding systems in typical ferromagnets.
Therefore, exploring further the role of symmetry and topology will remain a very fruitful research field in the foreseeable future. 

Clearly many open questions remain regarding the interplay of charge currents and magnetic structure for metallic antiferromagnets.  
In particular, a better understanding of the correlation of magnetic structure with charge transport is required as is indicated by recent results showing that electromigration can mimic transport signatures commonly associated with magnetic structure changes. 
Related to this is the challenge to identify materials where smaller current densities may result in sufficient spin-torques to manipulate antiferromagnetic spin order. 
This will help to identify and overcome thermal artifacts.
Another challenge is to obtain a clear understanding of the magnetization dynamics, especially in non-collinear and non-coplanar structures. 
Furthermore, an important open question is whether {\em dc} currents can efficiently manipulate the dynamics in antiferromagnets. 
Can we electrically change the damping in antiferromagnets to the point where they spontaneously start to oscillate? 
If so, then this may provide completely new perspectives for THz devices and technologies. 
Beyond the connection with THz radiation, it will also be important to better understand the interaction of antiferromagnetic spin structures with optical photons, which ultimately may enable more readily characterizations of the domain structures. 
Thus progress with new optical experimental approaches may be crucial for understanding the microscopic physics. 
Lastly, in terms of exploring material systems, the investigation of metallic antiferromagnets is really just in its infancy.
The research community is just starting to explore large areas of unusual material platforms, such as two-dimensional layered systems and topological semimetals.
Therefore, one can expect many new interesting phenomena to emerge, which will enrich our fundamental understanding of antiferromagnets, and also provide new technological solutions that are both robust and energy efficient.  

\begin{acknowledgments}
The preparation of this manuscript was primarily supported by the NSF through the University of Illinois at Urbana-Champaign Materials Research Science and Engineering Center DMR-1720633 and was carried out in part in the Materials Research Laboratory Central Research Facilities, University of Illinois.
This work also made use of the Illinois Campus Cluster, a computing resource that is operated by the Illinois Campus Cluster Program (ICCP) in conjunction with the National Center for Supercomputing Applications (NCSA) and which is supported by funds from the University of Illinois at Urbana-Champaign.
The authors thank Eric Huang for his help preparing the manuscript.  
\end{acknowledgments}

The data that support the findings of this study are available from the corresponding author upon reasonable request.

\nocite{*}
\bibliography{main}

\begin{thebibliography}{261}%
\makeatletter
\providecommand \@ifxundefined [1]{%
 \@ifx{#1\undefined}
}%
\providecommand \@ifnum [1]{%
 \ifnum #1\expandafter \@firstoftwo
 \else \expandafter \@secondoftwo
 \fi
}%
\providecommand \@ifx [1]{%
 \ifx #1\expandafter \@firstoftwo
 \else \expandafter \@secondoftwo
 \fi
}%
\providecommand \natexlab [1]{#1}%
\providecommand \enquote  [1]{``#1''}%
\providecommand \bibnamefont  [1]{#1}%
\providecommand \bibfnamefont [1]{#1}%
\providecommand \citenamefont [1]{#1}%
\providecommand \href@noop [0]{\@secondoftwo}%
\providecommand \href [0]{\begingroup \@sanitize@url \@href}%
\providecommand \@href[1]{\@@startlink{#1}\@@href}%
\providecommand \@@href[1]{\endgroup#1\@@endlink}%
\providecommand \@sanitize@url [0]{\catcode `\\12\catcode `\$12\catcode
  `\&12\catcode `\#12\catcode `\^12\catcode `\_12\catcode `\%12\relax}%
\providecommand \@@startlink[1]{}%
\providecommand \@@endlink[0]{}%
\providecommand \url  [0]{\begingroup\@sanitize@url \@url }%
\providecommand \@url [1]{\endgroup\@href {#1}{\urlprefix }}%
\providecommand \urlprefix  [0]{URL }%
\providecommand \Eprint [0]{\href }%
\providecommand \doibase [0]{http://dx.doi.org/}%
\providecommand \selectlanguage [0]{\@gobble}%
\providecommand \bibinfo  [0]{\@secondoftwo}%
\providecommand \bibfield  [0]{\@secondoftwo}%
\providecommand \translation [1]{[#1]}%
\providecommand \BibitemOpen [0]{}%
\providecommand \bibitemStop [0]{}%
\providecommand \bibitemNoStop [0]{.\EOS\space}%
\providecommand \EOS [0]{\spacefactor3000\relax}%
\providecommand \BibitemShut  [1]{\csname bibitem#1\endcsname}%
\let\auto@bib@innerbib\@empty
\bibitem [{\citenamefont {Meena}\ \emph {et~al.}(2014)\citenamefont {Meena},
  \citenamefont {Sze}, \citenamefont {Chand},\ and\ \citenamefont
  {Tseng}}]{Meena2014}%
  \BibitemOpen
  \bibfield  {author} {\bibinfo {author} {\bibfnamefont {J.~S.}\ \bibnamefont
  {Meena}}, \bibinfo {author} {\bibfnamefont {S.~M.}\ \bibnamefont {Sze}},
  \bibinfo {author} {\bibfnamefont {U.}~\bibnamefont {Chand}}, \ and\ \bibinfo
  {author} {\bibfnamefont {T.-Y.}\ \bibnamefont {Tseng}},\ }\href {\doibase
  10.1186/1556-276X-9-526} {\bibfield  {journal} {\bibinfo  {journal}
  {Nanoscale Res. Lett.}\ }\textbf {\bibinfo {volume} {9}},\ \bibinfo {pages}
  {526} (\bibinfo {year} {2014})}\BibitemShut {NoStop}%
\bibitem [{\citenamefont {Bossini}\ \emph {et~al.}(2016)\citenamefont
  {Bossini}, \citenamefont {Conte}, \citenamefont {Hashimoto}, \citenamefont
  {Secchi}, \citenamefont {Pisarev}, \citenamefont {Rasing}, \citenamefont
  {Cerullo},\ and\ \citenamefont {Kimel}}]{Bossini2016}%
  \BibitemOpen
  \bibfield  {author} {\bibinfo {author} {\bibfnamefont {D.}~\bibnamefont
  {Bossini}}, \bibinfo {author} {\bibfnamefont {S.~D.}\ \bibnamefont {Conte}},
  \bibinfo {author} {\bibfnamefont {Y.}~\bibnamefont {Hashimoto}}, \bibinfo
  {author} {\bibfnamefont {A.}~\bibnamefont {Secchi}}, \bibinfo {author}
  {\bibfnamefont {R.~V.}\ \bibnamefont {Pisarev}}, \bibinfo {author}
  {\bibfnamefont {T.}~\bibnamefont {Rasing}}, \bibinfo {author} {\bibfnamefont
  {G.}~\bibnamefont {Cerullo}}, \ and\ \bibinfo {author} {\bibfnamefont
  {A.~V.}\ \bibnamefont {Kimel}},\ }\href {\doibase 10.1038/ncomms10645}
  {\bibfield  {journal} {\bibinfo  {journal} {Nat. Comm.}\ }\textbf {\bibinfo
  {volume} {7}},\ \bibinfo {pages} {10645} (\bibinfo {year}
  {2016})}\BibitemShut {NoStop}%
\bibitem [{\citenamefont {Gomonay}\ \emph
  {et~al.}(2018{\natexlab{a}})\citenamefont {Gomonay}, \citenamefont {Baltz},
  \citenamefont {Brataas},\ and\ \citenamefont
  {Tserkovnyak}}]{GomonayNatPhys2018}%
  \BibitemOpen
  \bibfield  {author} {\bibinfo {author} {\bibfnamefont {O.}~\bibnamefont
  {Gomonay}}, \bibinfo {author} {\bibfnamefont {V.}~\bibnamefont {Baltz}},
  \bibinfo {author} {\bibfnamefont {A.}~\bibnamefont {Brataas}}, \ and\
  \bibinfo {author} {\bibfnamefont {Y.}~\bibnamefont {Tserkovnyak}},\
  }\href@noop {} {\bibfield  {journal} {\bibinfo  {journal} {Nat. Phys.}\
  }\textbf {\bibinfo {volume} {14}},\ \bibinfo {pages} {213} (\bibinfo {year}
  {2018}{\natexlab{a}})}\BibitemShut {NoStop}%
\bibitem [{\citenamefont {Wadley}\ \emph {et~al.}(2016)\citenamefont {Wadley},
  \citenamefont {Howells}, \citenamefont {{\v Z}elezn{\'y}}, \citenamefont
  {Andrews}, \citenamefont {Hills}, \citenamefont {Campion}, \citenamefont
  {Nov{\'a}k}, \citenamefont {Olejn{\'\i}k}, \citenamefont {Maccherozzi},
  \citenamefont {Dhesi}, \citenamefont {Martin}, \citenamefont {Wagner},
  \citenamefont {Wunderlich}, \citenamefont {Freimuth}, \citenamefont
  {Mokrousov}, \citenamefont {Kune{\v s}}, \citenamefont {Chauhan},
  \citenamefont {Grzybowski}, \citenamefont {Rushforth}, \citenamefont
  {Edmonds}, \citenamefont {Gallagher},\ and\ \citenamefont
  {Jungwirth}}]{WadleyScience2016}%
  \BibitemOpen
  \bibfield  {author} {\bibinfo {author} {\bibfnamefont {P.}~\bibnamefont
  {Wadley}}, \bibinfo {author} {\bibfnamefont {B.}~\bibnamefont {Howells}},
  \bibinfo {author} {\bibfnamefont {J.}~\bibnamefont {{\v Z}elezn{\'y}}},
  \bibinfo {author} {\bibfnamefont {C.}~\bibnamefont {Andrews}}, \bibinfo
  {author} {\bibfnamefont {V.}~\bibnamefont {Hills}}, \bibinfo {author}
  {\bibfnamefont {R.~P.}\ \bibnamefont {Campion}}, \bibinfo {author}
  {\bibfnamefont {V.}~\bibnamefont {Nov{\'a}k}}, \bibinfo {author}
  {\bibfnamefont {K.}~\bibnamefont {Olejn{\'\i}k}}, \bibinfo {author}
  {\bibfnamefont {F.}~\bibnamefont {Maccherozzi}}, \bibinfo {author}
  {\bibfnamefont {S.~S.}\ \bibnamefont {Dhesi}}, \bibinfo {author}
  {\bibfnamefont {S.~Y.}\ \bibnamefont {Martin}}, \bibinfo {author}
  {\bibfnamefont {T.}~\bibnamefont {Wagner}}, \bibinfo {author} {\bibfnamefont
  {J.}~\bibnamefont {Wunderlich}}, \bibinfo {author} {\bibfnamefont
  {F.}~\bibnamefont {Freimuth}}, \bibinfo {author} {\bibfnamefont
  {Y.}~\bibnamefont {Mokrousov}}, \bibinfo {author} {\bibfnamefont
  {J.}~\bibnamefont {Kune{\v s}}}, \bibinfo {author} {\bibfnamefont {J.~S.}\
  \bibnamefont {Chauhan}}, \bibinfo {author} {\bibfnamefont {M.~J.}\
  \bibnamefont {Grzybowski}}, \bibinfo {author} {\bibfnamefont {A.~W.}\
  \bibnamefont {Rushforth}}, \bibinfo {author} {\bibfnamefont {K.~W.}\
  \bibnamefont {Edmonds}}, \bibinfo {author} {\bibfnamefont {B.~L.}\
  \bibnamefont {Gallagher}}, \ and\ \bibinfo {author} {\bibfnamefont
  {T.}~\bibnamefont {Jungwirth}},\ }\href {\doibase 10.1126/science.aab1031}
  {\bibfield  {journal} {\bibinfo  {journal} {Science}\ }\textbf {\bibinfo
  {volume} {351}},\ \bibinfo {pages} {587} (\bibinfo {year}
  {2016})}\BibitemShut {NoStop}%
\bibitem [{\citenamefont {Olejník}\ \emph {et~al.}(2017)\citenamefont
  {Olejník}, \citenamefont {Schuler}, \citenamefont {Marti}, \citenamefont
  {Novák}, \citenamefont {Kašpar}, \citenamefont {Wadley}, \citenamefont
  {Campion}, \citenamefont {Edmonds}, \citenamefont {Gallagher}, \citenamefont
  {Garces},\ and\ \citenamefont {et~al.}}]{olejnikNatComm2017}%
  \BibitemOpen
  \bibfield  {author} {\bibinfo {author} {\bibfnamefont {K.}~\bibnamefont
  {Olejník}}, \bibinfo {author} {\bibfnamefont {V.}~\bibnamefont {Schuler}},
  \bibinfo {author} {\bibfnamefont {X.}~\bibnamefont {Marti}}, \bibinfo
  {author} {\bibfnamefont {V.}~\bibnamefont {Novák}}, \bibinfo {author}
  {\bibfnamefont {Z.}~\bibnamefont {Kašpar}}, \bibinfo {author} {\bibfnamefont
  {P.}~\bibnamefont {Wadley}}, \bibinfo {author} {\bibfnamefont {R.~P.}\
  \bibnamefont {Campion}}, \bibinfo {author} {\bibfnamefont {K.~W.}\
  \bibnamefont {Edmonds}}, \bibinfo {author} {\bibfnamefont {B.~L.}\
  \bibnamefont {Gallagher}}, \bibinfo {author} {\bibfnamefont {J.}~\bibnamefont
  {Garces}}, \ and\ \bibinfo {author} {\bibnamefont {et~al.}},\ }\href
  {\doibase 10.1038/ncomms15434} {\bibfield  {journal} {\bibinfo  {journal}
  {Nat. Comm.}\ }\textbf {\bibinfo {volume} {8}},\ \bibinfo {pages} {15434}
  (\bibinfo {year} {2017})}\BibitemShut {NoStop}%
\bibitem [{\citenamefont {Baldrati}\ \emph {et~al.}(2019)\citenamefont
  {Baldrati}, \citenamefont {Gomonay}, \citenamefont {Ross}, \citenamefont
  {Filianina}, \citenamefont {Lebrun}, \citenamefont {Ramos}, \citenamefont
  {Leveille}, \citenamefont {Fuhrmann}, \citenamefont {Forrest}, \citenamefont
  {Maccherozzi} \emph {et~al.}}]{baldrati2019mechanism}%
  \BibitemOpen
  \bibfield  {author} {\bibinfo {author} {\bibfnamefont {L.}~\bibnamefont
  {Baldrati}}, \bibinfo {author} {\bibfnamefont {O.}~\bibnamefont {Gomonay}},
  \bibinfo {author} {\bibfnamefont {A.}~\bibnamefont {Ross}}, \bibinfo {author}
  {\bibfnamefont {M.}~\bibnamefont {Filianina}}, \bibinfo {author}
  {\bibfnamefont {R.}~\bibnamefont {Lebrun}}, \bibinfo {author} {\bibfnamefont
  {R.}~\bibnamefont {Ramos}}, \bibinfo {author} {\bibfnamefont
  {C.}~\bibnamefont {Leveille}}, \bibinfo {author} {\bibfnamefont
  {F.}~\bibnamefont {Fuhrmann}}, \bibinfo {author} {\bibfnamefont
  {T.}~\bibnamefont {Forrest}}, \bibinfo {author} {\bibfnamefont
  {F.}~\bibnamefont {Maccherozzi}},  \emph {et~al.},\ }\href@noop {} {\bibfield
   {journal} {\bibinfo  {journal} {Phys. Rev. Lett.}\ }\textbf {\bibinfo
  {volume} {123}},\ \bibinfo {pages} {177201} (\bibinfo {year}
  {2019})}\BibitemShut {NoStop}%
\bibitem [{\citenamefont {Kirilyuk}, \citenamefont {Kimel},\ and\ \citenamefont
  {Rasing}(2010)}]{Kirilyuk2010}%
  \BibitemOpen
  \bibfield  {author} {\bibinfo {author} {\bibfnamefont {A.}~\bibnamefont
  {Kirilyuk}}, \bibinfo {author} {\bibfnamefont {A.~V.}\ \bibnamefont {Kimel}},
  \ and\ \bibinfo {author} {\bibfnamefont {T.}~\bibnamefont {Rasing}},\ }\href
  {\doibase 10.1103/revmodphys.82.2731} {\bibfield  {journal} {\bibinfo
  {journal} {Reviews of Modern Physics}\ }\textbf {\bibinfo {volume} {82}},\
  \bibinfo {pages} {2731} (\bibinfo {year} {2010})}\BibitemShut {NoStop}%
\bibitem [{\citenamefont {Chiang}\ \emph {et~al.}(2019)\citenamefont {Chiang},
  \citenamefont {Huang}, \citenamefont {Qu}, \citenamefont {Wu},\ and\
  \citenamefont {Chien}}]{chiang2019absence}%
  \BibitemOpen
  \bibfield  {author} {\bibinfo {author} {\bibfnamefont {C.~C.}\ \bibnamefont
  {Chiang}}, \bibinfo {author} {\bibfnamefont {S.~Y.}\ \bibnamefont {Huang}},
  \bibinfo {author} {\bibfnamefont {D.}~\bibnamefont {Qu}}, \bibinfo {author}
  {\bibfnamefont {P.~H.}\ \bibnamefont {Wu}}, \ and\ \bibinfo {author}
  {\bibfnamefont {C.~L.}\ \bibnamefont {Chien}},\ }\href@noop {} {\bibfield
  {journal} {\bibinfo  {journal} {Phys. Rev. Lett.}\ }\textbf {\bibinfo
  {volume} {123}},\ \bibinfo {pages} {227203} (\bibinfo {year}
  {2019})}\BibitemShut {NoStop}%
\bibitem [{\citenamefont {Matalla-Wagner}\ \emph {et~al.}(2019)\citenamefont
  {Matalla-Wagner}, \citenamefont {Schmalhorst}, \citenamefont {Reiss},
  \citenamefont {Tamura},\ and\ \citenamefont
  {Meinert}}]{matallawagner2019resistive}%
  \BibitemOpen
  \bibfield  {author} {\bibinfo {author} {\bibfnamefont {T.}~\bibnamefont
  {Matalla-Wagner}}, \bibinfo {author} {\bibfnamefont {J.-M.}\ \bibnamefont
  {Schmalhorst}}, \bibinfo {author} {\bibfnamefont {G.}~\bibnamefont {Reiss}},
  \bibinfo {author} {\bibfnamefont {N.}~\bibnamefont {Tamura}}, \ and\ \bibinfo
  {author} {\bibfnamefont {M.}~\bibnamefont {Meinert}},\ }\href@noop {}
  {\enquote {\bibinfo {title} {Resistive contribution in electrical switching
  experiments with antiferromagnets},}\ } (\bibinfo {year} {2019}),\ \Eprint
  {http://arxiv.org/abs/1910.08576} {arXiv:1910.08576 [cond-mat.mtrl-sci]}
  \BibitemShut {NoStop}%
\bibitem [{\citenamefont {Grzybowski}\ \emph {et~al.}(2017)\citenamefont
  {Grzybowski}, \citenamefont {Wadley}, \citenamefont {Edmonds}, \citenamefont
  {Beardsley}, \citenamefont {Hills}, \citenamefont {Campion}, \citenamefont
  {Gallagher}, \citenamefont {Chauhan}, \citenamefont {Novak}, \citenamefont
  {Jungwirth} \emph {et~al.}}]{grzybowski2017imaging}%
  \BibitemOpen
  \bibfield  {author} {\bibinfo {author} {\bibfnamefont {M.~J.}\ \bibnamefont
  {Grzybowski}}, \bibinfo {author} {\bibfnamefont {P.}~\bibnamefont {Wadley}},
  \bibinfo {author} {\bibfnamefont {K.~W.}\ \bibnamefont {Edmonds}}, \bibinfo
  {author} {\bibfnamefont {R.}~\bibnamefont {Beardsley}}, \bibinfo {author}
  {\bibfnamefont {V.}~\bibnamefont {Hills}}, \bibinfo {author} {\bibfnamefont
  {R.~P.}\ \bibnamefont {Campion}}, \bibinfo {author} {\bibfnamefont {B.~L.}\
  \bibnamefont {Gallagher}}, \bibinfo {author} {\bibfnamefont {J.~S.}\
  \bibnamefont {Chauhan}}, \bibinfo {author} {\bibfnamefont {V.}~\bibnamefont
  {Novak}}, \bibinfo {author} {\bibfnamefont {T.}~\bibnamefont {Jungwirth}},
  \emph {et~al.},\ }\href@noop {} {\bibfield  {journal} {\bibinfo  {journal}
  {Phys. Rev. Lett.}\ }\textbf {\bibinfo {volume} {118}},\ \bibinfo {pages}
  {057701} (\bibinfo {year} {2017})}\BibitemShut {NoStop}%
\bibitem [{\citenamefont {Nogu\'es}\ and\ \citenamefont
  {Schuller}(1999)}]{Nogues1999JMMM}%
  \BibitemOpen
  \bibfield  {author} {\bibinfo {author} {\bibfnamefont {J.}~\bibnamefont
  {Nogu\'es}}\ and\ \bibinfo {author} {\bibfnamefont {I.~K.}\ \bibnamefont
  {Schuller}},\ }\href {\doibase 10.1016/S0304-8853(98)00266-2} {\bibfield
  {journal} {\bibinfo  {journal} {J. Magn. Magn. Mater.}\ }\textbf {\bibinfo
  {volume} {192}},\ \bibinfo {pages} {203} (\bibinfo {year}
  {1999})}\BibitemShut {NoStop}%
\bibitem [{\citenamefont {Baltz}\ \emph {et~al.}(2018)\citenamefont {Baltz},
  \citenamefont {Manchon}, \citenamefont {Tsoi}, \citenamefont {Moriyama},
  \citenamefont {Ono},\ and\ \citenamefont {Tserkovnyak}}]{BaltzRMP2018}%
  \BibitemOpen
  \bibfield  {author} {\bibinfo {author} {\bibfnamefont {V.}~\bibnamefont
  {Baltz}}, \bibinfo {author} {\bibfnamefont {A.}~\bibnamefont {Manchon}},
  \bibinfo {author} {\bibfnamefont {M.}~\bibnamefont {Tsoi}}, \bibinfo {author}
  {\bibfnamefont {T.}~\bibnamefont {Moriyama}}, \bibinfo {author}
  {\bibfnamefont {T.}~\bibnamefont {Ono}}, \ and\ \bibinfo {author}
  {\bibfnamefont {Y.}~\bibnamefont {Tserkovnyak}},\ }\href {\doibase
  10.1103/RevModPhys.90.015005} {\bibfield  {journal} {\bibinfo  {journal}
  {Rev. Mod. Phys.}\ }\textbf {\bibinfo {volume} {90}},\ \bibinfo {pages}
  {015005} (\bibinfo {year} {2018})}\BibitemShut {NoStop}%
\bibitem [{\citenamefont {McGuire}\ and\ \citenamefont
  {Potter}(1975)}]{mcguire1975anisotropic}%
  \BibitemOpen
  \bibfield  {author} {\bibinfo {author} {\bibfnamefont {T.}~\bibnamefont
  {McGuire}}\ and\ \bibinfo {author} {\bibfnamefont {R.~L.}\ \bibnamefont
  {Potter}},\ }\href@noop {} {\bibfield  {journal} {\bibinfo  {journal} {IEEE
  Trans. Magn.}\ }\textbf {\bibinfo {volume} {11}},\ \bibinfo {pages} {1018}
  (\bibinfo {year} {1975})}\BibitemShut {NoStop}%
\bibitem [{\citenamefont {Juretschke}(1960)}]{JuretschkeJAP1960}%
  \BibitemOpen
  \bibfield  {author} {\bibinfo {author} {\bibfnamefont {H.~J.}\ \bibnamefont
  {Juretschke}},\ }\href {https://doi.org/10.1063/1.1735851} {\bibfield
  {journal} {\bibinfo  {journal} {J. Appl. Phys.}\ }\textbf {\bibinfo {volume}
  {31}},\ \bibinfo {pages} {1401} (\bibinfo {year} {1960})}\BibitemShut
  {NoStop}%
\bibitem [{\citenamefont {Juretschke}(1963)}]{JuretschkeJAP1963}%
  \BibitemOpen
  \bibfield  {author} {\bibinfo {author} {\bibfnamefont {H.~J.}\ \bibnamefont
  {Juretschke}},\ }\href {\doibase 10.1063/1.1729445} {\bibfield  {journal}
  {\bibinfo  {journal} {J. Appl. Phys.}\ }\textbf {\bibinfo {volume} {34}},\
  \bibinfo {pages} {1223} (\bibinfo {year} {1963})}\BibitemShut {NoStop}%
\bibitem [{\citenamefont {Liu}\ \emph {et~al.}(2011{\natexlab{a}})\citenamefont
  {Liu}, \citenamefont {Moriyama}, \citenamefont {Ralph},\ and\ \citenamefont
  {Buhrman}}]{liu2011spin}%
  \BibitemOpen
  \bibfield  {author} {\bibinfo {author} {\bibfnamefont {L.}~\bibnamefont
  {Liu}}, \bibinfo {author} {\bibfnamefont {T.}~\bibnamefont {Moriyama}},
  \bibinfo {author} {\bibfnamefont {D.~C.}\ \bibnamefont {Ralph}}, \ and\
  \bibinfo {author} {\bibfnamefont {R.~A.}\ \bibnamefont {Buhrman}},\
  }\href@noop {} {\bibfield  {journal} {\bibinfo  {journal} {Phys. Rev. Lett.}\
  }\textbf {\bibinfo {volume} {106}},\ \bibinfo {pages} {036601} (\bibinfo
  {year} {2011}{\natexlab{a}})}\BibitemShut {NoStop}%
\bibitem [{\citenamefont {Sklenar}\ \emph {et~al.}(2017)\citenamefont
  {Sklenar}, \citenamefont {Zhang}, \citenamefont {Jungfleisch}, \citenamefont
  {Saglam}, \citenamefont {Grudichak}, \citenamefont {Jiang}, \citenamefont
  {Pearson}, \citenamefont {Ketterson},\ and\ \citenamefont
  {Hoffmann}}]{sklenar2017unidirectional}%
  \BibitemOpen
  \bibfield  {author} {\bibinfo {author} {\bibfnamefont {J.}~\bibnamefont
  {Sklenar}}, \bibinfo {author} {\bibfnamefont {W.}~\bibnamefont {Zhang}},
  \bibinfo {author} {\bibfnamefont {M.~B.}\ \bibnamefont {Jungfleisch}},
  \bibinfo {author} {\bibfnamefont {H.}~\bibnamefont {Saglam}}, \bibinfo
  {author} {\bibfnamefont {S.}~\bibnamefont {Grudichak}}, \bibinfo {author}
  {\bibfnamefont {W.}~\bibnamefont {Jiang}}, \bibinfo {author} {\bibfnamefont
  {J.~E.}\ \bibnamefont {Pearson}}, \bibinfo {author} {\bibfnamefont {J.~B.}\
  \bibnamefont {Ketterson}}, \ and\ \bibinfo {author} {\bibfnamefont
  {A.}~\bibnamefont {Hoffmann}},\ }\href@noop {} {\bibfield  {journal}
  {\bibinfo  {journal} {Phys. Rev. B}\ }\textbf {\bibinfo {volume} {95}},\
  \bibinfo {pages} {224431} (\bibinfo {year} {2017})}\BibitemShut {NoStop}%
\bibitem [{\citenamefont {MacNeill}\ \emph {et~al.}(2017)\citenamefont
  {MacNeill}, \citenamefont {Stiehl}, \citenamefont {Guimaraes}, \citenamefont
  {Buhrman}, \citenamefont {Park},\ and\ \citenamefont
  {Ralph}}]{macneill2017control}%
  \BibitemOpen
  \bibfield  {author} {\bibinfo {author} {\bibfnamefont {D.}~\bibnamefont
  {MacNeill}}, \bibinfo {author} {\bibfnamefont {G.~M.}\ \bibnamefont
  {Stiehl}}, \bibinfo {author} {\bibfnamefont {M.~H.~D.}\ \bibnamefont
  {Guimaraes}}, \bibinfo {author} {\bibfnamefont {R.~A.}\ \bibnamefont
  {Buhrman}}, \bibinfo {author} {\bibfnamefont {J.}~\bibnamefont {Park}}, \
  and\ \bibinfo {author} {\bibfnamefont {D.~C.}\ \bibnamefont {Ralph}},\
  }\href@noop {} {\bibfield  {journal} {\bibinfo  {journal} {Nat. Phys.}\
  }\textbf {\bibinfo {volume} {13}},\ \bibinfo {pages} {300} (\bibinfo {year}
  {2017})}\BibitemShut {NoStop}%
\bibitem [{\citenamefont {Nan}\ \emph {et~al.}(2019)\citenamefont {Nan},
  \citenamefont {Quintela}, \citenamefont {Irwin}, \citenamefont {Gurung},
  \citenamefont {Shao}, \citenamefont {Gibbons}, \citenamefont {Campbell},
  \citenamefont {Song}, \citenamefont {Choi}, \citenamefont {Guo} \emph
  {et~al.}}]{nan2019controlling}%
  \BibitemOpen
  \bibfield  {author} {\bibinfo {author} {\bibfnamefont {T.}~\bibnamefont
  {Nan}}, \bibinfo {author} {\bibfnamefont {C.}~\bibnamefont {Quintela}},
  \bibinfo {author} {\bibfnamefont {J.}~\bibnamefont {Irwin}}, \bibinfo
  {author} {\bibfnamefont {G.}~\bibnamefont {Gurung}}, \bibinfo {author}
  {\bibfnamefont {D.}~\bibnamefont {Shao}}, \bibinfo {author} {\bibfnamefont
  {J.}~\bibnamefont {Gibbons}}, \bibinfo {author} {\bibfnamefont
  {N.}~\bibnamefont {Campbell}}, \bibinfo {author} {\bibfnamefont
  {K.}~\bibnamefont {Song}}, \bibinfo {author} {\bibfnamefont {S.}~\bibnamefont
  {Choi}}, \bibinfo {author} {\bibfnamefont {L.}~\bibnamefont {Guo}},  \emph
  {et~al.},\ }\href@noop {} {\bibfield  {journal} {\bibinfo  {journal} {arXiv
  preprint arXiv:1912.12586}\ } (\bibinfo {year} {2019})}\BibitemShut {NoStop}%
\bibitem [{\citenamefont {Langenfeld}\ \emph {et~al.}(2016)\citenamefont
  {Langenfeld}, \citenamefont {Tshitoyan}, \citenamefont {Fang}, \citenamefont
  {Wells}, \citenamefont {Moore},\ and\ \citenamefont
  {Ferguson}}]{langenfeld2016exchange}%
  \BibitemOpen
  \bibfield  {author} {\bibinfo {author} {\bibfnamefont {S.}~\bibnamefont
  {Langenfeld}}, \bibinfo {author} {\bibfnamefont {V.}~\bibnamefont
  {Tshitoyan}}, \bibinfo {author} {\bibfnamefont {Z.}~\bibnamefont {Fang}},
  \bibinfo {author} {\bibfnamefont {A.}~\bibnamefont {Wells}}, \bibinfo
  {author} {\bibfnamefont {T.~A.}\ \bibnamefont {Moore}}, \ and\ \bibinfo
  {author} {\bibfnamefont {A.~J.}\ \bibnamefont {Ferguson}},\ }\href@noop {}
  {\bibfield  {journal} {\bibinfo  {journal} {Appl. Phys. Lett.}\ }\textbf
  {\bibinfo {volume} {108}},\ \bibinfo {pages} {192402} (\bibinfo {year}
  {2016})}\BibitemShut {NoStop}%
\bibitem [{\citenamefont {Awad}\ \emph {et~al.}(2017)\citenamefont {Awad},
  \citenamefont {D{\"u}rrenfeld}, \citenamefont {Houshang}, \citenamefont
  {Dvornik}, \citenamefont {Iacocca}, \citenamefont {Dumas},\ and\
  \citenamefont {{\AA}kerman}}]{awad2017long}%
  \BibitemOpen
  \bibfield  {author} {\bibinfo {author} {\bibfnamefont {A.~A.}\ \bibnamefont
  {Awad}}, \bibinfo {author} {\bibfnamefont {P.}~\bibnamefont
  {D{\"u}rrenfeld}}, \bibinfo {author} {\bibfnamefont {A.}~\bibnamefont
  {Houshang}}, \bibinfo {author} {\bibfnamefont {M.}~\bibnamefont {Dvornik}},
  \bibinfo {author} {\bibfnamefont {E.}~\bibnamefont {Iacocca}}, \bibinfo
  {author} {\bibfnamefont {R.~K.}\ \bibnamefont {Dumas}}, \ and\ \bibinfo
  {author} {\bibfnamefont {J.}~\bibnamefont {{\AA}kerman}},\ }\href@noop {}
  {\bibfield  {journal} {\bibinfo  {journal} {Nat. Phys.}\ }\textbf {\bibinfo
  {volume} {13}},\ \bibinfo {pages} {292} (\bibinfo {year} {2017})}\BibitemShut
  {NoStop}%
\bibitem [{\citenamefont {Safranski}, \citenamefont {Montoya},\ and\
  \citenamefont {Krivorotov}(2019)}]{safranski2019spin}%
  \BibitemOpen
  \bibfield  {author} {\bibinfo {author} {\bibfnamefont {C.}~\bibnamefont
  {Safranski}}, \bibinfo {author} {\bibfnamefont {E.~A.}\ \bibnamefont
  {Montoya}}, \ and\ \bibinfo {author} {\bibfnamefont {I.~N.}\ \bibnamefont
  {Krivorotov}},\ }\href@noop {} {\bibfield  {journal} {\bibinfo  {journal}
  {Nat. Nanotechn.}\ }\textbf {\bibinfo {volume} {14}},\ \bibinfo {pages} {27}
  (\bibinfo {year} {2019})}\BibitemShut {NoStop}%
\bibitem [{\citenamefont {Zahedinejad}\ \emph {et~al.}(2019)\citenamefont
  {Zahedinejad}, \citenamefont {Awad}, \citenamefont {Muralidhar},
  \citenamefont {Khymyn}, \citenamefont {Fulara}, \citenamefont {Mazraati},
  \citenamefont {Dvornik},\ and\ \citenamefont
  {{\AA}kerman}}]{zahedinejad2019two}%
  \BibitemOpen
  \bibfield  {author} {\bibinfo {author} {\bibfnamefont {M.}~\bibnamefont
  {Zahedinejad}}, \bibinfo {author} {\bibfnamefont {A.~A.}\ \bibnamefont
  {Awad}}, \bibinfo {author} {\bibfnamefont {S.}~\bibnamefont {Muralidhar}},
  \bibinfo {author} {\bibfnamefont {R.}~\bibnamefont {Khymyn}}, \bibinfo
  {author} {\bibfnamefont {H.}~\bibnamefont {Fulara}}, \bibinfo {author}
  {\bibfnamefont {H.}~\bibnamefont {Mazraati}}, \bibinfo {author}
  {\bibfnamefont {M.}~\bibnamefont {Dvornik}}, \ and\ \bibinfo {author}
  {\bibfnamefont {J.}~\bibnamefont {{\AA}kerman}},\ }\href@noop {} {\bibfield
  {journal} {\bibinfo  {journal} {Nat. Nanotechn.}\ ,\ \bibinfo {pages} {1}}
  (\bibinfo {year} {2019})}\BibitemShut {NoStop}%
\bibitem [{\citenamefont {Marti}\ \emph {et~al.}(2014)\citenamefont {Marti},
  \citenamefont {Fina}, \citenamefont {Frontera}, \citenamefont {Liu},
  \citenamefont {Wadley}, \citenamefont {He}, \citenamefont {Paull},
  \citenamefont {Clarkson}, \citenamefont {Kudrnovsk{\`y}}, \citenamefont
  {Turek} \emph {et~al.}}]{marti2014room}%
  \BibitemOpen
  \bibfield  {author} {\bibinfo {author} {\bibfnamefont {X.}~\bibnamefont
  {Marti}}, \bibinfo {author} {\bibfnamefont {I.}~\bibnamefont {Fina}},
  \bibinfo {author} {\bibfnamefont {C.}~\bibnamefont {Frontera}}, \bibinfo
  {author} {\bibfnamefont {J.}~\bibnamefont {Liu}}, \bibinfo {author}
  {\bibfnamefont {P.}~\bibnamefont {Wadley}}, \bibinfo {author} {\bibfnamefont
  {Q.}~\bibnamefont {He}}, \bibinfo {author} {\bibfnamefont {R.~J.}\
  \bibnamefont {Paull}}, \bibinfo {author} {\bibfnamefont {J.~D.}\ \bibnamefont
  {Clarkson}}, \bibinfo {author} {\bibfnamefont {J.}~\bibnamefont
  {Kudrnovsk{\`y}}}, \bibinfo {author} {\bibfnamefont {I.}~\bibnamefont
  {Turek}},  \emph {et~al.},\ }\href@noop {} {\bibfield  {journal} {\bibinfo
  {journal} {Nat. Mater.}\ }\textbf {\bibinfo {volume} {13}},\ \bibinfo {pages}
  {367} (\bibinfo {year} {2014})}\BibitemShut {NoStop}%
\bibitem [{\citenamefont {Bodnar}\ \emph {et~al.}(2018)\citenamefont {Bodnar},
  \citenamefont {\ifmmode~\check{S}\else \v{S}\fi{}mejkal}, \citenamefont
  {Turek}, \citenamefont {Jungwirth}, \citenamefont {Gomonay}, \citenamefont
  {Sinova}, \citenamefont {Sapozhnik}, \citenamefont {Elmers}, \citenamefont
  {Kl\"aui},\ and\ \citenamefont {Jourdan}}]{BodnarNatComm2018}%
  \BibitemOpen
  \bibfield  {author} {\bibinfo {author} {\bibfnamefont {S.}~\bibnamefont
  {Bodnar}}, \bibinfo {author} {\bibfnamefont {L.}~\bibnamefont
  {\ifmmode~\check{S}\else \v{S}\fi{}mejkal}}, \bibinfo {author} {\bibfnamefont
  {I.}~\bibnamefont {Turek}}, \bibinfo {author} {\bibfnamefont
  {T.}~\bibnamefont {Jungwirth}}, \bibinfo {author} {\bibfnamefont
  {O.}~\bibnamefont {Gomonay}}, \bibinfo {author} {\bibfnamefont
  {J.}~\bibnamefont {Sinova}}, \bibinfo {author} {\bibfnamefont {A.~A.}\
  \bibnamefont {Sapozhnik}}, \bibinfo {author} {\bibfnamefont {J.-J.}\
  \bibnamefont {Elmers}}, \bibinfo {author} {\bibfnamefont {M.}~\bibnamefont
  {Kl\"aui}}, \ and\ \bibinfo {author} {\bibfnamefont {M.}~\bibnamefont
  {Jourdan}},\ }\href {\doibase 10.1038/s41467-017-02780-x} {\bibfield
  {journal} {\bibinfo  {journal} {Nat. Comm.}\ }\textbf {\bibinfo {volume}
  {9}},\ \bibinfo {pages} {348} (\bibinfo {year} {2018})}\BibitemShut {NoStop}%
\bibitem [{\citenamefont {Oh}\ \emph {et~al.}(2019)\citenamefont {Oh},
  \citenamefont {Humbard}, \citenamefont {Humbert}, \citenamefont {Sklenar},\
  and\ \citenamefont {Mason}}]{oh2019angular}%
  \BibitemOpen
  \bibfield  {author} {\bibinfo {author} {\bibfnamefont {J.}~\bibnamefont
  {Oh}}, \bibinfo {author} {\bibfnamefont {L.}~\bibnamefont {Humbard}},
  \bibinfo {author} {\bibfnamefont {V.}~\bibnamefont {Humbert}}, \bibinfo
  {author} {\bibfnamefont {J.}~\bibnamefont {Sklenar}}, \ and\ \bibinfo
  {author} {\bibfnamefont {N.}~\bibnamefont {Mason}},\ }\href@noop {}
  {\bibfield  {journal} {\bibinfo  {journal} {AIP Adv.}\ }\textbf {\bibinfo
  {volume} {9}},\ \bibinfo {pages} {045016} (\bibinfo {year}
  {2019})}\BibitemShut {NoStop}%
\bibitem [{\citenamefont {Moriyama}\ \emph {et~al.}(2018)\citenamefont
  {Moriyama}, \citenamefont {Zhou}, \citenamefont {Seki}, \citenamefont
  {Takanashi},\ and\ \citenamefont {Ono}}]{moriyama2018spin}%
  \BibitemOpen
  \bibfield  {author} {\bibinfo {author} {\bibfnamefont {T.}~\bibnamefont
  {Moriyama}}, \bibinfo {author} {\bibfnamefont {W.}~\bibnamefont {Zhou}},
  \bibinfo {author} {\bibfnamefont {T.}~\bibnamefont {Seki}}, \bibinfo {author}
  {\bibfnamefont {K.}~\bibnamefont {Takanashi}}, \ and\ \bibinfo {author}
  {\bibfnamefont {T.}~\bibnamefont {Ono}},\ }\href@noop {} {\bibfield
  {journal} {\bibinfo  {journal} {Phys. Rev. Lett.}\ }\textbf {\bibinfo
  {volume} {121}},\ \bibinfo {pages} {167202} (\bibinfo {year}
  {2018})}\BibitemShut {NoStop}%
\bibitem [{\citenamefont {Wang}\ \emph {et~al.}(2014)\citenamefont {Wang},
  \citenamefont {Seinige}, \citenamefont {Cao}, \citenamefont {Zhou},
  \citenamefont {Goodenough},\ and\ \citenamefont
  {Tsoi}}]{wang2014anisotropic}%
  \BibitemOpen
  \bibfield  {author} {\bibinfo {author} {\bibfnamefont {C.}~\bibnamefont
  {Wang}}, \bibinfo {author} {\bibfnamefont {H.}~\bibnamefont {Seinige}},
  \bibinfo {author} {\bibfnamefont {G.}~\bibnamefont {Cao}}, \bibinfo {author}
  {\bibfnamefont {J.-S.}\ \bibnamefont {Zhou}}, \bibinfo {author}
  {\bibfnamefont {J.~B.}\ \bibnamefont {Goodenough}}, \ and\ \bibinfo {author}
  {\bibfnamefont {M.}~\bibnamefont {Tsoi}},\ }\href@noop {} {\bibfield
  {journal} {\bibinfo  {journal} {Phys. Rev. X}\ }\textbf {\bibinfo {volume}
  {4}},\ \bibinfo {pages} {041034} (\bibinfo {year} {2014})}\BibitemShut
  {NoStop}%
\bibitem [{\citenamefont {Fina}\ \emph {et~al.}(2014)\citenamefont {Fina},
  \citenamefont {Marti}, \citenamefont {Yi}, \citenamefont {Liu}, \citenamefont
  {Chu}, \citenamefont {Rayan-Serrao}, \citenamefont {Suresha}, \citenamefont
  {Shick}, \citenamefont {{\v{Z}}elezn{\`y}}, \citenamefont {Jungwirth} \emph
  {et~al.}}]{fina2014anisotropic}%
  \BibitemOpen
  \bibfield  {author} {\bibinfo {author} {\bibfnamefont {I.}~\bibnamefont
  {Fina}}, \bibinfo {author} {\bibfnamefont {X.}~\bibnamefont {Marti}},
  \bibinfo {author} {\bibfnamefont {D.}~\bibnamefont {Yi}}, \bibinfo {author}
  {\bibfnamefont {J.}~\bibnamefont {Liu}}, \bibinfo {author} {\bibfnamefont
  {J.~H.}\ \bibnamefont {Chu}}, \bibinfo {author} {\bibfnamefont
  {C.}~\bibnamefont {Rayan-Serrao}}, \bibinfo {author} {\bibfnamefont
  {S.}~\bibnamefont {Suresha}}, \bibinfo {author} {\bibfnamefont {A.~B.}\
  \bibnamefont {Shick}}, \bibinfo {author} {\bibfnamefont {J.}~\bibnamefont
  {{\v{Z}}elezn{\`y}}}, \bibinfo {author} {\bibfnamefont {T.}~\bibnamefont
  {Jungwirth}},  \emph {et~al.},\ }\href@noop {} {\bibfield  {journal}
  {\bibinfo  {journal} {Nat. Comm.}\ }\textbf {\bibinfo {volume} {5}},\
  \bibinfo {pages} {1} (\bibinfo {year} {2014})}\BibitemShut {NoStop}%
\bibitem [{\citenamefont {Kriegner}\ \emph {et~al.}(2016)\citenamefont
  {Kriegner}, \citenamefont {V{\`y}born{\`y}}, \citenamefont {Olejn{\'\i}k},
  \citenamefont {Reichlov{\'a}}, \citenamefont {Nov{\'a}k}, \citenamefont
  {Marti}, \citenamefont {Gazquez}, \citenamefont {Saidl}, \citenamefont
  {N{\v{e}}mec}, \citenamefont {Volobuev} \emph
  {et~al.}}]{kriegner2016multiple}%
  \BibitemOpen
  \bibfield  {author} {\bibinfo {author} {\bibfnamefont {D.}~\bibnamefont
  {Kriegner}}, \bibinfo {author} {\bibfnamefont {K.}~\bibnamefont
  {V{\`y}born{\`y}}}, \bibinfo {author} {\bibfnamefont {K.}~\bibnamefont
  {Olejn{\'\i}k}}, \bibinfo {author} {\bibfnamefont {H.}~\bibnamefont
  {Reichlov{\'a}}}, \bibinfo {author} {\bibfnamefont {V.}~\bibnamefont
  {Nov{\'a}k}}, \bibinfo {author} {\bibfnamefont {X.}~\bibnamefont {Marti}},
  \bibinfo {author} {\bibfnamefont {J.}~\bibnamefont {Gazquez}}, \bibinfo
  {author} {\bibfnamefont {V.}~\bibnamefont {Saidl}}, \bibinfo {author}
  {\bibfnamefont {P.}~\bibnamefont {N{\v{e}}mec}}, \bibinfo {author}
  {\bibfnamefont {V.~V.}\ \bibnamefont {Volobuev}},  \emph {et~al.},\
  }\href@noop {} {\bibfield  {journal} {\bibinfo  {journal} {Nat. Comm.}\
  }\textbf {\bibinfo {volume} {7}},\ \bibinfo {pages} {1} (\bibinfo {year}
  {2016})}\BibitemShut {NoStop}%
\bibitem [{\citenamefont {Ahadi}\ \emph {et~al.}(2019)\citenamefont {Ahadi},
  \citenamefont {Lu}, \citenamefont {Salmani-Rezaie}, \citenamefont {Marshall},
  \citenamefont {Rondinelli},\ and\ \citenamefont
  {Stemmer}}]{ahadi2019anisotropic}%
  \BibitemOpen
  \bibfield  {author} {\bibinfo {author} {\bibfnamefont {K.}~\bibnamefont
  {Ahadi}}, \bibinfo {author} {\bibfnamefont {X.}~\bibnamefont {Lu}}, \bibinfo
  {author} {\bibfnamefont {S.}~\bibnamefont {Salmani-Rezaie}}, \bibinfo
  {author} {\bibfnamefont {P.~B.}\ \bibnamefont {Marshall}}, \bibinfo {author}
  {\bibfnamefont {J.~M.}\ \bibnamefont {Rondinelli}}, \ and\ \bibinfo {author}
  {\bibfnamefont {S.}~\bibnamefont {Stemmer}},\ }\href@noop {} {\bibfield
  {journal} {\bibinfo  {journal} {Phys. Rev. B}\ }\textbf {\bibinfo {volume}
  {99}},\ \bibinfo {pages} {041106} (\bibinfo {year} {2019})}\BibitemShut
  {NoStop}%
\bibitem [{\citenamefont {Moriyama}\ \emph {et~al.}(2015)\citenamefont
  {Moriyama}, \citenamefont {Matsuzaki}, \citenamefont {Kim}, \citenamefont
  {Suzuki}, \citenamefont {Taniyama},\ and\ \citenamefont
  {Ono}}]{moriyama2015sequential}%
  \BibitemOpen
  \bibfield  {author} {\bibinfo {author} {\bibfnamefont {T.}~\bibnamefont
  {Moriyama}}, \bibinfo {author} {\bibfnamefont {N.}~\bibnamefont {Matsuzaki}},
  \bibinfo {author} {\bibfnamefont {K.-J.}\ \bibnamefont {Kim}}, \bibinfo
  {author} {\bibfnamefont {I.}~\bibnamefont {Suzuki}}, \bibinfo {author}
  {\bibfnamefont {T.}~\bibnamefont {Taniyama}}, \ and\ \bibinfo {author}
  {\bibfnamefont {T.}~\bibnamefont {Ono}},\ }\href@noop {} {\bibfield
  {journal} {\bibinfo  {journal} {Appl. Phys. Lett.}\ }\textbf {\bibinfo
  {volume} {107}},\ \bibinfo {pages} {122403} (\bibinfo {year}
  {2015})}\BibitemShut {NoStop}%
\bibitem [{\citenamefont {McGrath}, \citenamefont {Camley},\ and\ \citenamefont
  {Livesey}(2020)}]{mcgrath2019self}%
  \BibitemOpen
  \bibfield  {author} {\bibinfo {author} {\bibfnamefont {B.~R.}\ \bibnamefont
  {McGrath}}, \bibinfo {author} {\bibfnamefont {R.~E.}\ \bibnamefont {Camley}},
  \ and\ \bibinfo {author} {\bibfnamefont {K.~L.}\ \bibnamefont {Livesey}},\
  }\href {\doibase 10.1103/PhysRevB.101.014444} {\bibfield  {journal} {\bibinfo
   {journal} {Phys. Rev. B}\ }\textbf {\bibinfo {volume} {101}},\ \bibinfo
  {pages} {014444} (\bibinfo {year} {2020})}\BibitemShut {NoStop}%
\bibitem [{\citenamefont {Nakayama}\ \emph {et~al.}(2013)\citenamefont
  {Nakayama}, \citenamefont {Althammer}, \citenamefont {Chen}, \citenamefont
  {Uchida}, \citenamefont {Kajiwara}, \citenamefont {Kikuchi}, \citenamefont
  {Ohtani}, \citenamefont {Gepr{\"a}gs}, \citenamefont {Opel}, \citenamefont
  {Takahashi} \emph {et~al.}}]{nakayama2013spin}%
  \BibitemOpen
  \bibfield  {author} {\bibinfo {author} {\bibfnamefont {H.}~\bibnamefont
  {Nakayama}}, \bibinfo {author} {\bibfnamefont {M.}~\bibnamefont {Althammer}},
  \bibinfo {author} {\bibfnamefont {Y.-T.}\ \bibnamefont {Chen}}, \bibinfo
  {author} {\bibfnamefont {K.}~\bibnamefont {Uchida}}, \bibinfo {author}
  {\bibfnamefont {Y.}~\bibnamefont {Kajiwara}}, \bibinfo {author}
  {\bibfnamefont {D.}~\bibnamefont {Kikuchi}}, \bibinfo {author} {\bibfnamefont
  {T.}~\bibnamefont {Ohtani}}, \bibinfo {author} {\bibfnamefont
  {S.}~\bibnamefont {Gepr{\"a}gs}}, \bibinfo {author} {\bibfnamefont
  {M.}~\bibnamefont {Opel}}, \bibinfo {author} {\bibfnamefont {S.}~\bibnamefont
  {Takahashi}},  \emph {et~al.},\ }\href@noop {} {\bibfield  {journal}
  {\bibinfo  {journal} {Phys. Rev. Lett.}\ }\textbf {\bibinfo {volume} {110}},\
  \bibinfo {pages} {206601} (\bibinfo {year} {2013})}\BibitemShut {NoStop}%
\bibitem [{\citenamefont {Avci}\ \emph {et~al.}(2015)\citenamefont {Avci},
  \citenamefont {Garello}, \citenamefont {Ghosh}, \citenamefont {Gabureac},
  \citenamefont {Alvarado},\ and\ \citenamefont
  {Gambardella}}]{avci2015unidirectional}%
  \BibitemOpen
  \bibfield  {author} {\bibinfo {author} {\bibfnamefont {C.~O.}\ \bibnamefont
  {Avci}}, \bibinfo {author} {\bibfnamefont {K.}~\bibnamefont {Garello}},
  \bibinfo {author} {\bibfnamefont {A.}~\bibnamefont {Ghosh}}, \bibinfo
  {author} {\bibfnamefont {M.}~\bibnamefont {Gabureac}}, \bibinfo {author}
  {\bibfnamefont {S.~F.}\ \bibnamefont {Alvarado}}, \ and\ \bibinfo {author}
  {\bibfnamefont {P.}~\bibnamefont {Gambardella}},\ }\href@noop {} {\bibfield
  {journal} {\bibinfo  {journal} {Nat. Phys.}\ }\textbf {\bibinfo {volume}
  {11}},\ \bibinfo {pages} {570} (\bibinfo {year} {2015})}\BibitemShut
  {NoStop}%
\bibitem [{\citenamefont {Kim}\ \emph {et~al.}(2016)\citenamefont {Kim},
  \citenamefont {Sheng}, \citenamefont {Takahashi}, \citenamefont {Mitani},\
  and\ \citenamefont {Hayashi}}]{kim2016spin}%
  \BibitemOpen
  \bibfield  {author} {\bibinfo {author} {\bibfnamefont {J.}~\bibnamefont
  {Kim}}, \bibinfo {author} {\bibfnamefont {P.}~\bibnamefont {Sheng}}, \bibinfo
  {author} {\bibfnamefont {S.}~\bibnamefont {Takahashi}}, \bibinfo {author}
  {\bibfnamefont {S.}~\bibnamefont {Mitani}}, \ and\ \bibinfo {author}
  {\bibfnamefont {M.}~\bibnamefont {Hayashi}},\ }\href@noop {} {\bibfield
  {journal} {\bibinfo  {journal} {Phys. Rev. Lett.}\ }\textbf {\bibinfo
  {volume} {116}},\ \bibinfo {pages} {097201} (\bibinfo {year}
  {2016})}\BibitemShut {NoStop}%
\bibitem [{\citenamefont {Hoogeboom}\ \emph {et~al.}(2017)\citenamefont
  {Hoogeboom}, \citenamefont {Aqeel}, \citenamefont {Kuschel}, \citenamefont
  {Palstra},\ and\ \citenamefont {van Wees}}]{hoogeboom2017negative}%
  \BibitemOpen
  \bibfield  {author} {\bibinfo {author} {\bibfnamefont {G.~R.}\ \bibnamefont
  {Hoogeboom}}, \bibinfo {author} {\bibfnamefont {A.}~\bibnamefont {Aqeel}},
  \bibinfo {author} {\bibfnamefont {T.}~\bibnamefont {Kuschel}}, \bibinfo
  {author} {\bibfnamefont {T.~T.~M.}\ \bibnamefont {Palstra}}, \ and\ \bibinfo
  {author} {\bibfnamefont {B.~J.}\ \bibnamefont {van Wees}},\ }\href@noop {}
  {\bibfield  {journal} {\bibinfo  {journal} {Appl. Phys. Lett.}\ }\textbf
  {\bibinfo {volume} {111}},\ \bibinfo {pages} {052409} (\bibinfo {year}
  {2017})}\BibitemShut {NoStop}%
\bibitem [{\citenamefont {Fischer}\ \emph {et~al.}(2018)\citenamefont
  {Fischer}, \citenamefont {Gomonay}, \citenamefont {Schlitz}, \citenamefont
  {Ganzhorn}, \citenamefont {Vlietstra}, \citenamefont {Althammer},
  \citenamefont {Huebl}, \citenamefont {Opel}, \citenamefont {Gross},
  \citenamefont {Goennenwein} \emph {et~al.}}]{fischer2018spin}%
  \BibitemOpen
  \bibfield  {author} {\bibinfo {author} {\bibfnamefont {J.}~\bibnamefont
  {Fischer}}, \bibinfo {author} {\bibfnamefont {O.}~\bibnamefont {Gomonay}},
  \bibinfo {author} {\bibfnamefont {R.}~\bibnamefont {Schlitz}}, \bibinfo
  {author} {\bibfnamefont {K.}~\bibnamefont {Ganzhorn}}, \bibinfo {author}
  {\bibfnamefont {N.}~\bibnamefont {Vlietstra}}, \bibinfo {author}
  {\bibfnamefont {M.}~\bibnamefont {Althammer}}, \bibinfo {author}
  {\bibfnamefont {H.}~\bibnamefont {Huebl}}, \bibinfo {author} {\bibfnamefont
  {M.}~\bibnamefont {Opel}}, \bibinfo {author} {\bibfnamefont {R.}~\bibnamefont
  {Gross}}, \bibinfo {author} {\bibfnamefont {S.~T.}\ \bibnamefont
  {Goennenwein}},  \emph {et~al.},\ }\href@noop {} {\bibfield  {journal}
  {\bibinfo  {journal} {Phys. Rev. B}\ }\textbf {\bibinfo {volume} {97}},\
  \bibinfo {pages} {014417} (\bibinfo {year} {2018})}\BibitemShut {NoStop}%
\bibitem [{\citenamefont {Baldrati}\ \emph {et~al.}(2018)\citenamefont
  {Baldrati}, \citenamefont {Ross}, \citenamefont {Niizeki}, \citenamefont
  {Schneider}, \citenamefont {Ramos}, \citenamefont {Cramer}, \citenamefont
  {Gomonay}, \citenamefont {Filianina}, \citenamefont {Savchenko},
  \citenamefont {Heinze} \emph {et~al.}}]{baldrati2018full}%
  \BibitemOpen
  \bibfield  {author} {\bibinfo {author} {\bibfnamefont {L.}~\bibnamefont
  {Baldrati}}, \bibinfo {author} {\bibfnamefont {A.}~\bibnamefont {Ross}},
  \bibinfo {author} {\bibfnamefont {T.}~\bibnamefont {Niizeki}}, \bibinfo
  {author} {\bibfnamefont {C.}~\bibnamefont {Schneider}}, \bibinfo {author}
  {\bibfnamefont {R.}~\bibnamefont {Ramos}}, \bibinfo {author} {\bibfnamefont
  {J.}~\bibnamefont {Cramer}}, \bibinfo {author} {\bibfnamefont
  {O.}~\bibnamefont {Gomonay}}, \bibinfo {author} {\bibfnamefont
  {M.}~\bibnamefont {Filianina}}, \bibinfo {author} {\bibfnamefont
  {T.}~\bibnamefont {Savchenko}}, \bibinfo {author} {\bibfnamefont
  {D.}~\bibnamefont {Heinze}},  \emph {et~al.},\ }\href@noop {} {\bibfield
  {journal} {\bibinfo  {journal} {Phys. Rev. B}\ }\textbf {\bibinfo {volume}
  {98}},\ \bibinfo {pages} {024422} (\bibinfo {year} {2018})}\BibitemShut
  {NoStop}%
\bibitem [{\citenamefont {Ji}\ \emph {et~al.}(2017)\citenamefont {Ji},
  \citenamefont {Miao}, \citenamefont {Meng}, \citenamefont {Ren},
  \citenamefont {Dong}, \citenamefont {Xu}, \citenamefont {Wu},\ and\
  \citenamefont {Jiang}}]{ji2017spin}%
  \BibitemOpen
  \bibfield  {author} {\bibinfo {author} {\bibfnamefont {Y.}~\bibnamefont
  {Ji}}, \bibinfo {author} {\bibfnamefont {J.}~\bibnamefont {Miao}}, \bibinfo
  {author} {\bibfnamefont {K.~K.}\ \bibnamefont {Meng}}, \bibinfo {author}
  {\bibfnamefont {Z.~Y.}\ \bibnamefont {Ren}}, \bibinfo {author} {\bibfnamefont
  {B.~W.}\ \bibnamefont {Dong}}, \bibinfo {author} {\bibfnamefont {X.~G.}\
  \bibnamefont {Xu}}, \bibinfo {author} {\bibfnamefont {Y.}~\bibnamefont {Wu}},
  \ and\ \bibinfo {author} {\bibfnamefont {Y.}~\bibnamefont {Jiang}},\
  }\href@noop {} {\bibfield  {journal} {\bibinfo  {journal} {Appl. Phys.
  Lett.}\ }\textbf {\bibinfo {volume} {110}},\ \bibinfo {pages} {262401}
  (\bibinfo {year} {2017})}\BibitemShut {NoStop}%
\bibitem [{\citenamefont {Chen}\ \emph
  {et~al.}(2018{\natexlab{a}})\citenamefont {Chen}, \citenamefont {Zarzuela},
  \citenamefont {Zhang}, \citenamefont {Song}, \citenamefont {Zhou},
  \citenamefont {Shi}, \citenamefont {Li}, \citenamefont {Zhou}, \citenamefont
  {Jiang}, \citenamefont {Pan} \emph {et~al.}}]{chen2018antidamping}%
  \BibitemOpen
  \bibfield  {author} {\bibinfo {author} {\bibfnamefont {X.~Z.}\ \bibnamefont
  {Chen}}, \bibinfo {author} {\bibfnamefont {R.}~\bibnamefont {Zarzuela}},
  \bibinfo {author} {\bibfnamefont {J.}~\bibnamefont {Zhang}}, \bibinfo
  {author} {\bibfnamefont {C.}~\bibnamefont {Song}}, \bibinfo {author}
  {\bibfnamefont {X.~F.}\ \bibnamefont {Zhou}}, \bibinfo {author}
  {\bibfnamefont {G.~Y.}\ \bibnamefont {Shi}}, \bibinfo {author} {\bibfnamefont
  {F.}~\bibnamefont {Li}}, \bibinfo {author} {\bibfnamefont {H.~A.}\
  \bibnamefont {Zhou}}, \bibinfo {author} {\bibfnamefont {W.~J.}\ \bibnamefont
  {Jiang}}, \bibinfo {author} {\bibfnamefont {F.}~\bibnamefont {Pan}},  \emph
  {et~al.},\ }\href@noop {} {\bibfield  {journal} {\bibinfo  {journal} {Phys.
  Rev. Lett.}\ }\textbf {\bibinfo {volume} {120}},\ \bibinfo {pages} {207204}
  (\bibinfo {year} {2018}{\natexlab{a}})}\BibitemShut {NoStop}%
\bibitem [{\citenamefont {Churikova}\ \emph {et~al.}(2020)\citenamefont
  {Churikova}, \citenamefont {Bono}, \citenamefont {Neltner}, \citenamefont
  {Wittmann}, \citenamefont {Scipioni}, \citenamefont {Shepard}, \citenamefont
  {Newhouse-Illige}, \citenamefont {Greer},\ and\ \citenamefont
  {Beach}}]{churikova2020non}%
  \BibitemOpen
  \bibfield  {author} {\bibinfo {author} {\bibfnamefont {A.}~\bibnamefont
  {Churikova}}, \bibinfo {author} {\bibfnamefont {D.}~\bibnamefont {Bono}},
  \bibinfo {author} {\bibfnamefont {B.}~\bibnamefont {Neltner}}, \bibinfo
  {author} {\bibfnamefont {A.}~\bibnamefont {Wittmann}}, \bibinfo {author}
  {\bibfnamefont {L.}~\bibnamefont {Scipioni}}, \bibinfo {author}
  {\bibfnamefont {A.}~\bibnamefont {Shepard}}, \bibinfo {author} {\bibfnamefont
  {T.}~\bibnamefont {Newhouse-Illige}}, \bibinfo {author} {\bibfnamefont
  {J.}~\bibnamefont {Greer}}, \ and\ \bibinfo {author} {\bibfnamefont
  {G.~S.~D.}\ \bibnamefont {Beach}},\ }\href@noop {} {\bibfield  {journal}
  {\bibinfo  {journal} {Appl. Phys. Lett.}\ }\textbf {\bibinfo {volume}
  {116}},\ \bibinfo {pages} {022410} (\bibinfo {year} {2020})}\BibitemShut
  {NoStop}%
\bibitem [{\citenamefont {Zhang}\ \emph {et~al.}(2019)\citenamefont {Zhang},
  \citenamefont {Finley}, \citenamefont {Safi},\ and\ \citenamefont
  {Liu}}]{zhang2019quantitative}%
  \BibitemOpen
  \bibfield  {author} {\bibinfo {author} {\bibfnamefont {P.}~\bibnamefont
  {Zhang}}, \bibinfo {author} {\bibfnamefont {J.}~\bibnamefont {Finley}},
  \bibinfo {author} {\bibfnamefont {T.}~\bibnamefont {Safi}}, \ and\ \bibinfo
  {author} {\bibfnamefont {L.}~\bibnamefont {Liu}},\ }\href@noop {} {\bibfield
  {journal} {\bibinfo  {journal} {Phys. Rev. Lett.}\ }\textbf {\bibinfo
  {volume} {123}},\ \bibinfo {pages} {247206} (\bibinfo {year}
  {2019})}\BibitemShut {NoStop}%
\bibitem [{\citenamefont {Cheng}\ \emph {et~al.}(2020)\citenamefont {Cheng},
  \citenamefont {Yu}, \citenamefont {Zhu}, \citenamefont {Hwang},\ and\
  \citenamefont {Yang}}]{cheng2020electrical}%
  \BibitemOpen
  \bibfield  {author} {\bibinfo {author} {\bibfnamefont {Y.}~\bibnamefont
  {Cheng}}, \bibinfo {author} {\bibfnamefont {S.}~\bibnamefont {Yu}}, \bibinfo
  {author} {\bibfnamefont {M.}~\bibnamefont {Zhu}}, \bibinfo {author}
  {\bibfnamefont {J.}~\bibnamefont {Hwang}}, \ and\ \bibinfo {author}
  {\bibfnamefont {F.}~\bibnamefont {Yang}},\ }\href@noop {} {\bibfield
  {journal} {\bibinfo  {journal} {Phys. Rev. Lett.}\ }\textbf {\bibinfo
  {volume} {124}},\ \bibinfo {pages} {027202} (\bibinfo {year}
  {2020})}\BibitemShut {NoStop}%
\bibitem [{\citenamefont {{Hoffmann}}(2013)}]{Hoffmann2013IEEETM}%
  \BibitemOpen
  \bibfield  {author} {\bibinfo {author} {\bibfnamefont {A.}~\bibnamefont
  {{Hoffmann}}},\ }\href {\doibase 10.1109/TMAG.2013.2262947} {\bibfield
  {journal} {\bibinfo  {journal} {IEEE Tran. Magn.}\ }\textbf {\bibinfo
  {volume} {49}},\ \bibinfo {pages} {5172} (\bibinfo {year}
  {2013})}\BibitemShut {NoStop}%
\bibitem [{\citenamefont {Zhang}\ \emph {et~al.}(2014)\citenamefont {Zhang},
  \citenamefont {Jungfleisch}, \citenamefont {Jiang}, \citenamefont {Pearson},
  \citenamefont {Hoffmann}, \citenamefont {Freimuth},\ and\ \citenamefont
  {Mokrousov}}]{ZhangPRL2014}%
  \BibitemOpen
  \bibfield  {author} {\bibinfo {author} {\bibfnamefont {W.}~\bibnamefont
  {Zhang}}, \bibinfo {author} {\bibfnamefont {M.~B.}\ \bibnamefont
  {Jungfleisch}}, \bibinfo {author} {\bibfnamefont {W.}~\bibnamefont {Jiang}},
  \bibinfo {author} {\bibfnamefont {J.~E.}\ \bibnamefont {Pearson}}, \bibinfo
  {author} {\bibfnamefont {A.}~\bibnamefont {Hoffmann}}, \bibinfo {author}
  {\bibfnamefont {F.}~\bibnamefont {Freimuth}}, \ and\ \bibinfo {author}
  {\bibfnamefont {Y.}~\bibnamefont {Mokrousov}},\ }\href {\doibase
  10.1103/PhysRevLett.113.196602} {\bibfield  {journal} {\bibinfo  {journal}
  {Phys. Rev. Lett.}\ }\textbf {\bibinfo {volume} {113}},\ \bibinfo {pages}
  {196602} (\bibinfo {year} {2014})}\BibitemShut {NoStop}%
\bibitem [{\citenamefont {Liu}\ \emph {et~al.}(2011{\natexlab{b}})\citenamefont
  {Liu}, \citenamefont {Moriyama}, \citenamefont {Ralph},\ and\ \citenamefont
  {Buhrman}}]{Li2011PRL}%
  \BibitemOpen
  \bibfield  {author} {\bibinfo {author} {\bibfnamefont {L.}~\bibnamefont
  {Liu}}, \bibinfo {author} {\bibfnamefont {T.}~\bibnamefont {Moriyama}},
  \bibinfo {author} {\bibfnamefont {D.~C.}\ \bibnamefont {Ralph}}, \ and\
  \bibinfo {author} {\bibfnamefont {R.~A.}\ \bibnamefont {Buhrman}},\ }\href
  {\doibase 10.1103/PhysRevLett.106.036601} {\bibfield  {journal} {\bibinfo
  {journal} {Phys. Rev. Lett.}\ }\textbf {\bibinfo {volume} {106}},\ \bibinfo
  {pages} {036601} (\bibinfo {year} {2011}{\natexlab{b}})}\BibitemShut
  {NoStop}%
\bibitem [{\citenamefont {Zhang}\ \emph {et~al.}(2015)\citenamefont {Zhang},
  \citenamefont {Jungfleisch}, \citenamefont {Freimuth}, \citenamefont {Jiang},
  \citenamefont {Sklenar}, \citenamefont {Pearson}, \citenamefont {Ketterson},
  \citenamefont {Mokrousov},\ and\ \citenamefont {Hoffmann}}]{ZhangPRB2015}%
  \BibitemOpen
  \bibfield  {author} {\bibinfo {author} {\bibfnamefont {W.}~\bibnamefont
  {Zhang}}, \bibinfo {author} {\bibfnamefont {M.~B.}\ \bibnamefont
  {Jungfleisch}}, \bibinfo {author} {\bibfnamefont {F.}~\bibnamefont
  {Freimuth}}, \bibinfo {author} {\bibfnamefont {W.}~\bibnamefont {Jiang}},
  \bibinfo {author} {\bibfnamefont {J.}~\bibnamefont {Sklenar}}, \bibinfo
  {author} {\bibfnamefont {J.~E.}\ \bibnamefont {Pearson}}, \bibinfo {author}
  {\bibfnamefont {J.~B.}\ \bibnamefont {Ketterson}}, \bibinfo {author}
  {\bibfnamefont {Y.}~\bibnamefont {Mokrousov}}, \ and\ \bibinfo {author}
  {\bibfnamefont {A.}~\bibnamefont {Hoffmann}},\ }\href {\doibase
  10.1103/PhysRevB.92.144405} {\bibfield  {journal} {\bibinfo  {journal} {Phys.
  Rev. B}\ }\textbf {\bibinfo {volume} {92}},\ \bibinfo {pages} {144405}
  (\bibinfo {year} {2015})}\BibitemShut {NoStop}%
\bibitem [{\citenamefont {Zhang}\ \emph {et~al.}(2016)\citenamefont {Zhang},
  \citenamefont {Han}, \citenamefont {Yang}, \citenamefont {Sun}, \citenamefont
  {Zhang}, \citenamefont {Yan},\ and\ \citenamefont
  {Parkin}}]{ZhangSciAdv2016}%
  \BibitemOpen
  \bibfield  {author} {\bibinfo {author} {\bibfnamefont {W.}~\bibnamefont
  {Zhang}}, \bibinfo {author} {\bibfnamefont {W.}~\bibnamefont {Han}}, \bibinfo
  {author} {\bibfnamefont {S.-H.}\ \bibnamefont {Yang}}, \bibinfo {author}
  {\bibfnamefont {Y.}~\bibnamefont {Sun}}, \bibinfo {author} {\bibfnamefont
  {Y.}~\bibnamefont {Zhang}}, \bibinfo {author} {\bibfnamefont
  {B.}~\bibnamefont {Yan}}, \ and\ \bibinfo {author} {\bibfnamefont {S.~S.~P.}\
  \bibnamefont {Parkin}},\ }\href {\doibase 10.1126/sciadv.1600759} {\bibfield
  {journal} {\bibinfo  {journal} {Sci. Adv.}\ }\textbf {\bibinfo {volume} {2}}
  (\bibinfo {year} {2016}),\ 10.1126/sciadv.1600759}\BibitemShut {NoStop}%
\bibitem [{\citenamefont {Culcer}\ and\ \citenamefont
  {Winkler}(2007)}]{CulcerPRL2007}%
  \BibitemOpen
  \bibfield  {author} {\bibinfo {author} {\bibfnamefont {D.}~\bibnamefont
  {Culcer}}\ and\ \bibinfo {author} {\bibfnamefont {R.}~\bibnamefont
  {Winkler}},\ }\href {\doibase 10.1103/PhysRevLett.99.226601} {\bibfield
  {journal} {\bibinfo  {journal} {Phys. Rev. Lett.}\ }\textbf {\bibinfo
  {volume} {99}},\ \bibinfo {pages} {226601} (\bibinfo {year}
  {2007})}\BibitemShut {NoStop}%
\bibitem [{\citenamefont {Sklenar}\ \emph {et~al.}(2016)\citenamefont
  {Sklenar}, \citenamefont {Zhang}, \citenamefont {Jungfleisch}, \citenamefont
  {Jiang}, \citenamefont {Saglam}, \citenamefont {Pearson}, \citenamefont
  {Ketterson},\ and\ \citenamefont {Hoffmann}}]{SklenarAIPAdv2016}%
  \BibitemOpen
  \bibfield  {author} {\bibinfo {author} {\bibfnamefont {J.}~\bibnamefont
  {Sklenar}}, \bibinfo {author} {\bibfnamefont {W.}~\bibnamefont {Zhang}},
  \bibinfo {author} {\bibfnamefont {M.~B.}\ \bibnamefont {Jungfleisch}},
  \bibinfo {author} {\bibfnamefont {W.}~\bibnamefont {Jiang}}, \bibinfo
  {author} {\bibfnamefont {H.}~\bibnamefont {Saglam}}, \bibinfo {author}
  {\bibfnamefont {J.~E.}\ \bibnamefont {Pearson}}, \bibinfo {author}
  {\bibfnamefont {J.~B.}\ \bibnamefont {Ketterson}}, \ and\ \bibinfo {author}
  {\bibfnamefont {A.}~\bibnamefont {Hoffmann}},\ }\href {\doibase
  10.1063/1.4943758} {\bibfield  {journal} {\bibinfo  {journal} {AIP Adv.}\
  }\textbf {\bibinfo {volume} {6}},\ \bibinfo {pages} {055603} (\bibinfo {year}
  {2016})}\BibitemShut {NoStop}%
\bibitem [{\citenamefont {Saglam}\ \emph {et~al.}(2018)\citenamefont {Saglam},
  \citenamefont {Rojas-Sanchez}, \citenamefont {Petit}, \citenamefont {Hehn},
  \citenamefont {Zhang}, \citenamefont {Pearson}, \citenamefont {Mangin},\ and\
  \citenamefont {Hoffmann}}]{SaglamPRB2018}%
  \BibitemOpen
  \bibfield  {author} {\bibinfo {author} {\bibfnamefont {H.}~\bibnamefont
  {Saglam}}, \bibinfo {author} {\bibfnamefont {J.~C.}\ \bibnamefont
  {Rojas-Sanchez}}, \bibinfo {author} {\bibfnamefont {S.}~\bibnamefont
  {Petit}}, \bibinfo {author} {\bibfnamefont {M.}~\bibnamefont {Hehn}},
  \bibinfo {author} {\bibfnamefont {W.}~\bibnamefont {Zhang}}, \bibinfo
  {author} {\bibfnamefont {J.~E.}\ \bibnamefont {Pearson}}, \bibinfo {author}
  {\bibfnamefont {S.}~\bibnamefont {Mangin}}, \ and\ \bibinfo {author}
  {\bibfnamefont {A.}~\bibnamefont {Hoffmann}},\ }\href {\doibase
  10.1103/PhysRevB.98.094407} {\bibfield  {journal} {\bibinfo  {journal} {Phys.
  Rev. B}\ }\textbf {\bibinfo {volume} {98}},\ \bibinfo {pages} {094407}
  (\bibinfo {year} {2018})}\BibitemShut {NoStop}%
\bibitem [{\citenamefont {Khodadadi}\ \emph {et~al.}(2019)\citenamefont
  {Khodadadi}, \citenamefont {Lim}, \citenamefont {Smith}, \citenamefont
  {Greening}, \citenamefont {Zheng}, \citenamefont {Diao}, \citenamefont
  {Kaiser},\ and\ \citenamefont {Emori}}]{KhodadadiPRB2019}%
  \BibitemOpen
  \bibfield  {author} {\bibinfo {author} {\bibfnamefont {B.}~\bibnamefont
  {Khodadadi}}, \bibinfo {author} {\bibfnamefont {Y.}~\bibnamefont {Lim}},
  \bibinfo {author} {\bibfnamefont {D.~A.}\ \bibnamefont {Smith}}, \bibinfo
  {author} {\bibfnamefont {R.~W.}\ \bibnamefont {Greening}}, \bibinfo {author}
  {\bibfnamefont {Y.}~\bibnamefont {Zheng}}, \bibinfo {author} {\bibfnamefont
  {Z.}~\bibnamefont {Diao}}, \bibinfo {author} {\bibfnamefont {C.}~\bibnamefont
  {Kaiser}}, \ and\ \bibinfo {author} {\bibfnamefont {S.}~\bibnamefont
  {Emori}},\ }\href {\doibase 10.1103/PhysRevB.99.024435} {\bibfield  {journal}
  {\bibinfo  {journal} {Phys. Rev. B}\ }\textbf {\bibinfo {volume} {99}},\
  \bibinfo {pages} {024435} (\bibinfo {year} {2019})}\BibitemShut {NoStop}%
\bibitem [{\citenamefont {Fukami}\ \emph {et~al.}(2016)\citenamefont {Fukami},
  \citenamefont {Zhang}, \citenamefont {DuttaGupta}, \citenamefont {Kurenkov},\
  and\ \citenamefont {Ohno}}]{FukamiNatMat2016}%
  \BibitemOpen
  \bibfield  {author} {\bibinfo {author} {\bibfnamefont {S.}~\bibnamefont
  {Fukami}}, \bibinfo {author} {\bibfnamefont {C.}~\bibnamefont {Zhang}},
  \bibinfo {author} {\bibfnamefont {S.}~\bibnamefont {DuttaGupta}}, \bibinfo
  {author} {\bibfnamefont {A.}~\bibnamefont {Kurenkov}}, \ and\ \bibinfo
  {author} {\bibfnamefont {H.}~\bibnamefont {Ohno}},\ }\href {\doibase
  10.1038/nmat4566} {\bibfield  {journal} {\bibinfo  {journal} {Nat. Mater.}\
  }\textbf {\bibinfo {volume} {15}},\ \bibinfo {pages} {535} (\bibinfo {year}
  {2016})}\BibitemShut {NoStop}%
\bibitem [{\citenamefont {Oh}\ \emph {et~al.}(2016)\citenamefont {Oh},
  \citenamefont {Chris~Baek}, \citenamefont {Kim}, \citenamefont {Lee},
  \citenamefont {Lee}, \citenamefont {Yang}, \citenamefont {Park},
  \citenamefont {Lee}, \citenamefont {Kim}, \citenamefont {Go}, \citenamefont
  {Jeong}, \citenamefont {Min}, \citenamefont {Lee}, \citenamefont {Lee},\ and\
  \citenamefont {Park}}]{OhNatNano2016}%
  \BibitemOpen
  \bibfield  {author} {\bibinfo {author} {\bibfnamefont {Y.-W.}\ \bibnamefont
  {Oh}}, \bibinfo {author} {\bibfnamefont {S.-h.}\ \bibnamefont {Chris~Baek}},
  \bibinfo {author} {\bibfnamefont {Y.~M.}\ \bibnamefont {Kim}}, \bibinfo
  {author} {\bibfnamefont {H.~Y.}\ \bibnamefont {Lee}}, \bibinfo {author}
  {\bibfnamefont {K.-D.}\ \bibnamefont {Lee}}, \bibinfo {author} {\bibfnamefont
  {C.-G.}\ \bibnamefont {Yang}}, \bibinfo {author} {\bibfnamefont {E.-S.}\
  \bibnamefont {Park}}, \bibinfo {author} {\bibfnamefont {K.-S.}\ \bibnamefont
  {Lee}}, \bibinfo {author} {\bibfnamefont {K.-W.}\ \bibnamefont {Kim}},
  \bibinfo {author} {\bibfnamefont {G.}~\bibnamefont {Go}}, \bibinfo {author}
  {\bibfnamefont {J.-R.}\ \bibnamefont {Jeong}}, \bibinfo {author}
  {\bibfnamefont {B.-C.}\ \bibnamefont {Min}}, \bibinfo {author} {\bibfnamefont
  {H.-W.}\ \bibnamefont {Lee}}, \bibinfo {author} {\bibfnamefont {K.-J.}\
  \bibnamefont {Lee}}, \ and\ \bibinfo {author} {\bibfnamefont {B.-G.}\
  \bibnamefont {Park}},\ }\href {\doibase 10.1038/nnano.2016.109} {\bibfield
  {journal} {\bibinfo  {journal} {Nat. Nanotechn.}\ }\textbf {\bibinfo {volume}
  {11}},\ \bibinfo {pages} {878} (\bibinfo {year} {2016})}\BibitemShut
  {NoStop}%
\bibitem [{\citenamefont {Borders}\ \emph {et~al.}(2017)\citenamefont
  {Borders}, \citenamefont {Akima}, \citenamefont {Fukami}, \citenamefont
  {Moriya}, \citenamefont {Kurihara}, \citenamefont {Horio}, \citenamefont
  {Sato},\ and\ \citenamefont {Ohno}}]{BordersAPE2017}%
  \BibitemOpen
  \bibfield  {author} {\bibinfo {author} {\bibfnamefont {W.~A.}\ \bibnamefont
  {Borders}}, \bibinfo {author} {\bibfnamefont {H.}~\bibnamefont {Akima}},
  \bibinfo {author} {\bibfnamefont {S.}~\bibnamefont {Fukami}}, \bibinfo
  {author} {\bibfnamefont {S.}~\bibnamefont {Moriya}}, \bibinfo {author}
  {\bibfnamefont {S.}~\bibnamefont {Kurihara}}, \bibinfo {author}
  {\bibfnamefont {Y.}~\bibnamefont {Horio}}, \bibinfo {author} {\bibfnamefont
  {S.}~\bibnamefont {Sato}}, \ and\ \bibinfo {author} {\bibfnamefont
  {H.}~\bibnamefont {Ohno}},\ }\href@noop {} {\bibfield  {journal} {\bibinfo
  {journal} {Appl. Phys. Exp.}\ }\textbf {\bibinfo {volume} {10}},\ \bibinfo
  {pages} {013007} (\bibinfo {year} {2017})}\BibitemShut {NoStop}%
\bibitem [{\citenamefont {Shindou}\ and\ \citenamefont
  {Nagaosa}(2001)}]{ShindouPRL2001}%
  \BibitemOpen
  \bibfield  {author} {\bibinfo {author} {\bibfnamefont {R.}~\bibnamefont
  {Shindou}}\ and\ \bibinfo {author} {\bibfnamefont {N.}~\bibnamefont
  {Nagaosa}},\ }\href {\doibase 10.1103/PhysRevLett.87.116801} {\bibfield
  {journal} {\bibinfo  {journal} {Phys. Rev. Lett.}\ }\textbf {\bibinfo
  {volume} {87}},\ \bibinfo {pages} {116801} (\bibinfo {year}
  {2001})}\BibitemShut {NoStop}%
\bibitem [{\citenamefont {Chen}, \citenamefont {Niu},\ and\ \citenamefont
  {MacDonald}(2014{\natexlab{a}})}]{ChenPRL2014}%
  \BibitemOpen
  \bibfield  {author} {\bibinfo {author} {\bibfnamefont {H.}~\bibnamefont
  {Chen}}, \bibinfo {author} {\bibfnamefont {Q.}~\bibnamefont {Niu}}, \ and\
  \bibinfo {author} {\bibfnamefont {A.~H.}\ \bibnamefont {MacDonald}},\ }\href
  {\doibase 10.1103/PhysRevLett.112.017205} {\bibfield  {journal} {\bibinfo
  {journal} {Phys. Rev. Lett.}\ }\textbf {\bibinfo {volume} {112}},\ \bibinfo
  {pages} {017205} (\bibinfo {year} {2014}{\natexlab{a}})}\BibitemShut
  {NoStop}%
\bibitem [{\citenamefont {K\"ubler}\ and\ \citenamefont
  {Felser}(2014)}]{KueblerEPL2014}%
  \BibitemOpen
  \bibfield  {author} {\bibinfo {author} {\bibfnamefont {J.}~\bibnamefont
  {K\"ubler}}\ and\ \bibinfo {author} {\bibfnamefont {C.}~\bibnamefont
  {Felser}},\ }\href {\doibase 10.1209/0295-5075/108/67001} {\bibfield
  {journal} {\bibinfo  {journal} {Europhys. Lett.}\ }\textbf {\bibinfo {volume}
  {108}},\ \bibinfo {pages} {67001} (\bibinfo {year} {2014})}\BibitemShut
  {NoStop}%
\bibitem [{\citenamefont {Nakatsuji}, \citenamefont {Kiyohara},\ and\
  \citenamefont {Higo}(2015)}]{Nakatsuji2015}%
  \BibitemOpen
  \bibfield  {author} {\bibinfo {author} {\bibfnamefont {S.}~\bibnamefont
  {Nakatsuji}}, \bibinfo {author} {\bibfnamefont {N.}~\bibnamefont {Kiyohara}},
  \ and\ \bibinfo {author} {\bibfnamefont {T.}~\bibnamefont {Higo}},\
  }\href@noop {} {\bibfield  {journal} {\bibinfo  {journal} {Nature}\ }\textbf
  {\bibinfo {volume} {527}},\ \bibinfo {pages} {212} (\bibinfo {year}
  {2015})}\BibitemShut {NoStop}%
\bibitem [{\citenamefont {Nayak}\ \emph {et~al.}(2016)\citenamefont {Nayak},
  \citenamefont {Fischer}, \citenamefont {Sun}, \citenamefont {Yan},
  \citenamefont {Karel}, \citenamefont {Komarek}, \citenamefont {Shekhar},
  \citenamefont {Kumar}, \citenamefont {Schnelle}, \citenamefont {K\"{u}bler},
  \citenamefont {Felser},\ and\ \citenamefont {Parkin}}]{Nayak2016}%
  \BibitemOpen
  \bibfield  {author} {\bibinfo {author} {\bibfnamefont {A.~K.}\ \bibnamefont
  {Nayak}}, \bibinfo {author} {\bibfnamefont {J.~E.}\ \bibnamefont {Fischer}},
  \bibinfo {author} {\bibfnamefont {Y.}~\bibnamefont {Sun}}, \bibinfo {author}
  {\bibfnamefont {B.}~\bibnamefont {Yan}}, \bibinfo {author} {\bibfnamefont
  {J.}~\bibnamefont {Karel}}, \bibinfo {author} {\bibfnamefont {A.~C.}\
  \bibnamefont {Komarek}}, \bibinfo {author} {\bibfnamefont {C.}~\bibnamefont
  {Shekhar}}, \bibinfo {author} {\bibfnamefont {N.}~\bibnamefont {Kumar}},
  \bibinfo {author} {\bibfnamefont {W.}~\bibnamefont {Schnelle}}, \bibinfo
  {author} {\bibfnamefont {J.}~\bibnamefont {K\"{u}bler}}, \bibinfo {author}
  {\bibfnamefont {C.}~\bibnamefont {Felser}}, \ and\ \bibinfo {author}
  {\bibfnamefont {S.~S.~P.}\ \bibnamefont {Parkin}},\ }\href@noop {} {\bibfield
   {journal} {\bibinfo  {journal} {Sci. Adv.}\ }\textbf {\bibinfo {volume}
  {2}},\ \bibinfo {pages} {1501870} (\bibinfo {year} {2016})}\BibitemShut
  {NoStop}%
\bibitem [{\citenamefont {Qin}\ \emph {et~al.}(2017)\citenamefont {Qin},
  \citenamefont {Chen}, \citenamefont {Cai}, \citenamefont {Kandaz},\ and\
  \citenamefont {Ji}}]{QinPRB2017}%
  \BibitemOpen
  \bibfield  {author} {\bibinfo {author} {\bibfnamefont {C.}~\bibnamefont
  {Qin}}, \bibinfo {author} {\bibfnamefont {S.}~\bibnamefont {Chen}}, \bibinfo
  {author} {\bibfnamefont {Y.}~\bibnamefont {Cai}}, \bibinfo {author}
  {\bibfnamefont {F.}~\bibnamefont {Kandaz}}, \ and\ \bibinfo {author}
  {\bibfnamefont {Y.}~\bibnamefont {Ji}},\ }\href {\doibase
  10.1103/PhysRevB.96.134418} {\bibfield  {journal} {\bibinfo  {journal} {Phys.
  Rev. B}\ }\textbf {\bibinfo {volume} {96}},\ \bibinfo {pages} {134418}
  (\bibinfo {year} {2017})}\BibitemShut {NoStop}%
\bibitem [{\citenamefont {Das}\ \emph {et~al.}(2017)\citenamefont {Das},
  \citenamefont {Schoemaker}, \citenamefont {van Wees},\ and\ \citenamefont
  {Vera-Marun}}]{DasPRB2017}%
  \BibitemOpen
  \bibfield  {author} {\bibinfo {author} {\bibfnamefont {K.~S.}\ \bibnamefont
  {Das}}, \bibinfo {author} {\bibfnamefont {W.~Y.}\ \bibnamefont {Schoemaker}},
  \bibinfo {author} {\bibfnamefont {B.~J.}\ \bibnamefont {van Wees}}, \ and\
  \bibinfo {author} {\bibfnamefont {I.~J.}\ \bibnamefont {Vera-Marun}},\ }\href
  {\doibase 10.1103/PhysRevB.96.220408} {\bibfield  {journal} {\bibinfo
  {journal} {Phys. Rev. B}\ }\textbf {\bibinfo {volume} {96}},\ \bibinfo
  {pages} {220408} (\bibinfo {year} {2017})}\BibitemShut {NoStop}%
\bibitem [{\citenamefont {Gibbons}\ \emph {et~al.}(2018)\citenamefont
  {Gibbons}, \citenamefont {MacNeill}, \citenamefont {Buhrman},\ and\
  \citenamefont {Ralph}}]{GibbonsPRAppl2018}%
  \BibitemOpen
  \bibfield  {author} {\bibinfo {author} {\bibfnamefont {J.~D.}\ \bibnamefont
  {Gibbons}}, \bibinfo {author} {\bibfnamefont {D.}~\bibnamefont {MacNeill}},
  \bibinfo {author} {\bibfnamefont {R.~A.}\ \bibnamefont {Buhrman}}, \ and\
  \bibinfo {author} {\bibfnamefont {D.~C.}\ \bibnamefont {Ralph}},\ }\href
  {\doibase 10.1103/PhysRevApplied.9.064033} {\bibfield  {journal} {\bibinfo
  {journal} {Phys. Rev. Applied}\ }\textbf {\bibinfo {volume} {9}},\ \bibinfo
  {pages} {064033} (\bibinfo {year} {2018})}\BibitemShut {NoStop}%
\bibitem [{\citenamefont {Kimata}\ \emph {et~al.}(2019)\citenamefont {Kimata},
  \citenamefont {Chen}, \citenamefont {Kondou}, \citenamefont {Sugimoto},
  \citenamefont {Muduli}, \citenamefont {Ikhlas}, \citenamefont {Omori},
  \citenamefont {Tomita}, \citenamefont {MacDonald}, \citenamefont
  {Nakatsuji},\ and\ \citenamefont {Otani}}]{KimataNature2019}%
  \BibitemOpen
  \bibfield  {author} {\bibinfo {author} {\bibfnamefont {M.}~\bibnamefont
  {Kimata}}, \bibinfo {author} {\bibfnamefont {H.}~\bibnamefont {Chen}},
  \bibinfo {author} {\bibfnamefont {K.}~\bibnamefont {Kondou}}, \bibinfo
  {author} {\bibfnamefont {S.}~\bibnamefont {Sugimoto}}, \bibinfo {author}
  {\bibfnamefont {P.~K.}\ \bibnamefont {Muduli}}, \bibinfo {author}
  {\bibfnamefont {M.}~\bibnamefont {Ikhlas}}, \bibinfo {author} {\bibfnamefont
  {Y.}~\bibnamefont {Omori}}, \bibinfo {author} {\bibfnamefont
  {T.}~\bibnamefont {Tomita}}, \bibinfo {author} {\bibfnamefont {A.~H.}\
  \bibnamefont {MacDonald}}, \bibinfo {author} {\bibfnamefont {S.}~\bibnamefont
  {Nakatsuji}}, \ and\ \bibinfo {author} {\bibfnamefont {Y.}~\bibnamefont
  {Otani}},\ }\href {\doibase 10.1038/s41586-018-0853-0} {\bibfield  {journal}
  {\bibinfo  {journal} {Nature}\ }\textbf {\bibinfo {volume} {565}},\ \bibinfo
  {pages} {627} (\bibinfo {year} {2019})}\BibitemShut {NoStop}%
\bibitem [{\citenamefont {Holanda}\ \emph {et~al.}(2020)\citenamefont
  {Holanda}, \citenamefont {Saglam}, \citenamefont {Karakas}, \citenamefont
  {Zang}, \citenamefont {Li}, \citenamefont {Divan}, \citenamefont {Liu},
  \citenamefont {Ozatay}, \citenamefont {Novosad}, \citenamefont {Pearson},\
  and\ \citenamefont {Hoffmann}}]{HolandaArxiv2019}%
  \BibitemOpen
  \bibfield  {author} {\bibinfo {author} {\bibfnamefont {J.}~\bibnamefont
  {Holanda}}, \bibinfo {author} {\bibfnamefont {H.}~\bibnamefont {Saglam}},
  \bibinfo {author} {\bibfnamefont {V.}~\bibnamefont {Karakas}}, \bibinfo
  {author} {\bibfnamefont {Z.}~\bibnamefont {Zang}}, \bibinfo {author}
  {\bibfnamefont {Y.}~\bibnamefont {Li}}, \bibinfo {author} {\bibfnamefont
  {R.}~\bibnamefont {Divan}}, \bibinfo {author} {\bibfnamefont
  {Y.}~\bibnamefont {Liu}}, \bibinfo {author} {\bibfnamefont {O.}~\bibnamefont
  {Ozatay}}, \bibinfo {author} {\bibfnamefont {V.}~\bibnamefont {Novosad}},
  \bibinfo {author} {\bibfnamefont {J.~E.}\ \bibnamefont {Pearson}}, \ and\
  \bibinfo {author} {\bibfnamefont {A.}~\bibnamefont {Hoffmann}},\ }\href
  {\doibase 10.1103/PhysRevLett.124.087204} {\enquote {\bibinfo {title}
  {Magnetic damping modulation in
  ${\mathrm{irmn}}_{3}/{\mathrm{ni}}_{80}{\mathrm{fe}}_{20}$ via the magnetic
  spin hall effect},}\ } (\bibinfo {year} {2020})\BibitemShut {NoStop}%
\bibitem [{\citenamefont {Liu}\ \emph {et~al.}(2019)\citenamefont {Liu},
  \citenamefont {Liu}, \citenamefont {Chen}, \citenamefont {Srivastava},
  \citenamefont {He}, \citenamefont {Teo}, \citenamefont {Phung}, \citenamefont
  {Yang},\ and\ \citenamefont {Yang}}]{LiuPRAppl2019}%
  \BibitemOpen
  \bibfield  {author} {\bibinfo {author} {\bibfnamefont {Y.}~\bibnamefont
  {Liu}}, \bibinfo {author} {\bibfnamefont {Y.}~\bibnamefont {Liu}}, \bibinfo
  {author} {\bibfnamefont {M.}~\bibnamefont {Chen}}, \bibinfo {author}
  {\bibfnamefont {S.}~\bibnamefont {Srivastava}}, \bibinfo {author}
  {\bibfnamefont {P.}~\bibnamefont {He}}, \bibinfo {author} {\bibfnamefont
  {K.~L.}\ \bibnamefont {Teo}}, \bibinfo {author} {\bibfnamefont
  {T.}~\bibnamefont {Phung}}, \bibinfo {author} {\bibfnamefont {S.-H.}\
  \bibnamefont {Yang}}, \ and\ \bibinfo {author} {\bibfnamefont
  {H.}~\bibnamefont {Yang}},\ }\href {\doibase
  10.1103/PhysRevApplied.12.064046} {\bibfield  {journal} {\bibinfo  {journal}
  {Phys. Rev. Appl.}\ }\textbf {\bibinfo {volume} {12}},\ \bibinfo {pages}
  {064046} (\bibinfo {year} {2019})}\BibitemShut {NoStop}%
\bibitem [{\citenamefont {Winkler}\ and\ \citenamefont
  {Zuelicke}(2019)}]{WinklerArxiv2019}%
  \BibitemOpen
  \bibfield  {author} {\bibinfo {author} {\bibfnamefont {R.}~\bibnamefont
  {Winkler}}\ and\ \bibinfo {author} {\bibfnamefont {U.}~\bibnamefont
  {Zuelicke}},\ }\href@noop {} {\enquote {\bibinfo {title} {Collinear orbital
  antiferromagnetic order and magnetoelectricity in quasi-2d itinerant-electron
  paramagnets, ferromagnets and antiferromagnets},}\ } (\bibinfo {year}
  {2019}),\ \Eprint {http://arxiv.org/abs/1912.09387} {arXiv:1912.09387
  [cond-mat.mes-hall]} \BibitemShut {NoStop}%
\bibitem [{\citenamefont {Seki}\ \emph {et~al.}(2015)\citenamefont {Seki},
  \citenamefont {Ideue}, \citenamefont {Kubota}, \citenamefont {Kozuka},
  \citenamefont {Takagi}, \citenamefont {Nakamura}, \citenamefont {Kaneko},
  \citenamefont {Kawasaki},\ and\ \citenamefont {Tokura}}]{SekiPRL2015}%
  \BibitemOpen
  \bibfield  {author} {\bibinfo {author} {\bibfnamefont {S.}~\bibnamefont
  {Seki}}, \bibinfo {author} {\bibfnamefont {T.}~\bibnamefont {Ideue}},
  \bibinfo {author} {\bibfnamefont {M.}~\bibnamefont {Kubota}}, \bibinfo
  {author} {\bibfnamefont {Y.}~\bibnamefont {Kozuka}}, \bibinfo {author}
  {\bibfnamefont {R.}~\bibnamefont {Takagi}}, \bibinfo {author} {\bibfnamefont
  {M.}~\bibnamefont {Nakamura}}, \bibinfo {author} {\bibfnamefont
  {Y.}~\bibnamefont {Kaneko}}, \bibinfo {author} {\bibfnamefont
  {M.}~\bibnamefont {Kawasaki}}, \ and\ \bibinfo {author} {\bibfnamefont
  {Y.}~\bibnamefont {Tokura}},\ }\href {\doibase
  10.1103/PhysRevLett.115.266601} {\bibfield  {journal} {\bibinfo  {journal}
  {Phys. Rev. Lett.}\ }\textbf {\bibinfo {volume} {115}},\ \bibinfo {pages}
  {266601} (\bibinfo {year} {2015})}\BibitemShut {NoStop}%
\bibitem [{\citenamefont {Wu}\ \emph {et~al.}(2016)\citenamefont {Wu},
  \citenamefont {Zhang}, \citenamefont {KC}, \citenamefont {Borisov},
  \citenamefont {Pearson}, \citenamefont {Jiang}, \citenamefont {Lederman},
  \citenamefont {Hoffmann},\ and\ \citenamefont {Bhattacharya}}]{WuPRL2019}%
  \BibitemOpen
  \bibfield  {author} {\bibinfo {author} {\bibfnamefont {S.~M.}\ \bibnamefont
  {Wu}}, \bibinfo {author} {\bibfnamefont {W.}~\bibnamefont {Zhang}}, \bibinfo
  {author} {\bibfnamefont {A.}~\bibnamefont {KC}}, \bibinfo {author}
  {\bibfnamefont {P.}~\bibnamefont {Borisov}}, \bibinfo {author} {\bibfnamefont
  {J.~E.}\ \bibnamefont {Pearson}}, \bibinfo {author} {\bibfnamefont {J.~S.}\
  \bibnamefont {Jiang}}, \bibinfo {author} {\bibfnamefont {D.}~\bibnamefont
  {Lederman}}, \bibinfo {author} {\bibfnamefont {A.}~\bibnamefont {Hoffmann}},
  \ and\ \bibinfo {author} {\bibfnamefont {A.}~\bibnamefont {Bhattacharya}},\
  }\href {\doibase 10.1103/PhysRevLett.116.097204} {\bibfield  {journal}
  {\bibinfo  {journal} {Phys. Rev. Lett.}\ }\textbf {\bibinfo {volume} {116}},\
  \bibinfo {pages} {097204} (\bibinfo {year} {2016})}\BibitemShut {NoStop}%
\bibitem [{\citenamefont {Rezende}, \citenamefont {Azevedo},\ and\
  \citenamefont {Rodríguez-Suárez}(2019)}]{RezendeJAP2019}%
  \BibitemOpen
  \bibfield  {author} {\bibinfo {author} {\bibfnamefont {S.~M.}\ \bibnamefont
  {Rezende}}, \bibinfo {author} {\bibfnamefont {A.}~\bibnamefont {Azevedo}}, \
  and\ \bibinfo {author} {\bibfnamefont {R.~L.}\ \bibnamefont
  {Rodríguez-Suárez}},\ }\href {\doibase 10.1063/1.5109132} {\bibfield
  {journal} {\bibinfo  {journal} {J. Appl. Phys.}\ }\textbf {\bibinfo {volume}
  {126}},\ \bibinfo {pages} {151101} (\bibinfo {year} {2019})}\BibitemShut
  {NoStop}%
\bibitem [{\citenamefont {Saglam}\ \emph {et~al.}(2016)\citenamefont {Saglam},
  \citenamefont {Zhang}, \citenamefont {Jungfleisch}, \citenamefont {Sklenar},
  \citenamefont {Pearson}, \citenamefont {Ketterson},\ and\ \citenamefont
  {Hoffmann}}]{SaglamPRB2016}%
  \BibitemOpen
  \bibfield  {author} {\bibinfo {author} {\bibfnamefont {H.}~\bibnamefont
  {Saglam}}, \bibinfo {author} {\bibfnamefont {W.}~\bibnamefont {Zhang}},
  \bibinfo {author} {\bibfnamefont {M.~B.}\ \bibnamefont {Jungfleisch}},
  \bibinfo {author} {\bibfnamefont {J.}~\bibnamefont {Sklenar}}, \bibinfo
  {author} {\bibfnamefont {J.~E.}\ \bibnamefont {Pearson}}, \bibinfo {author}
  {\bibfnamefont {J.~B.}\ \bibnamefont {Ketterson}}, \ and\ \bibinfo {author}
  {\bibfnamefont {A.}~\bibnamefont {Hoffmann}},\ }\href {\doibase
  10.1103/PhysRevB.94.140412} {\bibfield  {journal} {\bibinfo  {journal} {Phys.
  Rev. B}\ }\textbf {\bibinfo {volume} {94}},\ \bibinfo {pages} {140412}
  (\bibinfo {year} {2016})}\BibitemShut {NoStop}%
\bibitem [{\citenamefont {Gladii}\ \emph {et~al.}(2018)\citenamefont {Gladii},
  \citenamefont {Frangou}, \citenamefont {Forestier}, \citenamefont {Seeger},
  \citenamefont {Auffret}, \citenamefont {Joumard}, \citenamefont {Rubio-Roy},
  \citenamefont {Gambarelli},\ and\ \citenamefont {Baltz}}]{GladiiPRB2018}%
  \BibitemOpen
  \bibfield  {author} {\bibinfo {author} {\bibfnamefont {O.}~\bibnamefont
  {Gladii}}, \bibinfo {author} {\bibfnamefont {L.}~\bibnamefont {Frangou}},
  \bibinfo {author} {\bibfnamefont {G.}~\bibnamefont {Forestier}}, \bibinfo
  {author} {\bibfnamefont {R.~L.}\ \bibnamefont {Seeger}}, \bibinfo {author}
  {\bibfnamefont {S.}~\bibnamefont {Auffret}}, \bibinfo {author} {\bibfnamefont
  {I.}~\bibnamefont {Joumard}}, \bibinfo {author} {\bibfnamefont
  {M.}~\bibnamefont {Rubio-Roy}}, \bibinfo {author} {\bibfnamefont
  {S.}~\bibnamefont {Gambarelli}}, \ and\ \bibinfo {author} {\bibfnamefont
  {V.}~\bibnamefont {Baltz}},\ }\href {\doibase 10.1103/PhysRevB.98.094422}
  {\bibfield  {journal} {\bibinfo  {journal} {Phys. Rev. B}\ }\textbf {\bibinfo
  {volume} {98}},\ \bibinfo {pages} {094422} (\bibinfo {year}
  {2018})}\BibitemShut {NoStop}%
\bibitem [{\citenamefont {Huang}\ \emph {et~al.}(2011)\citenamefont {Huang},
  \citenamefont {Wang}, \citenamefont {Lee}, \citenamefont {Kwo},\ and\
  \citenamefont {Chien}}]{HuangPRL2011}%
  \BibitemOpen
  \bibfield  {author} {\bibinfo {author} {\bibfnamefont {S.~Y.}\ \bibnamefont
  {Huang}}, \bibinfo {author} {\bibfnamefont {W.~G.}\ \bibnamefont {Wang}},
  \bibinfo {author} {\bibfnamefont {S.~F.}\ \bibnamefont {Lee}}, \bibinfo
  {author} {\bibfnamefont {J.}~\bibnamefont {Kwo}}, \ and\ \bibinfo {author}
  {\bibfnamefont {C.~L.}\ \bibnamefont {Chien}},\ }\href {\doibase
  10.1103/PhysRevLett.107.216604} {\bibfield  {journal} {\bibinfo  {journal}
  {Phys. Rev. Lett.}\ }\textbf {\bibinfo {volume} {107}},\ \bibinfo {pages}
  {216604} (\bibinfo {year} {2011})}\BibitemShut {NoStop}%
\bibitem [{\citenamefont {Ikhlas}\ \emph {et~al.}(2017)\citenamefont {Ikhlas},
  \citenamefont {Tomita}, \citenamefont {Koretsune}, \citenamefont {Suzuki},
  \citenamefont {Nishio-Hamane}, \citenamefont {Arita}, \citenamefont {Otani},\
  and\ \citenamefont {Nakatsuji}}]{Ikhlas2017}%
  \BibitemOpen
  \bibfield  {author} {\bibinfo {author} {\bibfnamefont {M.}~\bibnamefont
  {Ikhlas}}, \bibinfo {author} {\bibfnamefont {T.}~\bibnamefont {Tomita}},
  \bibinfo {author} {\bibfnamefont {T.}~\bibnamefont {Koretsune}}, \bibinfo
  {author} {\bibfnamefont {M.~T.}\ \bibnamefont {Suzuki}}, \bibinfo {author}
  {\bibfnamefont {D.}~\bibnamefont {Nishio-Hamane}}, \bibinfo {author}
  {\bibfnamefont {R.}~\bibnamefont {Arita}}, \bibinfo {author} {\bibfnamefont
  {Y.}~\bibnamefont {Otani}}, \ and\ \bibinfo {author} {\bibfnamefont
  {S.}~\bibnamefont {Nakatsuji}},\ }\href@noop {} {\bibfield  {journal}
  {\bibinfo  {journal} {Nat. Phys.}\ }\textbf {\bibinfo {volume} {13}},\
  \bibinfo {pages} {1085} (\bibinfo {year} {2017})}\BibitemShut {NoStop}%
\bibitem [{\citenamefont {Cheng}\ \emph {et~al.}(2014)\citenamefont {Cheng},
  \citenamefont {Xiao}, \citenamefont {Niu},\ and\ \citenamefont
  {Brataas}}]{ChengPRL2014}%
  \BibitemOpen
  \bibfield  {author} {\bibinfo {author} {\bibfnamefont {R.}~\bibnamefont
  {Cheng}}, \bibinfo {author} {\bibfnamefont {J.}~\bibnamefont {Xiao}},
  \bibinfo {author} {\bibfnamefont {Q.}~\bibnamefont {Niu}}, \ and\ \bibinfo
  {author} {\bibfnamefont {A.}~\bibnamefont {Brataas}},\ }\href {\doibase
  10.1103/PhysRevLett.113.057601} {\bibfield  {journal} {\bibinfo  {journal}
  {Phys. Rev. Lett.}\ }\textbf {\bibinfo {volume} {113}},\ \bibinfo {pages}
  {057601} (\bibinfo {year} {2014})}\BibitemShut {NoStop}%
\bibitem [{\citenamefont {Tserkovnyak}, \citenamefont {Brataas},\ and\
  \citenamefont {Bauer}(2002)}]{TserkovnyakPRL2002}%
  \BibitemOpen
  \bibfield  {author} {\bibinfo {author} {\bibfnamefont {Y.}~\bibnamefont
  {Tserkovnyak}}, \bibinfo {author} {\bibfnamefont {A.}~\bibnamefont
  {Brataas}}, \ and\ \bibinfo {author} {\bibfnamefont {G.~E.~W.}\ \bibnamefont
  {Bauer}},\ }\href {\doibase 10.1103/PhysRevLett.88.117601} {\bibfield
  {journal} {\bibinfo  {journal} {Phys. Rev. Lett.}\ }\textbf {\bibinfo
  {volume} {88}},\ \bibinfo {pages} {117601} (\bibinfo {year}
  {2002})}\BibitemShut {NoStop}%
\bibitem [{\citenamefont {Heinrich}\ \emph {et~al.}(2003)\citenamefont
  {Heinrich}, \citenamefont {Tserkovnyak}, \citenamefont {Woltersdorf},
  \citenamefont {Brataas}, \citenamefont {Urban},\ and\ \citenamefont
  {Bauer}}]{HeinrichPRL2003}%
  \BibitemOpen
  \bibfield  {author} {\bibinfo {author} {\bibfnamefont {B.}~\bibnamefont
  {Heinrich}}, \bibinfo {author} {\bibfnamefont {Y.}~\bibnamefont
  {Tserkovnyak}}, \bibinfo {author} {\bibfnamefont {G.}~\bibnamefont
  {Woltersdorf}}, \bibinfo {author} {\bibfnamefont {A.}~\bibnamefont
  {Brataas}}, \bibinfo {author} {\bibfnamefont {R.}~\bibnamefont {Urban}}, \
  and\ \bibinfo {author} {\bibfnamefont {G.~E.~W.}\ \bibnamefont {Bauer}},\
  }\href {\doibase 10.1103/PhysRevLett.90.187601} {\bibfield  {journal}
  {\bibinfo  {journal} {Phys. Rev. Lett.}\ }\textbf {\bibinfo {volume} {90}},\
  \bibinfo {pages} {187601} (\bibinfo {year} {2003})}\BibitemShut {NoStop}%
\bibitem [{\citenamefont {Li}\ \emph {et~al.}(2020{\natexlab{a}})\citenamefont
  {Li}, \citenamefont {Wilson}, \citenamefont {Cheng}, \citenamefont {Lohmann},
  \citenamefont {Kavand}, \citenamefont {Yuan}, \citenamefont {Aldosary},
  \citenamefont {Agladze}, \citenamefont {Wei}, \citenamefont {Sherwin},\ and\
  \citenamefont {Shi}}]{Li2020Nature}%
  \BibitemOpen
  \bibfield  {author} {\bibinfo {author} {\bibfnamefont {J.}~\bibnamefont
  {Li}}, \bibinfo {author} {\bibfnamefont {B.}~\bibnamefont {Wilson}}, \bibinfo
  {author} {\bibfnamefont {R.}~\bibnamefont {Cheng}}, \bibinfo {author}
  {\bibfnamefont {M.}~\bibnamefont {Lohmann}}, \bibinfo {author} {\bibfnamefont
  {M.}~\bibnamefont {Kavand}}, \bibinfo {author} {\bibfnamefont
  {W.}~\bibnamefont {Yuan}}, \bibinfo {author} {\bibfnamefont {M.}~\bibnamefont
  {Aldosary}}, \bibinfo {author} {\bibfnamefont {N.}~\bibnamefont {Agladze}},
  \bibinfo {author} {\bibfnamefont {P.}~\bibnamefont {Wei}}, \bibinfo {author}
  {\bibfnamefont {M.~S.}\ \bibnamefont {Sherwin}}, \ and\ \bibinfo {author}
  {\bibfnamefont {J.}~\bibnamefont {Shi}},\ }\href@noop {} {\bibfield
  {journal} {\bibinfo  {journal} {Nature}\ }\textbf {\bibinfo {volume} {578}},\
  \bibinfo {pages} {70} (\bibinfo {year} {2020}{\natexlab{a}})}\BibitemShut
  {NoStop}%
\bibitem [{\citenamefont {Gomonay}\ and\ \citenamefont
  {Loktev}(2008)}]{GomonayJMSJ2008}%
  \BibitemOpen
  \bibfield  {author} {\bibinfo {author} {\bibfnamefont {H.}~\bibnamefont
  {Gomonay}}\ and\ \bibinfo {author} {\bibfnamefont {V.}~\bibnamefont
  {Loktev}},\ }\href {\doibase 10.3379/msjmag.32.535} {\bibfield  {journal}
  {\bibinfo  {journal} {J. Magn. Soc. Japan}\ }\textbf {\bibinfo {volume}
  {32}},\ \bibinfo {pages} {535} (\bibinfo {year} {2008})}\BibitemShut
  {NoStop}%
\bibitem [{\citenamefont {Cheng}\ \emph {et~al.}(2015)\citenamefont {Cheng},
  \citenamefont {Daniels}, \citenamefont {Zhu},\ and\ \citenamefont
  {Xiao}}]{ChengPRB2015}%
  \BibitemOpen
  \bibfield  {author} {\bibinfo {author} {\bibfnamefont {R.}~\bibnamefont
  {Cheng}}, \bibinfo {author} {\bibfnamefont {M.~W.}\ \bibnamefont {Daniels}},
  \bibinfo {author} {\bibfnamefont {J.-G.}\ \bibnamefont {Zhu}}, \ and\
  \bibinfo {author} {\bibfnamefont {D.}~\bibnamefont {Xiao}},\ }\href {\doibase
  10.1103/PhysRevB.91.064423} {\bibfield  {journal} {\bibinfo  {journal} {Phys.
  Rev. B}\ }\textbf {\bibinfo {volume} {91}},\ \bibinfo {pages} {064423}
  (\bibinfo {year} {2015})}\BibitemShut {NoStop}%
\bibitem [{\citenamefont {Khymyn}\ \emph {et~al.}(2017)\citenamefont {Khymyn},
  \citenamefont {Lisenkov}, \citenamefont {Tiberkevich}, \citenamefont
  {Ivanov},\ and\ \citenamefont {Slavin}}]{KhymynSciRep2017}%
  \BibitemOpen
  \bibfield  {author} {\bibinfo {author} {\bibfnamefont {R.}~\bibnamefont
  {Khymyn}}, \bibinfo {author} {\bibfnamefont {I.}~\bibnamefont {Lisenkov}},
  \bibinfo {author} {\bibfnamefont {V.}~\bibnamefont {Tiberkevich}}, \bibinfo
  {author} {\bibfnamefont {B.~A.}\ \bibnamefont {Ivanov}}, \ and\ \bibinfo
  {author} {\bibfnamefont {A.}~\bibnamefont {Slavin}},\ }\href {\doibase
  10.1038/srep43705} {\bibfield  {journal} {\bibinfo  {journal} {Sci. Rep.}\
  }\textbf {\bibinfo {volume} {7}},\ \bibinfo {pages} {43705} (\bibinfo {year}
  {2017})}\BibitemShut {NoStop}%
\bibitem [{\citenamefont {Acharyya}\ \emph {et~al.}(2011)\citenamefont
  {Acharyya}, \citenamefont {Nguyen}, \citenamefont {Pratt},\ and\
  \citenamefont {Bass}}]{AcharuyyaJAP2011}%
  \BibitemOpen
  \bibfield  {author} {\bibinfo {author} {\bibfnamefont {R.}~\bibnamefont
  {Acharyya}}, \bibinfo {author} {\bibfnamefont {H.~Y.~T.}\ \bibnamefont
  {Nguyen}}, \bibinfo {author} {\bibfnamefont {W.~P.}\ \bibnamefont {Pratt}}, \
  and\ \bibinfo {author} {\bibfnamefont {J.}~\bibnamefont {Bass}},\ }\href
  {\doibase 10.1063/1.3535340} {\bibfield  {journal} {\bibinfo  {journal} {J.
  Appl. Phys.}\ }\textbf {\bibinfo {volume} {109}},\ \bibinfo {pages} {07C503}
  (\bibinfo {year} {2011})}\BibitemShut {NoStop}%
\bibitem [{\citenamefont {Merodio}\ \emph {et~al.}(2014)\citenamefont
  {Merodio}, \citenamefont {Ghosh}, \citenamefont {Lemonias}, \citenamefont
  {Gautier}, \citenamefont {Ebels}, \citenamefont {Chshiev}, \citenamefont
  {Béa}, \citenamefont {Baltz},\ and\ \citenamefont
  {Bailey}}]{MerodioAPL2014}%
  \BibitemOpen
  \bibfield  {author} {\bibinfo {author} {\bibfnamefont {P.}~\bibnamefont
  {Merodio}}, \bibinfo {author} {\bibfnamefont {A.}~\bibnamefont {Ghosh}},
  \bibinfo {author} {\bibfnamefont {C.}~\bibnamefont {Lemonias}}, \bibinfo
  {author} {\bibfnamefont {E.}~\bibnamefont {Gautier}}, \bibinfo {author}
  {\bibfnamefont {U.}~\bibnamefont {Ebels}}, \bibinfo {author} {\bibfnamefont
  {M.}~\bibnamefont {Chshiev}}, \bibinfo {author} {\bibfnamefont
  {H.}~\bibnamefont {Béa}}, \bibinfo {author} {\bibfnamefont {V.}~\bibnamefont
  {Baltz}}, \ and\ \bibinfo {author} {\bibfnamefont {W.~E.}\ \bibnamefont
  {Bailey}},\ }\href {\doibase 10.1063/1.4862971} {\bibfield  {journal}
  {\bibinfo  {journal} {Appl. Phys. Lett.}\ }\textbf {\bibinfo {volume}
  {104}},\ \bibinfo {pages} {032406} (\bibinfo {year} {2014})}\BibitemShut
  {NoStop}%
\bibitem [{\citenamefont {Reichlov\'a}\ \emph {et~al.}(2015)\citenamefont
  {Reichlov\'a}, \citenamefont {Kriegner}, \citenamefont {Hol\'y},
  \citenamefont {Olejn\'{\i}k}, \citenamefont {Nov\'ak}, \citenamefont
  {Yamada}, \citenamefont {Miura}, \citenamefont {Ogawa}, \citenamefont
  {Takahashi}, \citenamefont {Jungwirth},\ and\ \citenamefont
  {Wunderlich}}]{ReichlovaPRB2015}%
  \BibitemOpen
  \bibfield  {author} {\bibinfo {author} {\bibfnamefont {H.}~\bibnamefont
  {Reichlov\'a}}, \bibinfo {author} {\bibfnamefont {D.}~\bibnamefont
  {Kriegner}}, \bibinfo {author} {\bibfnamefont {V.}~\bibnamefont {Hol\'y}},
  \bibinfo {author} {\bibfnamefont {K.}~\bibnamefont {Olejn\'{\i}k}}, \bibinfo
  {author} {\bibfnamefont {V.}~\bibnamefont {Nov\'ak}}, \bibinfo {author}
  {\bibfnamefont {M.}~\bibnamefont {Yamada}}, \bibinfo {author} {\bibfnamefont
  {K.}~\bibnamefont {Miura}}, \bibinfo {author} {\bibfnamefont
  {S.}~\bibnamefont {Ogawa}}, \bibinfo {author} {\bibfnamefont
  {H.}~\bibnamefont {Takahashi}}, \bibinfo {author} {\bibfnamefont
  {T.}~\bibnamefont {Jungwirth}}, \ and\ \bibinfo {author} {\bibfnamefont
  {J.}~\bibnamefont {Wunderlich}},\ }\href {\doibase
  10.1103/PhysRevB.92.165424} {\bibfield  {journal} {\bibinfo  {journal} {Phys.
  Rev. B}\ }\textbf {\bibinfo {volume} {92}},\ \bibinfo {pages} {165424}
  (\bibinfo {year} {2015})}\BibitemShut {NoStop}%
\bibitem [{\citenamefont {Shick}\ \emph {et~al.}(2010)\citenamefont {Shick},
  \citenamefont {Khmelevskyi}, \citenamefont {Mryasov}, \citenamefont
  {Wunderlich},\ and\ \citenamefont {Jungwirth}}]{ShickPRB2010}%
  \BibitemOpen
  \bibfield  {author} {\bibinfo {author} {\bibfnamefont {A.~B.}\ \bibnamefont
  {Shick}}, \bibinfo {author} {\bibfnamefont {S.}~\bibnamefont {Khmelevskyi}},
  \bibinfo {author} {\bibfnamefont {O.~N.}\ \bibnamefont {Mryasov}}, \bibinfo
  {author} {\bibfnamefont {J.}~\bibnamefont {Wunderlich}}, \ and\ \bibinfo
  {author} {\bibfnamefont {T.}~\bibnamefont {Jungwirth}},\ }\href {\doibase
  10.1103/PhysRevB.81.212409} {\bibfield  {journal} {\bibinfo  {journal} {Phys.
  Rev. B}\ }\textbf {\bibinfo {volume} {81}},\ \bibinfo {pages} {212409}
  (\bibinfo {year} {2010})}\BibitemShut {NoStop}%
\bibitem [{\citenamefont {\ifmmode~\check{Z}\else \v{Z}\fi{}elezn\'y}\ \emph
  {et~al.}(2014)\citenamefont {\ifmmode~\check{Z}\else \v{Z}\fi{}elezn\'y},
  \citenamefont {Gao}, \citenamefont {V\'yborn\'y}, \citenamefont {Zemen},
  \citenamefont {Ma\ifmmode~\check{s}\else \v{s}\fi{}ek}, \citenamefont
  {Manchon}, \citenamefont {Wunderlich}, \citenamefont {Sinova},\ and\
  \citenamefont {Jungwirth}}]{ZeleznyPRL2014}%
  \BibitemOpen
  \bibfield  {author} {\bibinfo {author} {\bibfnamefont {J.}~\bibnamefont
  {\ifmmode~\check{Z}\else \v{Z}\fi{}elezn\'y}}, \bibinfo {author}
  {\bibfnamefont {H.}~\bibnamefont {Gao}}, \bibinfo {author} {\bibfnamefont
  {K.}~\bibnamefont {V\'yborn\'y}}, \bibinfo {author} {\bibfnamefont
  {J.}~\bibnamefont {Zemen}}, \bibinfo {author} {\bibfnamefont
  {J.}~\bibnamefont {Ma\ifmmode~\check{s}\else \v{s}\fi{}ek}}, \bibinfo
  {author} {\bibfnamefont {A.}~\bibnamefont {Manchon}}, \bibinfo {author}
  {\bibfnamefont {J.}~\bibnamefont {Wunderlich}}, \bibinfo {author}
  {\bibfnamefont {J.}~\bibnamefont {Sinova}}, \ and\ \bibinfo {author}
  {\bibfnamefont {T.}~\bibnamefont {Jungwirth}},\ }\href {\doibase
  10.1103/PhysRevLett.113.157201} {\bibfield  {journal} {\bibinfo  {journal}
  {Phys. Rev. Lett.}\ }\textbf {\bibinfo {volume} {113}},\ \bibinfo {pages}
  {157201} (\bibinfo {year} {2014})}\BibitemShut {NoStop}%
\bibitem [{\citenamefont {Barthem}\ \emph {et~al.}(2013)\citenamefont
  {Barthem}, \citenamefont {Colin}, \citenamefont {H.Mayaffre}, \citenamefont
  {Julien},\ and\ \citenamefont {Givord}}]{BarthemNatComm2013}%
  \BibitemOpen
  \bibfield  {author} {\bibinfo {author} {\bibfnamefont {V.}~\bibnamefont
  {Barthem}}, \bibinfo {author} {\bibfnamefont {C.}~\bibnamefont {Colin}},
  \bibinfo {author} {\bibnamefont {H.Mayaffre}}, \bibinfo {author}
  {\bibfnamefont {M.-H.}\ \bibnamefont {Julien}}, \ and\ \bibinfo {author}
  {\bibfnamefont {D.}~\bibnamefont {Givord}},\ }\href {\doibase
  10.1038/ncomms3892} {\bibfield  {journal} {\bibinfo  {journal} {Nat. Comm.}\
  }\textbf {\bibinfo {volume} {4}},\ \bibinfo {pages} {2892} (\bibinfo {year}
  {2013})}\BibitemShut {NoStop}%
\bibitem [{\citenamefont {Meinert}, \citenamefont {Graulich},\ and\
  \citenamefont {Matalla-Wagner}(2018)}]{MeinertPRAppl2018}%
  \BibitemOpen
  \bibfield  {author} {\bibinfo {author} {\bibfnamefont {M.}~\bibnamefont
  {Meinert}}, \bibinfo {author} {\bibfnamefont {D.}~\bibnamefont {Graulich}}, \
  and\ \bibinfo {author} {\bibfnamefont {T.}~\bibnamefont {Matalla-Wagner}},\
  }\href {\doibase 10.1103/PhysRevApplied.9.064040} {\bibfield  {journal}
  {\bibinfo  {journal} {Phys. Rev. Appl.}\ }\textbf {\bibinfo {volume} {9}},\
  \bibinfo {pages} {064040} (\bibinfo {year} {2018})}\BibitemShut {NoStop}%
\bibitem [{\citenamefont {Lançon}\ \emph {et~al.}(2016)\citenamefont
  {Lançon}, \citenamefont {Walker}, \citenamefont {Ressouche}, \citenamefont
  {Ouladdiaf}, \citenamefont {Rule}, \citenamefont {Mcintyre}, \citenamefont
  {Hicks}, \citenamefont {Rønnow},\ and\ \citenamefont
  {Wildes}}]{lancon_PRB_2016}%
  \BibitemOpen
  \bibfield  {author} {\bibinfo {author} {\bibfnamefont {D.}~\bibnamefont
  {Lançon}}, \bibinfo {author} {\bibfnamefont {H.~C.}\ \bibnamefont {Walker}},
  \bibinfo {author} {\bibfnamefont {E.}~\bibnamefont {Ressouche}}, \bibinfo
  {author} {\bibfnamefont {B.}~\bibnamefont {Ouladdiaf}}, \bibinfo {author}
  {\bibfnamefont {K.~C.}\ \bibnamefont {Rule}}, \bibinfo {author}
  {\bibfnamefont {G.~J.}\ \bibnamefont {Mcintyre}}, \bibinfo {author}
  {\bibfnamefont {T.~J.}\ \bibnamefont {Hicks}}, \bibinfo {author}
  {\bibfnamefont {H.~M.}\ \bibnamefont {Rønnow}}, \ and\ \bibinfo {author}
  {\bibfnamefont {A.~R.}\ \bibnamefont {Wildes}},\ }\href {\doibase
  10.1103/physrevb.94.214407} {\bibfield  {journal} {\bibinfo  {journal} {Phys.
  Rev. B}\ }\textbf {\bibinfo {volume} {94}},\ \bibinfo {pages} {214407}
  (\bibinfo {year} {2016})}\BibitemShut {NoStop}%
\bibitem [{\citenamefont {Kim}\ \emph {et~al.}(2019{\natexlab{a}})\citenamefont
  {Kim}, \citenamefont {Yang}, \citenamefont {Li}, \citenamefont {Jiang},
  \citenamefont {Jin}, \citenamefont {Tao}, \citenamefont {Nichols},
  \citenamefont {Sfigakis}, \citenamefont {Zhong}, \citenamefont {Li},\ and\
  \citenamefont {et~al.}}]{kimPNAS2019}%
  \BibitemOpen
  \bibfield  {author} {\bibinfo {author} {\bibfnamefont {H.~H.}\ \bibnamefont
  {Kim}}, \bibinfo {author} {\bibfnamefont {B.}~\bibnamefont {Yang}}, \bibinfo
  {author} {\bibfnamefont {S.}~\bibnamefont {Li}}, \bibinfo {author}
  {\bibfnamefont {S.}~\bibnamefont {Jiang}}, \bibinfo {author} {\bibfnamefont
  {C.}~\bibnamefont {Jin}}, \bibinfo {author} {\bibfnamefont {Z.}~\bibnamefont
  {Tao}}, \bibinfo {author} {\bibfnamefont {G.}~\bibnamefont {Nichols}},
  \bibinfo {author} {\bibfnamefont {F.}~\bibnamefont {Sfigakis}}, \bibinfo
  {author} {\bibfnamefont {S.}~\bibnamefont {Zhong}}, \bibinfo {author}
  {\bibfnamefont {C.}~\bibnamefont {Li}}, \ and\ \bibinfo {author}
  {\bibnamefont {et~al.}},\ }\href {\doibase 10.1073/pnas.1902100116}
  {\bibfield  {journal} {\bibinfo  {journal} {Proc. Natl. Acad. Sci.}\ }\textbf
  {\bibinfo {volume} {116}},\ \bibinfo {pages} {11131} (\bibinfo {year}
  {2019}{\natexlab{a}})}\BibitemShut {NoStop}%
\bibitem [{\citenamefont {Gong}\ \emph {et~al.}(2017)\citenamefont {Gong},
  \citenamefont {Li}, \citenamefont {Li}, \citenamefont {Ji}, \citenamefont
  {Stern}, \citenamefont {Xia}, \citenamefont {Cao}, \citenamefont {Bao},
  \citenamefont {Wang}, \citenamefont {Wang} \emph
  {et~al.}}]{gong2017discovery}%
  \BibitemOpen
  \bibfield  {author} {\bibinfo {author} {\bibfnamefont {C.}~\bibnamefont
  {Gong}}, \bibinfo {author} {\bibfnamefont {L.}~\bibnamefont {Li}}, \bibinfo
  {author} {\bibfnamefont {Z.}~\bibnamefont {Li}}, \bibinfo {author}
  {\bibfnamefont {H.}~\bibnamefont {Ji}}, \bibinfo {author} {\bibfnamefont
  {A.}~\bibnamefont {Stern}}, \bibinfo {author} {\bibfnamefont
  {Y.}~\bibnamefont {Xia}}, \bibinfo {author} {\bibfnamefont {T.}~\bibnamefont
  {Cao}}, \bibinfo {author} {\bibfnamefont {W.}~\bibnamefont {Bao}}, \bibinfo
  {author} {\bibfnamefont {C.}~\bibnamefont {Wang}}, \bibinfo {author}
  {\bibfnamefont {Y.}~\bibnamefont {Wang}},  \emph {et~al.},\ }\href@noop {}
  {\bibfield  {journal} {\bibinfo  {journal} {Nature}\ }\textbf {\bibinfo
  {volume} {546}},\ \bibinfo {pages} {265} (\bibinfo {year}
  {2017})}\BibitemShut {NoStop}%
\bibitem [{\citenamefont {Wang}\ \emph {et~al.}(2016)\citenamefont {Wang},
  \citenamefont {Du}, \citenamefont {Liu}, \citenamefont {Hu}, \citenamefont
  {Zhang}, \citenamefont {Zhang}, \citenamefont {Owen}, \citenamefont {Lu},
  \citenamefont {Gan}, \citenamefont {Sengupta},\ and\ \citenamefont
  {et~al.}}]{wang_2Dmaterials_2016}%
  \BibitemOpen
  \bibfield  {author} {\bibinfo {author} {\bibfnamefont {X.}~\bibnamefont
  {Wang}}, \bibinfo {author} {\bibfnamefont {K.}~\bibnamefont {Du}}, \bibinfo
  {author} {\bibfnamefont {Y.~Y.~F.}\ \bibnamefont {Liu}}, \bibinfo {author}
  {\bibfnamefont {P.}~\bibnamefont {Hu}}, \bibinfo {author} {\bibfnamefont
  {J.}~\bibnamefont {Zhang}}, \bibinfo {author} {\bibfnamefont
  {Q.}~\bibnamefont {Zhang}}, \bibinfo {author} {\bibfnamefont {M.~H.~S.}\
  \bibnamefont {Owen}}, \bibinfo {author} {\bibfnamefont {X.}~\bibnamefont
  {Lu}}, \bibinfo {author} {\bibfnamefont {C.~K.}\ \bibnamefont {Gan}},
  \bibinfo {author} {\bibfnamefont {P.}~\bibnamefont {Sengupta}}, \ and\
  \bibinfo {author} {\bibnamefont {et~al.}},\ }\href {\doibase
  10.1088/2053-1583/3/3/031009} {\bibfield  {journal} {\bibinfo  {journal} {2D
  Mater.}\ }\textbf {\bibinfo {volume} {3}},\ \bibinfo {pages} {031009}
  (\bibinfo {year} {2016})}\BibitemShut {NoStop}%
\bibitem [{\citenamefont {Kim}\ \emph {et~al.}(2019{\natexlab{b}})\citenamefont
  {Kim}, \citenamefont {Lim}, \citenamefont {Kim}, \citenamefont {Lee},
  \citenamefont {Lee}, \citenamefont {Kim}, \citenamefont {Park}, \citenamefont
  {Son}, \citenamefont {Park}, \citenamefont {Park},\ and\ \citenamefont
  {et~al.}}]{kim_2Dmaterials_2019}%
  \BibitemOpen
  \bibfield  {author} {\bibinfo {author} {\bibfnamefont {K.}~\bibnamefont
  {Kim}}, \bibinfo {author} {\bibfnamefont {S.~Y.}\ \bibnamefont {Lim}},
  \bibinfo {author} {\bibfnamefont {J.}~\bibnamefont {Kim}}, \bibinfo {author}
  {\bibfnamefont {J.-U.}\ \bibnamefont {Lee}}, \bibinfo {author} {\bibfnamefont
  {S.}~\bibnamefont {Lee}}, \bibinfo {author} {\bibfnamefont {P.}~\bibnamefont
  {Kim}}, \bibinfo {author} {\bibfnamefont {K.}~\bibnamefont {Park}}, \bibinfo
  {author} {\bibfnamefont {S.}~\bibnamefont {Son}}, \bibinfo {author}
  {\bibfnamefont {C.-H.}\ \bibnamefont {Park}}, \bibinfo {author}
  {\bibfnamefont {J.-G.}\ \bibnamefont {Park}}, \ and\ \bibinfo {author}
  {\bibnamefont {et~al.}},\ }\href {\doibase 10.1088/2053-1583/ab27d5}
  {\bibfield  {journal} {\bibinfo  {journal} {2D Mater.}\ }\textbf {\bibinfo
  {volume} {6}},\ \bibinfo {pages} {041001} (\bibinfo {year}
  {2019}{\natexlab{b}})}\BibitemShut {NoStop}%
\bibitem [{\citenamefont {Ma}\ \emph {et~al.}(2012)\citenamefont {Ma},
  \citenamefont {Dai}, \citenamefont {Guo}, \citenamefont {Niu}, \citenamefont
  {Zhu},\ and\ \citenamefont {Huang}}]{ma_ACSNano_2012}%
  \BibitemOpen
  \bibfield  {author} {\bibinfo {author} {\bibfnamefont {Y.}~\bibnamefont
  {Ma}}, \bibinfo {author} {\bibfnamefont {Y.}~\bibnamefont {Dai}}, \bibinfo
  {author} {\bibfnamefont {M.}~\bibnamefont {Guo}}, \bibinfo {author}
  {\bibfnamefont {C.}~\bibnamefont {Niu}}, \bibinfo {author} {\bibfnamefont
  {Y.}~\bibnamefont {Zhu}}, \ and\ \bibinfo {author} {\bibfnamefont
  {B.}~\bibnamefont {Huang}},\ }\href {\doibase 10.1021/nn204667z} {\bibfield
  {journal} {\bibinfo  {journal} {ACS Nano}\ }\textbf {\bibinfo {volume} {6}},\
  \bibinfo {pages} {1695–1701} (\bibinfo {year} {2012})}\BibitemShut
  {NoStop}%
\bibitem [{\citenamefont {Weber}\ \emph {et~al.}(2016)\citenamefont {Weber},
  \citenamefont {Schoop}, \citenamefont {Duppel}, \citenamefont {Lippmann},
  \citenamefont {Nuss},\ and\ \citenamefont {Lotsch}}]{weber_NanoLett_2016}%
  \BibitemOpen
  \bibfield  {author} {\bibinfo {author} {\bibfnamefont {D.}~\bibnamefont
  {Weber}}, \bibinfo {author} {\bibfnamefont {L.~M.}\ \bibnamefont {Schoop}},
  \bibinfo {author} {\bibfnamefont {V.}~\bibnamefont {Duppel}}, \bibinfo
  {author} {\bibfnamefont {J.~M.}\ \bibnamefont {Lippmann}}, \bibinfo {author}
  {\bibfnamefont {J.}~\bibnamefont {Nuss}}, \ and\ \bibinfo {author}
  {\bibfnamefont {B.~V.}\ \bibnamefont {Lotsch}},\ }\href {\doibase
  10.1021/acs.nanolett.6b00701} {\bibfield  {journal} {\bibinfo  {journal}
  {Nano Lett.}\ }\textbf {\bibinfo {volume} {16}},\ \bibinfo {pages}
  {3578–3584} (\bibinfo {year} {2016})}\BibitemShut {NoStop}%
\bibitem [{\citenamefont {Du}\ \emph {et~al.}(2018)\citenamefont {Du},
  \citenamefont {Huang}, \citenamefont {Wang}, \citenamefont {Wang},
  \citenamefont {Yang}, \citenamefont {Tang}, \citenamefont {Liao},
  \citenamefont {Shi}, \citenamefont {Shi}, \citenamefont {Zhou},\ and\
  \citenamefont {et~al.}}]{du_2Dmaterials_2018}%
  \BibitemOpen
  \bibfield  {author} {\bibinfo {author} {\bibfnamefont {L.}~\bibnamefont
  {Du}}, \bibinfo {author} {\bibfnamefont {Y.}~\bibnamefont {Huang}}, \bibinfo
  {author} {\bibfnamefont {Y.}~\bibnamefont {Wang}}, \bibinfo {author}
  {\bibfnamefont {Q.}~\bibnamefont {Wang}}, \bibinfo {author} {\bibfnamefont
  {R.}~\bibnamefont {Yang}}, \bibinfo {author} {\bibfnamefont {J.}~\bibnamefont
  {Tang}}, \bibinfo {author} {\bibfnamefont {M.}~\bibnamefont {Liao}}, \bibinfo
  {author} {\bibfnamefont {D.}~\bibnamefont {Shi}}, \bibinfo {author}
  {\bibfnamefont {Y.}~\bibnamefont {Shi}}, \bibinfo {author} {\bibfnamefont
  {X.}~\bibnamefont {Zhou}}, \ and\ \bibinfo {author} {\bibnamefont {et~al.}},\
  }\href {\doibase 10.1088/2053-1583/aaee29} {\bibfield  {journal} {\bibinfo
  {journal} {2D Mater.}\ }\textbf {\bibinfo {volume} {6}},\ \bibinfo {pages}
  {015014} (\bibinfo {year} {2018})}\BibitemShut {NoStop}%
\bibitem [{\citenamefont {Zhou}\ \emph
  {et~al.}(2019{\natexlab{a}})\citenamefont {Zhou}, \citenamefont {Wang},
  \citenamefont {Osterhoudt}, \citenamefont {Lampen-Kelley}, \citenamefont
  {Mandrus}, \citenamefont {He}, \citenamefont {Burch},\ and\ \citenamefont
  {Henriksen}}]{zhou_JPCS_2019}%
  \BibitemOpen
  \bibfield  {author} {\bibinfo {author} {\bibfnamefont {B.}~\bibnamefont
  {Zhou}}, \bibinfo {author} {\bibfnamefont {Y.}~\bibnamefont {Wang}}, \bibinfo
  {author} {\bibfnamefont {G.~B.}\ \bibnamefont {Osterhoudt}}, \bibinfo
  {author} {\bibfnamefont {P.}~\bibnamefont {Lampen-Kelley}}, \bibinfo {author}
  {\bibfnamefont {D.}~\bibnamefont {Mandrus}}, \bibinfo {author} {\bibfnamefont
  {R.}~\bibnamefont {He}}, \bibinfo {author} {\bibfnamefont {K.~S.}\
  \bibnamefont {Burch}}, \ and\ \bibinfo {author} {\bibfnamefont {E.~A.}\
  \bibnamefont {Henriksen}},\ }\href {\doibase 10.1016/j.jpcs.2018.01.026}
  {\bibfield  {journal} {\bibinfo  {journal} {J. Phys. Chem. Solids}\ }\textbf
  {\bibinfo {volume} {128}},\ \bibinfo {pages} {291–295} (\bibinfo {year}
  {2019}{\natexlab{a}})}\BibitemShut {NoStop}%
\bibitem [{\citenamefont {Kurosawa}, \citenamefont {Saito},\ and\ \citenamefont
  {Yamaguchi}(1983)}]{kurosawa_JPSJ_1983}%
  \BibitemOpen
  \bibfield  {author} {\bibinfo {author} {\bibfnamefont {K.}~\bibnamefont
  {Kurosawa}}, \bibinfo {author} {\bibfnamefont {S.}~\bibnamefont {Saito}}, \
  and\ \bibinfo {author} {\bibfnamefont {Y.}~\bibnamefont {Yamaguchi}},\ }\href
  {\doibase 10.1143/jpsj.52.3919} {\bibfield  {journal} {\bibinfo  {journal}
  {J. Phys. Soc. Japan}\ }\textbf {\bibinfo {volume} {52}},\ \bibinfo {pages}
  {3919–3926} (\bibinfo {year} {1983})}\BibitemShut {NoStop}%
\bibitem [{\citenamefont {Lv}\ \emph {et~al.}(2015)\citenamefont {Lv},
  \citenamefont {Lu}, \citenamefont {Shao}, \citenamefont {Liu},\ and\
  \citenamefont {Sun}}]{LvPRB2015}%
  \BibitemOpen
  \bibfield  {author} {\bibinfo {author} {\bibfnamefont {H.~Y.}\ \bibnamefont
  {Lv}}, \bibinfo {author} {\bibfnamefont {W.~J.}\ \bibnamefont {Lu}}, \bibinfo
  {author} {\bibfnamefont {D.~F.}\ \bibnamefont {Shao}}, \bibinfo {author}
  {\bibfnamefont {Y.}~\bibnamefont {Liu}}, \ and\ \bibinfo {author}
  {\bibfnamefont {Y.~P.}\ \bibnamefont {Sun}},\ }\href {\doibase
  10.1103/PhysRevB.92.214419} {\bibfield  {journal} {\bibinfo  {journal} {Phys.
  Rev. B}\ }\textbf {\bibinfo {volume} {92}},\ \bibinfo {pages} {214419}
  (\bibinfo {year} {2015})}\BibitemShut {NoStop}%
\bibitem [{\citenamefont {Kumar}\ \emph {et~al.}(2017)\citenamefont {Kumar},
  \citenamefont {Frey}, \citenamefont {Dong}, \citenamefont {Anasori},
  \citenamefont {Gogotsi},\ and\ \citenamefont {Shenoy}}]{HemantACSNano2017}%
  \BibitemOpen
  \bibfield  {author} {\bibinfo {author} {\bibfnamefont {H.}~\bibnamefont
  {Kumar}}, \bibinfo {author} {\bibfnamefont {N.~C.}\ \bibnamefont {Frey}},
  \bibinfo {author} {\bibfnamefont {L.}~\bibnamefont {Dong}}, \bibinfo {author}
  {\bibfnamefont {B.}~\bibnamefont {Anasori}}, \bibinfo {author} {\bibfnamefont
  {Y.}~\bibnamefont {Gogotsi}}, \ and\ \bibinfo {author} {\bibfnamefont
  {V.~B.}\ \bibnamefont {Shenoy}},\ }\href {\doibase 10.1021/acsnano.7b02578}
  {\bibfield  {journal} {\bibinfo  {journal} {ACS Nano}\ }\textbf {\bibinfo
  {volume} {11}},\ \bibinfo {pages} {7648} (\bibinfo {year}
  {2017})}\BibitemShut {NoStop}%
\bibitem [{\citenamefont {Li}\ and\ \citenamefont {Guo}(2018)}]{LiPRB2018}%
  \BibitemOpen
  \bibfield  {author} {\bibinfo {author} {\bibfnamefont {Y.}~\bibnamefont
  {Li}}\ and\ \bibinfo {author} {\bibfnamefont {W.}~\bibnamefont {Guo}},\
  }\href {\doibase 10.1103/PhysRevB.97.104302} {\bibfield  {journal} {\bibinfo
  {journal} {Phys. Rev. B}\ }\textbf {\bibinfo {volume} {97}},\ \bibinfo
  {pages} {104302} (\bibinfo {year} {2018})}\BibitemShut {NoStop}%
\bibitem [{\citenamefont {Lei}\ \emph {et~al.}(2020)\citenamefont {Lei},
  \citenamefont {Lin}, \citenamefont {Jia}, \citenamefont {Gray}, \citenamefont
  {Topp}, \citenamefont {Farahi}, \citenamefont {Klemenz}, \citenamefont {Gao},
  \citenamefont {Rodolakis}, \citenamefont {McChesney}, \citenamefont {Ast},
  \citenamefont {Yazdani}, \citenamefont {Burch}, \citenamefont {Wu},
  \citenamefont {Ong},\ and\ \citenamefont {Schoop}}]{LeiSciAdv2020}%
  \BibitemOpen
  \bibfield  {author} {\bibinfo {author} {\bibfnamefont {S.}~\bibnamefont
  {Lei}}, \bibinfo {author} {\bibfnamefont {J.}~\bibnamefont {Lin}}, \bibinfo
  {author} {\bibfnamefont {Y.}~\bibnamefont {Jia}}, \bibinfo {author}
  {\bibfnamefont {M.}~\bibnamefont {Gray}}, \bibinfo {author} {\bibfnamefont
  {A.}~\bibnamefont {Topp}}, \bibinfo {author} {\bibfnamefont {G.}~\bibnamefont
  {Farahi}}, \bibinfo {author} {\bibfnamefont {S.}~\bibnamefont {Klemenz}},
  \bibinfo {author} {\bibfnamefont {T.}~\bibnamefont {Gao}}, \bibinfo {author}
  {\bibfnamefont {F.}~\bibnamefont {Rodolakis}}, \bibinfo {author}
  {\bibfnamefont {J.~L.}\ \bibnamefont {McChesney}}, \bibinfo {author}
  {\bibfnamefont {C.~R.}\ \bibnamefont {Ast}}, \bibinfo {author} {\bibfnamefont
  {A.}~\bibnamefont {Yazdani}}, \bibinfo {author} {\bibfnamefont {K.~S.}\
  \bibnamefont {Burch}}, \bibinfo {author} {\bibfnamefont {S.}~\bibnamefont
  {Wu}}, \bibinfo {author} {\bibfnamefont {N.~P.}\ \bibnamefont {Ong}}, \ and\
  \bibinfo {author} {\bibfnamefont {L.~M.}\ \bibnamefont {Schoop}},\ }\href
  {\doibase 10.1126/sciadv.aay6407} {\bibfield  {journal} {\bibinfo  {journal}
  {Sci. Adv.}\ }\textbf {\bibinfo {volume} {6}},\ \bibinfo {pages} {eaay6407}
  (\bibinfo {year} {2020})}\BibitemShut {NoStop}%
\bibitem [{\citenamefont {Jiao}\ \emph {et~al.}(2019)\citenamefont {Jiao},
  \citenamefont {Wu}, \citenamefont {Ma}, \citenamefont {Yu}, \citenamefont
  {Lu}, \citenamefont {Sheng}, \citenamefont {Zhang},\ and\ \citenamefont
  {Yang}}]{jiao_Nanoscale_2019}%
  \BibitemOpen
  \bibfield  {author} {\bibinfo {author} {\bibfnamefont {Y.}~\bibnamefont
  {Jiao}}, \bibinfo {author} {\bibfnamefont {W.}~\bibnamefont {Wu}}, \bibinfo
  {author} {\bibfnamefont {F.}~\bibnamefont {Ma}}, \bibinfo {author}
  {\bibfnamefont {Z.-M.}\ \bibnamefont {Yu}}, \bibinfo {author} {\bibfnamefont
  {Y.}~\bibnamefont {Lu}}, \bibinfo {author} {\bibfnamefont {X.-L.}\
  \bibnamefont {Sheng}}, \bibinfo {author} {\bibfnamefont {Y.}~\bibnamefont
  {Zhang}}, \ and\ \bibinfo {author} {\bibfnamefont {S.~A.}\ \bibnamefont
  {Yang}},\ }\href {\doibase 10.1039/c9nr04338a} {\bibfield  {journal}
  {\bibinfo  {journal} {Nanoscale}\ }\textbf {\bibinfo {volume} {11}},\
  \bibinfo {pages} {16508–16514} (\bibinfo {year} {2019})}\BibitemShut
  {NoStop}%
\bibitem [{\citenamefont {\u{S}mejkal}, \citenamefont {Jungwirth},\ and\
  \citenamefont {Sinova}(2017)}]{Smejkal2017}%
  \BibitemOpen
  \bibfield  {author} {\bibinfo {author} {\bibfnamefont {L.}~\bibnamefont
  {\u{S}mejkal}}, \bibinfo {author} {\bibfnamefont {T.}~\bibnamefont
  {Jungwirth}}, \ and\ \bibinfo {author} {\bibfnamefont {J.}~\bibnamefont
  {Sinova}},\ }\href@noop {} {\bibfield  {journal} {\bibinfo  {journal} {Phys.
  Stat. Sol. : Rap. Res. Lett.}\ }\textbf {\bibinfo {volume} {11}},\ \bibinfo
  {pages} {201700044} (\bibinfo {year} {2017})}\BibitemShut {NoStop}%
\bibitem [{\citenamefont {Qi}\ and\ \citenamefont {Zhang}(2011)}]{QiRMP}%
  \BibitemOpen
  \bibfield  {author} {\bibinfo {author} {\bibfnamefont {X.-L.}\ \bibnamefont
  {Qi}}\ and\ \bibinfo {author} {\bibfnamefont {S.-C.}\ \bibnamefont {Zhang}},\
  }\href@noop {} {\bibfield  {journal} {\bibinfo  {journal} {Rev. Mod. Phys.}\
  }\textbf {\bibinfo {volume} {83}},\ \bibinfo {pages} {1057} (\bibinfo {year}
  {2011})}\BibitemShut {NoStop}%
\bibitem [{\citenamefont {Hasan}\ and\ \citenamefont {Kane}(2010)}]{HasanRMP}%
  \BibitemOpen
  \bibfield  {author} {\bibinfo {author} {\bibfnamefont {M.~Z.}\ \bibnamefont
  {Hasan}}\ and\ \bibinfo {author} {\bibfnamefont {C.~L.}\ \bibnamefont
  {Kane}},\ }\href@noop {} {\bibfield  {journal} {\bibinfo  {journal} {Rev.
  Mod. Phys.}\ }\textbf {\bibinfo {volume} {82}},\ \bibinfo {pages} {3045}
  (\bibinfo {year} {2010})}\BibitemShut {NoStop}%
\bibitem [{\citenamefont {Armitage}, \citenamefont {Mele},\ and\ \citenamefont
  {Viswanath}(2018)}]{ArmitageRMP}%
  \BibitemOpen
  \bibfield  {author} {\bibinfo {author} {\bibfnamefont {N.~P.}\ \bibnamefont
  {Armitage}}, \bibinfo {author} {\bibfnamefont {E.~J.}\ \bibnamefont {Mele}},
  \ and\ \bibinfo {author} {\bibfnamefont {A.}~\bibnamefont {Viswanath}},\
  }\href@noop {} {\bibfield  {journal} {\bibinfo  {journal} {Rev. Mod. Phys.}\
  }\textbf {\bibinfo {volume} {90}},\ \bibinfo {pages} {015001} (\bibinfo
  {year} {2018})}\BibitemShut {NoStop}%
\bibitem [{\citenamefont {Zyuzin}, \citenamefont {Wu},\ and\ \citenamefont
  {Burkov}(2012)}]{Zyuzin2012A}%
  \BibitemOpen
  \bibfield  {author} {\bibinfo {author} {\bibfnamefont {A.~A.}\ \bibnamefont
  {Zyuzin}}, \bibinfo {author} {\bibfnamefont {S.}~\bibnamefont {Wu}}, \ and\
  \bibinfo {author} {\bibfnamefont {A.~A.}\ \bibnamefont {Burkov}},\
  }\href@noop {} {\bibfield  {journal} {\bibinfo  {journal} {Phys. Rev. B}\
  }\textbf {\bibinfo {volume} {85}},\ \bibinfo {pages} {165110} (\bibinfo
  {year} {2012})}\BibitemShut {NoStop}%
\bibitem [{\citenamefont {Zyuzin}\ and\ \citenamefont
  {Burkov}(2012)}]{Zyuzin2012B}%
  \BibitemOpen
  \bibfield  {author} {\bibinfo {author} {\bibfnamefont {A.~A.}\ \bibnamefont
  {Zyuzin}}\ and\ \bibinfo {author} {\bibfnamefont {A.~A.}\ \bibnamefont
  {Burkov}},\ }\href@noop {} {\bibfield  {journal} {\bibinfo  {journal} {Phys.
  Rev. B}\ }\textbf {\bibinfo {volume} {86}},\ \bibinfo {pages} {115133}
  (\bibinfo {year} {2012})}\BibitemShut {NoStop}%
\bibitem [{\citenamefont {K\"{u}bler}\ and\ \citenamefont
  {Felser}(2017)}]{Felser2017}%
  \BibitemOpen
  \bibfield  {author} {\bibinfo {author} {\bibfnamefont {J.}~\bibnamefont
  {K\"{u}bler}}\ and\ \bibinfo {author} {\bibfnamefont {C.}~\bibnamefont
  {Felser}},\ }\href@noop {} {\bibfield  {journal} {\bibinfo  {journal}
  {Europhys. Lett.}\ }\textbf {\bibinfo {volume} {120}},\ \bibinfo {pages}
  {47002} (\bibinfo {year} {2017})}\BibitemShut {NoStop}%
\bibitem [{\citenamefont {Nagaosa}\ \emph {et~al.}(2010)\citenamefont
  {Nagaosa}, \citenamefont {Sinova}, \citenamefont {Onoda}, \citenamefont
  {MacDonald},\ and\ \citenamefont {Ong}}]{NagaosaRMP}%
  \BibitemOpen
  \bibfield  {author} {\bibinfo {author} {\bibfnamefont {N.}~\bibnamefont
  {Nagaosa}}, \bibinfo {author} {\bibfnamefont {J.}~\bibnamefont {Sinova}},
  \bibinfo {author} {\bibfnamefont {S.}~\bibnamefont {Onoda}}, \bibinfo
  {author} {\bibfnamefont {A.~H.}\ \bibnamefont {MacDonald}}, \ and\ \bibinfo
  {author} {\bibfnamefont {N.~P.}\ \bibnamefont {Ong}},\ }\href@noop {}
  {\bibfield  {journal} {\bibinfo  {journal} {Rev. Mod. Phys.}\ }\textbf
  {\bibinfo {volume} {82}},\ \bibinfo {pages} {1539} (\bibinfo {year}
  {2010})}\BibitemShut {NoStop}%
\bibitem [{\citenamefont {Chen}, \citenamefont {Niu},\ and\ \citenamefont
  {MacDonald}(2014{\natexlab{b}})}]{MacDonald2014}%
  \BibitemOpen
  \bibfield  {author} {\bibinfo {author} {\bibfnamefont {H.}~\bibnamefont
  {Chen}}, \bibinfo {author} {\bibfnamefont {Q.}~\bibnamefont {Niu}}, \ and\
  \bibinfo {author} {\bibfnamefont {A.~H.}\ \bibnamefont {MacDonald}},\
  }\href@noop {} {\bibfield  {journal} {\bibinfo  {journal} {Phys. Rev. Lett.}\
  }\textbf {\bibinfo {volume} {112}},\ \bibinfo {pages} {017205} (\bibinfo
  {year} {2014}{\natexlab{b}})}\BibitemShut {NoStop}%
\bibitem [{\citenamefont {Kren}\ \emph {et~al.}(1975)\citenamefont {Kren},
  \citenamefont {Paitz}, \citenamefont {Zimmer},\ and\ \citenamefont
  {Zsoldos}}]{Kren1975}%
  \BibitemOpen
  \bibfield  {author} {\bibinfo {author} {\bibfnamefont {E.}~\bibnamefont
  {Kren}}, \bibinfo {author} {\bibfnamefont {J.}~\bibnamefont {Paitz}},
  \bibinfo {author} {\bibfnamefont {G.}~\bibnamefont {Zimmer}}, \ and\ \bibinfo
  {author} {\bibfnamefont {E.}~\bibnamefont {Zsoldos}},\ }\href@noop {}
  {\bibfield  {journal} {\bibinfo  {journal} {Phys. B}\ }\textbf {\bibinfo
  {volume} {80}},\ \bibinfo {pages} {226} (\bibinfo {year} {1975})}\BibitemShut
  {NoStop}%
\bibitem [{\citenamefont {Haldane}(2004)}]{Haldane2004}%
  \BibitemOpen
  \bibfield  {author} {\bibinfo {author} {\bibfnamefont {F.}~\bibnamefont
  {Haldane}},\ }\href@noop {} {\bibfield  {journal} {\bibinfo  {journal} {Phys.
  Rev. Lett.}\ }\textbf {\bibinfo {volume} {93}},\ \bibinfo {pages} {206602}
  (\bibinfo {year} {2004})}\BibitemShut {NoStop}%
\bibitem [{\citenamefont {Guo}\ \emph {et~al.}(2008)\citenamefont {Guo},
  \citenamefont {Murakami}, \citenamefont {Chen},\ and\ \citenamefont
  {Nagaosa}}]{Guo2008}%
  \BibitemOpen
  \bibfield  {author} {\bibinfo {author} {\bibfnamefont {G.~Y.}\ \bibnamefont
  {Guo}}, \bibinfo {author} {\bibfnamefont {S.}~\bibnamefont {Murakami}},
  \bibinfo {author} {\bibfnamefont {T.~W.}\ \bibnamefont {Chen}}, \ and\
  \bibinfo {author} {\bibfnamefont {N.}~\bibnamefont {Nagaosa}},\ }\href@noop
  {} {\bibfield  {journal} {\bibinfo  {journal} {Phys. Rev. Lett.}\ }\textbf
  {\bibinfo {volume} {100}},\ \bibinfo {pages} {096401} (\bibinfo {year}
  {2008})}\BibitemShut {NoStop}%
\bibitem [{\citenamefont {Shao}\ \emph {et~al.}(2019)\citenamefont {Shao},
  \citenamefont {Gurung}, \citenamefont {Zhang},\ and\ \citenamefont
  {Tsymbal}}]{Shao2019}%
  \BibitemOpen
  \bibfield  {author} {\bibinfo {author} {\bibfnamefont {D.-F.}\ \bibnamefont
  {Shao}}, \bibinfo {author} {\bibfnamefont {G.}~\bibnamefont {Gurung}},
  \bibinfo {author} {\bibfnamefont {S.~H.}\ \bibnamefont {Zhang}}, \ and\
  \bibinfo {author} {\bibfnamefont {E.~Y.}\ \bibnamefont {Tsymbal}},\
  }\href@noop {} {\bibfield  {journal} {\bibinfo  {journal} {Phys. Rev. Lett.}\
  }\textbf {\bibinfo {volume} {122}},\ \bibinfo {pages} {077203} (\bibinfo
  {year} {2019})}\BibitemShut {NoStop}%
\bibitem [{\citenamefont {Kim}\ \emph {et~al.}(2018)\citenamefont {Kim},
  \citenamefont {Kang}, \citenamefont {Schleife},\ and\ \citenamefont
  {Gilbert}}]{Kim2018}%
  \BibitemOpen
  \bibfield  {author} {\bibinfo {author} {\bibfnamefont {Y.}~\bibnamefont
  {Kim}}, \bibinfo {author} {\bibfnamefont {K.}~\bibnamefont {Kang}}, \bibinfo
  {author} {\bibfnamefont {A.}~\bibnamefont {Schleife}}, \ and\ \bibinfo
  {author} {\bibfnamefont {M.~J.}\ \bibnamefont {Gilbert}},\ }\href@noop {}
  {\bibfield  {journal} {\bibinfo  {journal} {Phys. Rev. B}\ }\textbf {\bibinfo
  {volume} {97}},\ \bibinfo {pages} {134415} (\bibinfo {year}
  {2018})}\BibitemShut {NoStop}%
\bibitem [{\citenamefont {Fang}, \citenamefont {Gilbert},\ and\ \citenamefont
  {Bernevig}(2013)}]{FangAFM2013}%
  \BibitemOpen
  \bibfield  {author} {\bibinfo {author} {\bibfnamefont {C.}~\bibnamefont
  {Fang}}, \bibinfo {author} {\bibfnamefont {M.~J.}\ \bibnamefont {Gilbert}}, \
  and\ \bibinfo {author} {\bibfnamefont {B.~A.}\ \bibnamefont {Bernevig}},\
  }\href@noop {} {\bibfield  {journal} {\bibinfo  {journal} {Phys. Rev. B}\
  }\textbf {\bibinfo {volume} {88}},\ \bibinfo {pages} {085406} (\bibinfo
  {year} {2013})}\BibitemShut {NoStop}%
\bibitem [{\citenamefont {Hoffmann}\ and\ \citenamefont
  {Bader}(2015)}]{hoffmann2015opportunities}%
  \BibitemOpen
  \bibfield  {author} {\bibinfo {author} {\bibfnamefont {A.}~\bibnamefont
  {Hoffmann}}\ and\ \bibinfo {author} {\bibfnamefont {S.~D.}\ \bibnamefont
  {Bader}},\ }\href@noop {} {\bibfield  {journal} {\bibinfo  {journal} {Phys.
  Rev. Appl.}\ }\textbf {\bibinfo {volume} {4}},\ \bibinfo {pages} {047001}
  (\bibinfo {year} {2015})}\BibitemShut {NoStop}%
\bibitem [{\citenamefont {Chumak}\ \emph {et~al.}(2015)\citenamefont {Chumak},
  \citenamefont {Vasyuchka}, \citenamefont {Serga},\ and\ \citenamefont
  {Hillebrands}}]{chumak2015magnon}%
  \BibitemOpen
  \bibfield  {author} {\bibinfo {author} {\bibfnamefont {A.}~\bibnamefont
  {Chumak}}, \bibinfo {author} {\bibfnamefont {V.}~\bibnamefont {Vasyuchka}},
  \bibinfo {author} {\bibfnamefont {A.}~\bibnamefont {Serga}}, \ and\ \bibinfo
  {author} {\bibfnamefont {B.}~\bibnamefont {Hillebrands}},\ }\href@noop {}
  {\bibfield  {journal} {\bibinfo  {journal} {Nat. Phys.}\ }\textbf {\bibinfo
  {volume} {11}},\ \bibinfo {pages} {453} (\bibinfo {year} {2015})}\BibitemShut
  {NoStop}%
\bibitem [{\citenamefont {Damon}\ and\ \citenamefont
  {Eshbach}(1961)}]{damon1961magnetostatic}%
  \BibitemOpen
  \bibfield  {author} {\bibinfo {author} {\bibfnamefont {R.~W.}\ \bibnamefont
  {Damon}}\ and\ \bibinfo {author} {\bibfnamefont {J.}~\bibnamefont
  {Eshbach}},\ }\href@noop {} {\bibfield  {journal} {\bibinfo  {journal} {J.
  Phys. Chem. Solids}\ }\textbf {\bibinfo {volume} {19}},\ \bibinfo {pages}
  {308} (\bibinfo {year} {1961})}\BibitemShut {NoStop}%
\bibitem [{\citenamefont {Sklenar}\ \emph {et~al.}(2012)\citenamefont
  {Sklenar}, \citenamefont {Bhat}, \citenamefont {Tsai}, \citenamefont
  {DeLong},\ and\ \citenamefont {Ketterson}}]{sklenar2012generating}%
  \BibitemOpen
  \bibfield  {author} {\bibinfo {author} {\bibfnamefont {J.}~\bibnamefont
  {Sklenar}}, \bibinfo {author} {\bibfnamefont {V.}~\bibnamefont {Bhat}},
  \bibinfo {author} {\bibfnamefont {C.}~\bibnamefont {Tsai}}, \bibinfo {author}
  {\bibfnamefont {L.}~\bibnamefont {DeLong}}, \ and\ \bibinfo {author}
  {\bibfnamefont {J.~B.}\ \bibnamefont {Ketterson}},\ }\href@noop {} {\bibfield
   {journal} {\bibinfo  {journal} {Appl. Phys. Lett.}\ }\textbf {\bibinfo
  {volume} {101}},\ \bibinfo {pages} {052404} (\bibinfo {year}
  {2012})}\BibitemShut {NoStop}%
\bibitem [{\citenamefont {Kalinikos}\ and\ \citenamefont
  {Slavin}(1986)}]{kalinikos1986theory}%
  \BibitemOpen
  \bibfield  {author} {\bibinfo {author} {\bibfnamefont {B.}~\bibnamefont
  {Kalinikos}}\ and\ \bibinfo {author} {\bibfnamefont {A.}~\bibnamefont
  {Slavin}},\ }\href@noop {} {\bibfield  {journal} {\bibinfo  {journal} {J.
  Phys. C: Solid State Phys.}\ }\textbf {\bibinfo {volume} {19}},\ \bibinfo
  {pages} {7013} (\bibinfo {year} {1986})}\BibitemShut {NoStop}%
\bibitem [{\citenamefont {Keffer}\ and\ \citenamefont
  {Kittel}(1952)}]{keffer1952theory}%
  \BibitemOpen
  \bibfield  {author} {\bibinfo {author} {\bibfnamefont {F.}~\bibnamefont
  {Keffer}}\ and\ \bibinfo {author} {\bibfnamefont {C.}~\bibnamefont
  {Kittel}},\ }\href@noop {} {\bibfield  {journal} {\bibinfo  {journal} {Phys.
  Rev.}\ }\textbf {\bibinfo {volume} {85}},\ \bibinfo {pages} {329} (\bibinfo
  {year} {1952})}\BibitemShut {NoStop}%
\bibitem [{\citenamefont {Gomonay}\ \emph
  {et~al.}(2018{\natexlab{b}})\citenamefont {Gomonay}, \citenamefont {Baltz},
  \citenamefont {Brataas},\ and\ \citenamefont
  {Tserkovnyak}}]{gomonay2018antiferromagnetic}%
  \BibitemOpen
  \bibfield  {author} {\bibinfo {author} {\bibfnamefont {O.}~\bibnamefont
  {Gomonay}}, \bibinfo {author} {\bibfnamefont {V.}~\bibnamefont {Baltz}},
  \bibinfo {author} {\bibfnamefont {A.}~\bibnamefont {Brataas}}, \ and\
  \bibinfo {author} {\bibfnamefont {Y.}~\bibnamefont {Tserkovnyak}},\
  }\href@noop {} {\bibfield  {journal} {\bibinfo  {journal} {Nat. Phys.}\
  }\textbf {\bibinfo {volume} {14}},\ \bibinfo {pages} {213} (\bibinfo {year}
  {2018}{\natexlab{b}})}\BibitemShut {NoStop}%
\bibitem [{\citenamefont {Kampfrath}\ \emph {et~al.}(2011)\citenamefont
  {Kampfrath}, \citenamefont {Sell}, \citenamefont {Klatt}, \citenamefont
  {Pashkin}, \citenamefont {M{\"a}hrlein}, \citenamefont {Dekorsy},
  \citenamefont {Wolf}, \citenamefont {Fiebig}, \citenamefont {Leitenstorfer},\
  and\ \citenamefont {Huber}}]{kampfrath2011coherent}%
  \BibitemOpen
  \bibfield  {author} {\bibinfo {author} {\bibfnamefont {T.}~\bibnamefont
  {Kampfrath}}, \bibinfo {author} {\bibfnamefont {A.}~\bibnamefont {Sell}},
  \bibinfo {author} {\bibfnamefont {G.}~\bibnamefont {Klatt}}, \bibinfo
  {author} {\bibfnamefont {A.}~\bibnamefont {Pashkin}}, \bibinfo {author}
  {\bibfnamefont {S.}~\bibnamefont {M{\"a}hrlein}}, \bibinfo {author}
  {\bibfnamefont {T.}~\bibnamefont {Dekorsy}}, \bibinfo {author} {\bibfnamefont
  {M.}~\bibnamefont {Wolf}}, \bibinfo {author} {\bibfnamefont {M.}~\bibnamefont
  {Fiebig}}, \bibinfo {author} {\bibfnamefont {A.}~\bibnamefont
  {Leitenstorfer}}, \ and\ \bibinfo {author} {\bibfnamefont {R.}~\bibnamefont
  {Huber}},\ }\href@noop {} {\bibfield  {journal} {\bibinfo  {journal} {Nat.
  Photon.}\ }\textbf {\bibinfo {volume} {5}},\ \bibinfo {pages} {31} (\bibinfo
  {year} {2011})}\BibitemShut {NoStop}%
\bibitem [{\citenamefont {Tzschaschel}\ \emph {et~al.}(2017)\citenamefont
  {Tzschaschel}, \citenamefont {Otani}, \citenamefont {Iida}, \citenamefont
  {Shimura}, \citenamefont {Ueda}, \citenamefont {G{\"u}nther}, \citenamefont
  {Fiebig},\ and\ \citenamefont {Satoh}}]{tzschaschel2017ultrafast}%
  \BibitemOpen
  \bibfield  {author} {\bibinfo {author} {\bibfnamefont {C.}~\bibnamefont
  {Tzschaschel}}, \bibinfo {author} {\bibfnamefont {K.}~\bibnamefont {Otani}},
  \bibinfo {author} {\bibfnamefont {R.}~\bibnamefont {Iida}}, \bibinfo {author}
  {\bibfnamefont {T.}~\bibnamefont {Shimura}}, \bibinfo {author} {\bibfnamefont
  {H.}~\bibnamefont {Ueda}}, \bibinfo {author} {\bibfnamefont {S.}~\bibnamefont
  {G{\"u}nther}}, \bibinfo {author} {\bibfnamefont {M.}~\bibnamefont {Fiebig}},
  \ and\ \bibinfo {author} {\bibfnamefont {T.}~\bibnamefont {Satoh}},\
  }\href@noop {} {\bibfield  {journal} {\bibinfo  {journal} {Phys. Rev. B}\
  }\textbf {\bibinfo {volume} {95}},\ \bibinfo {pages} {174407} (\bibinfo
  {year} {2017})}\BibitemShut {NoStop}%
\bibitem [{\citenamefont {Duine}\ \emph {et~al.}(2018)\citenamefont {Duine},
  \citenamefont {Lee}, \citenamefont {Parkin},\ and\ \citenamefont
  {Stiles}}]{duine2018synthetic}%
  \BibitemOpen
  \bibfield  {author} {\bibinfo {author} {\bibfnamefont {R.}~\bibnamefont
  {Duine}}, \bibinfo {author} {\bibfnamefont {K.-J.}\ \bibnamefont {Lee}},
  \bibinfo {author} {\bibfnamefont {S.~S.}\ \bibnamefont {Parkin}}, \ and\
  \bibinfo {author} {\bibfnamefont {M.~D.}\ \bibnamefont {Stiles}},\
  }\href@noop {} {\bibfield  {journal} {\bibinfo  {journal} {Nat. Phys.}\
  }\textbf {\bibinfo {volume} {14}},\ \bibinfo {pages} {217} (\bibinfo {year}
  {2018})}\BibitemShut {NoStop}%
\bibitem [{\citenamefont {Liu}\ \emph {et~al.}(2014)\citenamefont {Liu},
  \citenamefont {Nguyen}, \citenamefont {Ding}, \citenamefont {Cottam},\ and\
  \citenamefont {Adeyeye}}]{liu2014interlayer}%
  \BibitemOpen
  \bibfield  {author} {\bibinfo {author} {\bibfnamefont {X.}~\bibnamefont
  {Liu}}, \bibinfo {author} {\bibfnamefont {H.~T.}\ \bibnamefont {Nguyen}},
  \bibinfo {author} {\bibfnamefont {J.}~\bibnamefont {Ding}}, \bibinfo {author}
  {\bibfnamefont {M.}~\bibnamefont {Cottam}}, \ and\ \bibinfo {author}
  {\bibfnamefont {A.}~\bibnamefont {Adeyeye}},\ }\href@noop {} {\bibfield
  {journal} {\bibinfo  {journal} {Phys. Rev. B}\ }\textbf {\bibinfo {volume}
  {90}},\ \bibinfo {pages} {064428} (\bibinfo {year} {2014})}\BibitemShut
  {NoStop}%
\bibitem [{\citenamefont {Gomez-Perez}\ \emph {et~al.}(2018)\citenamefont
  {Gomez-Perez}, \citenamefont {V{\'e}lez}, \citenamefont {McKenzie-Sell},
  \citenamefont {Amado}, \citenamefont {Herrero-Mart{\'\i}n}, \citenamefont
  {L{\'o}pez-L{\'o}pez}, \citenamefont {Blanco-Canosa}, \citenamefont {Hueso},
  \citenamefont {Chuvilin}, \citenamefont {Robinson} \emph
  {et~al.}}]{gomez2018synthetic}%
  \BibitemOpen
  \bibfield  {author} {\bibinfo {author} {\bibfnamefont {J.~M.}\ \bibnamefont
  {Gomez-Perez}}, \bibinfo {author} {\bibfnamefont {S.}~\bibnamefont
  {V{\'e}lez}}, \bibinfo {author} {\bibfnamefont {L.}~\bibnamefont
  {McKenzie-Sell}}, \bibinfo {author} {\bibfnamefont {M.}~\bibnamefont
  {Amado}}, \bibinfo {author} {\bibfnamefont {J.}~\bibnamefont
  {Herrero-Mart{\'\i}n}}, \bibinfo {author} {\bibfnamefont {J.}~\bibnamefont
  {L{\'o}pez-L{\'o}pez}}, \bibinfo {author} {\bibfnamefont {S.}~\bibnamefont
  {Blanco-Canosa}}, \bibinfo {author} {\bibfnamefont {L.~E.}\ \bibnamefont
  {Hueso}}, \bibinfo {author} {\bibfnamefont {A.}~\bibnamefont {Chuvilin}},
  \bibinfo {author} {\bibfnamefont {J.~W.}\ \bibnamefont {Robinson}},  \emph
  {et~al.},\ }\href@noop {} {\bibfield  {journal} {\bibinfo  {journal} {Phys.
  Rev. Appl.}\ }\textbf {\bibinfo {volume} {10}},\ \bibinfo {pages} {044046}
  (\bibinfo {year} {2018})}\BibitemShut {NoStop}%
\bibitem [{\citenamefont {Demokritov}\ \emph {et~al.}(2006)\citenamefont
  {Demokritov}, \citenamefont {Demidov}, \citenamefont {Dzyapko}, \citenamefont
  {Melkov}, \citenamefont {Serga}, \citenamefont {Hillebrands},\ and\
  \citenamefont {Slavin}}]{demokritov2006bose}%
  \BibitemOpen
  \bibfield  {author} {\bibinfo {author} {\bibfnamefont {S.}~\bibnamefont
  {Demokritov}}, \bibinfo {author} {\bibfnamefont {V.}~\bibnamefont {Demidov}},
  \bibinfo {author} {\bibfnamefont {O.}~\bibnamefont {Dzyapko}}, \bibinfo
  {author} {\bibfnamefont {G.}~\bibnamefont {Melkov}}, \bibinfo {author}
  {\bibfnamefont {A.}~\bibnamefont {Serga}}, \bibinfo {author} {\bibfnamefont
  {B.}~\bibnamefont {Hillebrands}}, \ and\ \bibinfo {author} {\bibfnamefont
  {A.}~\bibnamefont {Slavin}},\ }\href@noop {} {\bibfield  {journal} {\bibinfo
  {journal} {Nature}\ }\textbf {\bibinfo {volume} {443}},\ \bibinfo {pages}
  {430} (\bibinfo {year} {2006})}\BibitemShut {NoStop}%
\bibitem [{\citenamefont {Radu}\ \emph {et~al.}(2005)\citenamefont {Radu},
  \citenamefont {Wilhelm}, \citenamefont {Yushankhai}, \citenamefont
  {Kovrizhin}, \citenamefont {Coldea}, \citenamefont {Tylczynski},
  \citenamefont {L{\"u}hmann},\ and\ \citenamefont {Steglich}}]{radu2005bose}%
  \BibitemOpen
  \bibfield  {author} {\bibinfo {author} {\bibfnamefont {T.}~\bibnamefont
  {Radu}}, \bibinfo {author} {\bibfnamefont {H.}~\bibnamefont {Wilhelm}},
  \bibinfo {author} {\bibfnamefont {V.}~\bibnamefont {Yushankhai}}, \bibinfo
  {author} {\bibfnamefont {D.}~\bibnamefont {Kovrizhin}}, \bibinfo {author}
  {\bibfnamefont {R.}~\bibnamefont {Coldea}}, \bibinfo {author} {\bibfnamefont
  {Z.}~\bibnamefont {Tylczynski}}, \bibinfo {author} {\bibfnamefont
  {T.}~\bibnamefont {L{\"u}hmann}}, \ and\ \bibinfo {author} {\bibfnamefont
  {F.}~\bibnamefont {Steglich}},\ }\href@noop {} {\bibfield  {journal}
  {\bibinfo  {journal} {Phys. Rev. Lett.}\ }\textbf {\bibinfo {volume} {95}},\
  \bibinfo {pages} {127202} (\bibinfo {year} {2005})}\BibitemShut {NoStop}%
\bibitem [{\citenamefont {Fj{\ae}rbu}, \citenamefont {Rohling},\ and\
  \citenamefont {Brataas}(2017)}]{fjaerbu2017electrically}%
  \BibitemOpen
  \bibfield  {author} {\bibinfo {author} {\bibfnamefont {E.~L.}\ \bibnamefont
  {Fj{\ae}rbu}}, \bibinfo {author} {\bibfnamefont {N.}~\bibnamefont {Rohling}},
  \ and\ \bibinfo {author} {\bibfnamefont {A.}~\bibnamefont {Brataas}},\
  }\href@noop {} {\bibfield  {journal} {\bibinfo  {journal} {Phys. Rev. B}\
  }\textbf {\bibinfo {volume} {95}},\ \bibinfo {pages} {144408} (\bibinfo
  {year} {2017})}\BibitemShut {NoStop}%
\bibitem [{\citenamefont {MacNeill}\ \emph {et~al.}(2019)\citenamefont
  {MacNeill}, \citenamefont {Hou}, \citenamefont {Klein}, \citenamefont
  {Zhang}, \citenamefont {Jarillo-Herrero},\ and\ \citenamefont
  {Liu}}]{macneill2019gigahertz}%
  \BibitemOpen
  \bibfield  {author} {\bibinfo {author} {\bibfnamefont {D.}~\bibnamefont
  {MacNeill}}, \bibinfo {author} {\bibfnamefont {J.~T.}\ \bibnamefont {Hou}},
  \bibinfo {author} {\bibfnamefont {D.~R.}\ \bibnamefont {Klein}}, \bibinfo
  {author} {\bibfnamefont {P.}~\bibnamefont {Zhang}}, \bibinfo {author}
  {\bibfnamefont {P.}~\bibnamefont {Jarillo-Herrero}}, \ and\ \bibinfo {author}
  {\bibfnamefont {L.}~\bibnamefont {Liu}},\ }\href {\doibase
  10.1103/PhysRevLett.123.047204} {\bibfield  {journal} {\bibinfo  {journal}
  {Phys. Rev. Lett.}\ }\textbf {\bibinfo {volume} {123}},\ \bibinfo {pages}
  {047204} (\bibinfo {year} {2019})}\BibitemShut {NoStop}%
\bibitem [{\citenamefont {Moriyama}\ \emph {et~al.}(2019)\citenamefont
  {Moriyama}, \citenamefont {Hayashi}, \citenamefont {Yamada}, \citenamefont
  {Shima}, \citenamefont {Ohya},\ and\ \citenamefont
  {Ono}}]{moriyama2019intrinsic}%
  \BibitemOpen
  \bibfield  {author} {\bibinfo {author} {\bibfnamefont {T.}~\bibnamefont
  {Moriyama}}, \bibinfo {author} {\bibfnamefont {K.}~\bibnamefont {Hayashi}},
  \bibinfo {author} {\bibfnamefont {K.}~\bibnamefont {Yamada}}, \bibinfo
  {author} {\bibfnamefont {M.}~\bibnamefont {Shima}}, \bibinfo {author}
  {\bibfnamefont {Y.}~\bibnamefont {Ohya}}, \ and\ \bibinfo {author}
  {\bibfnamefont {T.}~\bibnamefont {Ono}},\ }\href@noop {} {\bibfield
  {journal} {\bibinfo  {journal} {Phys. Rev. Mater.}\ }\textbf {\bibinfo
  {volume} {3}},\ \bibinfo {pages} {051402} (\bibinfo {year}
  {2019})}\BibitemShut {NoStop}%
\bibitem [{\citenamefont {Huang}\ \emph {et~al.}(2017)\citenamefont {Huang},
  \citenamefont {Clark}, \citenamefont {Navarro-Moratalla}, \citenamefont
  {Klein}, \citenamefont {Cheng}, \citenamefont {Seyler}, \citenamefont
  {Zhong}, \citenamefont {Schmidgall}, \citenamefont {McGuire}, \citenamefont
  {Cobden} \emph {et~al.}}]{huang2017layer}%
  \BibitemOpen
  \bibfield  {author} {\bibinfo {author} {\bibfnamefont {B.}~\bibnamefont
  {Huang}}, \bibinfo {author} {\bibfnamefont {G.}~\bibnamefont {Clark}},
  \bibinfo {author} {\bibfnamefont {E.}~\bibnamefont {Navarro-Moratalla}},
  \bibinfo {author} {\bibfnamefont {D.~R.}\ \bibnamefont {Klein}}, \bibinfo
  {author} {\bibfnamefont {R.}~\bibnamefont {Cheng}}, \bibinfo {author}
  {\bibfnamefont {K.~L.}\ \bibnamefont {Seyler}}, \bibinfo {author}
  {\bibfnamefont {D.}~\bibnamefont {Zhong}}, \bibinfo {author} {\bibfnamefont
  {E.}~\bibnamefont {Schmidgall}}, \bibinfo {author} {\bibfnamefont {M.~A.}\
  \bibnamefont {McGuire}}, \bibinfo {author} {\bibfnamefont {D.~H.}\
  \bibnamefont {Cobden}},  \emph {et~al.},\ }\href@noop {} {\bibfield
  {journal} {\bibinfo  {journal} {Nature}\ }\textbf {\bibinfo {volume} {546}},\
  \bibinfo {pages} {270} (\bibinfo {year} {2017})}\BibitemShut {NoStop}%
\bibitem [{\citenamefont {McGuire}\ \emph {et~al.}(2017)\citenamefont
  {McGuire}, \citenamefont {Clark}, \citenamefont {Santosh}, \citenamefont
  {Chance}, \citenamefont {Jellison~Jr}, \citenamefont {Cooper}, \citenamefont
  {Xu},\ and\ \citenamefont {Sales}}]{mcguire2017magnetic}%
  \BibitemOpen
  \bibfield  {author} {\bibinfo {author} {\bibfnamefont {M.~A.}\ \bibnamefont
  {McGuire}}, \bibinfo {author} {\bibfnamefont {G.}~\bibnamefont {Clark}},
  \bibinfo {author} {\bibfnamefont {K.}~\bibnamefont {Santosh}}, \bibinfo
  {author} {\bibfnamefont {W.~M.}\ \bibnamefont {Chance}}, \bibinfo {author}
  {\bibfnamefont {G.~E.}\ \bibnamefont {Jellison~Jr}}, \bibinfo {author}
  {\bibfnamefont {V.~R.}\ \bibnamefont {Cooper}}, \bibinfo {author}
  {\bibfnamefont {X.}~\bibnamefont {Xu}}, \ and\ \bibinfo {author}
  {\bibfnamefont {B.~C.}\ \bibnamefont {Sales}},\ }\href@noop {} {\bibfield
  {journal} {\bibinfo  {journal} {Phys. Rev. Mater.}\ }\textbf {\bibinfo
  {volume} {1}},\ \bibinfo {pages} {014001} (\bibinfo {year}
  {2017})}\BibitemShut {NoStop}%
\bibitem [{\citenamefont {Klein}\ \emph {et~al.}(2019)\citenamefont {Klein},
  \citenamefont {MacNeill}, \citenamefont {Song}, \citenamefont {Larson},
  \citenamefont {Fang}, \citenamefont {Xu}, \citenamefont {Ribeiro},
  \citenamefont {Canfield}, \citenamefont {Kaxiras}, \citenamefont {Comin}
  \emph {et~al.}}]{klein2019enhancement}%
  \BibitemOpen
  \bibfield  {author} {\bibinfo {author} {\bibfnamefont {D.~R.}\ \bibnamefont
  {Klein}}, \bibinfo {author} {\bibfnamefont {D.}~\bibnamefont {MacNeill}},
  \bibinfo {author} {\bibfnamefont {Q.}~\bibnamefont {Song}}, \bibinfo {author}
  {\bibfnamefont {D.~T.}\ \bibnamefont {Larson}}, \bibinfo {author}
  {\bibfnamefont {S.}~\bibnamefont {Fang}}, \bibinfo {author} {\bibfnamefont
  {M.}~\bibnamefont {Xu}}, \bibinfo {author} {\bibfnamefont {R.~A.}\
  \bibnamefont {Ribeiro}}, \bibinfo {author} {\bibfnamefont {P.~C.}\
  \bibnamefont {Canfield}}, \bibinfo {author} {\bibfnamefont {E.}~\bibnamefont
  {Kaxiras}}, \bibinfo {author} {\bibfnamefont {R.}~\bibnamefont {Comin}},
  \emph {et~al.},\ }\href@noop {} {\bibfield  {journal} {\bibinfo  {journal}
  {Nat. Phys.}\ }\textbf {\bibinfo {volume} {15}},\ \bibinfo {pages} {1255}
  (\bibinfo {year} {2019})}\BibitemShut {NoStop}%
\bibitem [{\citenamefont {Zhang}\ \emph {et~al.}(2020)\citenamefont {Zhang},
  \citenamefont {Li}, \citenamefont {Weber}, \citenamefont {Goldberger},
  \citenamefont {Mak},\ and\ \citenamefont {Shan}}]{zhang2020gate}%
  \BibitemOpen
  \bibfield  {author} {\bibinfo {author} {\bibfnamefont {X.-X.}\ \bibnamefont
  {Zhang}}, \bibinfo {author} {\bibfnamefont {L.}~\bibnamefont {Li}}, \bibinfo
  {author} {\bibfnamefont {D.}~\bibnamefont {Weber}}, \bibinfo {author}
  {\bibfnamefont {J.}~\bibnamefont {Goldberger}}, \bibinfo {author}
  {\bibfnamefont {K.~F.}\ \bibnamefont {Mak}}, \ and\ \bibinfo {author}
  {\bibfnamefont {J.}~\bibnamefont {Shan}},\ }\href@noop {} {\bibfield
  {journal} {\bibinfo  {journal} {arXiv preprint arXiv:2001.04044}\ } (\bibinfo
  {year} {2020})}\BibitemShut {NoStop}%
\bibitem [{\citenamefont {Cenker}\ \emph {et~al.}(2020)\citenamefont {Cenker},
  \citenamefont {Huang}, \citenamefont {Suri}, \citenamefont {Thijssen},
  \citenamefont {Miller}, \citenamefont {Song}, \citenamefont {Taniguchi},
  \citenamefont {Watanabe}, \citenamefont {McGuire}, \citenamefont {Xiao} \emph
  {et~al.}}]{cenker2020direct}%
  \BibitemOpen
  \bibfield  {author} {\bibinfo {author} {\bibfnamefont {J.}~\bibnamefont
  {Cenker}}, \bibinfo {author} {\bibfnamefont {B.}~\bibnamefont {Huang}},
  \bibinfo {author} {\bibfnamefont {N.}~\bibnamefont {Suri}}, \bibinfo {author}
  {\bibfnamefont {P.}~\bibnamefont {Thijssen}}, \bibinfo {author}
  {\bibfnamefont {A.}~\bibnamefont {Miller}}, \bibinfo {author} {\bibfnamefont
  {T.}~\bibnamefont {Song}}, \bibinfo {author} {\bibfnamefont {T.}~\bibnamefont
  {Taniguchi}}, \bibinfo {author} {\bibfnamefont {K.}~\bibnamefont {Watanabe}},
  \bibinfo {author} {\bibfnamefont {M.~A.}\ \bibnamefont {McGuire}}, \bibinfo
  {author} {\bibfnamefont {D.}~\bibnamefont {Xiao}},  \emph {et~al.},\
  }\href@noop {} {\bibfield  {journal} {\bibinfo  {journal} {arXiv preprint
  arXiv:2001.07025}\ } (\bibinfo {year} {2020})}\BibitemShut {NoStop}%
\bibitem [{\citenamefont {Huebl}\ \emph {et~al.}(2013)\citenamefont {Huebl},
  \citenamefont {Zollitsch}, \citenamefont {Lotze}, \citenamefont {Hocke},
  \citenamefont {Greifenstein}, \citenamefont {Marx}, \citenamefont {Gross},\
  and\ \citenamefont {Goennenwein}}]{huebl2013high}%
  \BibitemOpen
  \bibfield  {author} {\bibinfo {author} {\bibfnamefont {H.}~\bibnamefont
  {Huebl}}, \bibinfo {author} {\bibfnamefont {C.~W.}\ \bibnamefont
  {Zollitsch}}, \bibinfo {author} {\bibfnamefont {J.}~\bibnamefont {Lotze}},
  \bibinfo {author} {\bibfnamefont {F.}~\bibnamefont {Hocke}}, \bibinfo
  {author} {\bibfnamefont {M.}~\bibnamefont {Greifenstein}}, \bibinfo {author}
  {\bibfnamefont {A.}~\bibnamefont {Marx}}, \bibinfo {author} {\bibfnamefont
  {R.}~\bibnamefont {Gross}}, \ and\ \bibinfo {author} {\bibfnamefont {S.~T.}\
  \bibnamefont {Goennenwein}},\ }\href@noop {} {\bibfield  {journal} {\bibinfo
  {journal} {Phys. Rev. Lett.}\ }\textbf {\bibinfo {volume} {111}},\ \bibinfo
  {pages} {127003} (\bibinfo {year} {2013})}\BibitemShut {NoStop}%
\bibitem [{\citenamefont {Tabuchi}\ \emph {et~al.}(2014)\citenamefont
  {Tabuchi}, \citenamefont {Ishino}, \citenamefont {Ishikawa}, \citenamefont
  {Yamazaki}, \citenamefont {Usami},\ and\ \citenamefont
  {Nakamura}}]{tabuchi2014hybridizing}%
  \BibitemOpen
  \bibfield  {author} {\bibinfo {author} {\bibfnamefont {Y.}~\bibnamefont
  {Tabuchi}}, \bibinfo {author} {\bibfnamefont {S.}~\bibnamefont {Ishino}},
  \bibinfo {author} {\bibfnamefont {T.}~\bibnamefont {Ishikawa}}, \bibinfo
  {author} {\bibfnamefont {R.}~\bibnamefont {Yamazaki}}, \bibinfo {author}
  {\bibfnamefont {K.}~\bibnamefont {Usami}}, \ and\ \bibinfo {author}
  {\bibfnamefont {Y.}~\bibnamefont {Nakamura}},\ }\href@noop {} {\bibfield
  {journal} {\bibinfo  {journal} {Phys. Rev. Lett.}\ }\textbf {\bibinfo
  {volume} {113}},\ \bibinfo {pages} {083603} (\bibinfo {year}
  {2014})}\BibitemShut {NoStop}%
\bibitem [{\citenamefont {Bai}\ \emph {et~al.}(2015)\citenamefont {Bai},
  \citenamefont {Harder}, \citenamefont {Chen}, \citenamefont {Fan},
  \citenamefont {Xiao},\ and\ \citenamefont {Hu}}]{bai2015spin}%
  \BibitemOpen
  \bibfield  {author} {\bibinfo {author} {\bibfnamefont {L.}~\bibnamefont
  {Bai}}, \bibinfo {author} {\bibfnamefont {M.}~\bibnamefont {Harder}},
  \bibinfo {author} {\bibfnamefont {Y.}~\bibnamefont {Chen}}, \bibinfo {author}
  {\bibfnamefont {X.}~\bibnamefont {Fan}}, \bibinfo {author} {\bibfnamefont
  {J.}~\bibnamefont {Xiao}}, \ and\ \bibinfo {author} {\bibfnamefont {C.-M.}\
  \bibnamefont {Hu}},\ }\href@noop {} {\bibfield  {journal} {\bibinfo
  {journal} {Phys. Rev. Lett.}\ }\textbf {\bibinfo {volume} {114}},\ \bibinfo
  {pages} {227201} (\bibinfo {year} {2015})}\BibitemShut {NoStop}%
\bibitem [{\citenamefont {Li}\ \emph {et~al.}(2019)\citenamefont {Li},
  \citenamefont {Polakovic}, \citenamefont {Wang}, \citenamefont {Xu},
  \citenamefont {Lendinez}, \citenamefont {Zhang}, \citenamefont {Ding},
  \citenamefont {Khaire}, \citenamefont {Saglam}, \citenamefont {Divan} \emph
  {et~al.}}]{li2019strong}%
  \BibitemOpen
  \bibfield  {author} {\bibinfo {author} {\bibfnamefont {Y.}~\bibnamefont
  {Li}}, \bibinfo {author} {\bibfnamefont {T.}~\bibnamefont {Polakovic}},
  \bibinfo {author} {\bibfnamefont {Y.-L.}\ \bibnamefont {Wang}}, \bibinfo
  {author} {\bibfnamefont {J.}~\bibnamefont {Xu}}, \bibinfo {author}
  {\bibfnamefont {S.}~\bibnamefont {Lendinez}}, \bibinfo {author}
  {\bibfnamefont {Z.}~\bibnamefont {Zhang}}, \bibinfo {author} {\bibfnamefont
  {J.}~\bibnamefont {Ding}}, \bibinfo {author} {\bibfnamefont {T.}~\bibnamefont
  {Khaire}}, \bibinfo {author} {\bibfnamefont {H.}~\bibnamefont {Saglam}},
  \bibinfo {author} {\bibfnamefont {R.}~\bibnamefont {Divan}},  \emph
  {et~al.},\ }\href@noop {} {\bibfield  {journal} {\bibinfo  {journal} {Phys.
  Rev. Lett.}\ }\textbf {\bibinfo {volume} {123}},\ \bibinfo {pages} {107701}
  (\bibinfo {year} {2019})}\BibitemShut {NoStop}%
\bibitem [{\citenamefont {Klingler}\ \emph {et~al.}(2018)\citenamefont
  {Klingler}, \citenamefont {Amin}, \citenamefont {Gepr{\"a}gs}, \citenamefont
  {Ganzhorn}, \citenamefont {Maier-Flaig}, \citenamefont {Althammer},
  \citenamefont {Huebl}, \citenamefont {Gross}, \citenamefont {McMichael},
  \citenamefont {Stiles} \emph {et~al.}}]{klingler2018spin}%
  \BibitemOpen
  \bibfield  {author} {\bibinfo {author} {\bibfnamefont {S.}~\bibnamefont
  {Klingler}}, \bibinfo {author} {\bibfnamefont {V.}~\bibnamefont {Amin}},
  \bibinfo {author} {\bibfnamefont {S.}~\bibnamefont {Gepr{\"a}gs}}, \bibinfo
  {author} {\bibfnamefont {K.}~\bibnamefont {Ganzhorn}}, \bibinfo {author}
  {\bibfnamefont {H.}~\bibnamefont {Maier-Flaig}}, \bibinfo {author}
  {\bibfnamefont {M.}~\bibnamefont {Althammer}}, \bibinfo {author}
  {\bibfnamefont {H.}~\bibnamefont {Huebl}}, \bibinfo {author} {\bibfnamefont
  {R.}~\bibnamefont {Gross}}, \bibinfo {author} {\bibfnamefont {R.~D.}\
  \bibnamefont {McMichael}}, \bibinfo {author} {\bibfnamefont {M.~D.}\
  \bibnamefont {Stiles}},  \emph {et~al.},\ }\href@noop {} {\bibfield
  {journal} {\bibinfo  {journal} {Phys. Rev. Lett.}\ }\textbf {\bibinfo
  {volume} {120}},\ \bibinfo {pages} {127201} (\bibinfo {year}
  {2018})}\BibitemShut {NoStop}%
\bibitem [{\citenamefont {Chen}\ \emph
  {et~al.}(2018{\natexlab{b}})\citenamefont {Chen}, \citenamefont {Liu},
  \citenamefont {Liu}, \citenamefont {Xiao}, \citenamefont {Xia}, \citenamefont
  {Bauer}, \citenamefont {Wu},\ and\ \citenamefont {Yu}}]{chen2018strong}%
  \BibitemOpen
  \bibfield  {author} {\bibinfo {author} {\bibfnamefont {J.}~\bibnamefont
  {Chen}}, \bibinfo {author} {\bibfnamefont {C.}~\bibnamefont {Liu}}, \bibinfo
  {author} {\bibfnamefont {T.}~\bibnamefont {Liu}}, \bibinfo {author}
  {\bibfnamefont {Y.}~\bibnamefont {Xiao}}, \bibinfo {author} {\bibfnamefont
  {K.}~\bibnamefont {Xia}}, \bibinfo {author} {\bibfnamefont {G.~E.}\
  \bibnamefont {Bauer}}, \bibinfo {author} {\bibfnamefont {M.}~\bibnamefont
  {Wu}}, \ and\ \bibinfo {author} {\bibfnamefont {H.}~\bibnamefont {Yu}},\
  }\href@noop {} {\bibfield  {journal} {\bibinfo  {journal} {Phys. Rev. Lett.}\
  }\textbf {\bibinfo {volume} {120}},\ \bibinfo {pages} {217202} (\bibinfo
  {year} {2018}{\natexlab{b}})}\BibitemShut {NoStop}%
\bibitem [{\citenamefont {Li}\ \emph {et~al.}(2020{\natexlab{b}})\citenamefont
  {Li}, \citenamefont {Cao}, \citenamefont {Amin}, \citenamefont {Zhang},
  \citenamefont {Gibbons}, \citenamefont {Sklenar}, \citenamefont {Pearson},
  \citenamefont {Haney}, \citenamefont {Stiles}, \citenamefont {Bailey},
  \citenamefont {Novosad}, \citenamefont {Hoffmann},\ and\ \citenamefont
  {Zhang}}]{li2019coherent}%
  \BibitemOpen
  \bibfield  {author} {\bibinfo {author} {\bibfnamefont {Y.}~\bibnamefont
  {Li}}, \bibinfo {author} {\bibfnamefont {W.}~\bibnamefont {Cao}}, \bibinfo
  {author} {\bibfnamefont {V.~P.}\ \bibnamefont {Amin}}, \bibinfo {author}
  {\bibfnamefont {Z.}~\bibnamefont {Zhang}}, \bibinfo {author} {\bibfnamefont
  {J.}~\bibnamefont {Gibbons}}, \bibinfo {author} {\bibfnamefont
  {J.}~\bibnamefont {Sklenar}}, \bibinfo {author} {\bibfnamefont
  {J.}~\bibnamefont {Pearson}}, \bibinfo {author} {\bibfnamefont {P.~M.}\
  \bibnamefont {Haney}}, \bibinfo {author} {\bibfnamefont {M.~D.}\ \bibnamefont
  {Stiles}}, \bibinfo {author} {\bibfnamefont {W.~E.}\ \bibnamefont {Bailey}},
  \bibinfo {author} {\bibfnamefont {V.}~\bibnamefont {Novosad}}, \bibinfo
  {author} {\bibfnamefont {A.}~\bibnamefont {Hoffmann}}, \ and\ \bibinfo
  {author} {\bibfnamefont {W.}~\bibnamefont {Zhang}},\ }\href {\doibase
  10.1103/PhysRevLett.124.117202} {\bibfield  {journal} {\bibinfo  {journal}
  {Phys. Rev. Lett.}\ }\textbf {\bibinfo {volume} {124}},\ \bibinfo {pages}
  {117202} (\bibinfo {year} {2020}{\natexlab{b}})}\BibitemShut {NoStop}%
\bibitem [{\citenamefont {{\v{Z}}elezn{\`y}}\ \emph {et~al.}(2014)\citenamefont
  {{\v{Z}}elezn{\`y}}, \citenamefont {Gao}, \citenamefont {V{\`y}born{\`y}},
  \citenamefont {Zemen}, \citenamefont {Ma{\v{s}}ek}, \citenamefont {Manchon},
  \citenamefont {Wunderlich}, \citenamefont {Sinova},\ and\ \citenamefont
  {Jungwirth}}]{vzelezny2014relativistic}%
  \BibitemOpen
  \bibfield  {author} {\bibinfo {author} {\bibfnamefont {J.}~\bibnamefont
  {{\v{Z}}elezn{\`y}}}, \bibinfo {author} {\bibfnamefont {H.}~\bibnamefont
  {Gao}}, \bibinfo {author} {\bibfnamefont {K.}~\bibnamefont
  {V{\`y}born{\`y}}}, \bibinfo {author} {\bibfnamefont {J.}~\bibnamefont
  {Zemen}}, \bibinfo {author} {\bibfnamefont {J.}~\bibnamefont {Ma{\v{s}}ek}},
  \bibinfo {author} {\bibfnamefont {A.}~\bibnamefont {Manchon}}, \bibinfo
  {author} {\bibfnamefont {J.}~\bibnamefont {Wunderlich}}, \bibinfo {author}
  {\bibfnamefont {J.}~\bibnamefont {Sinova}}, \ and\ \bibinfo {author}
  {\bibfnamefont {T.}~\bibnamefont {Jungwirth}},\ }\href@noop {} {\bibfield
  {journal} {\bibinfo  {journal} {Phys. Rev. Lett.}\ }\textbf {\bibinfo
  {volume} {113}},\ \bibinfo {pages} {157201} (\bibinfo {year}
  {2014})}\BibitemShut {NoStop}%
\bibitem [{\citenamefont {Roy}, \citenamefont {Otxoa},\ and\ \citenamefont
  {Wunderlich}(2016)}]{roy2016robust}%
  \BibitemOpen
  \bibfield  {author} {\bibinfo {author} {\bibfnamefont {P.}~\bibnamefont
  {Roy}}, \bibinfo {author} {\bibfnamefont {R.}~\bibnamefont {Otxoa}}, \ and\
  \bibinfo {author} {\bibfnamefont {J.}~\bibnamefont {Wunderlich}},\
  }\href@noop {} {\bibfield  {journal} {\bibinfo  {journal} {Phys. Rev. B}\
  }\textbf {\bibinfo {volume} {94}},\ \bibinfo {pages} {014439} (\bibinfo
  {year} {2016})}\BibitemShut {NoStop}%
\bibitem [{\citenamefont {Gray}\ \emph {et~al.}(2019)\citenamefont {Gray},
  \citenamefont {Moriyama}, \citenamefont {Sivadas}, \citenamefont {Stiehl},
  \citenamefont {Heron}, \citenamefont {Need}, \citenamefont {Kirby},
  \citenamefont {Low}, \citenamefont {Nowack}, \citenamefont {Schlom} \emph
  {et~al.}}]{gray2019spin}%
  \BibitemOpen
  \bibfield  {author} {\bibinfo {author} {\bibfnamefont {I.}~\bibnamefont
  {Gray}}, \bibinfo {author} {\bibfnamefont {T.}~\bibnamefont {Moriyama}},
  \bibinfo {author} {\bibfnamefont {N.}~\bibnamefont {Sivadas}}, \bibinfo
  {author} {\bibfnamefont {G.~M.}\ \bibnamefont {Stiehl}}, \bibinfo {author}
  {\bibfnamefont {J.~T.}\ \bibnamefont {Heron}}, \bibinfo {author}
  {\bibfnamefont {R.}~\bibnamefont {Need}}, \bibinfo {author} {\bibfnamefont
  {B.~J.}\ \bibnamefont {Kirby}}, \bibinfo {author} {\bibfnamefont {D.~H.}\
  \bibnamefont {Low}}, \bibinfo {author} {\bibfnamefont {K.~C.}\ \bibnamefont
  {Nowack}}, \bibinfo {author} {\bibfnamefont {D.~G.}\ \bibnamefont {Schlom}},
  \emph {et~al.},\ }\href@noop {} {\bibfield  {journal} {\bibinfo  {journal}
  {Phys. Rev. X}\ }\textbf {\bibinfo {volume} {9}},\ \bibinfo {pages} {041016}
  (\bibinfo {year} {2019})}\BibitemShut {NoStop}%
\bibitem [{\citenamefont {Olejn{\'\i}k}\ \emph {et~al.}(2018)\citenamefont
  {Olejn{\'\i}k}, \citenamefont {Seifert}, \citenamefont {Ka{\v{s}}par},
  \citenamefont {Nov{\'a}k}, \citenamefont {Wadley}, \citenamefont {Campion},
  \citenamefont {Baumgartner}, \citenamefont {Gambardella}, \citenamefont
  {N{\v{e}}mec}, \citenamefont {Wunderlich} \emph
  {et~al.}}]{olejnik2018terahertz}%
  \BibitemOpen
  \bibfield  {author} {\bibinfo {author} {\bibfnamefont {K.}~\bibnamefont
  {Olejn{\'\i}k}}, \bibinfo {author} {\bibfnamefont {T.}~\bibnamefont
  {Seifert}}, \bibinfo {author} {\bibfnamefont {Z.}~\bibnamefont
  {Ka{\v{s}}par}}, \bibinfo {author} {\bibfnamefont {V.}~\bibnamefont
  {Nov{\'a}k}}, \bibinfo {author} {\bibfnamefont {P.}~\bibnamefont {Wadley}},
  \bibinfo {author} {\bibfnamefont {R.~P.}\ \bibnamefont {Campion}}, \bibinfo
  {author} {\bibfnamefont {M.}~\bibnamefont {Baumgartner}}, \bibinfo {author}
  {\bibfnamefont {P.}~\bibnamefont {Gambardella}}, \bibinfo {author}
  {\bibfnamefont {P.}~\bibnamefont {N{\v{e}}mec}}, \bibinfo {author}
  {\bibfnamefont {J.}~\bibnamefont {Wunderlich}},  \emph {et~al.},\ }\href@noop
  {} {\bibfield  {journal} {\bibinfo  {journal} {Sci. Adv.}\ }\textbf {\bibinfo
  {volume} {4}},\ \bibinfo {pages} {eaar3566} (\bibinfo {year}
  {2018})}\BibitemShut {NoStop}%
\bibitem [{\citenamefont {Simensen}\ \emph {et~al.}(2020)\citenamefont
  {Simensen}, \citenamefont {Kamra}, \citenamefont {Troncoso},\ and\
  \citenamefont {Brataas}}]{simensen2020magnon}%
  \BibitemOpen
  \bibfield  {author} {\bibinfo {author} {\bibfnamefont {H.~T.}\ \bibnamefont
  {Simensen}}, \bibinfo {author} {\bibfnamefont {A.}~\bibnamefont {Kamra}},
  \bibinfo {author} {\bibfnamefont {R.~E.}\ \bibnamefont {Troncoso}}, \ and\
  \bibinfo {author} {\bibfnamefont {A.}~\bibnamefont {Brataas}},\ }\href@noop
  {} {\bibfield  {journal} {\bibinfo  {journal} {Phys. Rev. B}\ }\textbf
  {\bibinfo {volume} {101}},\ \bibinfo {pages} {020403} (\bibinfo {year}
  {2020})}\BibitemShut {NoStop}%
\bibitem [{\citenamefont {Mancini}\ \emph {et~al.}(2013)\citenamefont
  {Mancini}, \citenamefont {Pressacco}, \citenamefont {Haertinger},
  \citenamefont {Fullerton}, \citenamefont {Suzuki}, \citenamefont
  {Woltersdorf},\ and\ \citenamefont {Back}}]{mancini2013magnetic}%
  \BibitemOpen
  \bibfield  {author} {\bibinfo {author} {\bibfnamefont {E.}~\bibnamefont
  {Mancini}}, \bibinfo {author} {\bibfnamefont {F.}~\bibnamefont {Pressacco}},
  \bibinfo {author} {\bibfnamefont {M.}~\bibnamefont {Haertinger}}, \bibinfo
  {author} {\bibfnamefont {E.~E.}\ \bibnamefont {Fullerton}}, \bibinfo {author}
  {\bibfnamefont {T.}~\bibnamefont {Suzuki}}, \bibinfo {author} {\bibfnamefont
  {G.}~\bibnamefont {Woltersdorf}}, \ and\ \bibinfo {author} {\bibfnamefont
  {C.~H.}\ \bibnamefont {Back}},\ }\href@noop {} {\bibfield  {journal}
  {\bibinfo  {journal} {J. Phys. D: Appl. Phys.}\ }\textbf {\bibinfo {volume}
  {46}},\ \bibinfo {pages} {245302} (\bibinfo {year} {2013})}\BibitemShut
  {NoStop}%
\bibitem [{\citenamefont {N{\v e}mec}\ \emph {et~al.}(2018)\citenamefont {N{\v
  e}mec}, \citenamefont {Fiebig}, \citenamefont {Kampfrath},\ and\
  \citenamefont {Kimel}}]{Nemec2018}%
  \BibitemOpen
  \bibfield  {author} {\bibinfo {author} {\bibfnamefont {P.}~\bibnamefont {N{\v
  e}mec}}, \bibinfo {author} {\bibfnamefont {M.}~\bibnamefont {Fiebig}},
  \bibinfo {author} {\bibfnamefont {T.}~\bibnamefont {Kampfrath}}, \ and\
  \bibinfo {author} {\bibfnamefont {A.~V.}\ \bibnamefont {Kimel}},\ }\href
  {\doibase 10.1038/s41567-018-0051-x} {\bibfield  {journal} {\bibinfo
  {journal} {Nat. Phys.}\ }\textbf {\bibinfo {volume} {14}},\ \bibinfo {pages}
  {229} (\bibinfo {year} {2018})}\BibitemShut {NoStop}%
\bibitem [{\citenamefont {Cheong}\ \emph {et~al.}(2020)\citenamefont {Cheong},
  \citenamefont {Fiebig}, \citenamefont {Wu}, \citenamefont {Chapon},\ and\
  \citenamefont {Kiryukhin}}]{Cheong2020}%
  \BibitemOpen
  \bibfield  {author} {\bibinfo {author} {\bibfnamefont {S.-W.}\ \bibnamefont
  {Cheong}}, \bibinfo {author} {\bibfnamefont {M.}~\bibnamefont {Fiebig}},
  \bibinfo {author} {\bibfnamefont {W.}~\bibnamefont {Wu}}, \bibinfo {author}
  {\bibfnamefont {L.}~\bibnamefont {Chapon}}, \ and\ \bibinfo {author}
  {\bibfnamefont {V.}~\bibnamefont {Kiryukhin}},\ }\href {\doibase
  10.1038/s41535-019-0204-x} {\bibfield  {journal} {\bibinfo  {journal} {npj
  Quan. Mater.}\ }\textbf {\bibinfo {volume} {5}},\ \bibinfo {pages} {3}
  (\bibinfo {year} {2020})}\BibitemShut {NoStop}%
\bibitem [{\citenamefont {Wadley}\ \emph {et~al.}(2015)\citenamefont {Wadley},
  \citenamefont {Hills}, \citenamefont {Shahedkhah}, \citenamefont {Edmonds},
  \citenamefont {Campion}, \citenamefont {Nov{\'a}k}, \citenamefont
  {Ouladdiaf}, \citenamefont {Khalyavin}, \citenamefont {Langridge},
  \citenamefont {Saidl}, \citenamefont {Nemec}, \citenamefont {Rushforth},
  \citenamefont {Gallagher}, \citenamefont {Dhesi}, \citenamefont
  {Maccherozzi}, \citenamefont {{\v Z}elezn{\'y}},\ and\ \citenamefont
  {Jungwirth}}]{Wadley2015}%
  \BibitemOpen
  \bibfield  {author} {\bibinfo {author} {\bibfnamefont {P.}~\bibnamefont
  {Wadley}}, \bibinfo {author} {\bibfnamefont {V.}~\bibnamefont {Hills}},
  \bibinfo {author} {\bibfnamefont {M.~R.}\ \bibnamefont {Shahedkhah}},
  \bibinfo {author} {\bibfnamefont {K.~W.}\ \bibnamefont {Edmonds}}, \bibinfo
  {author} {\bibfnamefont {R.~P.}\ \bibnamefont {Campion}}, \bibinfo {author}
  {\bibfnamefont {V.}~\bibnamefont {Nov{\'a}k}}, \bibinfo {author}
  {\bibfnamefont {B.}~\bibnamefont {Ouladdiaf}}, \bibinfo {author}
  {\bibfnamefont {D.}~\bibnamefont {Khalyavin}}, \bibinfo {author}
  {\bibfnamefont {S.}~\bibnamefont {Langridge}}, \bibinfo {author}
  {\bibfnamefont {V.}~\bibnamefont {Saidl}}, \bibinfo {author} {\bibfnamefont
  {P.}~\bibnamefont {Nemec}}, \bibinfo {author} {\bibfnamefont {A.~W.}\
  \bibnamefont {Rushforth}}, \bibinfo {author} {\bibfnamefont {B.~L.}\
  \bibnamefont {Gallagher}}, \bibinfo {author} {\bibfnamefont {S.~S.}\
  \bibnamefont {Dhesi}}, \bibinfo {author} {\bibfnamefont {F.}~\bibnamefont
  {Maccherozzi}}, \bibinfo {author} {\bibfnamefont {J.}~\bibnamefont {{\v
  Z}elezn{\'y}}}, \ and\ \bibinfo {author} {\bibfnamefont {T.}~\bibnamefont
  {Jungwirth}},\ }\href {\doibase 10.1038/srep17079} {\bibfield  {journal}
  {\bibinfo  {journal} {Sci. Rep.}\ }\textbf {\bibinfo {volume} {5}},\ \bibinfo
  {pages} {17079} (\bibinfo {year} {2015})}\BibitemShut {NoStop}%
\bibitem [{\citenamefont {Wadley}\ \emph {et~al.}(2017)\citenamefont {Wadley},
  \citenamefont {Edmonds}, \citenamefont {Shahedkhah}, \citenamefont {Campion},
  \citenamefont {Gallagher}, \citenamefont {{\v Z}elezn{\'y}}, \citenamefont
  {Kune{\v s}}, \citenamefont {Nov{\'a}k}, \citenamefont {Jungwirth},
  \citenamefont {Saidl}, \citenamefont {N{\v e}mec}, \citenamefont
  {Maccherozzi},\ and\ \citenamefont {Dhesi}}]{Wadley2017}%
  \BibitemOpen
  \bibfield  {author} {\bibinfo {author} {\bibfnamefont {P.}~\bibnamefont
  {Wadley}}, \bibinfo {author} {\bibfnamefont {K.~W.}\ \bibnamefont {Edmonds}},
  \bibinfo {author} {\bibfnamefont {M.~R.}\ \bibnamefont {Shahedkhah}},
  \bibinfo {author} {\bibfnamefont {R.~P.}\ \bibnamefont {Campion}}, \bibinfo
  {author} {\bibfnamefont {B.~L.}\ \bibnamefont {Gallagher}}, \bibinfo {author}
  {\bibfnamefont {J.}~\bibnamefont {{\v Z}elezn{\'y}}}, \bibinfo {author}
  {\bibfnamefont {J.}~\bibnamefont {Kune{\v s}}}, \bibinfo {author}
  {\bibfnamefont {V.}~\bibnamefont {Nov{\'a}k}}, \bibinfo {author}
  {\bibfnamefont {T.}~\bibnamefont {Jungwirth}}, \bibinfo {author}
  {\bibfnamefont {V.}~\bibnamefont {Saidl}}, \bibinfo {author} {\bibfnamefont
  {P.}~\bibnamefont {N{\v e}mec}}, \bibinfo {author} {\bibfnamefont
  {F.}~\bibnamefont {Maccherozzi}}, \ and\ \bibinfo {author} {\bibfnamefont
  {S.~S.}\ \bibnamefont {Dhesi}},\ }\href {\doibase 10.1038/s41598-017-11653-8}
  {\bibfield  {journal} {\bibinfo  {journal} {Sci. Rep.}\ }\textbf {\bibinfo
  {volume} {7}},\ \bibinfo {pages} {11147} (\bibinfo {year}
  {2017})}\BibitemShut {NoStop}%
\bibitem [{\citenamefont {Sapozhnik}\ \emph {et~al.}(2018)\citenamefont
  {Sapozhnik}, \citenamefont {Filianina}, \citenamefont {Bodnar}, \citenamefont
  {Lamirand}, \citenamefont {Mawass}, \citenamefont {Skourski}, \citenamefont
  {Elmers}, \citenamefont {Zabel}, \citenamefont {Kl\"aui},\ and\ \citenamefont
  {Jourdan}}]{Sapozhnik2018}%
  \BibitemOpen
  \bibfield  {author} {\bibinfo {author} {\bibfnamefont {A.~A.}\ \bibnamefont
  {Sapozhnik}}, \bibinfo {author} {\bibfnamefont {M.}~\bibnamefont
  {Filianina}}, \bibinfo {author} {\bibfnamefont {S.~Y.}\ \bibnamefont
  {Bodnar}}, \bibinfo {author} {\bibfnamefont {A.}~\bibnamefont {Lamirand}},
  \bibinfo {author} {\bibfnamefont {M.-A.}\ \bibnamefont {Mawass}}, \bibinfo
  {author} {\bibfnamefont {Y.}~\bibnamefont {Skourski}}, \bibinfo {author}
  {\bibfnamefont {H.-J.}\ \bibnamefont {Elmers}}, \bibinfo {author}
  {\bibfnamefont {H.}~\bibnamefont {Zabel}}, \bibinfo {author} {\bibfnamefont
  {M.}~\bibnamefont {Kl\"aui}}, \ and\ \bibinfo {author} {\bibfnamefont
  {M.}~\bibnamefont {Jourdan}},\ }\href {\doibase 10.1103/PhysRevB.97.134429}
  {\bibfield  {journal} {\bibinfo  {journal} {Phys. Rev. B}\ }\textbf {\bibinfo
  {volume} {97}},\ \bibinfo {pages} {134429} (\bibinfo {year}
  {2018})}\BibitemShut {NoStop}%
\bibitem [{\citenamefont {Loudon}(2012)}]{Loudon2012}%
  \BibitemOpen
  \bibfield  {author} {\bibinfo {author} {\bibfnamefont {J.~C.}\ \bibnamefont
  {Loudon}},\ }\href {\doibase 10.1103/PhysRevLett.109.267204} {\bibfield
  {journal} {\bibinfo  {journal} {Phys. Rev. Lett.}\ }\textbf {\bibinfo
  {volume} {109}},\ \bibinfo {pages} {267204} (\bibinfo {year}
  {2012})}\BibitemShut {NoStop}%
\bibitem [{\citenamefont {McCord}(2015)}]{McCord2015}%
  \BibitemOpen
  \bibfield  {author} {\bibinfo {author} {\bibfnamefont {J.}~\bibnamefont
  {McCord}},\ }\href {\doibase 10.1088/0022-3727/48/33/333001} {\bibfield
  {journal} {\bibinfo  {journal} {J. Phys. D: Appl. Phys.}\ }\textbf {\bibinfo
  {volume} {48}},\ \bibinfo {pages} {333001} (\bibinfo {year}
  {2015})}\BibitemShut {NoStop}%
\bibitem [{\citenamefont {Oppeneer}(2017)}]{Oppeneer2017}%
  \BibitemOpen
  \bibfield  {author} {\bibinfo {author} {\bibfnamefont {P.~M.}\ \bibnamefont
  {Oppeneer}},\ }\href {\doibase 10.1038/nphoton.2016.274} {\bibfield
  {journal} {\bibinfo  {journal} {Nat. Photon.}\ }\textbf {\bibinfo {volume}
  {11}},\ \bibinfo {pages} {74} (\bibinfo {year} {2017})}\BibitemShut {NoStop}%
\bibitem [{\citenamefont {Saidl}\ \emph {et~al.}(2017)\citenamefont {Saidl},
  \citenamefont {N{\v e}mec}, \citenamefont {Wadley}, \citenamefont {Hills},
  \citenamefont {Campion}, \citenamefont {Nov{\'a}k}, \citenamefont {Edmonds},
  \citenamefont {Maccherozzi}, \citenamefont {Dhesi}, \citenamefont
  {Gallagher}, \citenamefont {Troj{\'a}nek}, \citenamefont {Kune{\v s}},
  \citenamefont {{\v Z}elezn{\'y}}, \citenamefont {Mal{\'y}},\ and\
  \citenamefont {Jungwirth}}]{Saidl2017}%
  \BibitemOpen
  \bibfield  {author} {\bibinfo {author} {\bibfnamefont {V.}~\bibnamefont
  {Saidl}}, \bibinfo {author} {\bibfnamefont {P.}~\bibnamefont {N{\v e}mec}},
  \bibinfo {author} {\bibfnamefont {P.}~\bibnamefont {Wadley}}, \bibinfo
  {author} {\bibfnamefont {V.}~\bibnamefont {Hills}}, \bibinfo {author}
  {\bibfnamefont {R.~P.}\ \bibnamefont {Campion}}, \bibinfo {author}
  {\bibfnamefont {V.}~\bibnamefont {Nov{\'a}k}}, \bibinfo {author}
  {\bibfnamefont {K.~W.}\ \bibnamefont {Edmonds}}, \bibinfo {author}
  {\bibfnamefont {F.}~\bibnamefont {Maccherozzi}}, \bibinfo {author}
  {\bibfnamefont {S.~S.}\ \bibnamefont {Dhesi}}, \bibinfo {author}
  {\bibfnamefont {B.~L.}\ \bibnamefont {Gallagher}}, \bibinfo {author}
  {\bibfnamefont {F.}~\bibnamefont {Troj{\'a}nek}}, \bibinfo {author}
  {\bibfnamefont {J.}~\bibnamefont {Kune{\v s}}}, \bibinfo {author}
  {\bibfnamefont {J.}~\bibnamefont {{\v Z}elezn{\'y}}}, \bibinfo {author}
  {\bibfnamefont {P.}~\bibnamefont {Mal{\'y}}}, \ and\ \bibinfo {author}
  {\bibfnamefont {T.}~\bibnamefont {Jungwirth}},\ }\href {\doibase
  10.1038/nphoton.2016.255} {\bibfield  {journal} {\bibinfo  {journal} {Nat.
  Photon.}\ }\textbf {\bibinfo {volume} {11}},\ \bibinfo {pages} {91} (\bibinfo
  {year} {2017})}\BibitemShut {NoStop}%
\bibitem [{\citenamefont {Iida}\ \emph {et~al.}(2011)\citenamefont {Iida},
  \citenamefont {Satoh}, \citenamefont {Shimura}, \citenamefont {Kuroda},
  \citenamefont {Ivanov}, \citenamefont {Tokunaga},\ and\ \citenamefont
  {Tokura}}]{Iida2011}%
  \BibitemOpen
  \bibfield  {author} {\bibinfo {author} {\bibfnamefont {R.}~\bibnamefont
  {Iida}}, \bibinfo {author} {\bibfnamefont {T.}~\bibnamefont {Satoh}},
  \bibinfo {author} {\bibfnamefont {T.}~\bibnamefont {Shimura}}, \bibinfo
  {author} {\bibfnamefont {K.}~\bibnamefont {Kuroda}}, \bibinfo {author}
  {\bibfnamefont {B.~A.}\ \bibnamefont {Ivanov}}, \bibinfo {author}
  {\bibfnamefont {Y.}~\bibnamefont {Tokunaga}}, \ and\ \bibinfo {author}
  {\bibfnamefont {Y.}~\bibnamefont {Tokura}},\ }\href {\doibase
  10.1103/PhysRevB.84.064402} {\bibfield  {journal} {\bibinfo  {journal} {Phys.
  Rev. B}\ }\textbf {\bibinfo {volume} {84}},\ \bibinfo {pages} {064402}
  (\bibinfo {year} {2011})}\BibitemShut {NoStop}%
\bibitem [{\citenamefont {Eremenko}\ \emph {et~al.}(2012)\citenamefont
  {Eremenko}, \citenamefont {Kharchenko}, \citenamefont {Litvinenko},\ and\
  \citenamefont {Naumenko}}]{Eremenko2012}%
  \BibitemOpen
  \bibfield  {author} {\bibinfo {author} {\bibfnamefont {V.}~\bibnamefont
  {Eremenko}}, \bibinfo {author} {\bibfnamefont {N.}~\bibnamefont
  {Kharchenko}}, \bibinfo {author} {\bibfnamefont {Y.}~\bibnamefont
  {Litvinenko}}, \ and\ \bibinfo {author} {\bibfnamefont {V.}~\bibnamefont
  {Naumenko}},\ }\href {https://books.google.com/books?id=AxnSBwAAQBAJ} {\emph
  {\bibinfo {title} {Magneto-Optics and Spectroscopy of Antiferromagnets}}}\
  (\bibinfo  {publisher} {Springer New York},\ \bibinfo {year}
  {2012})\BibitemShut {NoStop}%
\bibitem [{\citenamefont {Yang}\ \emph {et~al.}(2019)\citenamefont {Yang},
  \citenamefont {Kang}, \citenamefont {Diao}, \citenamefont {Ramanathan},
  \citenamefont {Karigerasi}, \citenamefont {Shoemaker}, \citenamefont
  {Schleife},\ and\ \citenamefont {Cahill}}]{Yang2019}%
  \BibitemOpen
  \bibfield  {author} {\bibinfo {author} {\bibfnamefont {K.}~\bibnamefont
  {Yang}}, \bibinfo {author} {\bibfnamefont {K.}~\bibnamefont {Kang}}, \bibinfo
  {author} {\bibfnamefont {Z.}~\bibnamefont {Diao}}, \bibinfo {author}
  {\bibfnamefont {A.}~\bibnamefont {Ramanathan}}, \bibinfo {author}
  {\bibfnamefont {M.~H.}\ \bibnamefont {Karigerasi}}, \bibinfo {author}
  {\bibfnamefont {D.~P.}\ \bibnamefont {Shoemaker}}, \bibinfo {author}
  {\bibfnamefont {A.}~\bibnamefont {Schleife}}, \ and\ \bibinfo {author}
  {\bibfnamefont {D.~G.}\ \bibnamefont {Cahill}},\ }\href {\doibase
  10.1103/PhysRevMaterials.3.124408} {\bibfield  {journal} {\bibinfo  {journal}
  {Phys. Rev. Mater.}\ }\textbf {\bibinfo {volume} {3}},\ \bibinfo {pages}
  {124408} (\bibinfo {year} {2019})}\BibitemShut {NoStop}%
\bibitem [{\citenamefont {{Kerr LL. D.}}(1877)}]{John1877}%
  \BibitemOpen
  \bibfield  {author} {\bibinfo {author} {\bibfnamefont {J.}~\bibnamefont
  {{Kerr LL. D.}}},\ }\href {\doibase 10.1080/14786447708639245} {\bibfield
  {journal} {\bibinfo  {journal} {Lond. Edinb. Dubl. Phil. Mag.}\ }\textbf
  {\bibinfo {volume} {3}},\ \bibinfo {pages} {321} (\bibinfo {year}
  {1877})}\BibitemShut {NoStop}%
\bibitem [{\citenamefont {Faraday}(1855)}]{Faraday1855}%
  \BibitemOpen
  \bibfield  {author} {\bibinfo {author} {\bibfnamefont {M.}~\bibnamefont
  {Faraday}},\ }\href {https://books.google.com/books?id=Q7sKAAAAIAAJ} {\emph
  {\bibinfo {title} {Experimental Researches in Electricity}}},\ \bibinfo
  {series} {Experimental Researches in Electricity}\ No.\ \bibinfo {number} {v.
  3}\ (\bibinfo  {publisher} {Bernard Quaritch},\ \bibinfo {year}
  {1855})\BibitemShut {NoStop}%
\bibitem [{\citenamefont {Feng}\ \emph {et~al.}(2015)\citenamefont {Feng},
  \citenamefont {Guo}, \citenamefont {Zhou}, \citenamefont {Yao},\ and\
  \citenamefont {Niu}}]{Feng2015}%
  \BibitemOpen
  \bibfield  {author} {\bibinfo {author} {\bibfnamefont {W.}~\bibnamefont
  {Feng}}, \bibinfo {author} {\bibfnamefont {G.-Y.}\ \bibnamefont {Guo}},
  \bibinfo {author} {\bibfnamefont {J.}~\bibnamefont {Zhou}}, \bibinfo {author}
  {\bibfnamefont {Y.}~\bibnamefont {Yao}}, \ and\ \bibinfo {author}
  {\bibfnamefont {Q.}~\bibnamefont {Niu}},\ }\href {\doibase
  10.1103/PhysRevB.92.144426} {\bibfield  {journal} {\bibinfo  {journal} {Phys.
  Rev. B}\ }\textbf {\bibinfo {volume} {92}},\ \bibinfo {pages} {144426}
  (\bibinfo {year} {2015})}\BibitemShut {NoStop}%
\bibitem [{\citenamefont {Oppeneer}(2001)}]{Oppeneer2001}%
  \BibitemOpen
  \bibfield  {author} {\bibinfo {author} {\bibfnamefont {P.}~\bibnamefont
  {Oppeneer}},\ }\enquote {\bibinfo {title} {Chapter 3 magneto-optical kerr
  spectra},}\ in\ \href {\doibase
  https://doi.org/10.1016/S1567-2719(01)13007-6} {\emph {\bibinfo {booktitle}
  {Handbook of Magnetic Materials}}},\ Vol.~\bibinfo {volume} {13}\ (\bibinfo
  {publisher} {Elsevier},\ \bibinfo {year} {2001})\ pp.\ \bibinfo {pages} {229
  -- 422}\BibitemShut {NoStop}%
\bibitem [{\citenamefont {Higo}\ \emph {et~al.}(2018)\citenamefont {Higo},
  \citenamefont {Man}, \citenamefont {Gopman}, \citenamefont {Wu},
  \citenamefont {Koretsune}, \citenamefont {van~'t Erve}, \citenamefont
  {Kabanov}, \citenamefont {Rees}, \citenamefont {Li}, \citenamefont {Suzuki},
  \citenamefont {Patankar}, \citenamefont {Ikhlas}, \citenamefont {Chien},
  \citenamefont {Arita}, \citenamefont {Shull}, \citenamefont {Orenstein},\
  and\ \citenamefont {Nakatsuji}}]{Higo2018}%
  \BibitemOpen
  \bibfield  {author} {\bibinfo {author} {\bibfnamefont {T.}~\bibnamefont
  {Higo}}, \bibinfo {author} {\bibfnamefont {H.}~\bibnamefont {Man}}, \bibinfo
  {author} {\bibfnamefont {D.~B.}\ \bibnamefont {Gopman}}, \bibinfo {author}
  {\bibfnamefont {L.}~\bibnamefont {Wu}}, \bibinfo {author} {\bibfnamefont
  {T.}~\bibnamefont {Koretsune}}, \bibinfo {author} {\bibfnamefont {O.~M.~J.}\
  \bibnamefont {van~'t Erve}}, \bibinfo {author} {\bibfnamefont {Y.~P.}\
  \bibnamefont {Kabanov}}, \bibinfo {author} {\bibfnamefont {D.}~\bibnamefont
  {Rees}}, \bibinfo {author} {\bibfnamefont {Y.}~\bibnamefont {Li}}, \bibinfo
  {author} {\bibfnamefont {M.-T.}\ \bibnamefont {Suzuki}}, \bibinfo {author}
  {\bibfnamefont {S.}~\bibnamefont {Patankar}}, \bibinfo {author}
  {\bibfnamefont {M.}~\bibnamefont {Ikhlas}}, \bibinfo {author} {\bibfnamefont
  {C.~L.}\ \bibnamefont {Chien}}, \bibinfo {author} {\bibfnamefont
  {R.}~\bibnamefont {Arita}}, \bibinfo {author} {\bibfnamefont {R.~D.}\
  \bibnamefont {Shull}}, \bibinfo {author} {\bibfnamefont {J.}~\bibnamefont
  {Orenstein}}, \ and\ \bibinfo {author} {\bibfnamefont {S.}~\bibnamefont
  {Nakatsuji}},\ }\href {\doibase 10.1038/s41566-017-0086-z} {\bibfield
  {journal} {\bibinfo  {journal} {Nat. Photon.}\ }\textbf {\bibinfo {volume}
  {12}},\ \bibinfo {pages} {73} (\bibinfo {year} {2018})}\BibitemShut {NoStop}%
\bibitem [{\citenamefont {Suzuki}\ \emph {et~al.}(2017)\citenamefont {Suzuki},
  \citenamefont {Koretsune}, \citenamefont {Ochi},\ and\ \citenamefont
  {Arita}}]{Suzuki2017}%
  \BibitemOpen
  \bibfield  {author} {\bibinfo {author} {\bibfnamefont {M.-T.}\ \bibnamefont
  {Suzuki}}, \bibinfo {author} {\bibfnamefont {T.}~\bibnamefont {Koretsune}},
  \bibinfo {author} {\bibfnamefont {M.}~\bibnamefont {Ochi}}, \ and\ \bibinfo
  {author} {\bibfnamefont {R.}~\bibnamefont {Arita}},\ }\href {\doibase
  10.1103/PhysRevB.95.094406} {\bibfield  {journal} {\bibinfo  {journal} {Phys.
  Rev. B}\ }\textbf {\bibinfo {volume} {95}},\ \bibinfo {pages} {094406}
  (\bibinfo {year} {2017})}\BibitemShut {NoStop}%
\bibitem [{\citenamefont {Wimmer}\ \emph {et~al.}(2019)\citenamefont {Wimmer},
  \citenamefont {Mankovsky}, \citenamefont {Min\'ar}, \citenamefont {Yaresko},\
  and\ \citenamefont {Ebert}}]{Wimmer2019}%
  \BibitemOpen
  \bibfield  {author} {\bibinfo {author} {\bibfnamefont {S.}~\bibnamefont
  {Wimmer}}, \bibinfo {author} {\bibfnamefont {S.}~\bibnamefont {Mankovsky}},
  \bibinfo {author} {\bibfnamefont {J.}~\bibnamefont {Min\'ar}}, \bibinfo
  {author} {\bibfnamefont {A.~N.}\ \bibnamefont {Yaresko}}, \ and\ \bibinfo
  {author} {\bibfnamefont {H.}~\bibnamefont {Ebert}},\ }\href {\doibase
  10.1103/PhysRevB.100.214429} {\bibfield  {journal} {\bibinfo  {journal}
  {Phys. Rev. B}\ }\textbf {\bibinfo {volume} {100}},\ \bibinfo {pages}
  {214429} (\bibinfo {year} {2019})}\BibitemShut {NoStop}%
\bibitem [{\citenamefont {Feng}\ \emph {et~al.}(2020)\citenamefont {Feng},
  \citenamefont {Hanke}, \citenamefont {Zhou}, \citenamefont {Guo},
  \citenamefont {Bl{\"u}gel}, \citenamefont {Mokrousov},\ and\ \citenamefont
  {Yao}}]{Feng2020}%
  \BibitemOpen
  \bibfield  {author} {\bibinfo {author} {\bibfnamefont {W.}~\bibnamefont
  {Feng}}, \bibinfo {author} {\bibfnamefont {J.-P.}\ \bibnamefont {Hanke}},
  \bibinfo {author} {\bibfnamefont {X.}~\bibnamefont {Zhou}}, \bibinfo {author}
  {\bibfnamefont {G.-Y.}\ \bibnamefont {Guo}}, \bibinfo {author} {\bibfnamefont
  {S.}~\bibnamefont {Bl{\"u}gel}}, \bibinfo {author} {\bibfnamefont
  {Y.}~\bibnamefont {Mokrousov}}, \ and\ \bibinfo {author} {\bibfnamefont
  {Y.}~\bibnamefont {Yao}},\ }\href {\doibase 10.1038/s41467-019-13968-8}
  {\bibfield  {journal} {\bibinfo  {journal} {Nat. Comm.}\ }\textbf {\bibinfo
  {volume} {11}},\ \bibinfo {pages} {118} (\bibinfo {year} {2020})}\BibitemShut
  {NoStop}%
\bibitem [{\citenamefont {Balk}\ \emph {et~al.}(2019)\citenamefont {Balk},
  \citenamefont {Sung}, \citenamefont {Thomas}, \citenamefont {Rosa},
  \citenamefont {McDonald}, \citenamefont {Thompson}, \citenamefont {Bauer},
  \citenamefont {Ronning},\ and\ \citenamefont {Crooker}}]{Balk2019}%
  \BibitemOpen
  \bibfield  {author} {\bibinfo {author} {\bibfnamefont {A.~L.}\ \bibnamefont
  {Balk}}, \bibinfo {author} {\bibfnamefont {N.~H.}\ \bibnamefont {Sung}},
  \bibinfo {author} {\bibfnamefont {S.~M.}\ \bibnamefont {Thomas}}, \bibinfo
  {author} {\bibfnamefont {P.~F.~S.}\ \bibnamefont {Rosa}}, \bibinfo {author}
  {\bibfnamefont {R.~D.}\ \bibnamefont {McDonald}}, \bibinfo {author}
  {\bibfnamefont {J.~D.}\ \bibnamefont {Thompson}}, \bibinfo {author}
  {\bibfnamefont {E.~D.}\ \bibnamefont {Bauer}}, \bibinfo {author}
  {\bibfnamefont {F.}~\bibnamefont {Ronning}}, \ and\ \bibinfo {author}
  {\bibfnamefont {S.~A.}\ \bibnamefont {Crooker}},\ }\href {\doibase
  10.1063/1.5066557} {\bibfield  {journal} {\bibinfo  {journal} {Appl. Phys.
  Lett.}\ }\textbf {\bibinfo {volume} {114}},\ \bibinfo {pages} {032401}
  (\bibinfo {year} {2019})}\BibitemShut {NoStop}%
\bibitem [{\citenamefont {Dzyaloshinsky}(1958)}]{Dzyaloshinsky1958}%
  \BibitemOpen
  \bibfield  {author} {\bibinfo {author} {\bibfnamefont {I.}~\bibnamefont
  {Dzyaloshinsky}},\ }\href {\doibase
  https://doi.org/10.1016/0022-3697(58)90076-3} {\bibfield  {journal} {\bibinfo
   {journal} {J. Phys. Chem. Solids}\ }\textbf {\bibinfo {volume} {4}},\
  \bibinfo {pages} {241 } (\bibinfo {year} {1958})}\BibitemShut {NoStop}%
\bibitem [{\citenamefont {Moriya}(1960)}]{moriya1960}%
  \BibitemOpen
  \bibfield  {author} {\bibinfo {author} {\bibfnamefont {T.}~\bibnamefont
  {Moriya}},\ }\href {\doibase 10.1103/PhysRev.120.91} {\bibfield  {journal}
  {\bibinfo  {journal} {Phys. Rev.}\ }\textbf {\bibinfo {volume} {120}},\
  \bibinfo {pages} {91} (\bibinfo {year} {1960})}\BibitemShut {NoStop}%
\bibitem [{\citenamefont {Williams}, \citenamefont {Sherwood},\ and\
  \citenamefont {Remeika}(1958)}]{Williams1958}%
  \BibitemOpen
  \bibfield  {author} {\bibinfo {author} {\bibfnamefont {H.~J.}\ \bibnamefont
  {Williams}}, \bibinfo {author} {\bibfnamefont {R.~C.}\ \bibnamefont
  {Sherwood}}, \ and\ \bibinfo {author} {\bibfnamefont {J.~P.}\ \bibnamefont
  {Remeika}},\ }\href {\doibase 10.1063/1.1723049} {\bibfield  {journal}
  {\bibinfo  {journal} {J. Appl. Phys.}\ }\textbf {\bibinfo {volume} {29}},\
  \bibinfo {pages} {1772} (\bibinfo {year} {1958})}\BibitemShut {NoStop}%
\bibitem [{\citenamefont {Tabor}, \citenamefont {Anderson},\ and\ \citenamefont
  {Van~Uitert}(1970)}]{Tabor1970}%
  \BibitemOpen
  \bibfield  {author} {\bibinfo {author} {\bibfnamefont {W.~J.}\ \bibnamefont
  {Tabor}}, \bibinfo {author} {\bibfnamefont {A.~W.}\ \bibnamefont {Anderson}},
  \ and\ \bibinfo {author} {\bibfnamefont {L.~G.}\ \bibnamefont {Van~Uitert}},\
  }\href {\doibase 10.1063/1.1659357} {\bibfield  {journal} {\bibinfo
  {journal} {J. Appl. Phys.}\ }\textbf {\bibinfo {volume} {41}},\ \bibinfo
  {pages} {3018} (\bibinfo {year} {1970})}\BibitemShut {NoStop}%
\bibitem [{\citenamefont {Schmool}\ \emph {et~al.}(1999)\citenamefont
  {Schmool}, \citenamefont {Keller}, \citenamefont {Guyot}, \citenamefont
  {Krishnan},\ and\ \citenamefont {Tessier}}]{Schmool1999}%
  \BibitemOpen
  \bibfield  {author} {\bibinfo {author} {\bibfnamefont {D.~S.}\ \bibnamefont
  {Schmool}}, \bibinfo {author} {\bibfnamefont {N.}~\bibnamefont {Keller}},
  \bibinfo {author} {\bibfnamefont {M.}~\bibnamefont {Guyot}}, \bibinfo
  {author} {\bibfnamefont {R.}~\bibnamefont {Krishnan}}, \ and\ \bibinfo
  {author} {\bibfnamefont {M.}~\bibnamefont {Tessier}},\ }\href {\doibase
  10.1063/1.371583} {\bibfield  {journal} {\bibinfo  {journal} {J. Appl.
  Phys.}\ }\textbf {\bibinfo {volume} {86}},\ \bibinfo {pages} {5712} (\bibinfo
  {year} {1999})}\BibitemShut {NoStop}%
\bibitem [{\citenamefont {Saidl}\ \emph {et~al.}(2016)\citenamefont {Saidl},
  \citenamefont {Brajer}, \citenamefont {Hor{\'{a}}k}, \citenamefont
  {Reichlov{\'{a}}}, \citenamefont {V{\'{y}}born{\'{y}}}, \citenamefont {Veis},
  \citenamefont {Janda}, \citenamefont {Troj{\'{a}}nek}, \citenamefont
  {Mary{\v{s}}ko}, \citenamefont {Fina}, \citenamefont {Marti}, \citenamefont
  {Jungwirth},\ and\ \citenamefont {N{\v{e}}mec}}]{Saidl2016v2}%
  \BibitemOpen
  \bibfield  {author} {\bibinfo {author} {\bibfnamefont {V.}~\bibnamefont
  {Saidl}}, \bibinfo {author} {\bibfnamefont {M.}~\bibnamefont {Brajer}},
  \bibinfo {author} {\bibfnamefont {L.}~\bibnamefont {Hor{\'{a}}k}}, \bibinfo
  {author} {\bibfnamefont {H.}~\bibnamefont {Reichlov{\'{a}}}}, \bibinfo
  {author} {\bibfnamefont {K.}~\bibnamefont {V{\'{y}}born{\'{y}}}}, \bibinfo
  {author} {\bibfnamefont {M.}~\bibnamefont {Veis}}, \bibinfo {author}
  {\bibfnamefont {T.}~\bibnamefont {Janda}}, \bibinfo {author} {\bibfnamefont
  {F.}~\bibnamefont {Troj{\'{a}}nek}}, \bibinfo {author} {\bibfnamefont
  {M.}~\bibnamefont {Mary{\v{s}}ko}}, \bibinfo {author} {\bibfnamefont
  {I.}~\bibnamefont {Fina}}, \bibinfo {author} {\bibfnamefont {X.}~\bibnamefont
  {Marti}}, \bibinfo {author} {\bibfnamefont {T.}~\bibnamefont {Jungwirth}}, \
  and\ \bibinfo {author} {\bibfnamefont {P.}~\bibnamefont {N{\v{e}}mec}},\
  }\href {\doibase 10.1088/1367-2630/18/8/083017} {\bibfield  {journal}
  {\bibinfo  {journal} {New J. Phys.}\ }\textbf {\bibinfo {volume} {18}},\
  \bibinfo {pages} {083017} (\bibinfo {year} {2016})}\BibitemShut {NoStop}%
\bibitem [{\citenamefont {Lim}, \citenamefont {Schleife},\ and\ \citenamefont
  {Smith}(2017)}]{Lim2017}%
  \BibitemOpen
  \bibfield  {author} {\bibinfo {author} {\bibfnamefont {S.~J.}\ \bibnamefont
  {Lim}}, \bibinfo {author} {\bibfnamefont {A.}~\bibnamefont {Schleife}}, \
  and\ \bibinfo {author} {\bibfnamefont {A.~M.}\ \bibnamefont {Smith}},\ }\href
  {\doibase 10.1038/ncomms14849} {\bibfield  {journal} {\bibinfo  {journal}
  {Nat. Comm.}\ }\textbf {\bibinfo {volume} {8}},\ \bibinfo {pages} {14849}
  (\bibinfo {year} {2017})}\BibitemShut {NoStop}%
\bibitem [{\citenamefont {Zhang}\ and\ \citenamefont
  {Schleife}(2018)}]{Zhang2018}%
  \BibitemOpen
  \bibfield  {author} {\bibinfo {author} {\bibfnamefont {X.}~\bibnamefont
  {Zhang}}\ and\ \bibinfo {author} {\bibfnamefont {A.}~\bibnamefont
  {Schleife}},\ }\href {\doibase 10.1103/PhysRevB.97.125201} {\bibfield
  {journal} {\bibinfo  {journal} {Phys. Rev. B}\ }\textbf {\bibinfo {volume}
  {97}},\ \bibinfo {pages} {125201} (\bibinfo {year} {2018})}\BibitemShut
  {NoStop}%
\bibitem [{\citenamefont {Kresse}\ and\ \citenamefont
  {Furthm{\"u}ller}(1996)}]{Kresse:1996}%
  \BibitemOpen
  \bibfield  {author} {\bibinfo {author} {\bibfnamefont {G.}~\bibnamefont
  {Kresse}}\ and\ \bibinfo {author} {\bibfnamefont {J.}~\bibnamefont
  {Furthm{\"u}ller}},\ }\href {\doibase 10.1103/PhysRevB.54.11169} {\bibfield
  {journal} {\bibinfo  {journal} {Phys. Rev. B}\ }\textbf {\bibinfo {volume}
  {54}},\ \bibinfo {pages} {11169} (\bibinfo {year} {1996})}\BibitemShut
  {NoStop}%
\bibitem [{\citenamefont {Kresse}\ and\ \citenamefont
  {Joubert}(1999)}]{Kresse:1999}%
  \BibitemOpen
  \bibfield  {author} {\bibinfo {author} {\bibfnamefont {G.}~\bibnamefont
  {Kresse}}\ and\ \bibinfo {author} {\bibfnamefont {D.}~\bibnamefont
  {Joubert}},\ }\href {\doibase 10.1103/PhysRevB.59.1758} {\bibfield  {journal}
  {\bibinfo  {journal} {Phys. Rev. B}\ }\textbf {\bibinfo {volume} {59}},\
  \bibinfo {pages} {1758} (\bibinfo {year} {1999})}\BibitemShut {NoStop}%
\bibitem [{\citenamefont {Bl\"ochl}(1994)}]{Blochl:1994}%
  \BibitemOpen
  \bibfield  {author} {\bibinfo {author} {\bibfnamefont {P.~E.}\ \bibnamefont
  {Bl\"ochl}},\ }\href {\doibase 10.1103/PhysRevB.50.17953} {\bibfield
  {journal} {\bibinfo  {journal} {Phys. Rev. B}\ }\textbf {\bibinfo {volume}
  {50}},\ \bibinfo {pages} {17953} (\bibinfo {year} {1994})}\BibitemShut
  {NoStop}%
\bibitem [{\citenamefont {Gajdo\v{s}}\ \emph {et~al.}(2006)\citenamefont
  {Gajdo\v{s}}, \citenamefont {Hummer}, \citenamefont {Kresse}, \citenamefont
  {Furthm\"uller},\ and\ \citenamefont {Bechstedt}}]{Gajdos:2006}%
  \BibitemOpen
  \bibfield  {author} {\bibinfo {author} {\bibfnamefont {M.}~\bibnamefont
  {Gajdo\v{s}}}, \bibinfo {author} {\bibfnamefont {K.}~\bibnamefont {Hummer}},
  \bibinfo {author} {\bibfnamefont {G.}~\bibnamefont {Kresse}}, \bibinfo
  {author} {\bibfnamefont {J.}~\bibnamefont {Furthm\"uller}}, \ and\ \bibinfo
  {author} {\bibfnamefont {F.}~\bibnamefont {Bechstedt}},\ }\href {\doibase
  10.1103/PhysRevB.73.045112} {\bibfield  {journal} {\bibinfo  {journal} {Phys.
  Rev. B}\ }\textbf {\bibinfo {volume} {73}},\ \bibinfo {pages} {045112}
  (\bibinfo {year} {2006})}\BibitemShut {NoStop}%
\bibitem [{\citenamefont {Monkhorst}\ and\ \citenamefont
  {Pack}(1976)}]{Monkhorst:1976}%
  \BibitemOpen
  \bibfield  {author} {\bibinfo {author} {\bibfnamefont {H.~J.}\ \bibnamefont
  {Monkhorst}}\ and\ \bibinfo {author} {\bibfnamefont {J.~D.}\ \bibnamefont
  {Pack}},\ }\href {\doibase 10.1103/PhysRevB.13.5188} {\bibfield  {journal}
  {\bibinfo  {journal} {Phys. Rev. B}\ }\textbf {\bibinfo {volume} {13}},\
  \bibinfo {pages} {5188} (\bibinfo {year} {1976})}\BibitemShut {NoStop}%
\bibitem [{\citenamefont {Perdew}, \citenamefont {Burke},\ and\ \citenamefont
  {Ernzerhof}(1996)}]{Perdew:1997}%
  \BibitemOpen
  \bibfield  {author} {\bibinfo {author} {\bibfnamefont {J.~P.}\ \bibnamefont
  {Perdew}}, \bibinfo {author} {\bibfnamefont {K.}~\bibnamefont {Burke}}, \
  and\ \bibinfo {author} {\bibfnamefont {M.}~\bibnamefont {Ernzerhof}},\ }\href
  {\doibase 10.1103/PhysRevLett.77.3865} {\bibfield  {journal} {\bibinfo
  {journal} {Phys. Rev. Lett.}\ }\textbf {\bibinfo {volume} {77}},\ \bibinfo
  {pages} {3865} (\bibinfo {year} {1996})}\BibitemShut {NoStop}%
\bibitem [{\citenamefont {Pisarev}\ \emph {et~al.}(1972)\citenamefont
  {Pisarev}, \citenamefont {Ferre}, \citenamefont {Duran},\ and\ \citenamefont
  {Badoz}}]{Pisarev1972}%
  \BibitemOpen
  \bibfield  {author} {\bibinfo {author} {\bibfnamefont {R.}~\bibnamefont
  {Pisarev}}, \bibinfo {author} {\bibfnamefont {J.}~\bibnamefont {Ferre}},
  \bibinfo {author} {\bibfnamefont {J.}~\bibnamefont {Duran}}, \ and\ \bibinfo
  {author} {\bibfnamefont {J.}~\bibnamefont {Badoz}},\ }\href {\doibase
  https://doi.org/10.1016/0038-1098(72)91006-X} {\bibfield  {journal} {\bibinfo
   {journal} {Solid State Comm.}\ }\textbf {\bibinfo {volume} {11}},\ \bibinfo
  {pages} {913 } (\bibinfo {year} {1972})}\BibitemShut {NoStop}%
\bibitem [{\citenamefont {Kharchenko}, \citenamefont {Miloslavskaya},\ and\
  \citenamefont {Milner}(2005)}]{Kharchenko2005}%
  \BibitemOpen
  \bibfield  {author} {\bibinfo {author} {\bibfnamefont {N.~F.}\ \bibnamefont
  {Kharchenko}}, \bibinfo {author} {\bibfnamefont {O.~V.}\ \bibnamefont
  {Miloslavskaya}}, \ and\ \bibinfo {author} {\bibfnamefont {A.~A.}\
  \bibnamefont {Milner}},\ }\href {\doibase 10.1063/1.2008145} {\bibfield
  {journal} {\bibinfo  {journal} {Low Temp. Phys.}\ }\textbf {\bibinfo {volume}
  {31}},\ \bibinfo {pages} {825} (\bibinfo {year} {2005})}\BibitemShut
  {NoStop}%
\bibitem [{\citenamefont {Ferre}\ and\ \citenamefont
  {Gehring}(1984)}]{Ferre1984}%
  \BibitemOpen
  \bibfield  {author} {\bibinfo {author} {\bibfnamefont {J.}~\bibnamefont
  {Ferre}}\ and\ \bibinfo {author} {\bibfnamefont {G.~A.}\ \bibnamefont
  {Gehring}},\ }\href {\doibase 10.1088/0034-4885/47/5/002} {\bibfield
  {journal} {\bibinfo  {journal} {Rep. Prog. Phys.}\ }\textbf {\bibinfo
  {volume} {47}},\ \bibinfo {pages} {513} (\bibinfo {year} {1984})}\BibitemShut
  {NoStop}%
\bibitem [{\citenamefont {Silber}\ \emph {et~al.}(2019)\citenamefont {Silber},
  \citenamefont {Stejskal}, \citenamefont {Beran}, \citenamefont {Cejpek},
  \citenamefont {Anto\ifmmode~\check{s}\else \v{s}\fi{}}, \citenamefont
  {Matalla-Wagner}, \citenamefont {Thien}, \citenamefont {Kuschel},
  \citenamefont {Wollschl\"ager}, \citenamefont {Veis}, \citenamefont
  {Kuschel},\ and\ \citenamefont {Hamrle}}]{Silber2019}%
  \BibitemOpen
  \bibfield  {author} {\bibinfo {author} {\bibfnamefont {R.}~\bibnamefont
  {Silber}}, \bibinfo {author} {\bibfnamefont {O.~c.~v.}\ \bibnamefont
  {Stejskal}}, \bibinfo {author} {\bibfnamefont {L.~c.~v.}\ \bibnamefont
  {Beran}}, \bibinfo {author} {\bibfnamefont {P.}~\bibnamefont {Cejpek}},
  \bibinfo {author} {\bibfnamefont {R.}~\bibnamefont
  {Anto\ifmmode~\check{s}\else \v{s}\fi{}}}, \bibinfo {author} {\bibfnamefont
  {T.}~\bibnamefont {Matalla-Wagner}}, \bibinfo {author} {\bibfnamefont
  {J.}~\bibnamefont {Thien}}, \bibinfo {author} {\bibfnamefont
  {O.}~\bibnamefont {Kuschel}}, \bibinfo {author} {\bibfnamefont
  {J.}~\bibnamefont {Wollschl\"ager}}, \bibinfo {author} {\bibfnamefont
  {M.}~\bibnamefont {Veis}}, \bibinfo {author} {\bibfnamefont {T.}~\bibnamefont
  {Kuschel}}, \ and\ \bibinfo {author} {\bibfnamefont {J.}~\bibnamefont
  {Hamrle}},\ }\href {\doibase 10.1103/PhysRevB.100.064403} {\bibfield
  {journal} {\bibinfo  {journal} {Phys. Rev. B}\ }\textbf {\bibinfo {volume}
  {100}},\ \bibinfo {pages} {064403} (\bibinfo {year} {2019})}\BibitemShut
  {NoStop}%
\bibitem [{\citenamefont {Gnatchenko}\ \emph {et~al.}(1989)\citenamefont
  {Gnatchenko}, \citenamefont {Kharchenko}, \citenamefont {Lebedev},
  \citenamefont {Piotrowski}, \citenamefont {Szymczak},\ and\ \citenamefont
  {Szymczak}}]{Gnatchenko1989}%
  \BibitemOpen
  \bibfield  {author} {\bibinfo {author} {\bibfnamefont {S.}~\bibnamefont
  {Gnatchenko}}, \bibinfo {author} {\bibfnamefont {N.}~\bibnamefont
  {Kharchenko}}, \bibinfo {author} {\bibfnamefont {P.}~\bibnamefont {Lebedev}},
  \bibinfo {author} {\bibfnamefont {K.}~\bibnamefont {Piotrowski}}, \bibinfo
  {author} {\bibfnamefont {H.}~\bibnamefont {Szymczak}}, \ and\ \bibinfo
  {author} {\bibfnamefont {R.}~\bibnamefont {Szymczak}},\ }\href {\doibase
  https://doi.org/10.1016/0304-8853(89)90239-4} {\bibfield  {journal} {\bibinfo
   {journal} {J. Magn. Magn. Mater.}\ }\textbf {\bibinfo {volume} {81}},\
  \bibinfo {pages} {125 } (\bibinfo {year} {1989})}\BibitemShut {NoStop}%
\bibitem [{\citenamefont {Le~Gall}\ \emph {et~al.}(1976)\citenamefont
  {Le~Gall}, \citenamefont {Rudashewsky}, \citenamefont {Leycuras},\ and\
  \citenamefont {Minella}}]{LeGall1976}%
  \BibitemOpen
  \bibfield  {author} {\bibinfo {author} {\bibfnamefont {H.}~\bibnamefont
  {Le~Gall}}, \bibinfo {author} {\bibfnamefont {E.~G.}\ \bibnamefont
  {Rudashewsky}}, \bibinfo {author} {\bibfnamefont {C.}~\bibnamefont
  {Leycuras}}, \ and\ \bibinfo {author} {\bibfnamefont {D.}~\bibnamefont
  {Minella}},\ }\href {\doibase 10.1063/1.30504} {\bibfield  {journal}
  {\bibinfo  {journal} {AIP Conf. Proc.}\ }\textbf {\bibinfo {volume} {29}},\
  \bibinfo {pages} {656} (\bibinfo {year} {1976})}\BibitemShut {NoStop}%
\bibitem [{\citenamefont {D\"ahn}, \citenamefont {H\"ubner},\ and\
  \citenamefont {Bennemann}(1996)}]{Dahn1996}%
  \BibitemOpen
  \bibfield  {author} {\bibinfo {author} {\bibfnamefont {A.}~\bibnamefont
  {D\"ahn}}, \bibinfo {author} {\bibfnamefont {W.}~\bibnamefont {H\"ubner}}, \
  and\ \bibinfo {author} {\bibfnamefont {K.~H.}\ \bibnamefont {Bennemann}},\
  }\href {\doibase 10.1103/PhysRevLett.77.3929} {\bibfield  {journal} {\bibinfo
   {journal} {Phys. Rev. Lett.}\ }\textbf {\bibinfo {volume} {77}},\ \bibinfo
  {pages} {3929} (\bibinfo {year} {1996})}\BibitemShut {NoStop}%
\bibitem [{\citenamefont {Zvezdin}\ and\ \citenamefont
  {Kubrakov}(1999)}]{Zvezdin1999}%
  \BibitemOpen
  \bibfield  {author} {\bibinfo {author} {\bibfnamefont {A.~K.}\ \bibnamefont
  {Zvezdin}}\ and\ \bibinfo {author} {\bibfnamefont {N.~F.}\ \bibnamefont
  {Kubrakov}},\ }\href {\doibase 10.1134/1.558957} {\bibfield  {journal}
  {\bibinfo  {journal} {J. Exp. Theor. Phys.}\ }\textbf {\bibinfo {volume}
  {89}},\ \bibinfo {pages} {77} (\bibinfo {year} {1999})}\BibitemShut {NoStop}%
\bibitem [{\citenamefont {Fiebig}, \citenamefont {Pavlov},\ and\ \citenamefont
  {Pisarev}(2005)}]{Fiebig2005}%
  \BibitemOpen
  \bibfield  {author} {\bibinfo {author} {\bibfnamefont {M.}~\bibnamefont
  {Fiebig}}, \bibinfo {author} {\bibfnamefont {V.~V.}\ \bibnamefont {Pavlov}},
  \ and\ \bibinfo {author} {\bibfnamefont {R.~V.}\ \bibnamefont {Pisarev}},\
  }\href {\doibase 10.1364/JOSAB.22.000096} {\bibfield  {journal} {\bibinfo
  {journal} {J. Opt. Soc. Am. B}\ }\textbf {\bibinfo {volume} {22}},\ \bibinfo
  {pages} {96} (\bibinfo {year} {2005})}\BibitemShut {NoStop}%
\bibitem [{\citenamefont {Fiebig}\ \emph {et~al.}(1994)\citenamefont {Fiebig},
  \citenamefont {Fr\"ohlich}, \citenamefont {Krichevtsov},\ and\ \citenamefont
  {Pisarev}}]{Fiebig1994}%
  \BibitemOpen
  \bibfield  {author} {\bibinfo {author} {\bibfnamefont {M.}~\bibnamefont
  {Fiebig}}, \bibinfo {author} {\bibfnamefont {D.}~\bibnamefont {Fr\"ohlich}},
  \bibinfo {author} {\bibfnamefont {B.~B.}\ \bibnamefont {Krichevtsov}}, \ and\
  \bibinfo {author} {\bibfnamefont {R.~V.}\ \bibnamefont {Pisarev}},\ }\href
  {\doibase 10.1103/PhysRevLett.73.2127} {\bibfield  {journal} {\bibinfo
  {journal} {Phys. Rev. Lett.}\ }\textbf {\bibinfo {volume} {73}},\ \bibinfo
  {pages} {2127} (\bibinfo {year} {1994})}\BibitemShut {NoStop}%
\bibitem [{\citenamefont {Trzeciecki}, \citenamefont {D\"ahn},\ and\
  \citenamefont {H\"ubner}(1999)}]{Trzeciecki1999}%
  \BibitemOpen
  \bibfield  {author} {\bibinfo {author} {\bibfnamefont {M.}~\bibnamefont
  {Trzeciecki}}, \bibinfo {author} {\bibfnamefont {A.}~\bibnamefont {D\"ahn}},
  \ and\ \bibinfo {author} {\bibfnamefont {W.}~\bibnamefont {H\"ubner}},\
  }\href {\doibase 10.1103/PhysRevB.60.1144} {\bibfield  {journal} {\bibinfo
  {journal} {Phys. Rev. B}\ }\textbf {\bibinfo {volume} {60}},\ \bibinfo
  {pages} {1144} (\bibinfo {year} {1999})}\BibitemShut {NoStop}%
\bibitem [{\citenamefont {Satitkovitchai}, \citenamefont {Pavlyukh},\ and\
  \citenamefont {H\"ubner}(2003)}]{Satitkovitchai2003}%
  \BibitemOpen
  \bibfield  {author} {\bibinfo {author} {\bibfnamefont {K.}~\bibnamefont
  {Satitkovitchai}}, \bibinfo {author} {\bibfnamefont {Y.}~\bibnamefont
  {Pavlyukh}}, \ and\ \bibinfo {author} {\bibfnamefont {W.}~\bibnamefont
  {H\"ubner}},\ }\href {\doibase 10.1103/PhysRevB.67.165413} {\bibfield
  {journal} {\bibinfo  {journal} {Phys. Rev. B}\ }\textbf {\bibinfo {volume}
  {67}},\ \bibinfo {pages} {165413} (\bibinfo {year} {2003})}\BibitemShut
  {NoStop}%
\bibitem [{\citenamefont {Fr\"ohlich}\ \emph {et~al.}(1998)\citenamefont
  {Fr\"ohlich}, \citenamefont {Leute}, \citenamefont {Pavlov},\ and\
  \citenamefont {Pisarev}}]{Frohlich1998}%
  \BibitemOpen
  \bibfield  {author} {\bibinfo {author} {\bibfnamefont {D.}~\bibnamefont
  {Fr\"ohlich}}, \bibinfo {author} {\bibfnamefont {S.}~\bibnamefont {Leute}},
  \bibinfo {author} {\bibfnamefont {V.~V.}\ \bibnamefont {Pavlov}}, \ and\
  \bibinfo {author} {\bibfnamefont {R.~V.}\ \bibnamefont {Pisarev}},\ }\href
  {\doibase 10.1103/PhysRevLett.81.3239} {\bibfield  {journal} {\bibinfo
  {journal} {Phys. Rev. Lett.}\ }\textbf {\bibinfo {volume} {81}},\ \bibinfo
  {pages} {3239} (\bibinfo {year} {1998})}\BibitemShut {NoStop}%
\bibitem [{\citenamefont {Fiebig}\ \emph {et~al.}(2000)\citenamefont {Fiebig},
  \citenamefont {Fr\"ohlich}, \citenamefont {Kohn}, \citenamefont {Leute},
  \citenamefont {Lottermoser}, \citenamefont {Pavlov},\ and\ \citenamefont
  {Pisarev}}]{Fiebig2000}%
  \BibitemOpen
  \bibfield  {author} {\bibinfo {author} {\bibfnamefont {M.}~\bibnamefont
  {Fiebig}}, \bibinfo {author} {\bibfnamefont {D.}~\bibnamefont {Fr\"ohlich}},
  \bibinfo {author} {\bibfnamefont {K.}~\bibnamefont {Kohn}}, \bibinfo {author}
  {\bibfnamefont {S.}~\bibnamefont {Leute}}, \bibinfo {author} {\bibfnamefont
  {T.}~\bibnamefont {Lottermoser}}, \bibinfo {author} {\bibfnamefont {V.~V.}\
  \bibnamefont {Pavlov}}, \ and\ \bibinfo {author} {\bibfnamefont {R.~V.}\
  \bibnamefont {Pisarev}},\ }\href {\doibase 10.1103/PhysRevLett.84.5620}
  {\bibfield  {journal} {\bibinfo  {journal} {Phys. Rev. Lett.}\ }\textbf
  {\bibinfo {volume} {84}},\ \bibinfo {pages} {5620} (\bibinfo {year}
  {2000})}\BibitemShut {NoStop}%
\bibitem [{\citenamefont {Degenhardt}\ \emph {et~al.}(2001)\citenamefont
  {Degenhardt}, \citenamefont {Fiebig}, \citenamefont {Fr{\"o}hlich},
  \citenamefont {Lottermoser},\ and\ \citenamefont {Pisarev}}]{Degenhardt2001}%
  \BibitemOpen
  \bibfield  {author} {\bibinfo {author} {\bibfnamefont {C.}~\bibnamefont
  {Degenhardt}}, \bibinfo {author} {\bibfnamefont {M.}~\bibnamefont {Fiebig}},
  \bibinfo {author} {\bibfnamefont {D.}~\bibnamefont {Fr{\"o}hlich}}, \bibinfo
  {author} {\bibfnamefont {T.}~\bibnamefont {Lottermoser}}, \ and\ \bibinfo
  {author} {\bibfnamefont {R.~V.}\ \bibnamefont {Pisarev}},\ }\href {\doibase
  10.1007/s003400100617} {\bibfield  {journal} {\bibinfo  {journal} {Appl.
  Phys. B}\ }\textbf {\bibinfo {volume} {73}},\ \bibinfo {pages} {139}
  (\bibinfo {year} {2001})}\BibitemShut {NoStop}%
\bibitem [{\citenamefont {Fiebig}\ \emph {et~al.}(2002)\citenamefont {Fiebig},
  \citenamefont {Lottermoser}, \citenamefont {Fr{\"o}hlich}, \citenamefont
  {Goltsev},\ and\ \citenamefont {Pisarev}}]{Fiebig2002}%
  \BibitemOpen
  \bibfield  {author} {\bibinfo {author} {\bibfnamefont {M.}~\bibnamefont
  {Fiebig}}, \bibinfo {author} {\bibfnamefont {T.}~\bibnamefont {Lottermoser}},
  \bibinfo {author} {\bibfnamefont {D.}~\bibnamefont {Fr{\"o}hlich}}, \bibinfo
  {author} {\bibfnamefont {A.~V.}\ \bibnamefont {Goltsev}}, \ and\ \bibinfo
  {author} {\bibfnamefont {R.~V.}\ \bibnamefont {Pisarev}},\ }\href {\doibase
  10.1038/nature01077} {\bibfield  {journal} {\bibinfo  {journal} {Nature}\
  }\textbf {\bibinfo {volume} {419}},\ \bibinfo {pages} {818} (\bibinfo {year}
  {2002})}\BibitemShut {NoStop}%
\bibitem [{\citenamefont {Fiebig}\ \emph {et~al.}(2004)\citenamefont {Fiebig},
  \citenamefont {Lottermoser}, \citenamefont {Fr\"{o}hlich},\ and\
  \citenamefont {Kallenbach}}]{Fiebig2004}%
  \BibitemOpen
  \bibfield  {author} {\bibinfo {author} {\bibfnamefont {M.}~\bibnamefont
  {Fiebig}}, \bibinfo {author} {\bibfnamefont {T.}~\bibnamefont {Lottermoser}},
  \bibinfo {author} {\bibfnamefont {D.}~\bibnamefont {Fr\"{o}hlich}}, \ and\
  \bibinfo {author} {\bibfnamefont {S.}~\bibnamefont {Kallenbach}},\ }\href
  {\doibase 10.1364/OL.29.000041} {\bibfield  {journal} {\bibinfo  {journal}
  {Opt. Lett.}\ }\textbf {\bibinfo {volume} {29}},\ \bibinfo {pages} {41}
  (\bibinfo {year} {2004})}\BibitemShut {NoStop}%
\bibitem [{\citenamefont {Manz}\ \emph {et~al.}(2016)\citenamefont {Manz},
  \citenamefont {Matsubara}, \citenamefont {Lottermoser}, \citenamefont
  {B{\"u}chi}, \citenamefont {Iyama}, \citenamefont {Kimura}, \citenamefont
  {Meier},\ and\ \citenamefont {Fiebig}}]{Manz2016}%
  \BibitemOpen
  \bibfield  {author} {\bibinfo {author} {\bibfnamefont {S.}~\bibnamefont
  {Manz}}, \bibinfo {author} {\bibfnamefont {M.}~\bibnamefont {Matsubara}},
  \bibinfo {author} {\bibfnamefont {T.}~\bibnamefont {Lottermoser}}, \bibinfo
  {author} {\bibfnamefont {J.}~\bibnamefont {B{\"u}chi}}, \bibinfo {author}
  {\bibfnamefont {A.}~\bibnamefont {Iyama}}, \bibinfo {author} {\bibfnamefont
  {T.}~\bibnamefont {Kimura}}, \bibinfo {author} {\bibfnamefont
  {D.}~\bibnamefont {Meier}}, \ and\ \bibinfo {author} {\bibfnamefont
  {M.}~\bibnamefont {Fiebig}},\ }\href {\doibase 10.1038/nphoton.2016.146}
  {\bibfield  {journal} {\bibinfo  {journal} {Nat. Photon.}\ }\textbf {\bibinfo
  {volume} {10}},\ \bibinfo {pages} {653} (\bibinfo {year} {2016})}\BibitemShut
  {NoStop}%
\bibitem [{\citenamefont {Satoh}\ \emph
  {et~al.}(2010{\natexlab{a}})\citenamefont {Satoh}, \citenamefont {Cho},
  \citenamefont {Shimura}, \citenamefont {Kuroda}, \citenamefont {Ueda},
  \citenamefont {Ueda},\ and\ \citenamefont {Fiebig}}]{Satoh2010}%
  \BibitemOpen
  \bibfield  {author} {\bibinfo {author} {\bibfnamefont {T.}~\bibnamefont
  {Satoh}}, \bibinfo {author} {\bibfnamefont {S.-J.}\ \bibnamefont {Cho}},
  \bibinfo {author} {\bibfnamefont {T.}~\bibnamefont {Shimura}}, \bibinfo
  {author} {\bibfnamefont {K.}~\bibnamefont {Kuroda}}, \bibinfo {author}
  {\bibfnamefont {H.}~\bibnamefont {Ueda}}, \bibinfo {author} {\bibfnamefont
  {Y.}~\bibnamefont {Ueda}}, \ and\ \bibinfo {author} {\bibfnamefont
  {M.}~\bibnamefont {Fiebig}},\ }\href {\doibase 10.1364/JOSAB.27.001421}
  {\bibfield  {journal} {\bibinfo  {journal} {J. Opt. Soc. Am. B}\ }\textbf
  {\bibinfo {volume} {27}},\ \bibinfo {pages} {1421} (\bibinfo {year}
  {2010}{\natexlab{a}})}\BibitemShut {NoStop}%
\bibitem [{\citenamefont {Bennemann}(1998)}]{Bennemann1998}%
  \BibitemOpen
  \bibfield  {author} {\bibinfo {author} {\bibfnamefont {K.}~\bibnamefont
  {Bennemann}},\ }\href {https://books.google.com/books?id=XZhK7kEuzawC} {\emph
  {\bibinfo {title} {Non-linear Optics in Metals}}},\ International Series of
  Monographs on Physics\ (\bibinfo  {publisher} {Clarendon Press},\ \bibinfo
  {year} {1998})\BibitemShut {NoStop}%
\bibitem [{\citenamefont {Liao}\ \emph {et~al.}(1998)\citenamefont {Liao},
  \citenamefont {Xiao}, \citenamefont {Fu}, \citenamefont {Wang}, \citenamefont
  {Wong},\ and\ \citenamefont {Wong}}]{Liao1998}%
  \BibitemOpen
  \bibfield  {author} {\bibinfo {author} {\bibfnamefont {H.~B.}\ \bibnamefont
  {Liao}}, \bibinfo {author} {\bibfnamefont {R.~F.}\ \bibnamefont {Xiao}},
  \bibinfo {author} {\bibfnamefont {J.~S.}\ \bibnamefont {Fu}}, \bibinfo
  {author} {\bibfnamefont {H.}~\bibnamefont {Wang}}, \bibinfo {author}
  {\bibfnamefont {K.~S.}\ \bibnamefont {Wong}}, \ and\ \bibinfo {author}
  {\bibfnamefont {G.~K.~L.}\ \bibnamefont {Wong}},\ }\href {\doibase
  10.1364/OL.23.000388} {\bibfield  {journal} {\bibinfo  {journal} {Opt.
  Lett.}\ }\textbf {\bibinfo {volume} {23}},\ \bibinfo {pages} {388} (\bibinfo
  {year} {1998})}\BibitemShut {NoStop}%
\bibitem [{\citenamefont {Kimel}\ \emph {et~al.}(2004)\citenamefont {Kimel},
  \citenamefont {Kirilyuk}, \citenamefont {Tsvetkov}, \citenamefont {Pisarev},\
  and\ \citenamefont {Rasing}}]{Kimel2004}%
  \BibitemOpen
  \bibfield  {author} {\bibinfo {author} {\bibfnamefont {A.~V.}\ \bibnamefont
  {Kimel}}, \bibinfo {author} {\bibfnamefont {A.}~\bibnamefont {Kirilyuk}},
  \bibinfo {author} {\bibfnamefont {A.}~\bibnamefont {Tsvetkov}}, \bibinfo
  {author} {\bibfnamefont {R.~V.}\ \bibnamefont {Pisarev}}, \ and\ \bibinfo
  {author} {\bibfnamefont {T.}~\bibnamefont {Rasing}},\ }\href {\doibase
  10.1038/nature02659} {\bibfield  {journal} {\bibinfo  {journal} {Nature}\
  }\textbf {\bibinfo {volume} {429}},\ \bibinfo {pages} {850} (\bibinfo {year}
  {2004})}\BibitemShut {NoStop}%
\bibitem [{\citenamefont {Satoh}\ \emph {et~al.}(2015)\citenamefont {Satoh},
  \citenamefont {Iida}, \citenamefont {Higuchi}, \citenamefont {Fiebig},\ and\
  \citenamefont {Shimura}}]{Satoh2015}%
  \BibitemOpen
  \bibfield  {author} {\bibinfo {author} {\bibfnamefont {T.}~\bibnamefont
  {Satoh}}, \bibinfo {author} {\bibfnamefont {R.}~\bibnamefont {Iida}},
  \bibinfo {author} {\bibfnamefont {T.}~\bibnamefont {Higuchi}}, \bibinfo
  {author} {\bibfnamefont {M.}~\bibnamefont {Fiebig}}, \ and\ \bibinfo {author}
  {\bibfnamefont {T.}~\bibnamefont {Shimura}},\ }\href {\doibase
  10.1038/nphoton.2014.273} {\bibfield  {journal} {\bibinfo  {journal} {Nat.
  Photon.}\ }\textbf {\bibinfo {volume} {9}},\ \bibinfo {pages} {25} (\bibinfo
  {year} {2015})}\BibitemShut {NoStop}%
\bibitem [{\citenamefont {Kimel}\ \emph {et~al.}(2002)\citenamefont {Kimel},
  \citenamefont {Pisarev}, \citenamefont {Hohlfeld},\ and\ \citenamefont
  {Rasing}}]{Kimel2002}%
  \BibitemOpen
  \bibfield  {author} {\bibinfo {author} {\bibfnamefont {A.~V.}\ \bibnamefont
  {Kimel}}, \bibinfo {author} {\bibfnamefont {R.~V.}\ \bibnamefont {Pisarev}},
  \bibinfo {author} {\bibfnamefont {J.}~\bibnamefont {Hohlfeld}}, \ and\
  \bibinfo {author} {\bibfnamefont {T.}~\bibnamefont {Rasing}},\ }\href
  {\doibase 10.1103/PhysRevLett.89.287401} {\bibfield  {journal} {\bibinfo
  {journal} {Phys. Rev. Lett.}\ }\textbf {\bibinfo {volume} {89}},\ \bibinfo
  {pages} {287401} (\bibinfo {year} {2002})}\BibitemShut {NoStop}%
\bibitem [{\citenamefont {Trzeciecki}\ \emph {et~al.}(2001)\citenamefont
  {Trzeciecki}, \citenamefont {Ney}, \citenamefont {Zhang},\ and\ \citenamefont
  {H{\"u}bner}}]{Trzeciecki2001}%
  \BibitemOpen
  \bibfield  {author} {\bibinfo {author} {\bibfnamefont {M.}~\bibnamefont
  {Trzeciecki}}, \bibinfo {author} {\bibfnamefont {O.}~\bibnamefont {Ney}},
  \bibinfo {author} {\bibfnamefont {G.~P.}\ \bibnamefont {Zhang}}, \ and\
  \bibinfo {author} {\bibfnamefont {W.}~\bibnamefont {H{\"u}bner}},\ }\enquote
  {\bibinfo {title} {Laser-control of ferro- and antiferromagnetism},}\ in\
  \href {\doibase 10.1007/3-540-44946-9_43} {\emph {\bibinfo {booktitle}
  {Advances in Solid State Physics}}},\ \bibinfo {editor} {edited by\ \bibinfo
  {editor} {\bibfnamefont {B.}~\bibnamefont {Kramer}}}\ (\bibinfo  {publisher}
  {Springer Berlin Heidelberg},\ \bibinfo {address} {Berlin, Heidelberg},\
  \bibinfo {year} {2001})\ pp.\ \bibinfo {pages} {547--555}\BibitemShut
  {NoStop}%
\bibitem [{\citenamefont {Lefkidis}\ and\ \citenamefont
  {H\"ubner}(2007)}]{Lefkidis2007}%
  \BibitemOpen
  \bibfield  {author} {\bibinfo {author} {\bibfnamefont {G.}~\bibnamefont
  {Lefkidis}}\ and\ \bibinfo {author} {\bibfnamefont {W.}~\bibnamefont
  {H\"ubner}},\ }\href {\doibase 10.1103/PhysRevB.76.014418} {\bibfield
  {journal} {\bibinfo  {journal} {Phys. Rev. B}\ }\textbf {\bibinfo {volume}
  {76}},\ \bibinfo {pages} {014418} (\bibinfo {year} {2007})}\BibitemShut
  {NoStop}%
\bibitem [{\citenamefont {G\'omez-Abal}\ \emph {et~al.}(2004)\citenamefont
  {G\'omez-Abal}, \citenamefont {Ney}, \citenamefont {Satitkovitchai},\ and\
  \citenamefont {H\"ubner}}]{Gomez-Abal2004}%
  \BibitemOpen
  \bibfield  {author} {\bibinfo {author} {\bibfnamefont {R.}~\bibnamefont
  {G\'omez-Abal}}, \bibinfo {author} {\bibfnamefont {O.}~\bibnamefont {Ney}},
  \bibinfo {author} {\bibfnamefont {K.}~\bibnamefont {Satitkovitchai}}, \ and\
  \bibinfo {author} {\bibfnamefont {W.}~\bibnamefont {H\"ubner}},\ }\href
  {\doibase 10.1103/PhysRevLett.92.227402} {\bibfield  {journal} {\bibinfo
  {journal} {Phys. Rev. Lett.}\ }\textbf {\bibinfo {volume} {92}},\ \bibinfo
  {pages} {227402} (\bibinfo {year} {2004})}\BibitemShut {NoStop}%
\bibitem [{\citenamefont {Kimel}\ \emph {et~al.}(2009)\citenamefont {Kimel},
  \citenamefont {Ivanov}, \citenamefont {Pisarev}, \citenamefont {Usachev},
  \citenamefont {Kirilyuk},\ and\ \citenamefont {Rasing}}]{Kimel2009}%
  \BibitemOpen
  \bibfield  {author} {\bibinfo {author} {\bibfnamefont {A.~V.}\ \bibnamefont
  {Kimel}}, \bibinfo {author} {\bibfnamefont {B.~A.}\ \bibnamefont {Ivanov}},
  \bibinfo {author} {\bibfnamefont {R.~V.}\ \bibnamefont {Pisarev}}, \bibinfo
  {author} {\bibfnamefont {P.~A.}\ \bibnamefont {Usachev}}, \bibinfo {author}
  {\bibfnamefont {A.}~\bibnamefont {Kirilyuk}}, \ and\ \bibinfo {author}
  {\bibfnamefont {T.}~\bibnamefont {Rasing}},\ }\href {\doibase
  10.1038/nphys1369} {\bibfield  {journal} {\bibinfo  {journal} {Nat. Phys.}\
  }\textbf {\bibinfo {volume} {5}},\ \bibinfo {pages} {727} (\bibinfo {year}
  {2009})}\BibitemShut {NoStop}%
\bibitem [{\citenamefont {Satoh}\ \emph
  {et~al.}(2010{\natexlab{b}})\citenamefont {Satoh}, \citenamefont {Cho},
  \citenamefont {Iida}, \citenamefont {Shimura}, \citenamefont {Kuroda},
  \citenamefont {Ueda}, \citenamefont {Ueda}, \citenamefont {Ivanov},
  \citenamefont {Nori},\ and\ \citenamefont {Fiebig}}]{Satoh2010v2}%
  \BibitemOpen
  \bibfield  {author} {\bibinfo {author} {\bibfnamefont {T.}~\bibnamefont
  {Satoh}}, \bibinfo {author} {\bibfnamefont {S.-J.}\ \bibnamefont {Cho}},
  \bibinfo {author} {\bibfnamefont {R.}~\bibnamefont {Iida}}, \bibinfo {author}
  {\bibfnamefont {T.}~\bibnamefont {Shimura}}, \bibinfo {author} {\bibfnamefont
  {K.}~\bibnamefont {Kuroda}}, \bibinfo {author} {\bibfnamefont
  {H.}~\bibnamefont {Ueda}}, \bibinfo {author} {\bibfnamefont {Y.}~\bibnamefont
  {Ueda}}, \bibinfo {author} {\bibfnamefont {B.~A.}\ \bibnamefont {Ivanov}},
  \bibinfo {author} {\bibfnamefont {F.}~\bibnamefont {Nori}}, \ and\ \bibinfo
  {author} {\bibfnamefont {M.}~\bibnamefont {Fiebig}},\ }\href {\doibase
  10.1103/PhysRevLett.105.077402} {\bibfield  {journal} {\bibinfo  {journal}
  {Phys. Rev. Lett.}\ }\textbf {\bibinfo {volume} {105}},\ \bibinfo {pages}
  {077402} (\bibinfo {year} {2010}{\natexlab{b}})}\BibitemShut {NoStop}%
\bibitem [{\citenamefont {Kalashnikova}\ \emph {et~al.}(2007)\citenamefont
  {Kalashnikova}, \citenamefont {Kimel}, \citenamefont {Pisarev}, \citenamefont
  {Gridnev}, \citenamefont {Kirilyuk},\ and\ \citenamefont
  {Rasing}}]{Kalashnikova2007}%
  \BibitemOpen
  \bibfield  {author} {\bibinfo {author} {\bibfnamefont {A.~M.}\ \bibnamefont
  {Kalashnikova}}, \bibinfo {author} {\bibfnamefont {A.~V.}\ \bibnamefont
  {Kimel}}, \bibinfo {author} {\bibfnamefont {R.~V.}\ \bibnamefont {Pisarev}},
  \bibinfo {author} {\bibfnamefont {V.~N.}\ \bibnamefont {Gridnev}}, \bibinfo
  {author} {\bibfnamefont {A.}~\bibnamefont {Kirilyuk}}, \ and\ \bibinfo
  {author} {\bibfnamefont {T.}~\bibnamefont {Rasing}},\ }\href {\doibase
  10.1103/PhysRevLett.99.167205} {\bibfield  {journal} {\bibinfo  {journal}
  {Phys. Rev. Lett.}\ }\textbf {\bibinfo {volume} {99}},\ \bibinfo {pages}
  {167205} (\bibinfo {year} {2007})}\BibitemShut {NoStop}%
\bibitem [{\citenamefont {Kimel}\ \emph {et~al.}(2006)\citenamefont {Kimel},
  \citenamefont {Stanciu}, \citenamefont {Usachev}, \citenamefont {Pisarev},
  \citenamefont {Gridnev}, \citenamefont {Kirilyuk},\ and\ \citenamefont
  {Rasing}}]{Kimel2006}%
  \BibitemOpen
  \bibfield  {author} {\bibinfo {author} {\bibfnamefont {A.~V.}\ \bibnamefont
  {Kimel}}, \bibinfo {author} {\bibfnamefont {C.~D.}\ \bibnamefont {Stanciu}},
  \bibinfo {author} {\bibfnamefont {P.~A.}\ \bibnamefont {Usachev}}, \bibinfo
  {author} {\bibfnamefont {R.~V.}\ \bibnamefont {Pisarev}}, \bibinfo {author}
  {\bibfnamefont {V.~N.}\ \bibnamefont {Gridnev}}, \bibinfo {author}
  {\bibfnamefont {A.}~\bibnamefont {Kirilyuk}}, \ and\ \bibinfo {author}
  {\bibfnamefont {T.}~\bibnamefont {Rasing}},\ }\href {\doibase
  10.1103/PhysRevB.74.060403} {\bibfield  {journal} {\bibinfo  {journal} {Phys.
  Rev. B}\ }\textbf {\bibinfo {volume} {74}},\ \bibinfo {pages} {060403}
  (\bibinfo {year} {2006})}\BibitemShut {NoStop}%
\bibitem [{\citenamefont {Keffer}, \citenamefont {Sievers},\ and\ \citenamefont
  {Tinkham}(1961)}]{Keffer1961}%
  \BibitemOpen
  \bibfield  {author} {\bibinfo {author} {\bibfnamefont {F.}~\bibnamefont
  {Keffer}}, \bibinfo {author} {\bibfnamefont {A.~J.}\ \bibnamefont {Sievers}},
  \ and\ \bibinfo {author} {\bibfnamefont {M.}~\bibnamefont {Tinkham}},\ }\href
  {\doibase 10.1063/1.2000502} {\bibfield  {journal} {\bibinfo  {journal} {J.
  Appl. Phys.}\ }\textbf {\bibinfo {volume} {32}},\ \bibinfo {pages} {S65}
  (\bibinfo {year} {1961})}\BibitemShut {NoStop}%
\bibitem [{\citenamefont {Brunner}\ and\ \citenamefont
  {Renk}(1970)}]{Brunner1970}%
  \BibitemOpen
  \bibfield  {author} {\bibinfo {author} {\bibfnamefont {H.}~\bibnamefont
  {Brunner}}\ and\ \bibinfo {author} {\bibfnamefont {K.~F.}\ \bibnamefont
  {Renk}},\ }\href {\doibase 10.1063/1.1659199} {\bibfield  {journal} {\bibinfo
   {journal} {J. Appl. Phys.}\ }\textbf {\bibinfo {volume} {41}},\ \bibinfo
  {pages} {2250} (\bibinfo {year} {1970})}\BibitemShut {NoStop}%
\bibitem [{\citenamefont {Jongerden}\ \emph {et~al.}(1990)\citenamefont
  {Jongerden}, \citenamefont {Arts}, \citenamefont {Dijkhuis},\ and\
  \citenamefont {Wijn}}]{Jongerden1990}%
  \BibitemOpen
  \bibfield  {author} {\bibinfo {author} {\bibfnamefont {G.}~\bibnamefont
  {Jongerden}}, \bibinfo {author} {\bibfnamefont {A.}~\bibnamefont {Arts}},
  \bibinfo {author} {\bibfnamefont {J.}~\bibnamefont {Dijkhuis}}, \ and\
  \bibinfo {author} {\bibfnamefont {H.~D.}\ \bibnamefont {Wijn}},\ }\href
  {\doibase https://doi.org/10.1016/0022-2313(90)90123-S} {\bibfield  {journal}
  {\bibinfo  {journal} {J. Lumin.}\ }\textbf {\bibinfo {volume} {45}},\
  \bibinfo {pages} {126 } (\bibinfo {year} {1990})}\BibitemShut {NoStop}%
\bibitem [{\citenamefont {Holzrichter}, \citenamefont {Macfarlane},\ and\
  \citenamefont {Schawlow}(1971)}]{Holzrichter1971}%
  \BibitemOpen
  \bibfield  {author} {\bibinfo {author} {\bibfnamefont {J.~F.}\ \bibnamefont
  {Holzrichter}}, \bibinfo {author} {\bibfnamefont {R.~M.}\ \bibnamefont
  {Macfarlane}}, \ and\ \bibinfo {author} {\bibfnamefont {A.~L.}\ \bibnamefont
  {Schawlow}},\ }\href {\doibase 10.1103/PhysRevLett.26.652} {\bibfield
  {journal} {\bibinfo  {journal} {Phys. Rev. Lett.}\ }\textbf {\bibinfo
  {volume} {26}},\ \bibinfo {pages} {652} (\bibinfo {year} {1971})}\BibitemShut
  {NoStop}%
\bibitem [{\citenamefont {Essenberger}\ \emph {et~al.}(2011)\citenamefont
  {Essenberger}, \citenamefont {Sharma}, \citenamefont {Dewhurst},
  \citenamefont {Bersier}, \citenamefont {Cricchio}, \citenamefont
  {Nordstr\"om},\ and\ \citenamefont {Gross}}]{Essenberger:2011}%
  \BibitemOpen
  \bibfield  {author} {\bibinfo {author} {\bibfnamefont {F.}~\bibnamefont
  {Essenberger}}, \bibinfo {author} {\bibfnamefont {S.}~\bibnamefont {Sharma}},
  \bibinfo {author} {\bibfnamefont {J.~K.}\ \bibnamefont {Dewhurst}}, \bibinfo
  {author} {\bibfnamefont {C.}~\bibnamefont {Bersier}}, \bibinfo {author}
  {\bibfnamefont {F.}~\bibnamefont {Cricchio}}, \bibinfo {author}
  {\bibfnamefont {L.}~\bibnamefont {Nordstr\"om}}, \ and\ \bibinfo {author}
  {\bibfnamefont {E.~K.~U.}\ \bibnamefont {Gross}},\ }\href {\doibase
  10.1103/PhysRevB.84.174425} {\bibfield  {journal} {\bibinfo  {journal} {Phys.
  Rev. B}\ }\textbf {\bibinfo {volume} {84}},\ \bibinfo {pages} {174425}
  (\bibinfo {year} {2011})}\BibitemShut {NoStop}%
\bibitem [{\citenamefont {Krieger}\ \emph {et~al.}(2015)\citenamefont
  {Krieger}, \citenamefont {Dewhurst}, \citenamefont {Elliott}, \citenamefont
  {Sharma},\ and\ \citenamefont {Gross}}]{Krieger2015}%
  \BibitemOpen
  \bibfield  {author} {\bibinfo {author} {\bibfnamefont {K.}~\bibnamefont
  {Krieger}}, \bibinfo {author} {\bibfnamefont {J.~K.}\ \bibnamefont
  {Dewhurst}}, \bibinfo {author} {\bibfnamefont {P.}~\bibnamefont {Elliott}},
  \bibinfo {author} {\bibfnamefont {S.}~\bibnamefont {Sharma}}, \ and\ \bibinfo
  {author} {\bibfnamefont {E.~K.~U.}\ \bibnamefont {Gross}},\ }\href {\doibase
  10.1021/acs.jctc.5b00621} {\bibfield  {journal} {\bibinfo  {journal} {J.
  Chem. Theory Comput.}\ }\textbf {\bibinfo {volume} {11}},\ \bibinfo {pages}
  {4870} (\bibinfo {year} {2015})}\BibitemShut {NoStop}%
\bibitem [{\citenamefont {Elliott}\ \emph
  {et~al.}(2016{\natexlab{a}})\citenamefont {Elliott}, \citenamefont {Krieger},
  \citenamefont {Dewhurst}, \citenamefont {Sharma},\ and\ \citenamefont
  {Gross}}]{Elliott2016v2}%
  \BibitemOpen
  \bibfield  {author} {\bibinfo {author} {\bibfnamefont {P.}~\bibnamefont
  {Elliott}}, \bibinfo {author} {\bibfnamefont {K.}~\bibnamefont {Krieger}},
  \bibinfo {author} {\bibfnamefont {J.~K.}\ \bibnamefont {Dewhurst}}, \bibinfo
  {author} {\bibfnamefont {S.}~\bibnamefont {Sharma}}, \ and\ \bibinfo {author}
  {\bibfnamefont {E.~K.~U.}\ \bibnamefont {Gross}},\ }\href {\doibase
  10.1088/1367-2630/18/1/013014} {\bibfield  {journal} {\bibinfo  {journal}
  {New J. Phys.}\ }\textbf {\bibinfo {volume} {18}},\ \bibinfo {pages} {013014}
  (\bibinfo {year} {2016}{\natexlab{a}})}\BibitemShut {NoStop}%
\bibitem [{\citenamefont {Simoni}, \citenamefont {Stamenova},\ and\
  \citenamefont {Sanvito}(2017)}]{Simoni2017}%
  \BibitemOpen
  \bibfield  {author} {\bibinfo {author} {\bibfnamefont {J.}~\bibnamefont
  {Simoni}}, \bibinfo {author} {\bibfnamefont {M.}~\bibnamefont {Stamenova}}, \
  and\ \bibinfo {author} {\bibfnamefont {S.}~\bibnamefont {Sanvito}},\ }\href
  {\doibase 10.1103/PhysRevB.95.024412} {\bibfield  {journal} {\bibinfo
  {journal} {Phys. Rev. B}\ }\textbf {\bibinfo {volume} {95}},\ \bibinfo
  {pages} {024412} (\bibinfo {year} {2017})}\BibitemShut {NoStop}%
\bibitem [{\citenamefont {Chen}\ and\ \citenamefont {Wang}(2019)}]{Chen2019}%
  \BibitemOpen
  \bibfield  {author} {\bibinfo {author} {\bibfnamefont {Z.}~\bibnamefont
  {Chen}}\ and\ \bibinfo {author} {\bibfnamefont {L.-W.}\ \bibnamefont
  {Wang}},\ }\href {\doibase 10.1126/sciadv.aau8000} {\bibfield  {journal}
  {\bibinfo  {journal} {Sci. Adv.}\ }\textbf {\bibinfo {volume} {5}},\ \bibinfo
  {pages} {eaau8000} (\bibinfo {year} {2019})}\BibitemShut {NoStop}%
\bibitem [{\citenamefont {Elliott}\ \emph
  {et~al.}(2016{\natexlab{b}})\citenamefont {Elliott}, \citenamefont
  {M{\"u}ller}, \citenamefont {Dewhurst}, \citenamefont {Sharma},\ and\
  \citenamefont {Gross}}]{Elliott2016}%
  \BibitemOpen
  \bibfield  {author} {\bibinfo {author} {\bibfnamefont {P.}~\bibnamefont
  {Elliott}}, \bibinfo {author} {\bibfnamefont {T.}~\bibnamefont {M{\"u}ller}},
  \bibinfo {author} {\bibfnamefont {J.~K.}\ \bibnamefont {Dewhurst}}, \bibinfo
  {author} {\bibfnamefont {S.}~\bibnamefont {Sharma}}, \ and\ \bibinfo {author}
  {\bibfnamefont {E.~K.~U.}\ \bibnamefont {Gross}},\ }\href {\doibase
  10.1038/srep38911} {\bibfield  {journal} {\bibinfo  {journal} {Sci. Rep.}\
  }\textbf {\bibinfo {volume} {6}},\ \bibinfo {pages} {38911} (\bibinfo {year}
  {2016}{\natexlab{b}})}\BibitemShut {NoStop}%
\bibitem [{\citenamefont {Zhang}\ \emph {et~al.}(2017)\citenamefont {Zhang},
  \citenamefont {Babyak}, \citenamefont {Xue}, \citenamefont {Bai},\ and\
  \citenamefont {George}}]{Zhang2017}%
  \BibitemOpen
  \bibfield  {author} {\bibinfo {author} {\bibfnamefont {G.~P.}\ \bibnamefont
  {Zhang}}, \bibinfo {author} {\bibfnamefont {Z.}~\bibnamefont {Babyak}},
  \bibinfo {author} {\bibfnamefont {Y.}~\bibnamefont {Xue}}, \bibinfo {author}
  {\bibfnamefont {Y.~H.}\ \bibnamefont {Bai}}, \ and\ \bibinfo {author}
  {\bibfnamefont {T.~F.}\ \bibnamefont {George}},\ }\href {\doibase
  10.1103/PhysRevB.96.134407} {\bibfield  {journal} {\bibinfo  {journal} {Phys.
  Rev. B}\ }\textbf {\bibinfo {volume} {96}},\ \bibinfo {pages} {134407}
  (\bibinfo {year} {2017})}\BibitemShut {NoStop}%
\bibitem [{\citenamefont {Shokeen}\ \emph {et~al.}(2017)\citenamefont
  {Shokeen}, \citenamefont {Sanchez~Piaia}, \citenamefont {Bigot},
  \citenamefont {M\"uller}, \citenamefont {Elliott}, \citenamefont {Dewhurst},
  \citenamefont {Sharma},\ and\ \citenamefont {Gross}}]{Shokeen2017}%
  \BibitemOpen
  \bibfield  {author} {\bibinfo {author} {\bibfnamefont {V.}~\bibnamefont
  {Shokeen}}, \bibinfo {author} {\bibfnamefont {M.}~\bibnamefont
  {Sanchez~Piaia}}, \bibinfo {author} {\bibfnamefont {J.-Y.}\ \bibnamefont
  {Bigot}}, \bibinfo {author} {\bibfnamefont {T.}~\bibnamefont {M\"uller}},
  \bibinfo {author} {\bibfnamefont {P.}~\bibnamefont {Elliott}}, \bibinfo
  {author} {\bibfnamefont {J.~K.}\ \bibnamefont {Dewhurst}}, \bibinfo {author}
  {\bibfnamefont {S.}~\bibnamefont {Sharma}}, \ and\ \bibinfo {author}
  {\bibfnamefont {E.~K.~U.}\ \bibnamefont {Gross}},\ }\href {\doibase
  10.1103/PhysRevLett.119.107203} {\bibfield  {journal} {\bibinfo  {journal}
  {Phys. Rev. Lett.}\ }\textbf {\bibinfo {volume} {119}},\ \bibinfo {pages}
  {107203} (\bibinfo {year} {2017})}\BibitemShut {NoStop}%
\bibitem [{\citenamefont {Krieger}\ \emph {et~al.}(2017)\citenamefont
  {Krieger}, \citenamefont {Elliott}, \citenamefont {M\"uller}, \citenamefont
  {Singh}, \citenamefont {Dewhurst}, \citenamefont {Gross},\ and\ \citenamefont
  {Sharma}}]{Krieger2017}%
  \BibitemOpen
  \bibfield  {author} {\bibinfo {author} {\bibfnamefont {K.}~\bibnamefont
  {Krieger}}, \bibinfo {author} {\bibfnamefont {P.}~\bibnamefont {Elliott}},
  \bibinfo {author} {\bibfnamefont {T.}~\bibnamefont {M\"uller}}, \bibinfo
  {author} {\bibfnamefont {N.}~\bibnamefont {Singh}}, \bibinfo {author}
  {\bibfnamefont {J.~K.}\ \bibnamefont {Dewhurst}}, \bibinfo {author}
  {\bibfnamefont {E.~K.~U.}\ \bibnamefont {Gross}}, \ and\ \bibinfo {author}
  {\bibfnamefont {S.}~\bibnamefont {Sharma}},\ }\href {\doibase
  10.1088/1361-648x/aa66f2} {\bibfield  {journal} {\bibinfo  {journal} {J.
  Phys.: Conden. Matter}\ }\textbf {\bibinfo {volume} {29}},\ \bibinfo {pages}
  {224001} (\bibinfo {year} {2017})}\BibitemShut {NoStop}%
\bibitem [{\citenamefont {Dewhurst}\ \emph {et~al.}(2018)\citenamefont
  {Dewhurst}, \citenamefont {Elliott}, \citenamefont {Shallcross},
  \citenamefont {Gross},\ and\ \citenamefont {Sharma}}]{Dewhurst2018}%
  \BibitemOpen
  \bibfield  {author} {\bibinfo {author} {\bibfnamefont {J.~K.}\ \bibnamefont
  {Dewhurst}}, \bibinfo {author} {\bibfnamefont {P.}~\bibnamefont {Elliott}},
  \bibinfo {author} {\bibfnamefont {S.}~\bibnamefont {Shallcross}}, \bibinfo
  {author} {\bibfnamefont {E.~K.~U.}\ \bibnamefont {Gross}}, \ and\ \bibinfo
  {author} {\bibfnamefont {S.}~\bibnamefont {Sharma}},\ }\href {\doibase
  10.1021/acs.nanolett.7b05118} {\bibfield  {journal} {\bibinfo  {journal}
  {Nano Lett.}\ }\textbf {\bibinfo {volume} {18}},\ \bibinfo {pages} {1842}
  (\bibinfo {year} {2018})}\BibitemShut {NoStop}%
\bibitem [{\citenamefont {Chen}\ \emph {et~al.}(2019)\citenamefont {Chen},
  \citenamefont {Bovensiepen}, \citenamefont {Eschenlohr}, \citenamefont
  {M\"uller}, \citenamefont {Elliott}, \citenamefont {Gross}, \citenamefont
  {Dewhurst},\ and\ \citenamefont {Sharma}}]{Chen2019v2}%
  \BibitemOpen
  \bibfield  {author} {\bibinfo {author} {\bibfnamefont {J.}~\bibnamefont
  {Chen}}, \bibinfo {author} {\bibfnamefont {U.}~\bibnamefont {Bovensiepen}},
  \bibinfo {author} {\bibfnamefont {A.}~\bibnamefont {Eschenlohr}}, \bibinfo
  {author} {\bibfnamefont {T.}~\bibnamefont {M\"uller}}, \bibinfo {author}
  {\bibfnamefont {P.}~\bibnamefont {Elliott}}, \bibinfo {author} {\bibfnamefont
  {E.~K.~U.}\ \bibnamefont {Gross}}, \bibinfo {author} {\bibfnamefont {J.~K.}\
  \bibnamefont {Dewhurst}}, \ and\ \bibinfo {author} {\bibfnamefont
  {S.}~\bibnamefont {Sharma}},\ }\href {\doibase
  10.1103/PhysRevLett.122.067202} {\bibfield  {journal} {\bibinfo  {journal}
  {Phys. Rev. Lett.}\ }\textbf {\bibinfo {volume} {122}},\ \bibinfo {pages}
  {067202} (\bibinfo {year} {2019})}\BibitemShut {NoStop}%
\bibitem [{\citenamefont {Hofherr}\ \emph {et~al.}(2020)\citenamefont
  {Hofherr}, \citenamefont {H{\"a}user}, \citenamefont {Dewhurst},
  \citenamefont {Tengdin}, \citenamefont {Sakshath}, \citenamefont {Nembach},
  \citenamefont {Weber}, \citenamefont {Shaw}, \citenamefont {Silva},
  \citenamefont {Kapteyn}, \citenamefont {Cinchetti}, \citenamefont {Rethfeld},
  \citenamefont {Murnane}, \citenamefont {Steil}, \citenamefont
  {Stadtm{\"u}ller}, \citenamefont {Sharma}, \citenamefont {Aeschlimann},\ and\
  \citenamefont {Mathias}}]{Hofherr2020}%
  \BibitemOpen
  \bibfield  {author} {\bibinfo {author} {\bibfnamefont {M.}~\bibnamefont
  {Hofherr}}, \bibinfo {author} {\bibfnamefont {S.}~\bibnamefont {H{\"a}user}},
  \bibinfo {author} {\bibfnamefont {J.~K.}\ \bibnamefont {Dewhurst}}, \bibinfo
  {author} {\bibfnamefont {P.}~\bibnamefont {Tengdin}}, \bibinfo {author}
  {\bibfnamefont {S.}~\bibnamefont {Sakshath}}, \bibinfo {author}
  {\bibfnamefont {H.~T.}\ \bibnamefont {Nembach}}, \bibinfo {author}
  {\bibfnamefont {S.~T.}\ \bibnamefont {Weber}}, \bibinfo {author}
  {\bibfnamefont {J.~M.}\ \bibnamefont {Shaw}}, \bibinfo {author}
  {\bibfnamefont {T.~J.}\ \bibnamefont {Silva}}, \bibinfo {author}
  {\bibfnamefont {H.~C.}\ \bibnamefont {Kapteyn}}, \bibinfo {author}
  {\bibfnamefont {M.}~\bibnamefont {Cinchetti}}, \bibinfo {author}
  {\bibfnamefont {B.}~\bibnamefont {Rethfeld}}, \bibinfo {author}
  {\bibfnamefont {M.~M.}\ \bibnamefont {Murnane}}, \bibinfo {author}
  {\bibfnamefont {D.}~\bibnamefont {Steil}}, \bibinfo {author} {\bibfnamefont
  {B.}~\bibnamefont {Stadtm{\"u}ller}}, \bibinfo {author} {\bibfnamefont
  {S.}~\bibnamefont {Sharma}}, \bibinfo {author} {\bibfnamefont
  {M.}~\bibnamefont {Aeschlimann}}, \ and\ \bibinfo {author} {\bibfnamefont
  {S.}~\bibnamefont {Mathias}},\ }\href {\doibase 10.1126/sciadv.aay8717}
  {\bibfield  {journal} {\bibinfo  {journal} {Sci. Adv.}\ }\textbf {\bibinfo
  {volume} {6}},\ \bibinfo {pages} {eaay8717} (\bibinfo {year}
  {2020})}\BibitemShut {NoStop}%
\bibitem [{\citenamefont {Tancogne-Dejean}, \citenamefont {Eich},\ and\
  \citenamefont {Rubio}(2020)}]{Tancogne-Dejean2020}%
  \BibitemOpen
  \bibfield  {author} {\bibinfo {author} {\bibfnamefont {N.}~\bibnamefont
  {Tancogne-Dejean}}, \bibinfo {author} {\bibfnamefont {F.~G.}\ \bibnamefont
  {Eich}}, \ and\ \bibinfo {author} {\bibfnamefont {A.}~\bibnamefont {Rubio}},\
  }\href {\doibase 10.1021/acs.jctc.9b01064} {\bibfield  {journal} {\bibinfo
  {journal} {. Chem. Theory Comput.}\ }\textbf {\bibinfo {volume} {16}},\
  \bibinfo {pages} {1007} (\bibinfo {year} {2020})},\ \bibinfo {note} {pMID:
  31922758}\BibitemShut {NoStop}%
\bibitem [{\citenamefont {Balents}(2000)}]{BalentsNature2000}%
  \BibitemOpen
  \bibfield  {author} {\bibinfo {author} {\bibfnamefont {L.}~\bibnamefont
  {Balents}},\ }\href {\doibase 10.1038/nature08917} {\bibfield  {journal}
  {\bibinfo  {journal} {Nature}\ }\textbf {\bibinfo {volume} {464}},\ \bibinfo
  {pages} {199} (\bibinfo {year} {2000})}\BibitemShut {NoStop}%
\bibitem [{\citenamefont {Bhattacharjee}\ \emph {et~al.}(2018)\citenamefont
  {Bhattacharjee}, \citenamefont {Sapozhnik}, \citenamefont {Bodnar},
  \citenamefont {Grigorev}, \citenamefont {Agustsson}, \citenamefont {Cao},
  \citenamefont {Dominko}, \citenamefont {Obergfell}, \citenamefont {Gomonay},
  \citenamefont {Sinova} \emph {et~al.}}]{bhattacharjee2018neel}%
  \BibitemOpen
  \bibfield  {author} {\bibinfo {author} {\bibfnamefont {N.}~\bibnamefont
  {Bhattacharjee}}, \bibinfo {author} {\bibfnamefont {A.}~\bibnamefont
  {Sapozhnik}}, \bibinfo {author} {\bibfnamefont {S.~Y.}\ \bibnamefont
  {Bodnar}}, \bibinfo {author} {\bibfnamefont {V.~Y.}\ \bibnamefont
  {Grigorev}}, \bibinfo {author} {\bibfnamefont {S.~Y.}\ \bibnamefont
  {Agustsson}}, \bibinfo {author} {\bibfnamefont {J.}~\bibnamefont {Cao}},
  \bibinfo {author} {\bibfnamefont {D.}~\bibnamefont {Dominko}}, \bibinfo
  {author} {\bibfnamefont {M.}~\bibnamefont {Obergfell}}, \bibinfo {author}
  {\bibfnamefont {O.}~\bibnamefont {Gomonay}}, \bibinfo {author} {\bibfnamefont
  {J.}~\bibnamefont {Sinova}},  \emph {et~al.},\ }\href@noop {} {\bibfield
  {journal} {\bibinfo  {journal} {Phys. Rev. Lett.}\ }\textbf {\bibinfo
  {volume} {120}},\ \bibinfo {pages} {237201} (\bibinfo {year}
  {2018})}\BibitemShut {NoStop}%
\bibitem [{\citenamefont {Little}\ \emph {et~al.}(2017)\citenamefont {Little},
  \citenamefont {Wu}, \citenamefont {Lampen-Kelley}, \citenamefont {Banerjee},
  \citenamefont {Patankar}, \citenamefont {Rees}, \citenamefont {Bridges},
  \citenamefont {Yan}, \citenamefont {Mandrus}, \citenamefont {Nagler} \emph
  {et~al.}}]{little2017antiferromagnetic}%
  \BibitemOpen
  \bibfield  {author} {\bibinfo {author} {\bibfnamefont {A.}~\bibnamefont
  {Little}}, \bibinfo {author} {\bibfnamefont {L.}~\bibnamefont {Wu}}, \bibinfo
  {author} {\bibfnamefont {P.}~\bibnamefont {Lampen-Kelley}}, \bibinfo {author}
  {\bibfnamefont {A.}~\bibnamefont {Banerjee}}, \bibinfo {author}
  {\bibfnamefont {S.}~\bibnamefont {Patankar}}, \bibinfo {author}
  {\bibfnamefont {D.}~\bibnamefont {Rees}}, \bibinfo {author} {\bibfnamefont
  {C.}~\bibnamefont {Bridges}}, \bibinfo {author} {\bibfnamefont {J.-Q.}\
  \bibnamefont {Yan}}, \bibinfo {author} {\bibfnamefont {D.}~\bibnamefont
  {Mandrus}}, \bibinfo {author} {\bibfnamefont {S.}~\bibnamefont {Nagler}},
  \emph {et~al.},\ }\href@noop {} {\bibfield  {journal} {\bibinfo  {journal}
  {Phys. Rev. Lett.}\ }\textbf {\bibinfo {volume} {119}},\ \bibinfo {pages}
  {227201} (\bibinfo {year} {2017})}\BibitemShut {NoStop}%
\bibitem [{\citenamefont {Cherepov}\ \emph {et~al.}(2011)\citenamefont
  {Cherepov}, \citenamefont {Koop}, \citenamefont {Dzhezherya}, \citenamefont
  {Worledge},\ and\ \citenamefont {Korenivski}}]{cherepov2011resonant}%
  \BibitemOpen
  \bibfield  {author} {\bibinfo {author} {\bibfnamefont {S.}~\bibnamefont
  {Cherepov}}, \bibinfo {author} {\bibfnamefont {B.}~\bibnamefont {Koop}},
  \bibinfo {author} {\bibfnamefont {Y.~I.}\ \bibnamefont {Dzhezherya}},
  \bibinfo {author} {\bibfnamefont {D.}~\bibnamefont {Worledge}}, \ and\
  \bibinfo {author} {\bibfnamefont {V.}~\bibnamefont {Korenivski}},\
  }\href@noop {} {\bibfield  {journal} {\bibinfo  {journal} {Phys. Rev. Lett.}\
  }\textbf {\bibinfo {volume} {107}},\ \bibinfo {pages} {077202} (\bibinfo
  {year} {2011})}\BibitemShut {NoStop}%
\bibitem [{\citenamefont {Wadley}\ \emph {et~al.}(2018)\citenamefont {Wadley},
  \citenamefont {Reimers}, \citenamefont {Grzybowski}, \citenamefont {Andrews},
  \citenamefont {Wang}, \citenamefont {Chauhan}, \citenamefont {Gallagher},
  \citenamefont {Campion}, \citenamefont {Edmonds}, \citenamefont {Dhesi} \emph
  {et~al.}}]{wadley2018current}%
  \BibitemOpen
  \bibfield  {author} {\bibinfo {author} {\bibfnamefont {P.}~\bibnamefont
  {Wadley}}, \bibinfo {author} {\bibfnamefont {S.}~\bibnamefont {Reimers}},
  \bibinfo {author} {\bibfnamefont {M.~J.}\ \bibnamefont {Grzybowski}},
  \bibinfo {author} {\bibfnamefont {C.}~\bibnamefont {Andrews}}, \bibinfo
  {author} {\bibfnamefont {M.}~\bibnamefont {Wang}}, \bibinfo {author}
  {\bibfnamefont {J.~S.}\ \bibnamefont {Chauhan}}, \bibinfo {author}
  {\bibfnamefont {B.~L.}\ \bibnamefont {Gallagher}}, \bibinfo {author}
  {\bibfnamefont {R.~P.}\ \bibnamefont {Campion}}, \bibinfo {author}
  {\bibfnamefont {K.~W.}\ \bibnamefont {Edmonds}}, \bibinfo {author}
  {\bibfnamefont {S.~S.}\ \bibnamefont {Dhesi}},  \emph {et~al.},\ }\href@noop
  {} {\bibfield  {journal} {\bibinfo  {journal} {Nat. Nanotechn.}\ }\textbf
  {\bibinfo {volume} {13}},\ \bibinfo {pages} {362} (\bibinfo {year}
  {2018})}\BibitemShut {NoStop}%
\bibitem [{\citenamefont {Qian}\ \emph {et~al.}(2014)\citenamefont {Qian},
  \citenamefont {Nayak}, \citenamefont {Kreiner}, \citenamefont {Schnelle},\
  and\ \citenamefont {Felser}}]{Qian2014}%
  \BibitemOpen
  \bibfield  {author} {\bibinfo {author} {\bibfnamefont {J.~F.}\ \bibnamefont
  {Qian}}, \bibinfo {author} {\bibfnamefont {A.~K.}\ \bibnamefont {Nayak}},
  \bibinfo {author} {\bibfnamefont {G.}~\bibnamefont {Kreiner}}, \bibinfo
  {author} {\bibfnamefont {W.}~\bibnamefont {Schnelle}}, \ and\ \bibinfo
  {author} {\bibfnamefont {C.}~\bibnamefont {Felser}},\ }\href@noop {}
  {\bibfield  {journal} {\bibinfo  {journal} {J. Phys. D : Appl. Phys.}\
  }\textbf {\bibinfo {volume} {47}},\ \bibinfo {pages} {305001} (\bibinfo
  {year} {2014})}\BibitemShut {NoStop}%
\bibitem [{\citenamefont {Ohoyama}(1961)}]{Ohoyama1961}%
  \BibitemOpen
  \bibfield  {author} {\bibinfo {author} {\bibfnamefont {T.}~\bibnamefont
  {Ohoyama}},\ }\href@noop {} {\bibfield  {journal} {\bibinfo  {journal} {J.
  Phys. Soc. Jpn.}\ }\textbf {\bibinfo {volume} {16}},\ \bibinfo {pages} {1995}
  (\bibinfo {year} {1961})}\BibitemShut {NoStop}%
\bibitem [{\citenamefont {\u{Z}elez\'{n}y}\ \emph {et~al.}(2014)\citenamefont
  {\u{Z}elez\'{n}y}, \citenamefont {Gao}, \citenamefont {V\'{y}born\'{y}},
  \citenamefont {Zeman}, \citenamefont {Ma\u{s}ek}, \citenamefont {Manchon},
  \citenamefont {Wunderlich}, \citenamefont {Sinova},\ and\ \citenamefont
  {Jungwirth}}]{Zelezny2014}%
  \BibitemOpen
  \bibfield  {author} {\bibinfo {author} {\bibfnamefont {J.}~\bibnamefont
  {\u{Z}elez\'{n}y}}, \bibinfo {author} {\bibfnamefont {H.}~\bibnamefont
  {Gao}}, \bibinfo {author} {\bibnamefont {V\'{y}born\'{y}}}, \bibinfo {author}
  {\bibfnamefont {J.}~\bibnamefont {Zeman}}, \bibinfo {author} {\bibfnamefont
  {J.}~\bibnamefont {Ma\u{s}ek}}, \bibinfo {author} {\bibfnamefont
  {A.}~\bibnamefont {Manchon}}, \bibinfo {author} {\bibfnamefont
  {J.}~\bibnamefont {Wunderlich}}, \bibinfo {author} {\bibfnamefont
  {J.}~\bibnamefont {Sinova}}, \ and\ \bibinfo {author} {\bibfnamefont
  {T.}~\bibnamefont {Jungwirth}},\ }\href@noop {} {\bibfield  {journal}
  {\bibinfo  {journal} {Phys. Rev. Lett.}\ }\textbf {\bibinfo {volume} {113}},\
  \bibinfo {pages} {157201} (\bibinfo {year} {2014})}\BibitemShut {NoStop}%
\bibitem [{\citenamefont {Zhou}\ \emph
  {et~al.}(2019{\natexlab{b}})\citenamefont {Zhou}, \citenamefont {Hanke},
  \citenamefont {Feng}, \citenamefont {Li}, \citenamefont {Guo}, \citenamefont
  {Yao}, \citenamefont {Bl\"ugel},\ and\ \citenamefont {Mokrousov}}]{Zhou2019}%
  \BibitemOpen
  \bibfield  {author} {\bibinfo {author} {\bibfnamefont {X.}~\bibnamefont
  {Zhou}}, \bibinfo {author} {\bibfnamefont {J.-P.}\ \bibnamefont {Hanke}},
  \bibinfo {author} {\bibfnamefont {W.}~\bibnamefont {Feng}}, \bibinfo {author}
  {\bibfnamefont {F.}~\bibnamefont {Li}}, \bibinfo {author} {\bibfnamefont
  {G.-Y.}\ \bibnamefont {Guo}}, \bibinfo {author} {\bibfnamefont
  {Y.}~\bibnamefont {Yao}}, \bibinfo {author} {\bibfnamefont {S.}~\bibnamefont
  {Bl\"ugel}}, \ and\ \bibinfo {author} {\bibfnamefont {Y.}~\bibnamefont
  {Mokrousov}},\ }\href {\doibase 10.1103/PhysRevB.99.104428} {\bibfield
  {journal} {\bibinfo  {journal} {Phys. Rev. B}\ }\textbf {\bibinfo {volume}
  {99}},\ \bibinfo {pages} {104428} (\bibinfo {year}
  {2019}{\natexlab{b}})}\BibitemShut {NoStop}%
\bibitem [{\citenamefont {Mertins}\ \emph {et~al.}(2005)\citenamefont
  {Mertins}, \citenamefont {Valencia}, \citenamefont {Gaupp}, \citenamefont
  {Gudat}, \citenamefont {Oppeneer},\ and\ \citenamefont
  {Schneider}}]{Mertins2005}%
  \BibitemOpen
  \bibfield  {author} {\bibinfo {author} {\bibfnamefont {H.~C.}\ \bibnamefont
  {Mertins}}, \bibinfo {author} {\bibfnamefont {S.}~\bibnamefont {Valencia}},
  \bibinfo {author} {\bibfnamefont {A.}~\bibnamefont {Gaupp}}, \bibinfo
  {author} {\bibfnamefont {W.}~\bibnamefont {Gudat}}, \bibinfo {author}
  {\bibfnamefont {P.~M.}\ \bibnamefont {Oppeneer}}, \ and\ \bibinfo {author}
  {\bibfnamefont {C.~M.}\ \bibnamefont {Schneider}},\ }\href {\doibase
  10.1007/s00339-004-3129-5} {\bibfield  {journal} {\bibinfo  {journal} {Appl.
  Phys. A}\ }\textbf {\bibinfo {volume} {80}},\ \bibinfo {pages} {1011}
  (\bibinfo {year} {2005})}\BibitemShut {NoStop}%
\bibitem [{\citenamefont {Tomiyoshi}\ \emph {et~al.}(1986)\citenamefont
  {Tomiyoshi}, \citenamefont {Abe}, \citenamefont {Yamaguchi}, \citenamefont
  {Yamauchi},\ and\ \citenamefont {Yamamoto}}]{Tomiyoshi1986}%
  \BibitemOpen
  \bibfield  {author} {\bibinfo {author} {\bibfnamefont {S.}~\bibnamefont
  {Tomiyoshi}}, \bibinfo {author} {\bibfnamefont {S.}~\bibnamefont {Abe}},
  \bibinfo {author} {\bibfnamefont {Y.}~\bibnamefont {Yamaguchi}}, \bibinfo
  {author} {\bibfnamefont {H.}~\bibnamefont {Yamauchi}}, \ and\ \bibinfo
  {author} {\bibfnamefont {H.}~\bibnamefont {Yamamoto}},\ }\href {\doibase
  https://doi.org/10.1016/0304-8853(86)90353-7} {\bibfield  {journal} {\bibinfo
   {journal} {J. Magn. Magn. Mater.}\ }\textbf {\bibinfo {volume} {54-57}},\
  \bibinfo {pages} {1001 } (\bibinfo {year} {1986})}\BibitemShut {NoStop}%
\bibitem [{\citenamefont {Rout}\ \emph {et~al.}(2019)\citenamefont {Rout},
  \citenamefont {Madduri}, \citenamefont {Manna},\ and\ \citenamefont
  {Nayak}}]{Rout2019}%
  \BibitemOpen
  \bibfield  {author} {\bibinfo {author} {\bibfnamefont {P.~K.}\ \bibnamefont
  {Rout}}, \bibinfo {author} {\bibfnamefont {P.~V.~P.}\ \bibnamefont
  {Madduri}}, \bibinfo {author} {\bibfnamefont {S.~K.}\ \bibnamefont {Manna}},
  \ and\ \bibinfo {author} {\bibfnamefont {A.~K.}\ \bibnamefont {Nayak}},\
  }\href {\doibase 10.1103/PhysRevB.99.094430} {\bibfield  {journal} {\bibinfo
  {journal} {Phys. Rev. B}\ }\textbf {\bibinfo {volume} {99}},\ \bibinfo
  {pages} {094430} (\bibinfo {year} {2019})}\BibitemShut {NoStop}%
\bibitem [{\citenamefont {Chen}(1973)}]{Chen1973}%
  \BibitemOpen
  \bibfield  {author} {\bibinfo {author} {\bibfnamefont {D.}~\bibnamefont
  {Chen}},\ }in\ \href {\doibase 10.1117/12.953743} {\emph {\bibinfo
  {booktitle} {Electro-Optics Principles and Applications}}},\ Vol.\ \bibinfo
  {volume} {0038},\ \bibinfo {editor} {edited by\ \bibinfo {editor}
  {\bibfnamefont {J.~B.}\ \bibnamefont {DeVelis}}\ and\ \bibinfo {editor}
  {\bibfnamefont {B.~J.}\ \bibnamefont {Thompson}}},\ \bibinfo {organization}
  {International Society for Optics and Photonics}\ (\bibinfo  {publisher}
  {SPIE},\ \bibinfo {year} {1973})\ pp.\ \bibinfo {pages} {9 -- 16}\BibitemShut
  {NoStop}%
\bibitem [{\citenamefont {Sugano}\ and\ \citenamefont
  {Kojima}(2013)}]{Sugano2013}%
  \BibitemOpen
  \bibfield  {author} {\bibinfo {author} {\bibfnamefont {S.}~\bibnamefont
  {Sugano}}\ and\ \bibinfo {author} {\bibfnamefont {N.}~\bibnamefont
  {Kojima}},\ }\href {https://books.google.com/books?id=rj3zCAAAQBAJ} {\emph
  {\bibinfo {title} {Magneto-Optics}}},\ Springer Series in Solid-State
  Sciences\ (\bibinfo  {publisher} {Springer Berlin Heidelberg},\ \bibinfo
  {year} {2013})\BibitemShut {NoStop}%
\bibitem [{\citenamefont {Haider}(2017)}]{Haider2017}%
  \BibitemOpen
  \bibfield  {author} {\bibinfo {author} {\bibfnamefont {T.}~\bibnamefont
  {Haider}},\ }\href@noop {} {\bibfield  {journal} {\bibinfo  {journal}
  {International Journal of Electromagnetics and Applications}\ }\textbf
  {\bibinfo {volume} {7}},\ \bibinfo {pages} {17} (\bibinfo {year}
  {2017})}\BibitemShut {NoStop}%
\bibitem [{\citenamefont {Postava}\ \emph {et~al.}(1999)\citenamefont
  {Postava}, \citenamefont {Pistora}, \citenamefont {Ciprian}, \citenamefont
  {Hrabovsky}, \citenamefont {Lesnak},\ and\ \citenamefont
  {Fert}}]{Postava1999}%
  \BibitemOpen
  \bibfield  {author} {\bibinfo {author} {\bibfnamefont {K.}~\bibnamefont
  {Postava}}, \bibinfo {author} {\bibfnamefont {J.}~\bibnamefont {Pistora}},
  \bibinfo {author} {\bibfnamefont {D.}~\bibnamefont {Ciprian}}, \bibinfo
  {author} {\bibfnamefont {D.}~\bibnamefont {Hrabovsky}}, \bibinfo {author}
  {\bibfnamefont {M.}~\bibnamefont {Lesnak}}, \ and\ \bibinfo {author}
  {\bibfnamefont {A.~R.}\ \bibnamefont {Fert}},\ }in\ \href {\doibase
  10.1117/12.353094} {\emph {\bibinfo {booktitle} {11th Slovak-Czech-Polish
  Optical Conference on Wave and Quantum Aspects of Contemporary Optics}}},\
  Vol.\ \bibinfo {volume} {3820},\ \bibinfo {editor} {edited by\ \bibinfo
  {editor} {\bibfnamefont {M.}~\bibnamefont {Hrabovsky}}, \bibinfo {editor}
  {\bibfnamefont {A.}~\bibnamefont {Strba}}, \ and\ \bibinfo {editor}
  {\bibfnamefont {W.}~\bibnamefont {Urbanczyk}}},\ \bibinfo {organization}
  {International Society for Optics and Photonics}\ (\bibinfo  {publisher}
  {SPIE},\ \bibinfo {year} {1999})\ pp.\ \bibinfo {pages} {412 --
  422}\BibitemShut {NoStop}%
\bibitem [{\citenamefont {Vi{\v s}{\v n}ovsk{\'y}}(1986)}]{Visnovsky1986}%
  \BibitemOpen
  \bibfield  {author} {\bibinfo {author} {\bibfnamefont {{\v S}.}~\bibnamefont
  {Vi{\v s}{\v n}ovsk{\'y}}},\ }\href {\doibase 10.1007/BF01959567} {\bibfield
  {journal} {\bibinfo  {journal} {Czech. J. Phys. B}\ }\textbf {\bibinfo
  {volume} {36}},\ \bibinfo {pages} {1424} (\bibinfo {year}
  {1986})}\BibitemShut {NoStop}%
\bibitem [{\citenamefont {Tesa\ifmmode~\check{r}\else \v{r}\fi{}ov\'a}\ \emph
  {et~al.}(2014)\citenamefont {Tesa\ifmmode~\check{r}\else \v{r}\fi{}ov\'a},
  \citenamefont {Ostatnick\'y}, \citenamefont {Nov\'ak}, \citenamefont
  {Olejn\'{\i}k}, \citenamefont {\ifmmode~\check{S}\else \v{S}\fi{}ubrt},
  \citenamefont {Reichlov\'a}, \citenamefont {Ellis}, \citenamefont
  {Mukherjee}, \citenamefont {Lee}, \citenamefont {Sipahi}, \citenamefont
  {Sinova}, \citenamefont {Hamrle}, \citenamefont {Jungwirth}, \citenamefont
  {N\ifmmode~\check{e}\else \v{e}\fi{}mec}, \citenamefont
  {\ifmmode~\check{C}\else \v{C}\fi{}erne},\ and\ \citenamefont
  {V\'yborn\'y}}]{Tesarrova2014}%
  \BibitemOpen
  \bibfield  {author} {\bibinfo {author} {\bibfnamefont {N.}~\bibnamefont
  {Tesa\ifmmode~\check{r}\else \v{r}\fi{}ov\'a}}, \bibinfo {author}
  {\bibfnamefont {T.}~\bibnamefont {Ostatnick\'y}}, \bibinfo {author}
  {\bibfnamefont {V.}~\bibnamefont {Nov\'ak}}, \bibinfo {author} {\bibfnamefont
  {K.}~\bibnamefont {Olejn\'{\i}k}}, \bibinfo {author} {\bibfnamefont
  {J.}~\bibnamefont {\ifmmode~\check{S}\else \v{S}\fi{}ubrt}}, \bibinfo
  {author} {\bibfnamefont {H.}~\bibnamefont {Reichlov\'a}}, \bibinfo {author}
  {\bibfnamefont {C.~T.}\ \bibnamefont {Ellis}}, \bibinfo {author}
  {\bibfnamefont {A.}~\bibnamefont {Mukherjee}}, \bibinfo {author}
  {\bibfnamefont {J.}~\bibnamefont {Lee}}, \bibinfo {author} {\bibfnamefont
  {G.~M.}\ \bibnamefont {Sipahi}}, \bibinfo {author} {\bibfnamefont
  {J.}~\bibnamefont {Sinova}}, \bibinfo {author} {\bibfnamefont
  {J.}~\bibnamefont {Hamrle}}, \bibinfo {author} {\bibfnamefont
  {T.}~\bibnamefont {Jungwirth}}, \bibinfo {author} {\bibfnamefont
  {P.}~\bibnamefont {N\ifmmode~\check{e}\else \v{e}\fi{}mec}}, \bibinfo
  {author} {\bibfnamefont {J.}~\bibnamefont {\ifmmode~\check{C}\else
  \v{C}\fi{}erne}}, \ and\ \bibinfo {author} {\bibfnamefont {K.}~\bibnamefont
  {V\'yborn\'y}},\ }\href {\doibase 10.1103/PhysRevB.89.085203} {\bibfield
  {journal} {\bibinfo  {journal} {Phys. Rev. B}\ }\textbf {\bibinfo {volume}
  {89}},\ \bibinfo {pages} {085203} (\bibinfo {year} {2014})}\BibitemShut
  {NoStop}%
\bibitem [{\citenamefont {Palpant}(2006)}]{Palpant2006}%
  \BibitemOpen
  \bibfield  {author} {\bibinfo {author} {\bibfnamefont {B.}~\bibnamefont
  {Palpant}},\ }\enquote {\bibinfo {title} {Third-order nonlinear optical
  response of metal nanoparticles},}\ in\ \href {\doibase
  10.1007/1-4020-4850-5_15} {\emph {\bibinfo {booktitle} {Non-Linear Optical
  Properties of Matter: From Molecules to Condensed Phases}}},\ \bibinfo
  {editor} {edited by\ \bibinfo {editor} {\bibfnamefont {M.~G.}\ \bibnamefont
  {Papadopoulos}}, \bibinfo {editor} {\bibfnamefont {A.~J.}\ \bibnamefont
  {Sadlej}}, \ and\ \bibinfo {editor} {\bibfnamefont {J.}~\bibnamefont
  {Leszczynski}}}\ (\bibinfo  {publisher} {Springer Netherlands},\ \bibinfo
  {address} {Dordrecht},\ \bibinfo {year} {2006})\ pp.\ \bibinfo {pages}
  {461--508}\BibitemShut {NoStop}%
\bibitem [{\citenamefont {Lohner}\ and\ \citenamefont
  {Villaeys}(1999)}]{Lohner1999}%
  \BibitemOpen
  \bibfield  {author} {\bibinfo {author} {\bibfnamefont {F.}~\bibnamefont
  {Lohner}}\ and\ \bibinfo {author} {\bibfnamefont {A.}~\bibnamefont
  {Villaeys}},\ }\href {\doibase https://doi.org/10.1016/S0169-4332(98)00426-7}
  {\bibfield  {journal} {\bibinfo  {journal} {Applied Surface Science}\
  }\textbf {\bibinfo {volume} {138-139}},\ \bibinfo {pages} {325 } (\bibinfo
  {year} {1999})}\BibitemShut {NoStop}%
\bibitem [{\citenamefont {Dewhurst}, \citenamefont {Sanna},\ and\ \citenamefont
  {Sharma}(2018)}]{Dewhust2018v2}%
  \BibitemOpen
  \bibfield  {author} {\bibinfo {author} {\bibfnamefont {J.~K.}\ \bibnamefont
  {Dewhurst}}, \bibinfo {author} {\bibfnamefont {A.}~\bibnamefont {Sanna}}, \
  and\ \bibinfo {author} {\bibfnamefont {S.}~\bibnamefont {Sharma}},\ }\href
  {\doibase 10.1140/epjb/e2018-90146-1} {\bibfield  {journal} {\bibinfo
  {journal} {Euro. Phys. J. B}\ }\textbf {\bibinfo {volume} {91}},\ \bibinfo
  {pages} {218} (\bibinfo {year} {2018})}\BibitemShut {NoStop}%
\bibitem [{\citenamefont {Siegrist}\ \emph {et~al.}(2019)\citenamefont
  {Siegrist}, \citenamefont {Gessner}, \citenamefont {Ossiander}, \citenamefont
  {Denker}, \citenamefont {Chang}, \citenamefont {Schr{\"o}der}, \citenamefont
  {Guggenmos}, \citenamefont {Cui}, \citenamefont {Walowski}, \citenamefont
  {Martens}, \citenamefont {Dewhurst}, \citenamefont {Kleineberg},
  \citenamefont {M{\"u}nzenberg}, \citenamefont {Sharma},\ and\ \citenamefont
  {Schultze}}]{Siegrist2019}%
  \BibitemOpen
  \bibfield  {author} {\bibinfo {author} {\bibfnamefont {F.}~\bibnamefont
  {Siegrist}}, \bibinfo {author} {\bibfnamefont {J.~A.}\ \bibnamefont
  {Gessner}}, \bibinfo {author} {\bibfnamefont {M.}~\bibnamefont {Ossiander}},
  \bibinfo {author} {\bibfnamefont {C.}~\bibnamefont {Denker}}, \bibinfo
  {author} {\bibfnamefont {Y.-P.}\ \bibnamefont {Chang}}, \bibinfo {author}
  {\bibfnamefont {M.~C.}\ \bibnamefont {Schr{\"o}der}}, \bibinfo {author}
  {\bibfnamefont {A.}~\bibnamefont {Guggenmos}}, \bibinfo {author}
  {\bibfnamefont {Y.}~\bibnamefont {Cui}}, \bibinfo {author} {\bibfnamefont
  {J.}~\bibnamefont {Walowski}}, \bibinfo {author} {\bibfnamefont
  {U.}~\bibnamefont {Martens}}, \bibinfo {author} {\bibfnamefont {J.~K.}\
  \bibnamefont {Dewhurst}}, \bibinfo {author} {\bibfnamefont {U.}~\bibnamefont
  {Kleineberg}}, \bibinfo {author} {\bibfnamefont {M.}~\bibnamefont
  {M{\"u}nzenberg}}, \bibinfo {author} {\bibfnamefont {S.}~\bibnamefont
  {Sharma}}, \ and\ \bibinfo {author} {\bibfnamefont {M.}~\bibnamefont
  {Schultze}},\ }\href {\doibase 10.1038/s41586-019-1333-x} {\bibfield
  {journal} {\bibinfo  {journal} {Nature}\ }\textbf {\bibinfo {volume} {571}},\
  \bibinfo {pages} {240} (\bibinfo {year} {2019})}\BibitemShut {NoStop}%
\bibitem [{\citenamefont {N\'eel}(1953)}]{Neel1953}%
  \BibitemOpen
  \bibfield  {author} {\bibinfo {author} {\bibfnamefont {L.}~\bibnamefont
  {N\'eel}},\ }\href {\doibase 10.1103/RevModPhys.25.58} {\bibfield  {journal}
  {\bibinfo  {journal} {Rev. Mod. Phys.}\ }\textbf {\bibinfo {volume} {25}},\
  \bibinfo {pages} {58} (\bibinfo {year} {1953})}\BibitemShut {NoStop}%
\bibitem [{\citenamefont {{Okada}}\ \emph {et~al.}(2018)\citenamefont
  {{Okada}}, \citenamefont {{Kawasaki}}, \citenamefont {{Moriya}},\ and\
  \citenamefont {{Doi}}}]{Okada2018}%
  \BibitemOpen
  \bibfield  {author} {\bibinfo {author} {\bibfnamefont {H.}~\bibnamefont
  {{Okada}}}, \bibinfo {author} {\bibfnamefont {T.}~\bibnamefont {{Kawasaki}}},
  \bibinfo {author} {\bibfnamefont {K.}~\bibnamefont {{Moriya}}}, \ and\
  \bibinfo {author} {\bibfnamefont {M.}~\bibnamefont {{Doi}}},\ }\href
  {\doibase 10.1109/TMAG.2018.2841024} {\bibfield  {journal} {\bibinfo
  {journal} {IEEE Trans. Magn.}\ }\textbf {\bibinfo {volume} {54}},\ \bibinfo
  {pages} {1} (\bibinfo {year} {2018})}\BibitemShut {NoStop}%
\end{thebibliography}%

\end{document}